\documentclass[a4paper,fleqn]{cas-sc}

\usepackage[numbers]{natbib}

\def\tsc#1{\csdef{#1}{\textsc{\lowercase{#1}}\xspace}}
\tsc{WGM}
\tsc{QE}
\tsc{EP}
\tsc{PMS}
\tsc{BEC}
\tsc{DE}

\begin{document}
\let\WriteBookmarks\relax
\def\floatpagepagefraction{1}
\def\textpagefraction{.001}
\shorttitle{An Analytical Approach to Privacy and Performance Trade-Offs in Healthcare Data Sharing}
\shortauthors{Anonymous Authors}

\title [mode = title]{An Analytical Approach to Privacy and Performance Trade-Offs in Healthcare Data Sharing}

\author[1]{Yusi Wei}[orcid=0009-0008-4472-439X]
\cormark[1]
\ead{yw825@drexel.edu}
\affiliation[1]{organization={Decision Sciences \& MIS Department, Drexel University},
            city={Philadelphia},
            state={PA},
            country={United States}}
\cortext[cor1]{Corresponding author}

\author[1]{Hande Y. Benson}

\author[label3]{and Muge Capan}
\affiliation[label3]{organization={College of Engineering, University of Massachusetts Amherst},
            city={Amherst},
            state={MA},
            country={United States}}


\begin{abstract}
The secondary use of healthcare data is vital for research and clinical innovation, but it raises concerns about patient privacy. This study investigates how to balance privacy preservation and data utility in healthcare data sharing, considering the perspectives of both data providers and data users. Using a dataset of adult patients hospitalized between 2013 and 2015, we predict whether sepsis was present at admission or developed during the hospital stay. We identify subpopulations—such as older adults, frequently hospitalized patients, and racial minorities—that are especially vulnerable to privacy attacks due to their unique combinations of demographic and healthcare utilization attributes. These groups are also critical for machine learning (ML) model performance. We evaluate three anonymization methods—$k$-anonymity \citep{domingo2005ordinal}, the technique by Zheng et al. \cite{zheng2019effective}, and the MO-OBAM model \citep{wei2025multi}—based on their ability to reduce re-identification risk while maintaining ML utility. Results show that $k$-anonymity offers limited protection. The methods of Zheng et al. \cite{zheng2019effective} and MO-OBAM provide stronger privacy safeguards, with MO-OBAM yielding the best utility outcomes: only a 2\% change in precision and recall compared to the original dataset. This work provides actionable insights for healthcare organizations on how to share data responsibly. It highlights the need for anonymization methods that protect vulnerable populations without sacrificing the performance of data-driven models.
\end{abstract}

\begin{graphicalabstract}
\includegraphics[width=1\linewidth]{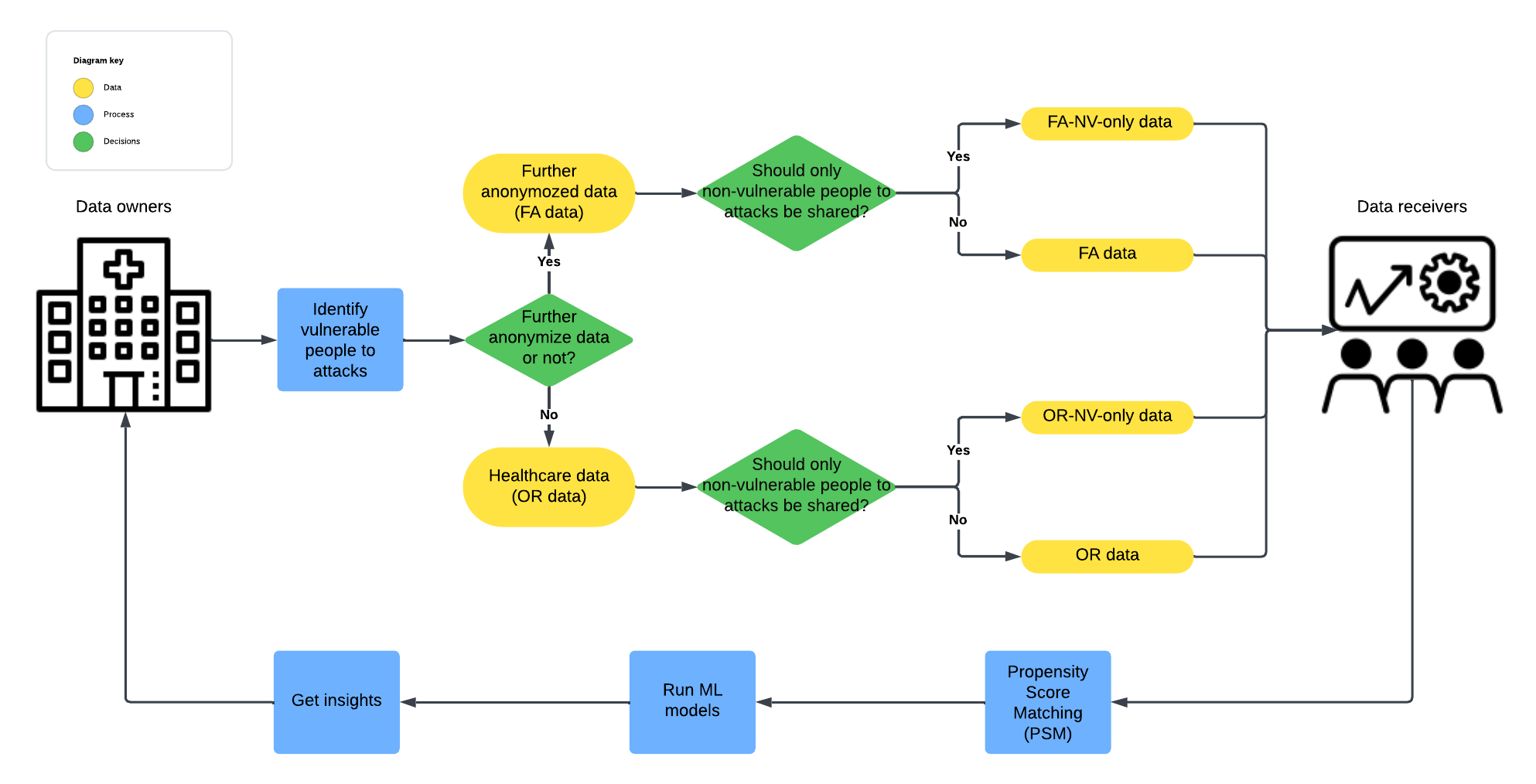}
\end{graphicalabstract}

\begin{highlights}
\item Identify vulnerable groups in health data based on demographics and utilization patterns.
\item Demonstrate that removing vulnerable patients lowers model prediction accuracy.
\item Show that our method best protects privacy with minimal performance loss.
\item Evaluate anonymization tools using both privacy and analytics criteria.
\item Provide guidance for ethical health data sharing in predictive applications.
\end{highlights}


\maketitle
\section{Introduction}
Healthcare is a complex and collaborative field where data sharing plays a crucial role in advancing knowledge and improving outcomes. Healthcare organizations are often motivated to share their data with experts, such as machine learning practitioners, who can provide valuable insights to help predict outcomes, guide decision-making, and identify individuals at risk of developing specific diseases. However, healthcare data contains both personally identifiable information and sensitive health details, making data sharing a process that increases individuals’ vulnerability to privacy attacks.

Healthcare organizations have a responsibility to preserve the privacy of patient information when sharing data for research purposes \citep{El2015}. To that end, the Health Insurance Portability and Accountability Act of 1996 (HIPAA) aims to ensure that healthcare providers and insurers maintain the confidentiality and integrity of medical records. All healthcare data for secondary use in research must be HIPAA de-identified, so in this paper, we refer to healthcare data compliant with HIPAA as {\em healthcare data} and data that is further anonymized beyond HIPAA requirements as {\em further anonymized data}. Further anonymization is a technique that often involves revising and recategorizing data values, particularly quasi-identifiers (QIs)—attributes like age, ZIP code, or gender that, when combined, can indirectly reveal identities. While further anonymization enhances privacy, it also introduces information loss for secondary use of healthcare data, which can reduce the ability of the ML to accurately identify patients at risk of adverse events. Therefore, a balance needs to be found between the level of privacy protection achieved and the utility of healthcare data for predictive analytics purposes. 

Therefore, studying the trade-off between privacy preservation and the utility of healthcare data is essential. However, finding a balance between maintaining high levels of privacy and ensuring healthcare data utility is a complex task. It relies on understanding how the shared healthcare data will be used for research purposes by data receivers \citep{im2024exploring}. Much of the existing research utilizes simple algorithms such as $k$-anonymity, $l$-diversity, and $t$-closeness \citep{karagiannis2024mastering,wu2013utility,li2009tradeoff}, or has employed ARX, a publicly accessible data anonymization tool \citep{jeon2020proposal,im2024exploring}. Others, such as that by Zamani et al. \cite{zamani2023privacy}, have primarily focused on the theoretical aspects of the trade-off between privacy and utility. Optimization models can effectively address this trade-off, as well, but have not been widely applied to data from a healthcare setting.

There is a need for application-oriented research that accurately reflects the complexities of real-world healthcare data analytics tasks and takes into account the multifaceted nature of data utility and privacy in healthcare applications \cite{MACKAY2023100155}. To address this need, this paper examines the anonymization process from two perspectives: that of the data owners and the data receivers. From the perspective of healthcare organizations as data owners, the primary concern is addressing the privacy risks associated with sharing sensitive healthcare data. Conversely, from the perspective of data receivers, the focus is on ensuring that the shared healthcare data retains sufficient information to generate meaningful and accurate results. By considering these dual perspectives, we aim to explore how the anonymization process can balance privacy protection and data utility to facilitate healthcare data sharing. Therefore, in this paper, we aim to answer the following research questions:
\begin{itemize}
    \item Research Question 1: Which patient subpopulations defined by demographic and healthcare utilization characteristics in healthcare data are more vulnerable to privacy attacks?
    \item Research Question 2: a) Do anonymization models reduce the number of patients vulnerable to attacks in healthcare data? b) How do anonymization models affect the percentage of patients vulnerable to attacks within each patient subpopulation?
    \item Research Question 3: a) How does the decision to include or exclude the vulnerable populations' data impact the performance of ML models that aim to predict the presence of sepsis for hospitalized patients? b) How can the anonymization process help protect these vulnerable populations to privacy while developing ML models that work well for both majority and minority groups based on their representation in the EHR data?
\end{itemize}


\section{Related Work}\label{sec:related work}

\subsection{Overview of healthcare data sharing and privacy concerns}

The sharing of patients’ health information across organizations is increasingly common and essential for healthcare advancements \citep{hulsen2020sharing}. Whiddett et al. \cite{whiddett2006patients} studied that patients’ willingness to share their health information varies depending on the type of data and the intended recipient. They observed that patients became increasingly hesitant to share their health data with other stakeholders, such as administrators, researchers, or government departments. Second, the study highlighted that patients were more likely to agree to share their data if their identity could be anonymized. Finally, patients were more willing to share general demographic information, like age or gender, with researchers or administrators, but were increasingly reluctant to share sensitive details, such as medical history, with anyone other than their doctor. This study highlights an increasing concern among patients about the privacy of their healthcare data. This also underscores a balance between protecting privacy to address patient concerns and ensuring the shared data remains accurate and useful for secondary use, such as research or treatment development \citep{nowrozy2024privacy}. 

One of the foundational techniques to preserve privacy is $k$-anonymity, which ensures that each record—defined as a row of data containing all variables—is indistinguishable from at least $k-1$ other records, thus reducing re-identification risks \citep{sweeney2002k}. Various algorithms have been developed to achieve $k$-anonymity, for instance, Mondrian \citep{lefevre2006mondrian}, Incognito \citep{lefevre2005incognito}, and others \citep{bayardo2005data, domingo2005ordinal, cao2012publishing, khan2020theta} adapt the principles of $k$-anonymity. In this study, we utilize the $k$-anonymity algorithm proposed by Domingo-Ferrer and Torra \cite{domingo2005ordinal}, which clusters data and replaces original values with the cluster’s central value. However, $k$-anonymity alone cannot mitigate all privacy risks. To solve these limitations, advanced extensions such as \citep{machanavajjhala2007diversity, li2006t, cao2012publishing, khan2020theta} have been proposed. Additionally, differential privacy has emerged as a robust alternative \citep{10.1145/3706584,10.1145/3643821,10.1145/3639411}. Given that healthcare data is intended for secondary use, such as machine learning applications, achieving a balance between privacy preservation and data utility is paramount \citep{tertulino2024privacy, ahammed2024privacy}. Optimization-based anonymization models have emerged as robust tools due to their ability to effectively manage this trade-off. These models, such as \citep{doka2015k, liang2020optimization, aminifar2021diversity, zheng2019effective, wei2025multi}, provide a flexible framework to incorporate diverse privacy and utility constraints. In this paper, we adopt the advanced optimization-based anonymization model proposed by Zheng et al. \cite{zheng2019effective} and the Multi-Objective Optimization-Based Anonymization Model (MO-OBAM) proposed by us \cite{wei2025multi}. These models have demonstrated superior performance in anonymizing datasets containing both numeric and categorical QIs while offering comprehensive protection against diverse privacy attacks. Their effectiveness in preserving data utility and mitigating privacy risks makes them suitable choices for further anonymization in healthcare datasets.

It is imperative to strengthen regulations and adopt advanced models to safeguard patient data privacy. However, even with advancements in technology and legal frameworks, healthcare data remains vulnerable to privacy risks \citep{el2011systematic,kwok2011harder,lafky2010safe,fernandez2013security}. The most common privacy attacks in healthcare data are linkage attacks and homogeneity attacks. Linkage attacks refer to a situation where an attacker re-identifies individuals in de-identified data by combining information in the dataset with external information, such as publicly available data \citep{wood2018differential, halvorsen2022attacker}. Homogeneity attacks happen when all the values for a sensitive attribute within a group are identical. In this case, an attacker is able to infer the value of a sensitive attribute of an individual based on other information in the data. This situation can occur even after applying $k$-anonymization. Machanavajjhala et al. \cite{machanavajjhala2007diversity} described the scenario of the homogeneity attack in their paper. Therefore, both linkage and homogeneity attacks have significant threats to healthcare data due to the inclusion of highly sensitive personal information. As a result, these concerns have driven extensive research efforts focused on improving privacy preservation in healthcare data.

While privacy preservation is essential for safeguarding patient data, it is equally important to ensure that the shared data remains valuable for secondary use, such as machine learning-driven disease prediction. The increasing adoption of ML in healthcare has significantly improved disease diagnosis, outcome prediction, and resource optimization \cite{DONG2025100397}. Therefore, the secondary use of further anonymized healthcare data in the context of utilizing ML models has received attention in the literature. These investigations aim to understand how anonymization affects the performance of predictive models, such as changes in accuracy, precision, or recall, when applied to tasks such as disease classification or patient outcome predictions. Senavirathne and Torra emphasized that achieving substantial privacy protection requires applying high levels of anonymization, which often leads to a noticeable decline in ML model utility \citep{senavirathne2020role}. This trade-off underscores the need to investigate which patient subpopulations are at greater risk of privacy breaches and whether providing enhanced privacy for these groups impacts ML model performance. Moreover, much of the existing research focuses solely on assessing ML performance before and after anonymization. For example, Im et al. \cite{im2024exploring} used ARX to anonymize healthcare data and compared the results of logistic regression models applied to the original and anonymized datasets, finding that model performance varied significantly with different anonymization levels. Additionally, studies such as Wimmer and Powell \cite{wimmer2014comparison} and Pitoglou et al.\cite{pitoglou2022exploring} highlight that the impact of anonymization on healthcare data depends on the ML model used, with some exhibiting stable or even improved performance and others being more adversely affected. 

\subsection{Machine Learning for Sepsis Detection and Prediction}

Sepsis is a life-threatening condition characterized by a dysregulated immune response to infection, leading to systemic inflammation and organ dysfunction \citep{rackow1986clinical}. Clinically, sepsis is significant due to its high mortality rates, with severe cases progressing to septic shock \citep{dellinger2008surviving}. Beyond its immediate threat to life, sepsis is a leading cause of long-term disability, as many survivors experience persistent physical, cognitive, and psychological impairments that reduce their quality of life \citep{iwashyna2012population}. Additionally, sepsis imposes a substantial economic burden, accounting for billions in annual healthcare costs due to prolonged hospitalizations, intensive care requirements, and post-discharge complications \citep{henry2015targeted}. Therefore, early detection is crucial, as timely intervention can reduce mortality and improve patient outcomes \citep{dellinger2013surviving}. Consequently, numerous studies have focused on early sepsis prediction, such as \citep{10.1145/3637528.3671586, 10.1145/3582099.3582129,lin2018early}. 

Sepsis detection has been guided by various clinical definitions and scoring systems, each with distinct advantages and limitations. The Sepsis-3 criteria \citep{singer2016third} define sepsis as life-threatening organ dysfunction due to a dysregulated response to infection, assessed primarily through the SOFA (Sepsis-related Organ Failure Assessment) score \citep{vincent1996sofa}. While SOFA effectively quantifies organ dysfunction, it was originally developed for research purposes, and its clinical utility is often limited \citep{toker2021comparison}. Additionally, its reliance on extensive laboratory data makes it less practical for real-time decision-making, and its sensitivity for early sepsis detection remains a concern \citep{reddy2024navigating}. Earlier definitions, such as SIRS (Systemic Inflammatory Response Syndrome) \citep{bone1992definitions}, classified sepsis based on systemic inflammatory markers but failed to account for organ dysfunction, making it unsuitable for identifying at-risk patients or predicting mortality risk \citep{churpek2015incidence, kaukonen2015systemic}. To improve early detection, the qSOFA (Quick SOFA) score \citep{singer2016third} was introduced as a simplified bedside tool using vital sign-based criteria. However, while qSOFA offers ease of use in non-ICU settings, its lower sensitivity limits its reliability as a standalone diagnostic method \citep{goulden2018qsofa, anand2019epidemiology}. Another approach, the PIRO (Predisposition, Infection, Response, Organ Dysfunction) model \citep{howell2011proof}, provides a more comprehensive framework by incorporating patient risk factors and disease progression. These challenges highlight the limitations of traditional rule-based scoring systems, underscoring the need for advanced approaches, such as machine learning models, that can leverage large-scale EHR for more accurate and timely sepsis prediction.

Zhang et al. \cite{10.1145/3644116.3644330} employed multiple machine learning models, including Support Vector Machine (SVM), Decision Tree, Random Forest, Gradient Boosting, Logistic Regression, and Gaussian Naïve Bayes, to analyze ICU patient data. Their study aimed to predict sepsis-related mortality and identify personalized medication recommendations for sepsis patients. Similarly, van Wyk et al. \cite{van2019minimal} applied Random Forest classification to various physiological datasets to identify patients at risk of developing sepsis, while Tsoukalas et al.\cite{tsoukalas2015data} developed SVM classifiers to predict sepsis patients’ length of stay and mortality risk. Beyond machine learning-based approaches, researchers have also explored other modeling techniques for sepsis prediction and progression analysis. Saka et al. \cite{saka2007use} designed an empirically based Monte Carlo model to simulate the progression of sepsis in hospitalized patients from the GenIMS study, predicting changes in health status (SOFA score) as a function of previous health state and hospital length of stay. Shapiro et al.\cite{shapiro2007mortality}, Ribas et al. \cite{ribas2012severe}, and Chen et al. \cite{chen2017development} applied statistical approaches, such as logistic regression, to predict sepsis-related mortality.

In addition to mortality prediction, researchers have focused on optimizing clinical decision-making for sepsis treatment. Rosenstrom et al.\cite{rosenstrom2022optimizing} investigated the challenges of selecting the initial treatment for sepsis, incorporating organ dysfunction type and treatment choices (e.g., fluids or anti-infectives). Using electronic health record (EHR) data, they developed a Markov Decision Process (MDP) model to determine the optimal first-line treatment policy for sepsis. Capan et al. \cite{capan2018not} employed multivariate logistic regression to identify the organ dysfunction measures most strongly associated with mortality in sepsis patients, highlighting that certain dysfunctions have a stronger impact on adverse outcomes than others. Additionally, several studies have developed models to analyze patterns of simultaneous cellular and physiological dysfunctions to enhance sepsis prediction \citep{jazayeri2019network,jazayeri2021proximity}.

While privacy-preserving techniques for healthcare data have advanced significantly, there is still a lack of research that translates these methods into practical applications. Addressing the challenges of real-world healthcare data analytics requires considering both the data owner’s need for privacy and the user’s demand for predictive accuracy. A balanced approach must account for the objectives of both decision-makers to ensure effective and ethical data utilization. Current studies often simplify the evaluation of the trade-offs between data utility and privacy. This paper addresses this gap by identifying patient subpopulations vulnerable to attacks and evaluating how their inclusion impacts ML performance. Our focus, on the one hand, is on the perspective of data receivers concerned with data utility for ML applications, and, on the other hand, it is to provide actionable recommendations to data owners. We aim to guide healthcare organizations on what types of data to share in a way that balances privacy protection with maintaining meaningful ML model performance.

\section{Experiments}\label{sec:experiment}
\subsection{Data}
The healthcare data used in this study, as detailed in Table~\ref{tab:sample data}, is composed of retrospectively collected Electronic Health Record (EHR) data from two hospitals of a single tertiary-care healthcare system in the United States (in total, 1100 in-hospital beds). The data collection was performed from patients admitted to these hospitals between July 2013 and December 2015. The inclusion criteria consist of patient age $\geq$ 18 at admission and visit types of in-patient, Emergency Department only, or observational visits. 

The dataset includes 119,871 unique patients and 106 variables. Each patient corresponds to a record including all variables. The variables consist of 6 quasi-identifiers (QIs), 30 sensitive attributes (SAs), and 70 non-sensitive attributes (NSAs). The QIs in the dataset include age, the number of days spent in the hospital (LOS), the number of visits to the hospital observed during the study time period (\# of Visits), gender, race, and ethnicity. These QIs have been carefully chosen based on the criterion that any variable, when combined with others, could uniquely identify people. Variables such as age, gender, race, and ethnicity have been widely recognized as QIs \citep{el2009evaluating}. Additionally, LOS and \# of Visits in our data further contribute to the unique identification of a patient when combined with the aforementioned QIs. The dataset includes 30 SAs, each denoted by ``HX'', which stands for ``history of'', meaning this condition was observed and documented on the patient's chart before the current admission to the hospital. The ``HX'' is followed by an abbreviation of the disease. The NSAs are variables that are less prone to privacy concerns. In our study, NSAs include disease flags, which indicate the presence of a condition (e.g., sepsis) diagnosed at the end of a hospitalization and are encoded using the International Classification of Diseases (ICD) system. The ICD codes are, while not sensitive themselves, are utilized for financial purposes such as healthcare billing. Among them is the sepsis flag which is the target variable for the ML models in this paper. The sepsis flag of 1 indicates that the patient was discharged with a sepsis-related ICD code in their chart based on meeting clinical sepsis criteria during their hospitalization.

\begin{table}
    \caption{Sample of healthcare data used in this study}
   {\resizebox{\columnwidth}{!}{
    \begin{tabular}{>{\centering\arraybackslash}m{1.2em}>
    {\centering\arraybackslash}m{1.5em}>
    {\centering\arraybackslash}m{3.0em}>{\centering\arraybackslash}m{3em}>{\centering\arraybackslash}m{4.0em}>{\centering\arraybackslash}m{5.8em}>{\centering\arraybackslash}m{3.1em}>{\centering\arraybackslash}m{1em}>{\centering\arraybackslash}m{3.5em}>{\centering\arraybackslash}m{3.0em}>{\centering\arraybackslash}m{1em}>{\centering\arraybackslash}m{3.8em}} \hline
    \multicolumn{6}{c}{Quasi-identifers}  & \multicolumn{3}{c}{Sensitive attributes} & \multicolumn{3}{c}{Non-sensitive attributes} \\\hline
    Age & LOS & \#ofVisits  & Gender & Race  & Ethnicity & HX\_DM &   \ldots & HX\_HTN  & UtiFlag  & \ldots  & SepsisFlag\\
    \hline
    50  & 2 & 1  & Female & White & Non-Hispanic or Latino   & 0  &   \ldots  & 1  &  0 & \ldots     & 0 \\ \hline
    29 & 19 & 2  & Female & Asian & Non-Hispanic or Latino    & 0     &    \ldots   & 0  & P  &   \ldots   & 0   \\ \hline
    38  & 1 & 1  & Male & White & Non-Hispanic or Latino    & 1     &    \ldots   & 0 & 0 &\ldots   & 0   \\ \hline
    86 & 30 & 10   & Male  & Black or African American & Hispanic or Latino & 1     &   \ldots    & 1   & P   &   \ldots      & 1   \\ \hline
    \end{tabular}%
   }}
  \label{tab:sample data}%
\end{table}

\subsection{Experiment Settings}\label{sec:experiment settings}

\begin{figure}
    \centering
    \includegraphics[width=1\linewidth]{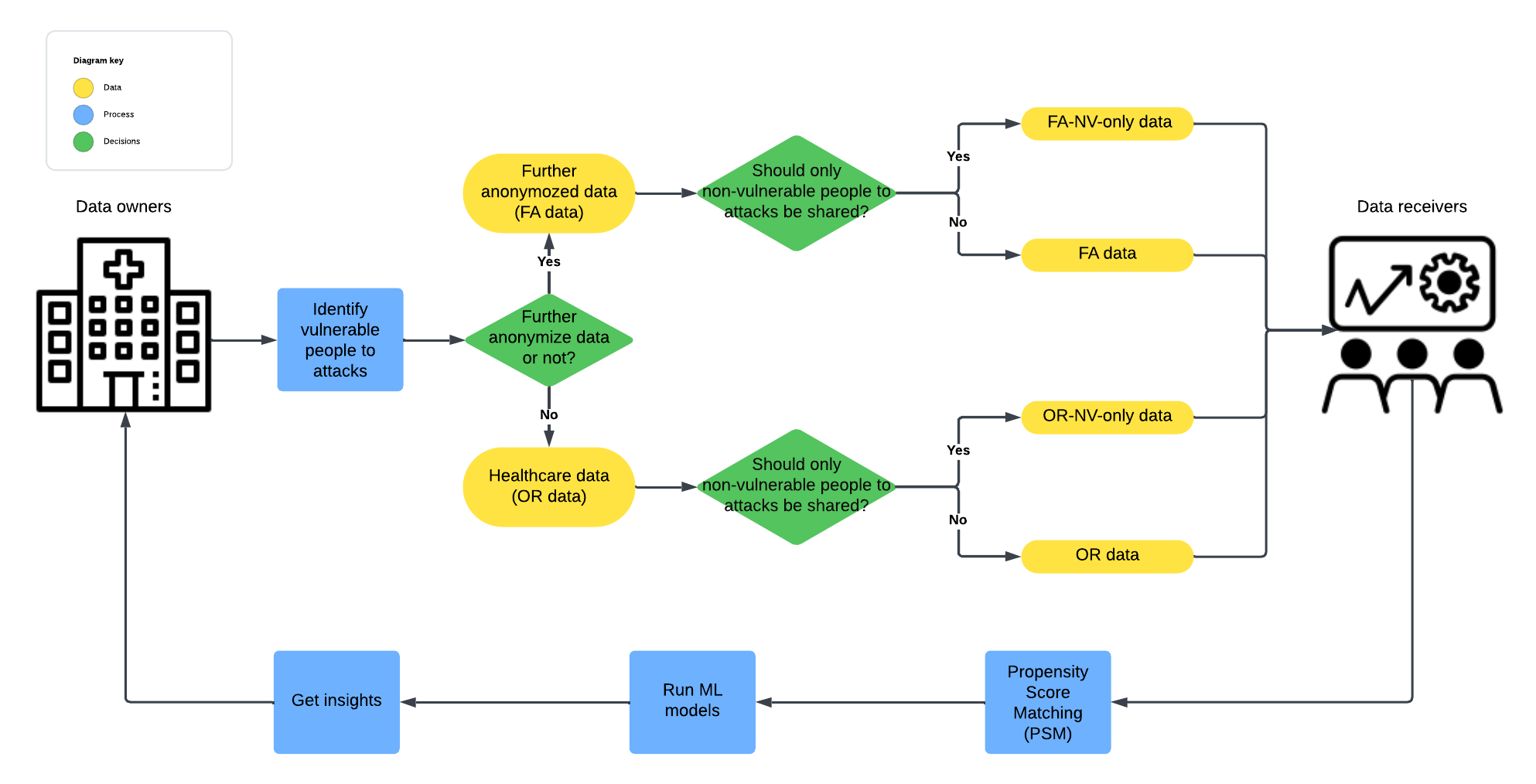}
    \caption{The process of sharing healthcare data between data owners and data receivers}
    \label{fig:flow chart}
\end{figure}

Figure~\ref{fig:flow chart} illustrates the process of sharing healthcare data between data owners and data receivers. This study adopts this framework to design the experiments. The first stage involves identifying people who are vulnerable to privacy attacks, as detailed in Section~\ref{sec:baseline attack evaluation}. Section~\ref{sec:further anonymization} outlines the anonymization models employed and the subsequent processing of further anonymized (FA) datasets. At the conclusion of this stage, four datasets are generated based on different decisions. We refer to these four datasets as {\em candidate datasets}. Subsequently, candidate datasets are used to classify the presence of a sepsis flag. Due to the inherent imbalance in the data with regards to patients who have a sepsis flag, propensity score matching (PSM) is applied to achieve a more balanced dataset. Our PSM implementation is described in Section~\ref{sec: PSM}. Following this, ML models are employed to perform the classification task, as discussed in Section~\ref{sec:ML performance}. The primary contribution of this study is assessing how ML performance varies across candidate datasets and guiding data owners in selecting which dataset to share to effectively balance privacy preservation with data utility and ensure optimal outcomes for data receivers.

\subsubsection{Baseline Attack Evaluation}\label{sec:baseline attack evaluation}
To gain insights into the initial privacy attack risk levels in the healthcare data, we assess the number of people who are vulnerable to linkage and homogeneity attacks. The process consists of the following steps:
\begin{enumerate}
    \item Data Grouping:
    \begin{itemize}
        \item Construct a new column by concatenating the values of QIs for each individual in the dataset. For instance, if the QIs include age = 26, gender = female, and race = white, the resulting value in this column will be represented as ''26/female/white.''
        \item Group the dataset based on the concatenated QI values to form clusters, ensuring that individuals with identical QI combinations are clustered together.
    \end{itemize}
    \item Evaluating Linkage Attack Risk:
    \begin{itemize}
        \item Calculate the size of each cluster, denoted as $|C_i|$ , where $C_i$ represents the $i$-th cluster.
        \item Determine the risk value for each individual as $1/|C_i|$. 
        \item Identify individuals vulnerable to linkage attacks by applying threshold values $\tau = 0.05, 0.075, 0.1$, as suggested by El Emam \cite{ElEmamKhaled2013MtPo, ElEmamKhaled2013CMT}. An individual is classified as vulnerable if their risk value  $1/|C_i|$  exceeds the specified threshold.
    \end{itemize}
    \item Evaluating Homogeneity Attack Risk:
    \begin{itemize}
        \item Assess each cluster to determine whether all individuals within the cluster have an identical sensitive attribute value.
        \item Classify individuals in these clusters as vulnerable to homogeneity attacks \citep{machanavajjhala2007diversity}.
    \end{itemize}
\end{enumerate}

\subsubsection{Further Anonymization and Attack Evaluation}\label{sec:further anonymization}
In the subsequent phase of the experiment, we apply MO-OBAM \citep{wei2025multi}, $k$-anonymity algorithm proposed by Domingo-Ferrer and Torra \cite{domingo2005ordinal} and algorithm introduced by Zheng et al. \cite{zheng2019effective} to further anonymize the healthcare data with different protection levels. The anonymization process modifies the QIs as follows:
\begin{itemize}
    \item For numeric QIs, replace original values with their mean within each cluster.
    \item For categorical QIs, replace original values with their mode within each cluster.
\end{itemize}
After applying these transformations, we obtain some FA datasets for each model. The FA datasets will be clustered based on the new QI values. The subsequent steps—forming clusters, computing individual risk values, and identifying individuals vulnerable to linkage and homogeneity attacks—follow the same procedure described earlier. Therefore, we identify people subject to linkage and homogeneity attacks in these FA datasets and split data regarding privacy attacks into vulnerable and non-vulnerable groups.

\subsubsection{Propensity Score Matching}\label{sec: PSM}
An additional challenge frequently encountered in healthcare data is imbalance, where positive outcomes outnumber negative outcomes. For example, considering the entire hospitalized patient population, usually a small proportion of patients experience a negative outcome (e.g. presence of sepsis during hospitalization), whereas a larger proportion does not experience the negative outcome of interest. Therefore, propensity score matching (PSM) is a widely used approach for balancing healthcare data \citep{rosenbaum1983central}. This technique aims to establish similarity in the distributions of key variables between groups with negative and positive outcomes, thereby enhancing the comparability of the studied cohorts. Valojerdi and Janani emphasized that utilizing five strata in PSM significantly reduces bias when investigating imbalanced groups \citep{valojerdi2018brief}. In this study, we use demographic variables, namely age, gender, race, and ethnicity to perform a PSM and apply five strata to the healthcare data to reduce the imbalance with regards to the binary indicator (referred as a "flag") that indicates the presence of sepsis during a hospitalization visit. Table~\ref{tab:summary of data}(a) provides descriptive statistics for the demographic variables in the healthcare data. Initially, the dataset consists of 116,020 people without a sepsis diagnosis and 3,851 people with a sepsis diagnosis. After applying PSM with a ratio of 1:5 \citep{valojerdi2018brief, cochran1968effectiveness}, we obtain a matched dataset comprising 19,255 people without sepsis and 3,851 people with sepsis. The descriptive statistics for the PSM-adjusted data are presented in Table~\ref{tab:summary of data}(b). We also assessed whether the distribution of each variable in the control group matches that in the case group using statistical tests. The results indicate that the distributions of variables in both groups are statistically equivalent.

\subsubsection{ML Performance}\label{sec:ML performance}
We utilize the candidate datasets generated in the data-sharing process to train and evaluate ML models,  specifically, Decision Trees (DT), Logistic Regression (LR), Gaussian Naive Bayes (NB), Random Forests (RF), Neural Networks (NN), and Support Vector Machines (SVM) to predict sepsis flag. In this stage of the experiment, we perform 100 iterations of training and testing using a consistent division of the datasets into training and test sets, with an 80:20 split ratio. This iterative approach aims to comprehensively assess the performance of the ML models across the candidate datasets. To ensure robust evaluation, we conduct statistical analyses of the precision and recall metrics obtained from the 100 iterations. Depending on whether the assumption of normality is satisfied, we apply either t-tests or Mann-Whitney U tests to compare model performance. In addition, our primary goal in ML is to accurately classify sepsis status, making it vital to examine the effect of including or excluding vulnerable populations to privacy on model performance.


\section{Results and Analysis}\label{sec:results}
\subsection*{Research Question 1: Which patient subpopulations defined by demographic and healthcare utilization characteristics in healthcare data are more vulnerable to privacy attacks?}

\begin{figure}
    \centering
    \includegraphics[scale=0.355]{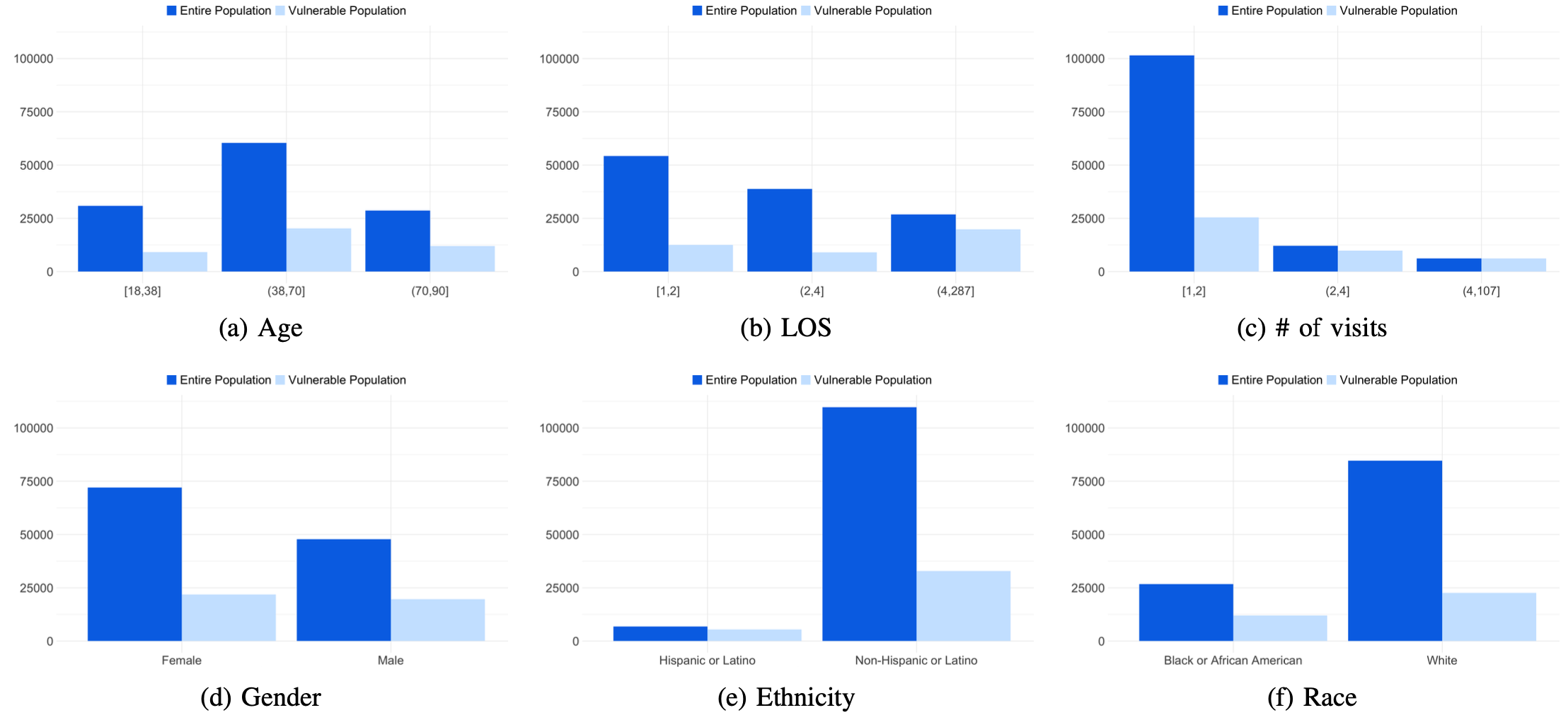}
    \caption{Distribution of the entire dataset vs the subset of individuals vulnerable to linkage attacks ($\tau=0.1$) for each QI.}
    \label{fig:initial linkage attacks}
\end{figure}

\begin{figure}
    \centering
    \includegraphics[scale=0.355]{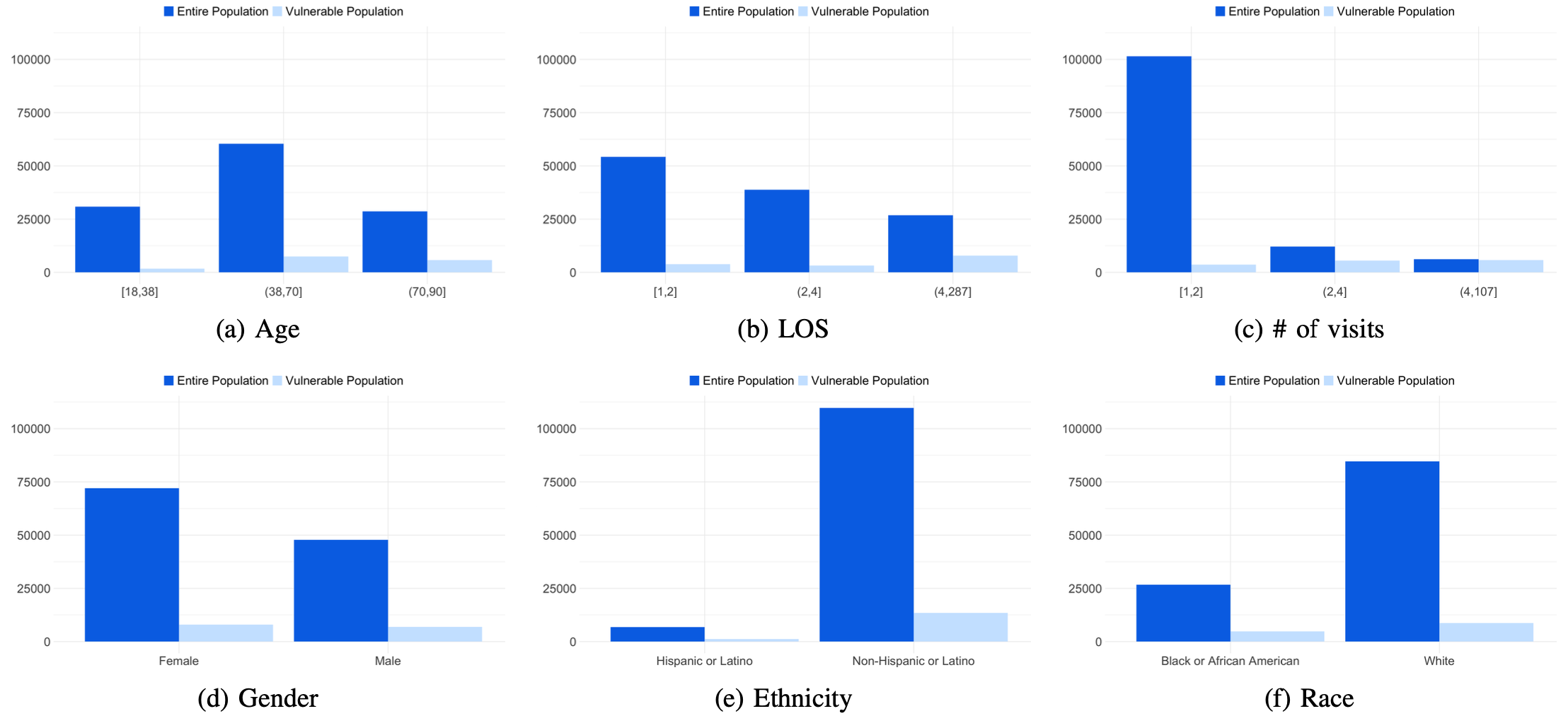}
    \caption{Distribution of the entire dataset vs the subset of individuals vulnerable to homogeneity attacks for each QI.}
    \label{fig:initial homogeneity attacks}
\end{figure}

To address this research question, we present two figures: Figure~\ref{fig:initial linkage attacks} and Figure~\ref{fig:initial homogeneity attacks}. These figures compare the distribution of the vulnerable population regarding privacy attacks with the overall distribution of the entire population across each QI. For simplification of visualization and analysis purposes, we grouped numeric QIs into three quartile-based ranges: from the minimum to the 1st quartile, the 1st quartile to the 3rd quartile, and the 3rd quartile to the maximum. Additionally, only the most prevalent categories are shown for categorical QIs. Detailed information for all QIs is provided in Appendix~\ref{app:attack evaluation}.

In the healthcare data, we identified 56,113 people who were vulnerable to linkage attacks at a risk level of $\tau=0.05$, 48,396 people who were vulnerable to linkage attacks at a risk level of $\tau=0.075$, and 41,445 people who were vulnerable to linkage attacks at a risk level of $\tau=0.1$. Figure~\ref{fig:initial linkage attacks} shows the distribution of QIs for individuals identified as vulnerable to linkage attacks at a risk level of $\tau=0.1$. 

\begin{itemize}
    \item Age: The distribution of individuals at risk of linkage attacks at the risk level of $\tau=0.1$ closely aligns with the overall patient population. However, nearly half of individuals in the age group (70, 90] are vulnerable, indicating that older individuals, particularly those over 70, face a significantly higher risk of linkage attacks.
    \item LOS: The distribution of individuals vulnerable to linkage attacks at $\tau=0.1$ contrasts sharply with that of the general population, suggesting that extended hospitalizations significantly increase susceptibility to linkage attacks.
    \item \# of Visits: While the overall distribution of at-risk individuals at $\tau=0.1$ mirrors that of the general population, nearly all individuals in the groups (2, 4] and (4, 107] are vulnerable.  This indicates that high healthcare utilization increase the risk of linkage attacks.
    \item Gender: The distribution of vulnerable individuals at $\tau=0.1$ is more balanced across genders compared to the general population. However, nearly half of the male population is vulnerable, highlighting that males face a higher risk of linkage attacks.
    \item Ethnicity: The overall distribution of at-risk individuals at $\tau=0.1$ is similar to that of the general population. However, nearly half of Black or African American individuals are vulnerable, suggesting they are at a higher risk of linkage attacks.
    \item Race: While the distribution of individuals vulnerable at $\tau=0.1$ largely aligns with the general population, almost all individuals identifying as Hispanic or Latino are vulnerable, underscoring their particularly high risk of linkage attacks.
\end{itemize}

Out of 119,871 individuals in the healthcare data, 14,937 (12\% of the study population) were identified as vulnerable to homogeneity attacks. Figure~\ref{fig:initial homogeneity attacks} compares the distribution of the entire dataset with that of people at risk of homogeneity attacks. For most QIs, the distribution of people vulnerable to homogeneity attacks closely resembles the distribution observed for linkage attacks in Figure~\ref{fig:initial linkage attacks}. This alignment suggests that groups identified as having a higher risk of linkage attacks also face a higher risk of homogeneity attacks. However, the distribution of vulnerable people by the number of visits diverges notably from that of the general population, highlighting that people with more frequent hospital visits are at an increased risk of homogeneity attacks.

\subsection*{Research Question 2a) Do anonymization models reduce the number of patients vulnerable to attacks in healthcare data?}

To address sub-question 2a), we refer to Table~\ref{tab: sepsis model evaluations2}, which shows the number of individuals at risk of linkage and homogeneity attacks across the three anonymization models. Different $k$ values correspond to varying levels of protection, demonstrating that all three models can effectively reduce the number of individuals vulnerable to both linkage and homogeneity attacks compared to the original healthcare data. Furthermore, when compared to the basic $k$-anonymity model, the model proposed by Zheng \emph{et al.} and MO-OBAM provide superior protection against homogeneity attacks, reducing the number of individuals at risk to zero. Comparing these two anonymization models reveals that MO-OBAM offers enhanced protection against linkage attacks.

\begin{table}
\caption{Number of people at risk of linkage attacks and homogeneity attacks by model and level of protection. The $\tau$ columns indicate the number of people at risk of linkage attacks for each value of $\tau$. The HA column gives the number of people at risk of homogeneity attacks. The values for ``Baseline'' are for the original healthcare data.}
\begin{tabular}{lcllll}\hline
Model    & Level of Protection                   & $\tau$=0.05 & $\tau$=0.075 & $\tau$=0.1 & HA     \\ \hline
Baseline      &                                          & 56113       & 48396        & 41445      & 14937 \\ \hline
$k$-anonymity   & \multirow{3}{*}{$k$=5}                                & 53225       & 46250        & 35480      & 6905   \\
Zheng et al     &                        & 18174       & 11656        & 7486       & 0      \\
MO-OBAM      &       & 15680       & 7375         & 2954       & 0      \\ \hline
$k$-anonymity   & \multirow{3}{*}{$k$=10}                               & 47990       & 47990        & 0          & 2260   \\
Zheng et al    &                         & 17068       & 12929        & 0          & 0      \\
MO-OBAM     &       & 6939        & 1960         & 0          & 0      \\ \hline
$k$-anonymity   & \multirow{3}{*}{$k$=15}                              & 46355       & 0            & 0          & 780    \\
Zheng et al     &                       & 21607       & 0           & 0          & 0      \\
MO-OBAM       &      & 2020        & 0           & 0         & 0      \\ \hline
$k$-anonymity   & \multirow{3}{*}{$k$=20}                              & 0           & 0            & 0          & 320    \\
Zheng et al    &                         & 0           & 0            & 0          & 0      \\
MO-OBAM      &   & 0           & 0            & 0          & 0      \\ \hline
\end{tabular}%
\label{tab: sepsis model evaluations2}
\end{table}

\subsection*{Research Question 2b) How do anonymization models affect the percentage of patients vulnerable to attacks within each patient subpopulation?}
\begin{figure}
    \centering
    \includegraphics[scale=0.355]{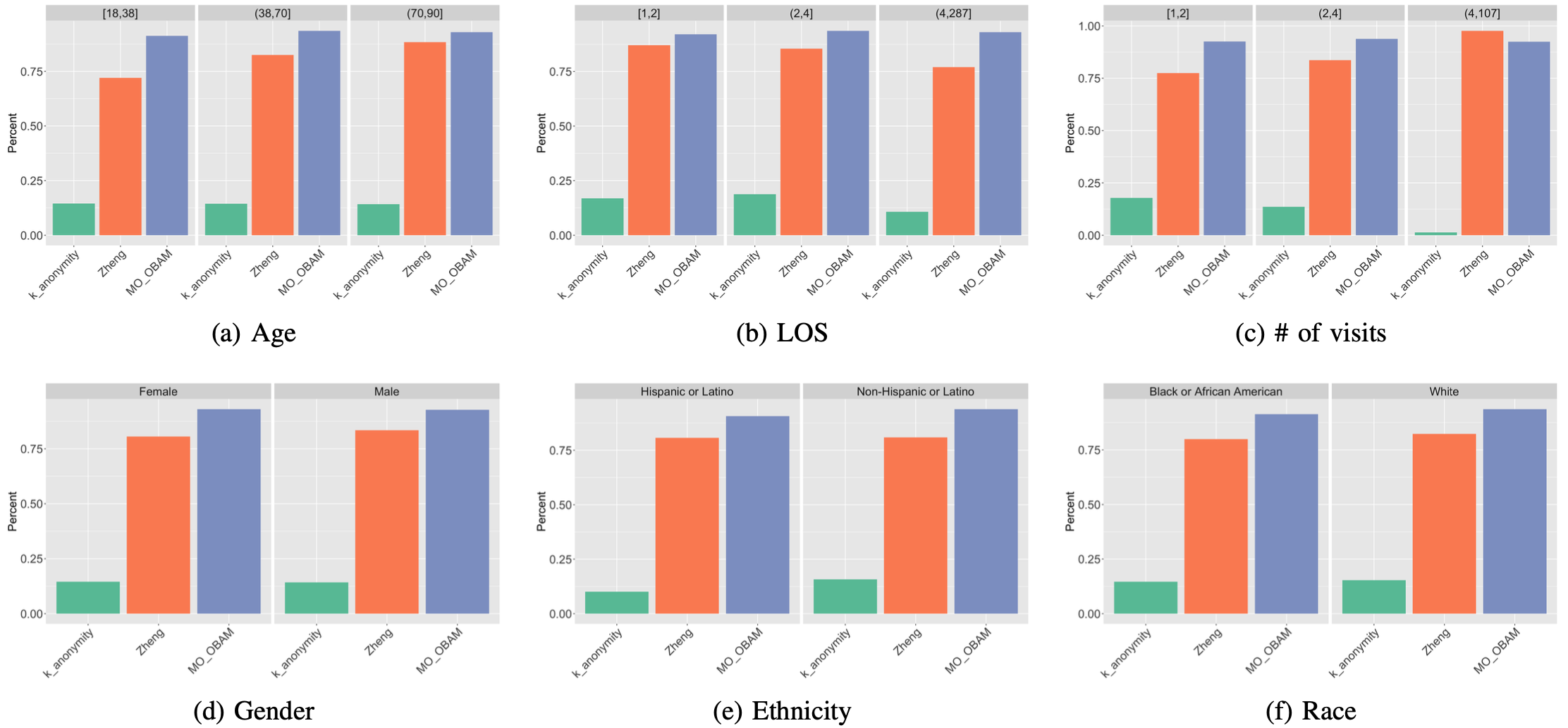}
    \caption{Percentage decrease in the number of people who are vulnerable to linkage attacks ($\tau=0.1$) after applying three anonymization methods, relative to the original vulnerable population for each QI with a protection level of $k=5$}
    \label{fig:RQ2.2}
\end{figure}

For sub-question 2b), we evaluate how the three anonymization models affect the percentage of individuals vulnerable to linkage attacks. Specifically, we focus on the risk level of $\tau=0.1$ and a protection level of $k=5$ as a representative example. Figure~\ref{fig:RQ2.2} illustrates the percentage decreases in vulnerability after applying anonymization across categories for each QI, with additional details provided in Appendix~\ref{app:attack evaluation}. Higher bars in the figure represent greater reductions relative to the original vulnerable population before anonymization. The results in Figure~\ref{fig:RQ2.2} reveal that the basic $k$-anonymity model achieves a percentage decrease of less than 20\% in the vulnerable population to linkage attacks with the risk level of $\tau=0.1$. In contrast, the algorithm proposed by Zheng \emph{et al.} results in a 70\% to 80\% decrease, while MO-OBAM decreases the vulnerable population to linkage attack with the risk level of $\tau=0.1$ by nearly 90\% across most cases, demonstrating its superior performance in protecting vulnerable groups. These findings underscore that both advanced anonymization models, particularly MO-OBAM, significantly outperform the basic $k$-anonymity model, emphasizing that $k$-anonymity alone is insufficient to provide robust protection against linkage attacks.

Before implementing ML models, we aimed to investigate whether individuals identified as vulnerable to privacy attacks exhibit a higher likelihood of being diagnosed with sepsis (denoted by the variable sepsisFlag). To address this, we analyzed the proportions of individuals with sepsisFlag in two distinct groups: the entire population and the subset identified as vulnerable to privacy attacks. Specifically, we focused on vulnerabilities stemming from linkage and homogeneity attacks, utilizing the PSM-adjusted healthcare data. Table~\ref{tab: statistical test between entirepop and vulnerable pop} presents these comparisons. All reported p-values are below 0.05, indicating statistically significant differences between the two groups. Therefore, these results suggest a higher incidence of sepsisFlag among the vulnerable population, hinting at a potential link between vulnerability to privacy attacks and increased risk of being diagnosed with sepsis. Given our primary ML objective—accurate classification of sepsis status—it is crucial to assess how the presence of vulnerable individuals may influence ML model performance.

\begin{table}
\caption{Percentage of population with sepsis diagnosis and results of two-proportion z-tests assessing whether the proportion of individuals with sepsis diagnoses statistically significantly differs between the entire population and vulnerable groups (for linkage attack with $\tau=0.05$, $\tau=0.075$, $\tau=0.1$, and homogeneity attack (HA)).}
\begin{tabular}{llll}\hline
Risk Indicator & Entire Population & Vulnerable Population & p value  \\ \hline
$\tau=0.05$    & 16.67\%           & 18.19\%               & 2.84E-05 \\ \hline
$\tau=0.075$   & 16.67\%           & 19.49\%               & 9.39E-14 \\ \hline
$\tau=0.1$     & 16.67\%           & 20.85\%               & 4.70E-26 \\ \hline
HA             & 16.67\%           & 25.40\%               & 1.17E-54 \\ \hline
\end{tabular}%
\label{tab: statistical test between entirepop and vulnerable pop}
\end{table}

Detailed results for each evaluation scenario are provided in Appendix~\ref{app:ml model evaluation}. Tables~\ref{tab:average precision full}(a) and~\ref{tab:average recall full}(a) report the average precision and recall values calculated across all scenarios for each ML model and anonymization model. Tables~\ref{tab:average precision full}(b) and~\ref{tab:average recall full}(b) indicate the percentage change of precision and recall between two different datasets, with the dataset appearing before “vs.” serving as the baseline for comparison. For instance, when comparing “OR vs. OR-NV-only,” the percentage change is calculated relative to the metrics of the “OR” dataset.

\subsection*{Research Question 3a) How does the decision to include or exclude the vulnerable populations' data impact the performance of ML models that aim to predict the presence of sepsis for hospitalized patients?}
To address subquestion 3a), we compared precision and recall obtained using the OR and OR-NV-only datasets, as well as the FA and FA-NV-only datasets. The results demonstrate a significant decline in model performance when excluding vulnerable populations. Using the OR-NV-only dataset resulted in a reduction in both precision and recall compared to the OR dataset. Precision decreased by 8\% to 23\%, while recall showed a more substantial decline of 17\% to 43\% across all ML models. This underscores the critical contribution of vulnerable populations to the overall performance of ML models. A similar trend was observed when comparing the FA and FA-NV-only datasets. Excluding vulnerable populations in the FA-NV-only dataset led to decreased precision and recall, though the extent of the decline varied across anonymization models:
\begin{itemize}
    \item The $k$-anonymity algorithm resulted in an average reduction of 4\% in precision and 8\% in recall. \item The method proposed by Zheng \emph{et al.} exhibited smaller average reductions of 2\% in precision and 3\% in recall.
    \item MO-OBAM showed the least reductions in all ML models compared to two other anonymization models, with average reductions of 1\% in precision and 2\% in recall, highlighting its effectiveness in minimizing performance degradation while protecting vulnerable populations.
\end{itemize}
Therefore, including vulnerable populations is essential for achieving better ML model performance. Also, this is important for developing inclusive models that have good predictive performance for all subgroups and not just perform well for the majority group - which would result in disparities in predicting sepsis in underrepresented populations. The results highlight that vulnerable populations contribute significantly to both precision and recall, and their exclusion leads to a marked reduction in model effectiveness. Furthermore, among the anonymization models examined, MO-OBAM stands out as the most effective approach, minimizing the performance trade-offs while providing protection for vulnerable groups.

\subsection*{Research Question 3b) How can the anonymization process help protect these vulnerable populations to privacy while developing ML models that work well for both majority and minority groups based on their representation in the EHR data?}
To answer sub-question 3b), we compared the results obtained from the OR and FA datasets, and from the OR-NV-only and FA-NV-only datasets. The comparison between the OR and FA datasets revealed that across most ML models, both precision and recall values remained similar, with percentage changes of less than 2\%. An exception was observed with SVM, where precision obtained using the FA dataset was approximately 3\% higher than that of the OR dataset, while recall was approximately 5\% lower. This indicates that anonymization is able to preserve the utility of the data for most ML models but there may be some variability in the effect of anonymization on SVM. When comparing the OR-NV-only and FA-NV-only datasets, the FA-NV-only dataset consistently demonstrated better precision across all ML models, and recall improved for five out of six models, with the exception of NB. A closer analysis of the improvements achieved by using the FA-NV-only dataset reveals notable differences across anonymization methods:
\begin{itemize}
    \item The $k$-anonymity algorithm resulted in an average improvement of 17\% in precision and 27\% in recall.
    \item The method proposed by Zheng \emph{et al.} showed even greater improvements, with an average increase of 20\% in precision and 35\% in recall.
    \item MO-OBAM demonstrated the most improvements compared to other two models, achieving an average increase of 22\% in precision and 37\% in recall. These results underscore its superior effectiveness in maintaining useful information while protecting vulnerable populations.
\end{itemize}
When comparing the OR and FA datasets, precision and recall exhibited minimal changes, indicating that the anonymization process effectively preserves data utility for ML models. When comparing the OR-NV-only and FA-NV-only datasets, precision and recall showed notable improvements for most ML models when using the FA-NV-only dataset. Therefore, the results demonstrate that all three anonymization models are effective in preserving the utility of the data for ML model performance while safeguarding vulnerable populations. Among these methods, MO-OBAM stands out, achieving superior precision and recall improvements when applied to the FA-NV-only dataset. This highlights its advanced capability to maintain useful information in anonymized data while providing robust protection for vulnerable groups.

\begin{table}[]
    \caption{Average precision values for six ML models under different datasets and anonymization methods.}
    {\begin{tabular}{c}
        (a) Precision Values \\
\begin{tabular}{llllllll}
\hline
Data &
  \begin{tabular}[c]{@{}l@{}}Anonymization \\ Model\end{tabular} &
  \multicolumn{1}{c}{DT} &
  \multicolumn{1}{c}{LR} &
  \multicolumn{1}{c}{NB} &
  \multicolumn{1}{c}{NN} &
  \multicolumn{1}{c}{RF} &
  \multicolumn{1}{c}{SVM} \\ \hline
OR &
   &
  \multicolumn{1}{r}{0.5793} &
  \multicolumn{1}{r}{0.7085} &
  \multicolumn{1}{r}{0.6001} &
  \multicolumn{1}{r}{0.6735} &
  \multicolumn{1}{r}{0.7092} &
  \multicolumn{1}{r}{0.6594} \\ \hline
OR-NV-only                  &               & 0.4775 & 0.6269 & 0.3981 & 0.6141 & 0.6495 & 0.5122 \\ \hline
\multirow{3}{*}{FA}         & $k$-anonymity & 0.5798 & 0.7044 & 0.6021 & 0.6675 & 0.7041 & 0.6805 \\ \cline{2-8} 
                            & Zheng et al   & 0.5802 & 0.7077 & 0.6006 & 0.6680 & 0.7045 & 0.6763 \\ \cline{2-8} 
                            & MO-OBAM       & 0.5786 & 0.7082 & 0.6019 & 0.6662 & 0.7052 & 0.6850 \\ \hline
\multirow{3}{*}{FA-NV-only} & $k$-anonymity & 0.5451 & 0.6892 & 0.5428 & 0.6499 & 0.6825 & 0.6606 \\ \cline{2-8} 
                            & Zheng et al   & 0.5630 & 0.6963 & 0.5733 & 0.6555 & 0.6944 & 0.6801 \\ \cline{2-8} 
                            & MO-OBAM       & 0.5737 & 0.7034 & 0.5935 & 0.6642 & 0.7006 & 0.6831 \\ \hline
\end{tabular}%
\\ \\
        (b) Percentage Changes of Precision Values \\
\begin{tabular}{llrrrrrr}
\hline
Pairwise Comparison &
  \begin{tabular}[c]{@{}l@{}}Anonymization \\ Model\end{tabular} &
  \multicolumn{1}{c}{DT} &
  \multicolumn{1}{c}{LR} &
  \multicolumn{1}{c}{NB} &
  \multicolumn{1}{c}{NN} &
  \multicolumn{1}{c}{RF} &
  \multicolumn{1}{c}{SVM} \\ \hline
OR vs. OR-NV-only                          &               & -17.57\% & -11.52\% & -33.67\% & -8.82\% & -8.42\% & -22.32\% \\ \hline
\multirow{3}{*}{OR vs. FA}                 & $k$-anonymity & 0.08\%   & -0.59\%  & 0.34\%   & -0.90\% & -0.73\% & 3.19\%   \\ \cline{2-8} 
                                           & Zheng et al   & 0.16\%   & -0.11\%  & 0.08\%   & -0.82\% & -0.66\% & 2.56\%   \\ \cline{2-8} 
                                           & MO-OBAM       & -0.13\%  & -0.04\%  & 0.30\%   & -1.08\% & -0.57\% & 3.88\%   \\ \hline
\multirow{3}{*}{OR-NV-only vs. FA-NV-only} & $k$-anonymity & 14.16\%  & 9.94\%   & 36.37\%  & 5.83\%  & 5.08\%  & 28.98\%  \\ \cline{2-8} 
                                           & Zheng et al   & 17.91\%  & 11.08\%  & 44.03\%  & 6.74\%  & 6.91\%  & 32.79\%  \\ \cline{2-8} 
                                           & MO-OBAM       & 20.16\%  & 12.20\%  & 49.10\%  & 8.16\%  & 7.87\%  & 33.37\%  \\ \hline
\multirow{3}{*}{FA vs. FA-NV-only}         & $k$-anonymity & -5.97\%  & -2.16\%  & -9.85\%  & -2.63\% & -3.06\% & -2.92\%  \\ \cline{2-8} 
                                           & Zheng et al   & -2.96\%  & -1.61\%  & -4.54\%  & -1.86\% & -1.44\% & 0.57\%   \\ \cline{2-8} 
                                           & MO-OBAM       & -0.83\%  & -0.68\%  & -1.40\%  & -0.31\% & -0.64\% & -0.27\%  \\ \hline
\end{tabular}%
    \end{tabular}}
    \label{tab:average precision full}
\end{table}

\begin{table}[]
    \caption{Average recall values for six ML models under different datasets and anonymization methods}
    {\begin{tabular}{c}
         (a) Recall Values \\
\begin{tabular}{llllllll}
\hline
Data &
  \begin{tabular}[c]{@{}l@{}}Anonymization \\ Model\end{tabular} &
  \multicolumn{1}{c}{DT} &
  \multicolumn{1}{c}{LR} &
  \multicolumn{1}{c}{NB} &
  \multicolumn{1}{c}{NN} &
  \multicolumn{1}{c}{RF} &
  \multicolumn{1}{c}{SVM} \\ \hline
OR &
   &
  \multicolumn{1}{r}{0.5748} &
  \multicolumn{1}{r}{0.6197} &
  \multicolumn{1}{r}{0.4370} &
  \multicolumn{1}{r}{0.6160} &
  \multicolumn{1}{r}{0.6021} &
  \multicolumn{1}{r}{0.6097} \\ \hline
OR-NV-only                  &               & 0.4761 & 0.4340 & 0.6177 & 0.4665 & 0.3488 & 0.4350 \\ \hline
\multirow{3}{*}{FA}         & $k$-anonymity & 0.5749 & 0.6169 & 0.4409 & 0.6150 & 0.6033 & 0.5752 \\ \cline{2-8} 
                            & Zheng et al   & 0.5759 & 0.6204 & 0.4451 & 0.6167 & 0.6007 & 0.5885 \\ \cline{2-8} 
                            & MO-OBAM       & 0.5727 & 0.6174 & 0.4408 & 0.6201 & 0.6009 & 0.5732 \\ \hline
\multirow{3}{*}{FA-NV-only} & $k$-anonymity & 0.5406 & 0.5592 & 0.4561 & 0.5694 & 0.5218 & 0.5304 \\ \cline{2-8} 
                            & Zheng et al   & 0.5605 & 0.6000 & 0.4387 & 0.5978 & 0.5754 & 0.5561 \\ \cline{2-8} 
                            & MO-OBAM       & 0.5650 & 0.6063 & 0.4397 & 0.6028 & 0.5830 & 0.5570 \\ \hline
\end{tabular}%
\\  \\
    (b) Percentage Changes of Recall Values \\
\begin{tabular}{llrrrrrr}
\hline
Pairwise Comparison &
  \begin{tabular}[c]{@{}l@{}}Anonymization \\ Model\end{tabular} &
  \multicolumn{1}{c}{DT} &
  \multicolumn{1}{c}{LR} &
  \multicolumn{1}{c}{NB} &
  \multicolumn{1}{c}{NN} &
  \multicolumn{1}{c}{RF} &
  \multicolumn{1}{c}{SVM} \\ \hline
OR vs. OR-NV-only                          &               & -17.18\% & -29.96\% & 41.34\%  & -24.28\% & -42.07\% & -28.65\% \\ \hline
\multirow{3}{*}{OR vs. FA}                 & $k$-anonymity & 0.01\%   & -0.46\%  & 0.88\%   & -0.17\%  & 0.19\%   & -5.66\%  \\ \cline{2-8} 
                                           & Zheng et al   & 0.20\%   & 0.12\%   & 1.84\%   & 0.12\%   & -0.24\%  & -3.48\%  \\ \cline{2-8} 
                                           & MO-OBAM       & -0.36\%  & -0.38\%  & 0.86\%   & 0.67\%   & -0.20\%  & -5.98\%  \\ \hline
\multirow{3}{*}{OR-NV-only vs. FA-NV-only} & $k$-anonymity & 13.56\%  & 28.84\%  & -26.15\% & 22.06\%  & 49.58\%  & 21.92\%  \\ \cline{2-8} 
                                           & Zheng et al   & 17.74\%  & 38.24\%  & -28.97\% & 28.17\%  & 64.98\%  & 27.84\%  \\ \cline{2-8} 
                                           & MO-OBAM       & 18.68\%  & 39.70\%  & -28.81\% & 29.24\%  & 67.13\%  & 28.03\%  \\ \hline
\multirow{3}{*}{FA vs. FA-NV-only}         & $k$-anonymity & -5.96\%  & -9.35\%  & 3.46\%   & -7.42\%  & -13.51\% & -7.80\%  \\ \cline{2-8} 
                                           & Zheng et al   & -2.68\%  & -3.30\%  & -1.42\%  & -3.06\%  & -4.20\%  & -5.50\%  \\ \cline{2-8} 
                                           & MO-OBAM       & -1.36\%  & -1.79\%  & -0.24\%  & -2.78\%  & -2.99\%  & -2.84\%  \\ \hline
\end{tabular}%
    \end{tabular}}
    \label{tab:average recall full}
\end{table}

\section{Discussion}\label{sec:conclusion}
This study examines the data-sharing process outlined in Figure~\ref{fig:flow chart} and addresses key research questions to guide data owners in making informed decisions. Data owners, such as hospitals and healthcare organizations, often aim to share healthcare data for secondary use, such as research and analytics. However, they are obligated to ensure the protection of individuals’ privacy within the data before sharing it. Our findings highlight that healthcare datasets remain vulnerable to privacy attacks, even when the datasets meet HIPAA requirements. Specifically, older individuals, those with prolonged hospital stays, high healthcare utilization, and members of minority groups by gender, ethnicity, and race face higher risks of privacy attacks. By identifying these vulnerable populations, the study underscores a critical first decision for data owners: healthcare data must undergo additional anonymization processes to enhance privacy protection before it is shared.

The next decision for data owners is whether to share only the data of individuals who are not vulnerable to privacy attacks or to include the entire population. In our case, the healthcare data is intended for training ML models to classify sepsis. If individuals vulnerable to privacy attacks also exhibit a higher risk of being diagnosed with sepsis, excluding this group could significantly impact the effectiveness of ML applications. Our study reveals that vulnerable populations have a higher proportion of individuals diagnosed with sepsis, suggesting a potential connection between vulnerability to privacy attacks and an increased risk of sepsis. These findings emphasize that excluding vulnerable individuals would compromise the utility of the data. To evaluate this, we compared the precision and recall obtained from ML models using the FA dataset (which includes the entire population) and the FA-NV-only dataset (which excludes vulnerable individuals). The results demonstrate that precision and recall are consistently higher when using the FA dataset. This underscores the importance of sharing further anonymized data that includes vulnerable populations, as it ensures better ML model performance and retains the utility of the dataset for its intended purpose. Therefore, we recommend that the further anonymized data that includes the entire population be shared with data receivers. 

Since healthcare data will undergo further anonymization, the choice of anonymization technique is crucial. This study compared three anonymization models and found that the basic $k$-anonymity algorithm is less effective at providing robust protection against privacy attacks compared to two advanced anonymization models. Specifically, $k$-anonymity has significant limitations: (1) it fails to fully protect individuals from homogeneity attacks, leaving some at risk, and (2) it achieves only a marginal reduction in the number of individuals vulnerable to privacy attacks, highlighting its limitations as an anonymization approach. When comparing the two advanced models—the method proposed by Zheng \emph{et al.} and MO-OBAM—we observed that both are capable of completely mitigating homogeneity attacks, ensuring no individuals remain at risk after their application. However, MO-OBAM consistently outperforms in terms of protecting against linkage attacks, achieving the greatest reduction in the number of individuals at risk across most scenarios. From the perspective of privacy protection, MO-OBAM is the most effective choice. Furthermore, when evaluating the utility of the further anonymized data, MO-OBAM also proves superior. It minimizes performance reductions across all ML models when comparing FA and FA-NV-only datasets and demonstrates the highest performance improvements when comparing OR-NV-only and FA-NV-only datasets. These findings underscore MO-OBAM’s dual strengths: it excels at safeguarding privacy while preserving the utility of the data for ML applications. 

\section{Conclusion}
This study explores and answers critical research questions about the healthcare data-sharing process. The objective is to guide data owners in selecting datasets that balance patients’ privacy concerns with the utility required for secondary use. Our experimental results demonstrate that populations vulnerable to privacy attacks positively contribute to robust ML performance. Therefore, further anonymizing healthcare data and sharing the entire population within it is a promising approach to ensuring both privacy and data utility.

However, this study has several limitations. For instance, the potential link observed between vulnerability to privacy attacks and sepsis diagnosis may not generalize to other diseases. This limitation stems from using a single dataset designed for sepsis classification. Moreover, ICD codes, primarily intended for reimbursement, may not fully capture clinical severity or disease trajectory, meaning the sepsis flag may represent only a subset of individuals at risk. Additionally, our study relied on labeled data with an available sepsis flag; applying these techniques to unlabeled datasets may necessitate modifications. Furthermore, this study focused only on two common types of privacy attacks—linkage and homogeneity attacks—and employed three anonymization models to mitigate them. Other privacy attacks, such as semantic skewness attacks \citep{khan2020theta} and semantic similarity attacks \citep{anjum2018efficient}, were not addressed.

Future work aims to address these limitations. We plan to extend the study to datasets targeting other diseases and investigate the effects of different privacy attacks on ML models. Additionally, we intend to evaluate the impact of a broader range of anonymization algorithms and models to strengthen privacy protection and maintain data utility.

\section{Acknowledgment}

This work was supported by the National Science Foundation Smart and Connected Health under Award Number 1833538; the National Library of Medicine of the National Institutes of Health under Grant Number 1R01LM012300-01A1 and Award Number R01LM012300. We would also like to acknowledge the S.E.P.S.I.S (Sepsis Early Prediction Support Implementation System) Collaborative. Finally, we thank the providers and data scientists at the study hospital who facilitated data access, cleaning, and provided insights into the sepsis-related clinical workflow.


\pagebreak
\appendix
\section{Descriptive Statistics of the Healthcare Data}\label{app: descriptive stat}

\begin{table}[h!]
\caption{Demographic variables before and after PSM}{\begin{tabular}{c}
     (a) Summary of demographic variables in original data \\
    \resizebox{\columnwidth}{!}{%
\begin{tabular}{l|lrr|llr}
\hline
\multicolumn{1}{c|}{\multirow{2}{*}{\textbf{Variables}}} &
  \multicolumn{3}{c|}{\multirow{2}{*}{\textbf{Controls (SepsisFlag = 0)}}} &
  \multicolumn{3}{c}{\multirow{2}{*}{\textbf{Cases (SepsisFlag = 1)}}} \\
\multicolumn{1}{c|}{} &
  \multicolumn{3}{c|}{} &
  \multicolumn{3}{c}{} \\ \hline
\multicolumn{1}{c|}{\textbf{}} &
  \multicolumn{1}{l|}{\textbf{Mean}} &
  \multicolumn{2}{l|}{\textbf{Std. Dev}} &
  \multicolumn{1}{l|}{\textbf{Mean}} &
  \multicolumn{2}{l}{\textbf{Std. Dev}} \\ \hline
AgeCategory &
  \multicolumn{1}{l|}{54.29807} &
  \multicolumn{2}{l|}{19.43443} &
  \multicolumn{1}{l|}{65.14775} &
  \multicolumn{2}{l}{17.67968} \\ \hline
 &
  \multicolumn{1}{l|}{\textbf{Values}} &
  \multicolumn{1}{c|}{\textbf{Count}} &
  \multicolumn{1}{c|}{\textbf{Percentage}} &
  \multicolumn{1}{l|}{\textbf{Values}} &
  \multicolumn{1}{c|}{\textbf{Count}} &
  \multicolumn{1}{c}{\textbf{Percentage}} \\ \hline
\multirow{2}{*}{GenderDescription} &
  \multicolumn{1}{l|}{Female} &
  \multicolumn{1}{r|}{70174} &
  60.484\% &
  \multicolumn{1}{l|}{Female} &
  \multicolumn{1}{r|}{1921} &
  49.883\% \\ \cline{2-7} 
 &
  \multicolumn{1}{l|}{Male} &
  \multicolumn{1}{r|}{45846} &
  39.516\% &
  \multicolumn{1}{l|}{Male} &
  \multicolumn{1}{r|}{1930} &
  50.117\% \\ \hline
\multirow{7}{*}{RaceDescription} &
  \multicolumn{1}{l|}{White} &
  \multicolumn{1}{r|}{81761} &
  70.471\% &
  \multicolumn{1}{l|}{White} &
  \multicolumn{1}{r|}{2829} &
  73.461\% \\ \cline{2-7} 
 &
  \multicolumn{1}{p{2.5cm}|}{Black or African American} &
  \multicolumn{1}{r|}{25937} &
  22.356\% &
  \multicolumn{1}{l|}{Black or African   American} &
  \multicolumn{1}{r|}{809} &
  21.008\% \\ \cline{2-7} 
 &
  \multicolumn{1}{l|}{Other Rance} &
  \multicolumn{1}{r|}{4377} &
  3.773\% &
  \multicolumn{1}{l|}{Other Rance} &
  \multicolumn{1}{r|}{107} &
  2.778\% \\ \cline{2-7} 
 &
  \multicolumn{1}{l|}{Asian} &
  \multicolumn{1}{r|}{2717} &
  2.342\% &
  \multicolumn{1}{l|}{Asian} &
  \multicolumn{1}{r|}{69} &
  1.792\% \\ \cline{2-7} 
 &
  \multicolumn{1}{l|}{Unavailable} &
  \multicolumn{1}{r|}{947} &
  0.816\% &
  \multicolumn{1}{l|}{Unavailable} &
  \multicolumn{1}{r|}{24} &
  0.623\% \\ \cline{2-7} 
 &
  \multicolumn{1}{p{2.5cm}|}{American Indian or Alaska Native} &
  \multicolumn{1}{r|}{217} &
  0.187\% &
  \multicolumn{1}{l|}{American Indian or   Alaska Native} &
  \multicolumn{1}{r|}{13} &
  0.338\% \\ \cline{2-7} 
 &
  \multicolumn{1}{l|}{Multiple Racial} &
  \multicolumn{1}{r|}{64} &
  0.055\% &
  \multicolumn{1}{l|}{Multiple} &
  \multicolumn{1}{r|}{0} &
  0.000\% \\ \hline
\multirow{4}{*}{EthnicGroupDescription} &
  \multicolumn{1}{l|}{Not Hispanic or   Latino} &
  \multicolumn{1}{r|}{106173} &
  91.513\% &
  \multicolumn{1}{l|}{Not Hispanic or   Latino} &
  \multicolumn{1}{r|}{3566} &
  92.599\% \\ \cline{2-7} 
 &
  \multicolumn{1}{l|}{Hispanic} &
  \multicolumn{1}{r|}{6698} &
  5.773\% &
  \multicolumn{1}{l|}{Hispanic} &
  \multicolumn{1}{r|}{157} &
  4.077\% \\ \cline{2-7} 
 &
  \multicolumn{1}{l|}{Unknown} &
  \multicolumn{1}{r|}{2910} &
  2.508\% &
  \multicolumn{1}{l|}{Unknown} &
  \multicolumn{1}{r|}{122} &
  3.168\% \\ \cline{2-7} 
 &
  \multicolumn{1}{l|}{Declined} &
  \multicolumn{1}{r|}{239} &
  0.206\% &
  \multicolumn{1}{l|}{Declined} &
  \multicolumn{1}{r|}{6} &
  0.156\% \\ \hline
  \end{tabular}
} \\ \\
(b) Summary of demographic variables in PSM-adjusted data \\
\resizebox{\columnwidth}{!}{%
\begin{tabular}{l|lrr|llr}
\hline
\multicolumn{1}{c|}{\multirow{2}{*}{\textbf{Variables}}} &
  \multicolumn{3}{c|}{\multirow{2}{*}{\textbf{Controls (SepsisFlag = 0)}}} &
  \multicolumn{3}{c}{\multirow{2}{*}{\textbf{Cases (SepsisFlag = 1)}}} \\
\multicolumn{1}{c|}{} &
  \multicolumn{3}{c|}{} &
  \multicolumn{3}{c}{} \\ \hline
\multicolumn{1}{c|}{\textbf{}} &
  \multicolumn{1}{l|}{\textbf{Mean}} &
  \multicolumn{2}{l|}{\textbf{Std. Dev}} &
  \multicolumn{1}{l|}{\textbf{Mean}} &
  \multicolumn{2}{l}{\textbf{Std. Dev}} \\ \hline
AgeCategory &
  \multicolumn{1}{l|}{65.15674} &
  \multicolumn{2}{l|}{17.65043} &
  \multicolumn{1}{l|}{65.14775} &
  \multicolumn{2}{l}{17.67968} \\ \hline
 &
  \multicolumn{1}{l|}{\textbf{Values}} &
  \multicolumn{1}{c|}{\textbf{Count}} &
  \multicolumn{1}{c|}{\textbf{Percentage}} &
  \multicolumn{1}{l|}{\textbf{Values}} &
  \multicolumn{1}{c|}{\textbf{Count}} &
  \multicolumn{1}{c}{\textbf{Percentage}} \\ \hline
\multirow{2}{*}{GenderDescription} &
  \multicolumn{1}{l|}{Female} &
  \multicolumn{1}{r|}{9596} &
  49.836\% &
  \multicolumn{1}{l|}{Female} &
  \multicolumn{1}{r|}{1921} &
  49.883\% \\ \cline{2-7} 
 &
  \multicolumn{1}{l|}{Male} &
  \multicolumn{1}{r|}{9659} &
  50.164\% &
  \multicolumn{1}{l|}{Male} &
  \multicolumn{1}{r|}{1930} &
  50.117\% \\ \hline
\multirow{7}{*}{RaceDescription} &
  \multicolumn{1}{l|}{White} &
  \multicolumn{1}{r|}{14151} &
  73.493\% &
  \multicolumn{1}{l|}{White} &
  \multicolumn{1}{r|}{2829} &
  73.461\% \\ \cline{2-7} 
 &
  \multicolumn{1}{p{2.5cm}|}{Black or African American} &
  \multicolumn{1}{r|}{4065} &
  21.111\% &
  \multicolumn{1}{l|}{Black or African   American} &
  \multicolumn{1}{r|}{809} &
  21.008\% \\ \cline{2-7} 
 &
  \multicolumn{1}{l|}{Other Rance} &
  \multicolumn{1}{r|}{573} &
  2.976\% &
  \multicolumn{1}{l|}{Other Rance} &
  \multicolumn{1}{r|}{107} &
  2.778\% \\ \cline{2-7} 
 &
  \multicolumn{1}{l|}{Asian} &
  \multicolumn{1}{r|}{339} &
  1.761\% &
  \multicolumn{1}{l|}{Asian} &
  \multicolumn{1}{r|}{69} &
  1.792\% \\ \cline{2-7} 
 &
  \multicolumn{1}{l|}{Unavailable} &
  \multicolumn{1}{r|}{80} &
  0.415\% &
  \multicolumn{1}{l|}{Unavailable} &
  \multicolumn{1}{r|}{24} &
  0.623\% \\ \cline{2-7} 
 &
  \multicolumn{1}{p{2.5cm}|}{American Indian or Alaska Native} &
  \multicolumn{1}{r|}{47} &
  0.244\% &
  \multicolumn{1}{l|}{American Indian or   Alaska Native} &
  \multicolumn{1}{r|}{13} &
  0.338\% \\ \cline{2-7} 
 &
  \multicolumn{1}{l|}{Multiple Racial} &
  \multicolumn{1}{r|}{0} &
  0.000\% &
  \multicolumn{1}{l|}{Multiple} &
  \multicolumn{1}{r|}{0} &
  0.000\% \\ \hline
\multirow{4}{*}{EthnicGroupDescription} &
  \multicolumn{1}{l|}{Not Hispanic or   Latino} &
  \multicolumn{1}{r|}{17887} &
  92.895\% &
  \multicolumn{1}{l|}{Not Hispanic or   Latino} &
  \multicolumn{1}{r|}{3566} &
  92.599\% \\ \cline{2-7} 
 &
  \multicolumn{1}{l|}{Hispanic} &
  \multicolumn{1}{r|}{795} &
  4.129\% &
  \multicolumn{1}{l|}{Hispanic} &
  \multicolumn{1}{r|}{157} &
  4.077\% \\ \cline{2-7} 
 &
  \multicolumn{1}{l|}{Unknown} &
  \multicolumn{1}{r|}{557} &
  2.893\% &
  \multicolumn{1}{l|}{Unknown} &
  \multicolumn{1}{r|}{122} &
  3.168\% \\ \cline{2-7} 
 &
  \multicolumn{1}{l|}{Declined} &
  \multicolumn{1}{r|}{16} &
  0.083\% &
  \multicolumn{1}{l|}{Declined} &
  \multicolumn{1}{r|}{6} &
  0.156\% \\ \hline\end{tabular}%
}
\end{tabular}}
\label{tab:summary of data}
\end{table}

\section{Attack Evaluation for Entire Healthcare Data}\label{app:attack evaluation}

\begin{table}[h!]
\caption{Number of people at risk of linkage attack before and after applying anonymization, by QIs}
     (a) Age \\ 
\resizebox{\columnwidth}{!}{
\begin{tabular}{ccl|lll|lll|lll}
\hline
\multicolumn{3}{l|}{Attack   Level} &
  \multicolumn{3}{c|}{$\tau=0.05$} &
  \multicolumn{3}{c|}{$\tau=0.075$} &
  \multicolumn{3}{c}{$\tau=0.1$} \\ \hline
\multicolumn{3}{l|}{Group} &
  \multicolumn{1}{l|}{{[}18,38{]}} &
  \multicolumn{1}{l|}{(38,70{]}} &
  (70,90{]} &
  \multicolumn{1}{l|}{{[}18,38{]}} &
  \multicolumn{1}{l|}{(38,70{]}} &
  (70,90{]} &
  \multicolumn{1}{l|}{{[}18,38{]}} &
  \multicolumn{1}{l|}{(38,70{]}} &
  (70,90{]} \\ \hline
\multicolumn{3}{l|}{Entire Population} &
  \multicolumn{1}{l|}{30815} &
  \multicolumn{1}{l|}{60445} &
  28611 &
  \multicolumn{1}{l|}{30815} &
  \multicolumn{1}{l|}{60445} &
  28611 &
  \multicolumn{1}{l|}{30815} &
  \multicolumn{1}{l|}{60445} &
  28611 \\ \hline
\multicolumn{3}{l|}{Vulnerable   Population in Entire Population} &
  \multicolumn{1}{l|}{12608} &
  \multicolumn{1}{l|}{26883} &
  16622 &
  \multicolumn{1}{l|}{10567} &
  \multicolumn{1}{l|}{23611} &
  14218 &
  \multicolumn{1}{l|}{9156} &
  \multicolumn{1}{l|}{20239} &
  12050 \\ \hline
\multicolumn{1}{p{2.5cm}|}{\multirow{12}{*}{\parbox{2.5cm}{Vulnerable Population after applying anonymization}}} &
  \multicolumn{1}{c|}{\multirow{3}{*}{$k$=5}} &
  $k$-anonymity &
  \multicolumn{1}{l|}{11765} &
  \multicolumn{1}{l|}{25710} &
  15750 &
  \multicolumn{1}{l|}{10055} &
  \multicolumn{1}{l|}{22635} &
  13560 &
  \multicolumn{1}{l|}{7825} &
  \multicolumn{1}{l|}{17315} &
  10340 \\ \cline{3-12} 
\multicolumn{1}{c|}{} &
  \multicolumn{1}{c|}{} &
  Zheng et al &
  \multicolumn{1}{l|}{5156} &
  \multicolumn{1}{l|}{9298} &
  3720 &
  \multicolumn{1}{l|}{3747} &
  \multicolumn{1}{l|}{5723} &
  2186 &
  \multicolumn{1}{l|}{2566} &
  \multicolumn{1}{l|}{3526} &
  1394 \\ \cline{3-12} 
\multicolumn{1}{c|}{} &
  \multicolumn{1}{c|}{} &
  MO-OBAM &
  \multicolumn{1}{l|}{3747} &
  \multicolumn{1}{l|}{7732} &
  4201 &
  \multicolumn{1}{l|}{1928} &
  \multicolumn{1}{l|}{3495} &
  1952 &
  \multicolumn{1}{l|}{801} &
  \multicolumn{1}{l|}{1311} &
  842 \\ \cline{2-12} 
\multicolumn{1}{c|}{} &
  \multicolumn{1}{c|}{\multirow{3}{*}{$k$=10}} &
  $k$-anonymity &
  \multicolumn{1}{l|}{10480} &
  \multicolumn{1}{l|}{23440} &
  14070 &
  \multicolumn{1}{l|}{10480} &
  \multicolumn{1}{l|}{23440} &
  14070 &
  \multicolumn{1}{l|}{0} &
  \multicolumn{1}{l|}{0} &
  0 \\ \cline{3-12} 
\multicolumn{1}{c|}{} &
  \multicolumn{1}{c|}{} &
  Zheng et al &
  \multicolumn{1}{l|}{4712} &
  \multicolumn{1}{l|}{8431} &
  3925 &
  \multicolumn{1}{l|}{3998} &
  \multicolumn{1}{l|}{6242} &
  2689 &
  \multicolumn{1}{l|}{0} &
  \multicolumn{1}{l|}{0} &
  0 \\ \cline{3-12} 
\multicolumn{1}{c|}{} &
  \multicolumn{1}{c|}{} &
  MO-OBAM &
  \multicolumn{1}{l|}{2040} &
  \multicolumn{1}{l|}{3428} &
  1471 &
  \multicolumn{1}{l|}{714} &
  \multicolumn{1}{l|}{894} &
  352 &
  \multicolumn{1}{l|}{0} &
  \multicolumn{1}{l|}{0} &
  0 \\ \cline{2-12} 
\multicolumn{1}{c|}{} &
  \multicolumn{1}{c|}{\multirow{3}{*}{$k$=15}} &
  $k$-anonymity &
  \multicolumn{1}{l|}{12510} &
  \multicolumn{1}{l|}{17825} &
  16020 &
  \multicolumn{1}{l|}{0} &
  \multicolumn{1}{l|}{0} &
  0 &
  \multicolumn{1}{l|}{0} &
  \multicolumn{1}{l|}{0} &
  0 \\ \cline{3-12} 
\multicolumn{1}{c|}{} &
  \multicolumn{1}{c|}{} &
  Zheng et al &
  \multicolumn{1}{l|}{6019} &
  \multicolumn{1}{l|}{10630} &
  4958 &
  \multicolumn{1}{l|}{0} &
  \multicolumn{1}{l|}{0} &
  0 &
  \multicolumn{1}{l|}{0} &
  \multicolumn{1}{l|}{0} &
  0 \\ \cline{3-12} 
\multicolumn{1}{c|}{} &
  \multicolumn{1}{c|}{} &
  MO-OBAM &
  \multicolumn{1}{l|}{470} &
  \multicolumn{1}{l|}{1133} &
  417 &
  \multicolumn{1}{l|}{0} &
  \multicolumn{1}{l|}{0} &
  0 &
  \multicolumn{1}{l|}{0} &
  \multicolumn{1}{l|}{0} &
  0 \\ \cline{2-12} 
\multicolumn{1}{c|}{} &
  \multicolumn{1}{c|}{\multirow{3}{*}{$k$=20}} &
  $k$-anonymity &
  \multicolumn{1}{l|}{0} &
  \multicolumn{1}{l|}{0} &
  0 &
  \multicolumn{1}{l|}{0} &
  \multicolumn{1}{l|}{0} &
  0 &
  \multicolumn{1}{l|}{0} &
  \multicolumn{1}{l|}{0} &
  0 \\ \cline{3-12} 
\multicolumn{1}{c|}{} &
  \multicolumn{1}{c|}{} &
  Zheng et al &
  \multicolumn{1}{l|}{0} &
  \multicolumn{1}{l|}{0} &
  0 &
  \multicolumn{1}{l|}{0} &
  \multicolumn{1}{l|}{0} &
  0 &
  \multicolumn{1}{l|}{0} &
  \multicolumn{1}{l|}{0} &
  0 \\ \cline{3-12} 
\multicolumn{1}{c|}{} &
  \multicolumn{1}{c|}{} &
  MO-OBAM &
  \multicolumn{1}{l|}{0} &
  \multicolumn{1}{l|}{0} &
  0 &
  \multicolumn{1}{l|}{0} &
  \multicolumn{1}{l|}{0} &
  0 &
  \multicolumn{1}{l|}{0} &
  \multicolumn{1}{l|}{0} &
  0 \\ \hline
\end{tabular}
} 
\label{tab: attack eval-LA}
\end{table}
\begin{table}[h!]
(b) Length of Stay \\
\resizebox{\columnwidth}{!}{
\begin{tabular}{ccl|lll|lll|lll}
\hline
\multicolumn{3}{l|}{Attack   Level} &
  \multicolumn{3}{c|}{$\tau=0.05$} &
  \multicolumn{3}{c|}{$\tau=0.075$} &
  \multicolumn{3}{c}{$\tau=0.1$} \\ \hline
\multicolumn{3}{l|}{Group} &
  \multicolumn{1}{l|}{{[}1,2{]}} &
  \multicolumn{1}{l|}{(2,4{]}} &
  (4,287{]} &
  \multicolumn{1}{l|}{{[}1,2{]}} &
  \multicolumn{1}{l|}{(2,4{]}} &
  (4,287{]} &
  \multicolumn{1}{l|}{{[}1,2{]}} &
  \multicolumn{1}{l|}{(2,4{]}} &
  (4,287{]} \\ \hline
\multicolumn{3}{l|}{Entire Population} &
  \multicolumn{1}{l|}{54242} &
  \multicolumn{1}{l|}{38747} &
  26882 &
  \multicolumn{1}{l|}{54242} &
  \multicolumn{1}{l|}{38747} &
  26882 &
  \multicolumn{1}{l|}{54242} &
  \multicolumn{1}{l|}{38747} &
  26882 \\ \hline
\multicolumn{3}{l|}{Vulnerable   Population in Entire Population} &
  \multicolumn{1}{l|}{17729} &
  \multicolumn{1}{l|}{14169} &
  24215 &
  \multicolumn{1}{l|}{14966} &
  \multicolumn{1}{l|}{11322} &
  22108 &
  \multicolumn{1}{l|}{12540} &
  \multicolumn{1}{l|}{9065} &
  19840 \\ \hline
\multicolumn{1}{p{2.5cm}|}{\multirow{12}{*}{\parbox{2.5cm}{Vulnerable Population after applying anonymization}}} &
  \multicolumn{1}{c|}{\multirow{3}{*}{$k$=5}} &
  $k$-anonymity &
  \multicolumn{1}{l|}{16525} &
  \multicolumn{1}{l|}{13115} &
  23585 &
  \multicolumn{1}{l|}{14215} &
  \multicolumn{1}{l|}{10640} &
  21395 &
  \multicolumn{1}{l|}{10415} &
  \multicolumn{1}{l|}{7360} &
  17705 \\ \cline{3-12} 
\multicolumn{1}{c|}{} &
  \multicolumn{1}{c|}{} &
  Zheng et al &
  \multicolumn{1}{l|}{4669} &
  \multicolumn{1}{l|}{4036} &
  9469 &
  \multicolumn{1}{l|}{2725} &
  \multicolumn{1}{l|}{2279} &
  6652 &
  \multicolumn{1}{l|}{1618} &
  \multicolumn{1}{l|}{1316} &
  4552 \\ \cline{3-12} 
\multicolumn{1}{c|}{} &
  \multicolumn{1}{c|}{} &
  MO-OBAM &
  \multicolumn{1}{l|}{4682} &
  \multicolumn{1}{l|}{3472} &
  7526 &
  \multicolumn{1}{l|}{2206} &
  \multicolumn{1}{l|}{1428} &
  3741 &
  \multicolumn{1}{l|}{996} &
  \multicolumn{1}{l|}{580} &
  1378 \\ \cline{2-12} 
\multicolumn{1}{c|}{} &
  \multicolumn{1}{c|}{\multirow{3}{*}{$k$=10}} &
  $k$-anonymity &
  \multicolumn{1}{l|}{14940} &
  \multicolumn{1}{l|}{11270} &
  21780 &
  \multicolumn{1}{l|}{14940} &
  \multicolumn{1}{l|}{11270} &
  21780 &
  \multicolumn{1}{l|}{0} &
  \multicolumn{1}{l|}{0} &
  0 \\ \cline{3-12} 
\multicolumn{1}{c|}{} &
  \multicolumn{1}{c|}{} &
  Zheng et al &
  \multicolumn{1}{l|}{4016} &
  \multicolumn{1}{l|}{3684} &
  9368 &
  \multicolumn{1}{l|}{3023} &
  \multicolumn{1}{l|}{2824} &
  7082 &
  \multicolumn{1}{l|}{0} &
  \multicolumn{1}{l|}{0} &
  0 \\ \cline{3-12} 
\multicolumn{1}{c|}{} &
  \multicolumn{1}{c|}{} &
  MO-OBAM &
  \multicolumn{1}{l|}{2089} &
  \multicolumn{1}{l|}{1682} &
  3168 &
  \multicolumn{1}{l|}{606} &
  \multicolumn{1}{l|}{428} &
  926 &
  \multicolumn{1}{l|}{0} &
  \multicolumn{1}{l|}{0} &
  0 \\ \cline{2-12} 
\multicolumn{1}{c|}{} &
  \multicolumn{1}{c|}{\multirow{3}{*}{$k$=15}} &
  $k$-anonymity &
  \multicolumn{1}{l|}{13090} &
  \multicolumn{1}{l|}{9190} &
  24075 &
  \multicolumn{1}{l|}{0} &
  \multicolumn{1}{l|}{0} &
  0 &
  \multicolumn{1}{l|}{0} &
  \multicolumn{1}{l|}{0} &
  0 \\ \cline{3-12} 
\multicolumn{1}{c|}{} &
  \multicolumn{1}{c|}{} &
  Zheng et al &
  \multicolumn{1}{l|}{5488} &
  \multicolumn{1}{l|}{5347} &
  10772 &
  \multicolumn{1}{l|}{0} &
  \multicolumn{1}{l|}{0} &
  0 &
  \multicolumn{1}{l|}{0} &
  \multicolumn{1}{l|}{0} &
  0 \\ \cline{3-12} 
\multicolumn{1}{c|}{} &
  \multicolumn{1}{c|}{} &
  MO-OBAM &
  \multicolumn{1}{l|}{611} &
  \multicolumn{1}{l|}{435} &
  974 &
  \multicolumn{1}{l|}{0} &
  \multicolumn{1}{l|}{0} &
  0 &
  \multicolumn{1}{l|}{0} &
  \multicolumn{1}{l|}{0} &
  0 \\ \cline{2-12} 
\multicolumn{1}{c|}{} &
  \multicolumn{1}{c|}{\multirow{3}{*}{$k$=20}} &
  $k$-anonymity &
  \multicolumn{1}{l|}{0} &
  \multicolumn{1}{l|}{0} &
  0 &
  \multicolumn{1}{l|}{0} &
  \multicolumn{1}{l|}{0} &
  0 &
  \multicolumn{1}{l|}{0} &
  \multicolumn{1}{l|}{0} &
  0 \\ \cline{3-12} 
\multicolumn{1}{c|}{} &
  \multicolumn{1}{c|}{} &
  Zheng et al &
  \multicolumn{1}{l|}{0} &
  \multicolumn{1}{l|}{0} &
  0 &
  \multicolumn{1}{l|}{0} &
  \multicolumn{1}{l|}{0} &
  0 &
  \multicolumn{1}{l|}{0} &
  \multicolumn{1}{l|}{0} &
  0 \\ \cline{3-12} 
\multicolumn{1}{c|}{} &
  \multicolumn{1}{c|}{} &
  MO-OBAM &
  \multicolumn{1}{l|}{0} &
  \multicolumn{1}{l|}{0} &
  0 &
  \multicolumn{1}{l|}{0} &
  \multicolumn{1}{l|}{0} &
  0 &
  \multicolumn{1}{l|}{0} &
  \multicolumn{1}{l|}{0} &
  0 \\ \hline
\end{tabular}
} 
\end{table}
\begin{table}[h!]
(c) \# of Visits \\
\resizebox{\columnwidth}{!}{
\begin{tabular}{ccl|lll|lll|lll}
\hline
\multicolumn{3}{l|}{Attack   Level} &
  \multicolumn{3}{c|}{$\tau=0.05$} &
  \multicolumn{3}{c|}{$\tau=0.075$} &
  \multicolumn{3}{c}{$\tau=0.1$} \\ \hline
\multicolumn{3}{l|}{Group} &
  \multicolumn{1}{l|}{[1,2]} &
  \multicolumn{1}{l|}{(2,4]} &
  (4,107] &
  \multicolumn{1}{l|}{[1,2]} &
  \multicolumn{1}{l|}{(2,4]} &
  (4,107] &
  \multicolumn{1}{l|}{[1,2]} &
  \multicolumn{1}{l|}{(2,4]} &
  (4,107] \\ \hline
\multicolumn{3}{l|}{Entire Population} &
  \multicolumn{1}{l|}{101524} &
  \multicolumn{1}{l|}{12155} &
  6192 &
  \multicolumn{1}{l|}{101524} &
  \multicolumn{1}{l|}{12155} &
  6192 &
  \multicolumn{1}{l|}{101524} &
  \multicolumn{1}{l|}{12155} &
  6192 \\ \hline
\multicolumn{3}{l|}{Vulnerable   Population in Entire Population} &
  \multicolumn{1}{l|}{38006} &
  \multicolumn{1}{l|}{11915} &
  6192 &
  \multicolumn{1}{l|}{31094} &
  \multicolumn{1}{l|}{11110} &
  6192 &
  \multicolumn{1}{l|}{25495} &
  \multicolumn{1}{l|}{9758} &
  6192 \\ \hline
\multicolumn{1}{p{2.5cm}|}{\multirow{12}{*}{\parbox{2.5cm}{Vulnerable Population after applying anonymization}}} &
  \multicolumn{1}{c|}{\multirow{3}{*}{$k$=5}} &
  $k$-anonymity &
  \multicolumn{1}{l|}{35465} &
  \multicolumn{1}{l|}{11580} &
  6180 &
  \multicolumn{1}{l|}{29330} &
  \multicolumn{1}{l|}{10740} &
  6180 &
  \multicolumn{1}{l|}{20940} &
  \multicolumn{1}{l|}{8430} &
  6110 \\ \cline{3-12} 
\multicolumn{1}{c|}{} &
  \multicolumn{1}{c|}{} &
  Zheng et al &
  \multicolumn{1}{l|}{14098} &
  \multicolumn{1}{l|}{3511} &
  565 &
  \multicolumn{1}{l|}{8977} &
  \multicolumn{1}{l|}{2356} &
  323 &
  \multicolumn{1}{l|}{5749} &
  \multicolumn{1}{l|}{1591} &
  146 \\ \cline{3-12} 
\multicolumn{1}{c|}{} &
  \multicolumn{1}{c|}{} &
  MO-OBAM &
  \multicolumn{1}{l|}{9064} &
  \multicolumn{1}{l|}{3922} &
  2694 &
  \multicolumn{1}{l|}{4442} &
  \multicolumn{1}{l|}{1718} &
  1215 &
  \multicolumn{1}{l|}{1891} &
  \multicolumn{1}{l|}{598} &
  465 \\ \cline{2-12} 
\multicolumn{1}{c|}{} &
  \multicolumn{1}{c|}{\multirow{3}{*}{$k$=10}} &
  $k$-anonymity &
  \multicolumn{1}{l|}{31000} &
  \multicolumn{1}{l|}{10900} &
  6090 &
  \multicolumn{1}{l|}{31000} &
  \multicolumn{1}{l|}{10900} &
  6090 &
  \multicolumn{1}{l|}{0} &
  \multicolumn{1}{l|}{0} &
  0 \\ \cline{3-12} 
\multicolumn{1}{c|}{} &
  \multicolumn{1}{c|}{} &
  Zheng et al &
  \multicolumn{1}{l|}{13291} &
  \multicolumn{1}{l|}{3114} &
  663 &
  \multicolumn{1}{l|}{10019} &
  \multicolumn{1}{l|}{2477} &
  433 &
  \multicolumn{1}{l|}{0} &
  \multicolumn{1}{l|}{0} &
  0 \\ \cline{3-12} 
\multicolumn{1}{c|}{} &
  \multicolumn{1}{c|}{} &
  MO-OBAM &
  \multicolumn{1}{l|}{3826} &
  \multicolumn{1}{l|}{1511} &
  1602 &
  \multicolumn{1}{l|}{1212} &
  \multicolumn{1}{l|}{358} &
  390 &
  \multicolumn{1}{l|}{0} &
  \multicolumn{1}{l|}{0} &
  0 \\ \cline{2-12} 
\multicolumn{1}{c|}{} &
  \multicolumn{1}{c|}{\multirow{3}{*}{$k$=15}} &
  $k$-anonymity &
  \multicolumn{1}{l|}{28370} &
  \multicolumn{1}{l|}{11775} &
  6210 &
  \multicolumn{1}{l|}{0} &
  \multicolumn{1}{l|}{0} &
  0 &
  \multicolumn{1}{l|}{0} &
  \multicolumn{1}{l|}{0} &
  0 \\ \cline{3-12} 
\multicolumn{1}{c|}{} &
  \multicolumn{1}{c|}{} &
  Zheng et al &
  \multicolumn{1}{l|}{17045} &
  \multicolumn{1}{l|}{3913} &
  649 &
  \multicolumn{1}{l|}{0} &
  \multicolumn{1}{l|}{0} &
  0 &
  \multicolumn{1}{l|}{0} &
  \multicolumn{1}{l|}{0} &
  0 \\ \cline{3-12} 
\multicolumn{1}{c|}{} &
  \multicolumn{1}{c|}{} &
  MO-OBAM &
  \multicolumn{1}{l|}{1163} &
  \multicolumn{1}{l|}{422} &
  435 &
  \multicolumn{1}{l|}{0} &
  \multicolumn{1}{l|}{0} &
  0 &
  \multicolumn{1}{l|}{0} &
  \multicolumn{1}{l|}{0} &
  0 \\ \cline{2-12} 
\multicolumn{1}{c|}{} &
  \multicolumn{1}{c|}{\multirow{3}{*}{$k$=20}} &
  $k$-anonymity &
  \multicolumn{1}{l|}{0} &
  \multicolumn{1}{l|}{0} &
  0 &
  \multicolumn{1}{l|}{0} &
  \multicolumn{1}{l|}{0} &
  0 &
  \multicolumn{1}{l|}{0} &
  \multicolumn{1}{l|}{0} &
  0 \\ \cline{3-12} 
\multicolumn{1}{c|}{} &
  \multicolumn{1}{c|}{} &
  Zheng et al &
  \multicolumn{1}{l|}{0} &
  \multicolumn{1}{l|}{0} &
  0 &
  \multicolumn{1}{l|}{0} &
  \multicolumn{1}{l|}{0} &
  0 &
  \multicolumn{1}{l|}{0} &
  \multicolumn{1}{l|}{0} &
  0 \\ \cline{3-12} 
\multicolumn{1}{c|}{} &
  \multicolumn{1}{c|}{} &
  MO-OBAM &
  \multicolumn{1}{l|}{0} &
  \multicolumn{1}{l|}{0} &
  0 &
  \multicolumn{1}{l|}{0} &
  \multicolumn{1}{l|}{0} &
  0 &
  \multicolumn{1}{l|}{0} &
  \multicolumn{1}{l|}{0} &
  0 \\ \hline
\end{tabular}
} 
\end{table}
\begin{table}[h!]
(d) GenderDescription \\
\resizebox{\columnwidth}{!}{
\begin{tabular}{ccl|ll|ll|ll}
\hline
\multicolumn{3}{l|}{Attack   Level} &
  \multicolumn{2}{c|}{$\tau=0.05$} &
  \multicolumn{2}{c|}{$\tau=0.075$} &
  \multicolumn{2}{c}{$\tau=0.1$} \\ \hline
\multicolumn{3}{l|}{Group} &
  \multicolumn{1}{l|}{Female} &
  Male &
  \multicolumn{1}{l|}{Female} &
  Male &
  \multicolumn{1}{l|}{Female} &
  Male \\ \hline
\multicolumn{3}{l|}{Entire Population} &
  \multicolumn{1}{l|}{72095} &
  47776 &
  \multicolumn{1}{l|}{72095} &
  47776 &
  \multicolumn{1}{l|}{72095} &
  47776 \\ \hline
\multicolumn{3}{l|}{Vulnerable   Population in Entire Population} &
  \multicolumn{1}{l|}{30194} &
  25919 &
  \multicolumn{1}{l|}{25671} &
  22725 &
  \multicolumn{1}{l|}{21813} &
  19632 \\ \hline
\multicolumn{1}{p{2.5cm}|}{\multirow{12}{*}{\parbox{2.5cm}{Vulnerable Population after applying anonymization}}} &
  \multicolumn{1}{c|}{\multirow{3}{*}{$k$=5}} &
  $k$-anonymity &
  \multicolumn{1}{l|}{28575} &
  24650 &
  \multicolumn{1}{l|}{24420} &
  21830 &
  \multicolumn{1}{l|}{18650} &
  16830 \\ \cline{3-9} 
\multicolumn{1}{c|}{} &
  \multicolumn{1}{c|}{} &
  Zheng et al &
  \multicolumn{1}{l|}{10081} &
  8093 &
  \multicolumn{1}{l|}{6543} &
  5113 &
  \multicolumn{1}{l|}{4232} &
  3254 \\ \cline{3-9} 
\multicolumn{1}{c|}{} &
  \multicolumn{1}{c|}{} &
  MO-OBAM &
  \multicolumn{1}{l|}{8686} &
  6994 &
  \multicolumn{1}{l|}{3812} &
  3563 &
  \multicolumn{1}{l|}{1532} &
  1422 \\ \cline{2-9} 
\multicolumn{1}{c|}{} &
  \multicolumn{1}{c|}{\multirow{3}{*}{$k$=10}} &
  $k$-anonymity &
  \multicolumn{1}{l|}{25230} &
  22760 &
  \multicolumn{1}{l|}{25230} &
  22760 &
  \multicolumn{1}{l|}{0} &
  0 \\ \cline{3-9} 
\multicolumn{1}{c|}{} &
  \multicolumn{1}{c|}{} &
  Zheng et al &
  \multicolumn{1}{l|}{9080} &
  7988 &
  \multicolumn{1}{l|}{7014} &
  5915 &
  \multicolumn{1}{l|}{0} &
  0 \\ \cline{3-9} 
\multicolumn{1}{c|}{} &
  \multicolumn{1}{c|}{} &
  MO-OBAM &
  \multicolumn{1}{l|}{3698} &
  3241 &
  \multicolumn{1}{l|}{998} &
  962 &
  \multicolumn{1}{l|}{0} &
  0 \\ \cline{2-9} 
\multicolumn{1}{c|}{} &
  \multicolumn{1}{c|}{\multirow{3}{*}{$k$=15}} &
  $k$-anonymity &
  \multicolumn{1}{l|}{25330} &
  21025 &
  \multicolumn{1}{l|}{0} &
  0 &
  \multicolumn{1}{l|}{0} &
  0 \\ \cline{3-9} 
\multicolumn{1}{c|}{} &
  \multicolumn{1}{c|}{} &
  Zheng et al &
  \multicolumn{1}{l|}{12126} &
  9481 &
  \multicolumn{1}{l|}{0} &
  0 &
  \multicolumn{1}{l|}{0} &
  0 \\ \cline{3-9} 
\multicolumn{1}{c|}{} &
  \multicolumn{1}{c|}{} &
  MO-OBAM &
  \multicolumn{1}{l|}{907} &
  1113 &
  \multicolumn{1}{l|}{0} &
  0 &
  \multicolumn{1}{l|}{0} &
  0 \\ \cline{2-9} 
\multicolumn{1}{c|}{} &
  \multicolumn{1}{c|}{\multirow{3}{*}{$k$=20}} &
  $k$-anonymity &
  \multicolumn{1}{l|}{0} &
  0 &
  \multicolumn{1}{l|}{0} &
  0 &
  \multicolumn{1}{l|}{0} &
  0 \\ \cline{3-9} 
\multicolumn{1}{c|}{} &
  \multicolumn{1}{c|}{} &
  Zheng et al &
  \multicolumn{1}{l|}{0} &
  0 &
  \multicolumn{1}{l|}{0} &
  0 &
  \multicolumn{1}{l|}{0} &
  0 \\ \cline{3-9} 
\multicolumn{1}{c|}{} &
  \multicolumn{1}{c|}{} &
  MO-OBAM &
  \multicolumn{1}{l|}{0} &
  0 &
  \multicolumn{1}{l|}{0} &
  0 &
  \multicolumn{1}{l|}{0} &
  0 \\ \hline
\end{tabular}
} 
\end{table}
\begin{table}[h!]
(e) RaceDescription \\
\resizebox{\columnwidth}{!}{
\begin{tabular}{ccl|lllllll|lllllll|lllllll}
\hline
\multicolumn{3}{l|}{Attack   Level} &
  \multicolumn{7}{c|}{$\tau=0.05$} &
  \multicolumn{7}{c|}{$\tau=0.075$} &
  \multicolumn{7}{c}{$\tau=0.1$} \\ \hline
\multicolumn{3}{l|}{Group} &
  \multicolumn{1}{p{2.5cm}|}{American Indian or Alaska Native} &
  \multicolumn{1}{l|}{Asian} &
  \multicolumn{1}{p{2.5cm}|}{Black or African American} &
  \multicolumn{1}{l|}{Multi Racial} &
  \multicolumn{1}{l|}{Other Race} &
  \multicolumn{1}{l|}{Unavailable} &
  White &
  \multicolumn{1}{p{2.5cm}|}{American Indian or Alaska Native} &
  \multicolumn{1}{l|}{Asian} &
  \multicolumn{1}{p{2.5cm}|}{Black or African American} &
  \multicolumn{1}{l|}{Multi Racial} &
  \multicolumn{1}{l|}{Other Race} &
  \multicolumn{1}{l|}{Unavailable} &
  White &
  \multicolumn{1}{p{2.5cm}|}{American Indian or Alaska Native} &
  \multicolumn{1}{l|}{Asian} &
  \multicolumn{1}{p{2.5cm}|}{Black or African American} &
  \multicolumn{1}{l|}{Multi Racial} &
  \multicolumn{1}{l|}{Other Race} &
  \multicolumn{1}{l|}{Unavailable} &
  White \\ \hline
\multicolumn{3}{l|}{Entire Population} &
  \multicolumn{1}{l|}{230} &
  \multicolumn{1}{l|}{2786} &
  \multicolumn{1}{l|}{26746} &
  \multicolumn{1}{l|}{64} &
  \multicolumn{1}{l|}{4484} &
  \multicolumn{1}{l|}{971} &
  84590 &
  \multicolumn{1}{l|}{230} &
  \multicolumn{1}{l|}{2786} &
  \multicolumn{1}{l|}{26746} &
  \multicolumn{1}{l|}{64} &
  \multicolumn{1}{l|}{4484} &
  \multicolumn{1}{l|}{971} &
  84590 &
  \multicolumn{1}{l|}{230} &
  \multicolumn{1}{l|}{2786} &
  \multicolumn{1}{l|}{26746} &
  \multicolumn{1}{l|}{64} &
  \multicolumn{1}{l|}{4484} &
  \multicolumn{1}{l|}{971} &
  84590 \\ \hline
\multicolumn{3}{l|}{Vulnerable Population in Entire Population} &
  \multicolumn{1}{l|}{230} &
  \multicolumn{1}{l|}{2177} &
  \multicolumn{1}{l|}{16335} &
  \multicolumn{1}{l|}{64} &
  \multicolumn{1}{l|}{4034} &
  \multicolumn{1}{l|}{971} &
  32302 &
  \multicolumn{1}{l|}{230} &
  \multicolumn{1}{l|}{2063} &
  \multicolumn{1}{l|}{13911} &
  \multicolumn{1}{l|}{64} &
  \multicolumn{1}{l|}{3891} &
  \multicolumn{1}{l|}{971} &
  27266 &
  \multicolumn{1}{l|}{230} &
  \multicolumn{1}{l|}{1965} &
  \multicolumn{1}{l|}{12055} &
  \multicolumn{1}{l|}{64} &
  \multicolumn{1}{l|}{3551} &
  \multicolumn{1}{l|}{971} &
  22609 \\ \hline
\multicolumn{1}{p{2.5cm}|}{\multirow{12}{*}{\parbox{2.5cm}{Vulnerable Population after applying anonymization}}} &
  \multicolumn{1}{c|}{\multirow{3}{*}{$k$=5}} &
  $k$-anonymity &
  \multicolumn{1}{l|}{226} &
  \multicolumn{1}{l|}{2085} &
  \multicolumn{1}{l|}{15530} &
  \multicolumn{1}{l|}{64} &
  \multicolumn{1}{l|}{4030} &
  \multicolumn{1}{l|}{971} &
  30319 &
  \multicolumn{1}{l|}{145} &
  \multicolumn{1}{l|}{1980} &
  \multicolumn{1}{l|}{13410} &
  \multicolumn{1}{l|}{64} &
  \multicolumn{1}{l|}{3855} &
  \multicolumn{1}{l|}{945} &
  25851 &
  \multicolumn{1}{l|}{226} &
  \multicolumn{1}{l|}{1640} &
  \multicolumn{1}{l|}{10290} &
  \multicolumn{1}{l|}{64} &
  \multicolumn{1}{l|}{3195} &
  \multicolumn{1}{l|}{925} &
  19140 \\ \cline{3-24} 
\multicolumn{1}{c|}{} &
  \multicolumn{1}{c|}{} &
  Zheng et al &
  \multicolumn{1}{l|}{82} &
  \multicolumn{1}{l|}{563} &
  \multicolumn{1}{l|}{6030} &
  \multicolumn{1}{l|}{0} &
  \multicolumn{1}{l|}{1071} &
  \multicolumn{1}{l|}{77} &
  10351 &
  \multicolumn{1}{l|}{66} &
  \multicolumn{1}{l|}{430} &
  \multicolumn{1}{l|}{3835} &
  \multicolumn{1}{l|}{0} &
  \multicolumn{1}{l|}{909} &
  \multicolumn{1}{l|}{59} &
  6357 &
  \multicolumn{1}{l|}{24} &
  \multicolumn{1}{l|}{305} &
  \multicolumn{1}{l|}{2414} &
  \multicolumn{1}{l|}{0} &
  \multicolumn{1}{l|}{693} &
  \multicolumn{1}{l|}{49} &
  4001 \\ \cline{3-24} 
\multicolumn{1}{c|}{} &
  \multicolumn{1}{c|}{} &
  MO-OBAM &
  \multicolumn{1}{l|}{66} &
  \multicolumn{1}{l|}{436} &
  \multicolumn{1}{l|}{4894} &
  \multicolumn{1}{l|}{0} &
  \multicolumn{1}{l|}{1085} &
  \multicolumn{1}{l|}{225} &
  8974 &
  \multicolumn{1}{l|}{5} &
  \multicolumn{1}{l|}{354} &
  \multicolumn{1}{l|}{2522} &
  \multicolumn{1}{l|}{13} &
  \multicolumn{1}{l|}{546} &
  \multicolumn{1}{l|}{149} &
  3786 &
  \multicolumn{1}{l|}{5} &
  \multicolumn{1}{l|}{140} &
  \multicolumn{1}{l|}{1038} &
  \multicolumn{1}{l|}{0} &
  \multicolumn{1}{l|}{261} &
  \multicolumn{1}{l|}{67} &
  1443 \\ \cline{2-24} 
\multicolumn{1}{c|}{} &
  \multicolumn{1}{c|}{\multirow{3}{*}{$k$=10}} &
  $k$-anonymity &
  \multicolumn{1}{l|}{170} &
  \multicolumn{1}{l|}{2080} &
  \multicolumn{1}{l|}{14018} &
  \multicolumn{1}{l|}{64} &
  \multicolumn{1}{l|}{3926} &
  \multicolumn{1}{l|}{902} &
  26830 &
  \multicolumn{1}{l|}{200} &
  \multicolumn{1}{l|}{2050} &
  \multicolumn{1}{l|}{13910} &
  \multicolumn{1}{l|}{64} &
  \multicolumn{1}{l|}{3890} &
  \multicolumn{1}{l|}{902} &
  26974 &
  \multicolumn{1}{l|}{0} &
  \multicolumn{1}{l|}{0} &
  \multicolumn{1}{l|}{0} &
  \multicolumn{1}{l|}{0} &
  \multicolumn{1}{l|}{0} &
  \multicolumn{1}{l|}{0} &
  0 \\ \cline{3-24} 
\multicolumn{1}{c|}{} &
  \multicolumn{1}{c|}{} &
  Zheng et al &
  \multicolumn{1}{l|}{71} &
  \multicolumn{1}{l|}{584} &
  \multicolumn{1}{l|}{5455} &
  \multicolumn{1}{l|}{10} &
  \multicolumn{1}{l|}{1257} &
  \multicolumn{1}{l|}{174} &
  9517 &
  \multicolumn{1}{l|}{71} &
  \multicolumn{1}{l|}{460} &
  \multicolumn{1}{l|}{4100} &
  \multicolumn{1}{l|}{10} &
  \multicolumn{1}{l|}{1074} &
  \multicolumn{1}{l|}{146} &
  7068 &
  \multicolumn{1}{l|}{0} &
  \multicolumn{1}{l|}{0} &
  \multicolumn{1}{l|}{0} &
  \multicolumn{1}{l|}{0} &
  \multicolumn{1}{l|}{0} &
  \multicolumn{1}{l|}{0} &
  0 \\ \cline{3-24} 
\multicolumn{1}{c|}{} &
  \multicolumn{1}{c|}{} &
  MO-OBAM &
  \multicolumn{1}{l|}{30} &
  \multicolumn{1}{l|}{242} &
  \multicolumn{1}{l|}{2701} &
  \multicolumn{1}{l|}{26} &
  \multicolumn{1}{l|}{511} &
  \multicolumn{1}{l|}{153} &
  3276 &
  \multicolumn{1}{l|}{12} &
  \multicolumn{1}{l|}{144} &
  \multicolumn{1}{l|}{750} &
  \multicolumn{1}{l|}{12} &
  \multicolumn{1}{l|}{210} &
  \multicolumn{1}{l|}{55} &
  777 &
  \multicolumn{1}{l|}{0} &
  \multicolumn{1}{l|}{0} &
  \multicolumn{1}{l|}{0} &
  \multicolumn{1}{l|}{0} &
  \multicolumn{1}{l|}{0} &
  \multicolumn{1}{l|}{0} &
  0 \\ \cline{2-24} 
\multicolumn{1}{c|}{} &
  \multicolumn{1}{c|}{\multirow{3}{*}{$k$=15}} &
  $k$-anonymity &
  \multicolumn{1}{l|}{180} &
  \multicolumn{1}{l|}{2080} &
  \multicolumn{1}{l|}{15180} &
  \multicolumn{1}{l|}{0} &
  \multicolumn{1}{l|}{3615} &
  \multicolumn{1}{l|}{915} &
  24385 &
  \multicolumn{1}{l|}{0} &
  \multicolumn{1}{l|}{0} &
  \multicolumn{1}{l|}{0} &
  \multicolumn{1}{l|}{0} &
  \multicolumn{1}{l|}{0} &
  \multicolumn{1}{l|}{0} &
  0 &
  \multicolumn{1}{l|}{0} &
  \multicolumn{1}{l|}{0} &
  \multicolumn{1}{l|}{0} &
  \multicolumn{1}{l|}{0} &
  \multicolumn{1}{l|}{0} &
  \multicolumn{1}{l|}{0} &
  0 \\ \cline{3-24} 
\multicolumn{1}{c|}{} &
  \multicolumn{1}{c|}{} &
  Zheng et al &
  \multicolumn{1}{l|}{15} &
  \multicolumn{1}{l|}{686} &
  \multicolumn{1}{l|}{6711} &
  \multicolumn{1}{l|}{0} &
  \multicolumn{1}{l|}{1572} &
  \multicolumn{1}{l|}{226} &
  12397 &
  \multicolumn{1}{l|}{0} &
  \multicolumn{1}{l|}{0} &
  \multicolumn{1}{l|}{0} &
  \multicolumn{1}{l|}{0} &
  \multicolumn{1}{l|}{0} &
  \multicolumn{1}{l|}{0} &
  0 &
  \multicolumn{1}{l|}{0} &
  \multicolumn{1}{l|}{0} &
  \multicolumn{1}{l|}{0} &
  \multicolumn{1}{l|}{0} &
  \multicolumn{1}{l|}{0} &
  \multicolumn{1}{l|}{0} &
  0 \\ \cline{3-24} 
\multicolumn{1}{c|}{} &
  \multicolumn{1}{c|}{} &
  MO-OBAM &
  \multicolumn{1}{l|}{0} &
  \multicolumn{1}{l|}{53} &
  \multicolumn{1}{l|}{676} &
  \multicolumn{1}{l|}{0} &
  \multicolumn{1}{l|}{207} &
  \multicolumn{1}{l|}{} &
  1084 &
  \multicolumn{1}{l|}{0} &
  \multicolumn{1}{l|}{0} &
  \multicolumn{1}{l|}{0} &
  \multicolumn{1}{l|}{0} &
  \multicolumn{1}{l|}{0} &
  \multicolumn{1}{l|}{0} &
  0 &
  \multicolumn{1}{l|}{0} &
  \multicolumn{1}{l|}{0} &
  \multicolumn{1}{l|}{0} &
  \multicolumn{1}{l|}{0} &
  \multicolumn{1}{l|}{0} &
  \multicolumn{1}{l|}{0} &
  0 \\ \cline{2-24} 
\multicolumn{1}{c|}{} &
  \multicolumn{1}{c|}{\multirow{3}{*}{$k$=20}} &
  $k$-anonymity &
  \multicolumn{1}{l|}{0} &
  \multicolumn{1}{l|}{0} &
  \multicolumn{1}{l|}{0} &
  \multicolumn{1}{l|}{0} &
  \multicolumn{1}{l|}{0} &
  \multicolumn{1}{l|}{0} &
  0 &
  \multicolumn{1}{l|}{0} &
  \multicolumn{1}{l|}{0} &
  \multicolumn{1}{l|}{0} &
  \multicolumn{1}{l|}{0} &
  \multicolumn{1}{l|}{0} &
  \multicolumn{1}{l|}{0} &
  0 &
  \multicolumn{1}{l|}{0} &
  \multicolumn{1}{l|}{0} &
  \multicolumn{1}{l|}{0} &
  \multicolumn{1}{l|}{0} &
  \multicolumn{1}{l|}{0} &
  \multicolumn{1}{l|}{0} &
  0 \\ \cline{3-24} 
\multicolumn{1}{c|}{} &
  \multicolumn{1}{c|}{} &
  Zheng et al &
  \multicolumn{1}{l|}{0} &
  \multicolumn{1}{l|}{0} &
  \multicolumn{1}{l|}{0} &
  \multicolumn{1}{l|}{0} &
  \multicolumn{1}{l|}{0} &
  \multicolumn{1}{l|}{0} &
  0 &
  \multicolumn{1}{l|}{0} &
  \multicolumn{1}{l|}{0} &
  \multicolumn{1}{l|}{0} &
  \multicolumn{1}{l|}{0} &
  \multicolumn{1}{l|}{0} &
  \multicolumn{1}{l|}{0} &
  0 &
  \multicolumn{1}{l|}{0} &
  \multicolumn{1}{l|}{0} &
  \multicolumn{1}{l|}{0} &
  \multicolumn{1}{l|}{0} &
  \multicolumn{1}{l|}{0} &
  \multicolumn{1}{l|}{0} &
  0 \\ \cline{3-24} 
\multicolumn{1}{c|}{} &
  \multicolumn{1}{c|}{} &
  MO-OBAM &
  \multicolumn{1}{l|}{0} &
  \multicolumn{1}{l|}{0} &
  \multicolumn{1}{l|}{0} &
  \multicolumn{1}{l|}{0} &
  \multicolumn{1}{l|}{0} &
  \multicolumn{1}{l|}{0} &
  0 &
  \multicolumn{1}{l|}{0} &
  \multicolumn{1}{l|}{0} &
  \multicolumn{1}{l|}{0} &
  \multicolumn{1}{l|}{0} &
  \multicolumn{1}{l|}{0} &
  \multicolumn{1}{l|}{0} &
  0 &
  \multicolumn{1}{l|}{0} &
  \multicolumn{1}{l|}{0} &
  \multicolumn{1}{l|}{0} &
  \multicolumn{1}{l|}{0} &
  \multicolumn{1}{l|}{0} &
  \multicolumn{1}{l|}{0} &
  0 \\ \hline
\end{tabular}
}
\end{table}
\begin{table}[h!]
(f) EthnicGroupDescription \\
\resizebox{\columnwidth}{!}{%
\begin{tabular}{ccl|llll|llll|llll}
\hline
\multicolumn{3}{l|}{Attack   Level} &
  \multicolumn{4}{c|}{$\tau=0.05$} &
  \multicolumn{4}{c|}{$\tau=0.075$} &
  \multicolumn{4}{c}{$\tau=0.1$} \\ \hline
\multicolumn{3}{l|}{Group} &
  \multicolumn{1}{l|}{Declined} &
  \multicolumn{1}{l|}{Hispanic or Latino} &
  \multicolumn{1}{l|}{Non-Hispanic or Latino} &
  Unknown &
  \multicolumn{1}{l|}{Declined} &
  \multicolumn{1}{l|}{Hispanic or Latino} &
  \multicolumn{1}{l|}{Non-Hispanic or Latino} &
  Unknown &
  \multicolumn{1}{l|}{Declined} &
  \multicolumn{1}{l|}{Hispanic or Latino} &
  \multicolumn{1}{l|}{Non-Hispanic or Latino} &
  Unknown \\ \hline
\multicolumn{3}{l|}{Entire Population} &
  \multicolumn{1}{l|}{245} &
  \multicolumn{1}{l|}{6855} &
  \multicolumn{1}{l|}{109739} &
  3032 &
  \multicolumn{1}{l|}{245} &
  \multicolumn{1}{l|}{6855} &
  \multicolumn{1}{l|}{109739} &
  3032 &
  \multicolumn{1}{l|}{245} &
  \multicolumn{1}{l|}{6855} &
  \multicolumn{1}{l|}{109739} &
  3032 \\ \hline
\multicolumn{3}{l|}{Vulnerable   Population in Entire Population} &
  \multicolumn{1}{l|}{245} &
  \multicolumn{1}{l|}{6136} &
  \multicolumn{1}{l|}{46700} &
  3032 &
  \multicolumn{1}{l|}{245} &
  \multicolumn{1}{l|}{5905} &
  \multicolumn{1}{l|}{39245} &
  3001 &
  \multicolumn{1}{l|}{245} &
  \multicolumn{1}{l|}{5402} &
  \multicolumn{1}{l|}{32899} &
  2899 \\ \hline
\multicolumn{1}{p{2.5cm}|}{\multirow{12}{*}{\parbox{2.5cm}{Vulnerable Population after applying anonymization}}} &
  \multicolumn{1}{c|}{\multirow{3}{*}{$k$=5}} &
  $k$-anonymity &
  \multicolumn{1}{l|}{245} &
  \multicolumn{1}{l|}{6060} &
  \multicolumn{1}{l|}{43890} &
  3030 &
  \multicolumn{1}{l|}{245} &
  \multicolumn{1}{l|}{5760} &
  \multicolumn{1}{l|}{37245} &
  3000 &
  \multicolumn{1}{l|}{245} &
  \multicolumn{1}{l|}{4860} &
  \multicolumn{1}{l|}{27725} &
  2650 \\ \cline{3-15} 
\multicolumn{1}{c|}{} &
  \multicolumn{1}{c|}{} &
  Zheng et al &
  \multicolumn{1}{l|}{56} &
  \multicolumn{1}{l|}{1775} &
  \multicolumn{1}{l|}{16023} &
  320 &
  \multicolumn{1}{l|}{38} &
  \multicolumn{1}{l|}{1350} &
  \multicolumn{1}{l|}{10055} &
  213 &
  \multicolumn{1}{l|}{28} &
  \multicolumn{1}{l|}{1044} &
  \multicolumn{1}{l|}{6282} &
  132 \\ \cline{3-15} 
\multicolumn{1}{c|}{} &
  \multicolumn{1}{c|}{} &
  MO-OBAM &
  \multicolumn{1}{l|}{22} &
  \multicolumn{1}{l|}{1775} &
  \multicolumn{1}{l|}{13072} &
  811 &
  \multicolumn{1}{l|}{22} &
  \multicolumn{1}{l|}{1040} &
  \multicolumn{1}{l|}{5704} &
  609 &
  \multicolumn{1}{l|}{0} &
  \multicolumn{1}{l|}{505} &
  \multicolumn{1}{l|}{2040} &
  409 \\ \cline{2-15} 
\multicolumn{1}{c|}{} &
  \multicolumn{1}{c|}{\multirow{3}{*}{$k$=10}} &
  $k$-anonymity &
  \multicolumn{1}{l|}{245} &
  \multicolumn{1}{l|}{5710} &
  \multicolumn{1}{l|}{39020} &
  3015 &
  \multicolumn{1}{l|}{245} &
  \multicolumn{1}{l|}{5710} &
  \multicolumn{1}{l|}{39035} &
  3000 &
  \multicolumn{1}{l|}{0} &
  \multicolumn{1}{l|}{0} &
  \multicolumn{1}{l|}{0} &
  0 \\ \cline{3-15} 
\multicolumn{1}{c|}{} &
  \multicolumn{1}{c|}{} &
  Zheng et al &
  \multicolumn{1}{l|}{77} &
  \multicolumn{1}{l|}{1923} &
  \multicolumn{1}{l|}{14645} &
  423 &
  \multicolumn{1}{l|}{46} &
  \multicolumn{1}{l|}{1743} &
  \multicolumn{1}{l|}{10862} &
  278 &
  \multicolumn{1}{l|}{0} &
  \multicolumn{1}{l|}{0} &
  \multicolumn{1}{l|}{0} &
  0 \\ \cline{3-15} 
\multicolumn{1}{c|}{} &
  \multicolumn{1}{c|}{} &
  MO-OBAM &
  \multicolumn{1}{l|}{14} &
  \multicolumn{1}{l|}{820} &
  \multicolumn{1}{l|}{5819} &
  286 &
  \multicolumn{1}{l|}{0} &
  \multicolumn{1}{l|}{288} &
  \multicolumn{1}{l|}{1518} &
  154 &
  \multicolumn{1}{l|}{0} &
  \multicolumn{1}{l|}{0} &
  \multicolumn{1}{l|}{0} &
  0 \\ \cline{2-15} 
\multicolumn{1}{c|}{} &
  \multicolumn{1}{c|}{\multirow{3}{*}{$k$=15}} &
  $k$-anonymity &
  \multicolumn{1}{l|}{240} &
  \multicolumn{1}{l|}{5195} &
  \multicolumn{1}{l|}{37890} &
  3030 &
  \multicolumn{1}{l|}{0} &
  \multicolumn{1}{l|}{0} &
  \multicolumn{1}{l|}{0} &
  0 &
  \multicolumn{1}{l|}{0} &
  \multicolumn{1}{l|}{0} &
  \multicolumn{1}{l|}{0} &
  0 \\ \cline{3-15} 
\multicolumn{1}{c|}{} &
  \multicolumn{1}{c|}{} &
  Zheng et al &
  \multicolumn{1}{l|}{75} &
  \multicolumn{1}{l|}{2315} &
  \multicolumn{1}{l|}{18579} &
  638 &
  \multicolumn{1}{l|}{0} &
  \multicolumn{1}{l|}{0} &
  \multicolumn{1}{l|}{0} &
  0 &
  \multicolumn{1}{l|}{0} &
  \multicolumn{1}{l|}{0} &
  \multicolumn{1}{l|}{0} &
  0 \\ \cline{3-15} 
\multicolumn{1}{c|}{} &
  \multicolumn{1}{c|}{} &
  MO-OBAM &
  \multicolumn{1}{l|}{0} &
  \multicolumn{1}{l|}{292} &
  \multicolumn{1}{l|}{1606} &
  122 &
  \multicolumn{1}{l|}{0} &
  \multicolumn{1}{l|}{0} &
  \multicolumn{1}{l|}{0} &
  0 &
  \multicolumn{1}{l|}{0} &
  \multicolumn{1}{l|}{0} &
  \multicolumn{1}{l|}{0} &
  0 \\ \cline{2-15} 
\multicolumn{1}{c|}{} &
  \multicolumn{1}{c|}{\multirow{3}{*}{$k$=20}} &
  $k$-anonymity &
  \multicolumn{1}{l|}{0} &
  \multicolumn{1}{l|}{0} &
  \multicolumn{1}{l|}{0} &
  0 &
  \multicolumn{1}{l|}{0} &
  \multicolumn{1}{l|}{0} &
  \multicolumn{1}{l|}{0} &
  0 &
  \multicolumn{1}{l|}{0} &
  \multicolumn{1}{l|}{0} &
  \multicolumn{1}{l|}{0} &
  0 \\ \cline{3-15} 
\multicolumn{1}{c|}{} &
  \multicolumn{1}{c|}{} &
  Zheng et al &
  \multicolumn{1}{l|}{0} &
  \multicolumn{1}{l|}{0} &
  \multicolumn{1}{l|}{0} &
  0 &
  \multicolumn{1}{l|}{0} &
  \multicolumn{1}{l|}{0} &
  \multicolumn{1}{l|}{0} &
  0 &
  \multicolumn{1}{l|}{0} &
  \multicolumn{1}{l|}{0} &
  \multicolumn{1}{l|}{0} &
  0 \\ \cline{3-15} 
\multicolumn{1}{c|}{} &
  \multicolumn{1}{c|}{} &
  MO-OBAM &
  \multicolumn{1}{l|}{0} &
  \multicolumn{1}{l|}{0} &
  \multicolumn{1}{l|}{0} &
  0 &
  \multicolumn{1}{l|}{0} &
  \multicolumn{1}{l|}{0} &
  \multicolumn{1}{l|}{0} &
  0 &
  \multicolumn{1}{l|}{0} &
  \multicolumn{1}{l|}{0} &
  \multicolumn{1}{l|}{0} &
  0 \\ \hline
\end{tabular}%
}
\end{table}

\begin{table}[h!]
\caption{Number of people at risk of homogeneity attack before and after applying anonymization, by QIs}
\label{tab: attack eval-HA}
(a) Age  \\
\begin{tabular}{cll|l|l|l}
\hline
\multicolumn{3}{l|}{Group}                                        & {[}18,38{]} & (38,70{]} & (70,90{]} \\ \hline
\multicolumn{3}{l|}{Entire Population}                            & 30815       & 60445     & 28611     \\ \hline
\multicolumn{3}{l|}{Vulnerable   Population in Entire Population} & 1740        & 7415      & 5782      \\ \hline
\multicolumn{1}{p{2.5cm}|}{\multirow{4}{*}{\parbox{2.5cm}{Vulnerable Population after applying anonymization}}} &
  \multicolumn{1}{l|}{\multirow{4}{*}{$k$-anonymity}} &
  $k$=5 &
  135 &
  3205 &
  3565 \\ \cline{3-6} 
\multicolumn{1}{c|}{}     & \multicolumn{1}{l|}{}     & $k$=10    & 10          & 930       & 1320      \\ \cline{3-6} 
\multicolumn{1}{c|}{}     & \multicolumn{1}{l|}{}     & $k$=15    & 0           & 285       & 495       \\ \cline{3-6} 
\multicolumn{1}{c|}{}     & \multicolumn{1}{l|}{}     & $k$=20    & 0           & 100       & 220       \\ \hline
\end{tabular}
\end{table}

\begin{table}[h!]
(b) Length of Stay \\
\begin{tabular}{cll|l|l|l}
\hline
\multicolumn{3}{l|}{Group}                                        & {[}1,2{]} & (2,4{]} & (4,287{]} \\ \hline
\multicolumn{3}{l|}{Entire Population}                            & 54242     & 38747   & 26882     \\ \hline
\multicolumn{3}{l|}{Vulnerable   Population in Entire Population} & 3896      & 3205    & 7836      \\ \hline
\multicolumn{1}{p{2.5cm}|}{\multirow{4}{*}{\parbox{2.5cm}{Vulnerable Population after applying anonymization}}} &
  \multicolumn{1}{l|}{\multirow{4}{*}{$k$-anonymity}} &
  $k$=5 &
  1960 &
  1655 &
  3290 \\ \cline{3-6} 
\multicolumn{1}{c|}{}     & \multicolumn{1}{l|}{}     & $k$=10    & 750       & 550     & 960       \\ \cline{3-6} 
\multicolumn{1}{c|}{}     & \multicolumn{1}{l|}{}     & $k$=15    & 255       & 105     & 420       \\ \cline{3-6} 
\multicolumn{1}{c|}{}     & \multicolumn{1}{l|}{}     & $k$=20    & 160       & 40      & 120       \\ \hline
\end{tabular}
\end{table}

\begin{table}[h!]
(c) \# of Visits \\
\begin{tabular}{lll|l|l|l}
\hline
\multicolumn{3}{l|}{Group}                                        & {[}1,2{]} & (2,4{]} & (4,107{]} \\ \hline
\multicolumn{3}{l|}{Entire Population}                            & 101524	&12155	&6192     \\ \hline
\multicolumn{3}{l|}{Vulnerable   Population in Entire Population} & 3606 &5577	&5754     \\ \hline
\multicolumn{1}{p{2.5cm}|}{\multirow{4}{*}{\parbox{2.5cm}{Vulnerable Population after applying anonymization}}} & \multicolumn{1}{l|}{\multirow{4}{*}{$k$-anonymity}} & $k$=5 & 880	&2440	&3585 \\ \cline{3-6} 
\multicolumn{1}{l|}{}     & \multicolumn{1}{l|}{}     & $k$=10    &180	&620 &1460      \\ \cline{3-6} 
\multicolumn{1}{l|}{}     & \multicolumn{1}{l|}{}     & $k$=15    & 90	&165 &525       \\ \cline{3-6} 
\multicolumn{1}{l|}{}     & \multicolumn{1}{l|}{}     & $k$=20    & 20	&60	&240      \\ \hline
\end{tabular}
\end{table}

\begin{table}[h!]
(d) GenderDescription \\
\begin{tabular}{lll|l|l}
\hline
\multicolumn{3}{l|}{Group}                                        & Female & Male  \\ \hline
\multicolumn{3}{l|}{Entire Population}                            & 72095  & 47776 \\ \hline
\multicolumn{3}{l|}{Vulnerable   Population in Entire Population} & 7984   & 6953  \\ \hline
\multicolumn{1}{p{2.5cm}|}{\multirow{4}{*}{\parbox{2.5cm}{Vulnerable Population after applying anonymization}}} & \multicolumn{1}{l|}{\multirow{4}{*}{$k$-anonymity}} & $k$=5 & 3805 & 3100 \\ \cline{3-5} 
\multicolumn{1}{l|}{}     & \multicolumn{1}{l|}{}     & $k$=10    & 1330   & 930   \\ \cline{3-5} 
\multicolumn{1}{l|}{}     & \multicolumn{1}{l|}{}     & $k$=15    & 465    & 315   \\ \cline{3-5} 
\multicolumn{1}{l|}{}     & \multicolumn{1}{l|}{}     & $k$=20    & 180    & 140   \\ \hline
\end{tabular}
\end{table}

\begin{table}[h!]
(e) RaceDescription \\
\resizebox{\textwidth}{!}{%
\begin{tabular}{cll|l|l|l|l|l|l|l}
\hline
\multicolumn{3}{l|}{Group} &
  \multicolumn{1}{p{2.5cm}|}{American Indian or Alaska Native} &
  Asian &
  \multicolumn{1}{p{2.5cm}|}{Black or African American} &
  Multi Racial &
  Other Race &
  Unavailable &
  White \\ \hline
\multicolumn{3}{l|}{Entire Population}                            & 230 & 2786 & 26746 & 64 & 4484 & 971 & 84590 \\ \hline
\multicolumn{3}{l|}{Vulnerable   Population in Entire Population} & 60  & 369  & 4831  & 2  & 802  & 99  & 8774  \\ \hline
\multicolumn{1}{p{2.5cm}|}{\multirow{4}{*}{\parbox{2.5cm}{Vulnerable Population after applying anonymization}}} &
  \multicolumn{1}{l|}{\multirow{4}{*}{$k$-anonymity}} &
  $k$=5 &
  0 &
  55 &
  2180 &
  25 &
  170 &
  40 &
  4435 \\ \cline{3-10} 
\multicolumn{1}{c|}{}     & \multicolumn{1}{l|}{}     & $k$=10    & 0   & 10   & 600   & 10 & 60   & 30  & 1550  \\ \cline{3-10} 
\multicolumn{1}{c|}{}     & \multicolumn{1}{l|}{}     & $k$=15    & 0   & 0    & 165   & 0  & 15   & 30  & 570   \\ \cline{3-10} 
\multicolumn{1}{c|}{}     & \multicolumn{1}{l|}{}     & $k$=20    & 0   & 0    & 40    & 0  & 20   & 0   & 260   \\ \hline
\end{tabular}%
}
\end{table}

\begin{table}[h!]
(f) EthnicGroupDescription \\
\resizebox{\textwidth}{!}{%
\begin{tabular}{lll|l|l|l|l}
\hline
\multicolumn{3}{l|}{Group}                                        & Declined & Hispanic or Latino & Non-Hispanic or Latino & Unknown \\ \hline
\multicolumn{3}{l|}{Entire Population}                            & 245      & 6855               & 109739                 & 3032    \\ \hline
\multicolumn{3}{l|}{Vulnerable   Population in Entire Population} & 31       & 1234               & 13534                  & 138     \\ \hline
\multicolumn{1}{p{2.5cm}|}{\multirow{4}{*}{\parbox{2.5cm}{Vulnerable Population after applying anonymization}}} &
  \multicolumn{1}{l|}{\multirow{4}{*}{$k$-anonymity}} &
  $k$=5 &
  0 &
  200 &
  6695 &
  10 \\ \cline{3-7} 
\multicolumn{1}{l|}{}     & \multicolumn{1}{l|}{}     & $k$=10    & 0        & 60                 & 2200                   & 0       \\ \cline{3-7} 
\multicolumn{1}{l|}{}     & \multicolumn{1}{l|}{}     & $k$=15    & 0        & 0                  & 780                    & 0       \\ \cline{3-7} 
\multicolumn{1}{l|}{}     & \multicolumn{1}{l|}{}     & $k$=20    & 0        & 0                  & 320                    & 0       \\ \hline
\end{tabular}%
}
\end{table}

\section{ML Model Evaluation}\label{app:ml model evaluation}

In all tables, we show average precision and recall values across different ML models for a pair of datasets. The percentage difference between the pair is also provided.  The highlighted cells indicate a statistically significant difference between the averages from the two datasets.


\begin{table}[h!]
    \centering
    \caption{OR vs FA}
    \label{tab:OR vs FA in app}
    \begin{tabular}{c}
    (a) Precision \\
    \resizebox{\columnwidth}{!}{%
\begin{tabular}{l|l|c|rr|rr|rr|rr|rl|rr}
\hline
Data &
  \begin{tabular}[c]{@{}l@{}}Anonymization \\ Model\end{tabular} &
  \multicolumn{1}{l|}{$k$} &
  \multicolumn{2}{c|}{DT} &
  \multicolumn{2}{c|}{LR} &
  \multicolumn{2}{c|}{NB} &
  \multicolumn{2}{c|}{NN} &
  \multicolumn{2}{c|}{RF} &
  \multicolumn{2}{c}{SVM} \\ \hline
OR data &
   &
  \multicolumn{1}{l|}{} &
  \multicolumn{1}{r|}{0.5793} &
   &
  \multicolumn{1}{r|}{0.7085} &
   &
  \multicolumn{1}{r|}{0.6001} &
   &
  \multicolumn{1}{r|}{0.6735} &
   &
  \multicolumn{1}{r|}{0.7092} &
   &
  \multicolumn{1}{r|}{0.6594} &
   \\ \hline
 &
  $k$-anonymity &
   &
  \multicolumn{1}{r|}{0.5793} &
  0.00\% &
  \multicolumn{1}{r|}{\cellcolor[RGB]{242, 206, 239}0.7027} &
  \cellcolor[RGB]{242, 206, 239}-0.82\% &
  \multicolumn{1}{r|}{0.6016} &
  0.25\% &
  \multicolumn{1}{r|}{0.6680} &
  -0.82\% &
  \multicolumn{1}{r|}{0.7045} &
  -0.66\% &
  \multicolumn{1}{r|}{\cellcolor[RGB]{242, 206, 239}0.7042} &
  \cellcolor[RGB]{242, 206, 239}6.79\% \\ \cline{2-2} \cline{4-15} 
 &
  Zheng et al &
   &
  \multicolumn{1}{r|}{0.5822} &
  0.50\% &
  \multicolumn{1}{r|}{0.7090} &
  0.07\% &
  \multicolumn{1}{r|}{0.5978} &
  -0.38\% &
  \multicolumn{1}{r|}{0.6736} &
  0.01\% &
  \multicolumn{1}{r|}{\cellcolor[RGB]{242, 206, 239}0.7025} &
  \cellcolor[RGB]{242, 206, 239}-0.94\% &
  \multicolumn{1}{r|}{0.6576} &
  -0.27\% \\ \cline{2-2} \cline{4-15} 
 &
  MO-OBAM &
  \multirow{-3}{*}{5} &
  \multicolumn{1}{r|}{0.5809} &
  0.28\% &
  \multicolumn{1}{r|}{0.7110} &
  0.35\% &
  \multicolumn{1}{r|}{0.6029} &
  0.47\% &
  \multicolumn{1}{r|}{0.6671} &
  -0.95\% &
  \multicolumn{1}{r|}{0.7087} &
  -0.07\% &
  \multicolumn{1}{r|}{\cellcolor[RGB]{242, 206, 239}0.6999} &
  \cellcolor[RGB]{242, 206, 239}6.14\% \\ \cline{2-15} 
 &
  $k$-anonymity &
   &
  \multicolumn{1}{r|}{0.5833} &
  0.69\% &
  \multicolumn{1}{r|}{0.7050} &
  -0.49\% &
  \multicolumn{1}{r|}{0.5995} &
  -0.10\% &
  \multicolumn{1}{r|}{0.6682} &
  -0.79\% &
  \multicolumn{1}{r|}{0.7053} &
  -0.55\% &
  \multicolumn{1}{r|}{0.6905} &
  4.72\% \\ \cline{2-2} \cline{4-15} 
 &
  Zheng et al &
   &
  \multicolumn{1}{r|}{0.5777} &
  -0.28\% &
  \multicolumn{1}{r|}{0.7045} &
  -0.56\% &
  \multicolumn{1}{r|}{0.6005} &
  0.07\% &
  \multicolumn{1}{r|}{0.6703} &
  -0.48\% &
  \multicolumn{1}{r|}{0.7049} &
  -0.61\% &
  \multicolumn{1}{r|}{0.6823} &
  3.47\% \\ \cline{2-2} \cline{4-15} 
 &
  MO-OBAM &
  \multirow{-3}{*}{10} &
  \multicolumn{1}{r|}{0.5777} &
  -0.28\% &
  \multicolumn{1}{r|}{0.7082} &
  -0.04\% &
  \multicolumn{1}{r|}{0.6006} &
  0.08\% &
  \multicolumn{1}{r|}{0.6690} &
  -0.67\% &
  \multicolumn{1}{r|}{\cellcolor[RGB]{242, 206, 239}0.7031} &
  \cellcolor[RGB]{242, 206, 239}-0.86\% &
  \multicolumn{1}{r|}{\cellcolor[RGB]{242, 206, 239}0.7087} &
  \cellcolor[RGB]{242, 206, 239}7.48\% \\ \cline{2-15} 
 &
  $k$-anonymity &
   &
  \multicolumn{1}{r|}{0.5771} &
  -0.38\% &
  \multicolumn{1}{r|}{0.7060} &
  -0.35\% &
  \multicolumn{1}{r|}{0.6037} &
  0.60\% &
  \multicolumn{1}{r|}{0.6663} &
  -1.07\% &
  \multicolumn{1}{r|}{\cellcolor[RGB]{242, 206, 239}0.7024} &
  \cellcolor[RGB]{242, 206, 239}-0.96\% &
  \multicolumn{1}{r|}{0.6610} &
  0.24\% \\ \cline{2-2} \cline{4-15} 
 &
  Zheng et al &
   &
  \multicolumn{1}{r|}{0.5801} &
  0.14\% &
  \multicolumn{1}{r|}{0.7102} &
  0.24\% &
  \multicolumn{1}{r|}{0.6024} &
  0.38\% &
  \multicolumn{1}{r|}{0.6648} &
  -1.29\% &
  \multicolumn{1}{r|}{0.7069} &
  -0.32\% &
  \multicolumn{1}{r|}{0.6589} &
  -0.08\% \\ \cline{2-2} \cline{4-15} 
 &
  MO-OBAM &
  \multirow{-3}{*}{15} &
  \multicolumn{1}{r|}{0.5785} &
  -0.14\% &
  \multicolumn{1}{r|}{0.7061} &
  -0.34\% &
  \multicolumn{1}{r|}{0.6037} &
  0.60\% &
  \multicolumn{1}{r|}{0.6676} &
  -0.88\% &
  \multicolumn{1}{r|}{0.7061} &
  -0.44\% &
  \multicolumn{1}{r|}{0.6798} &
  3.09\% \\ \cline{2-15} 
 &
  $k$-anonymity &
   &
  \multicolumn{1}{r|}{0.5793} &
  0.00\% &
  \multicolumn{1}{r|}{0.7037} &
  -0.68\% &
  \multicolumn{1}{r|}{0.6037} &
  0.60\% &
  \multicolumn{1}{r|}{0.6673} &
  -0.92\% &
  \multicolumn{1}{r|}{0.7040} &
  -0.73\% &
  \multicolumn{1}{r|}{0.6661} &
  1.02\% \\ \cline{2-2} \cline{4-15} 
 &
  Zheng et al &
   &
  \multicolumn{1}{r|}{0.5808} &
  0.26\% &
  \multicolumn{1}{r|}{0.7071} &
  -0.20\% &
  \multicolumn{1}{r|}{0.6017} &
  0.27\% &
  \multicolumn{1}{r|}{\cellcolor[RGB]{242, 206, 239}0.6631} &
  \cellcolor[RGB]{242, 206, 239}-1.54\% &
  \multicolumn{1}{r|}{\cellcolor[RGB]{242, 206, 239}0.7037} &
  \cellcolor[RGB]{242, 206, 239}-0.78\% &
  \multicolumn{1}{r|}{\cellcolor[RGB]{242, 206, 239}0.7063} &
  \cellcolor[RGB]{242, 206, 239}7.11\% \\ \cline{2-2} \cline{4-15} 
\multirow{-12}{*}{FA data} &
  MO-OBAM &
  \multirow{-3}{*}{20} &
  \multicolumn{1}{r|}{0.5771} &
  -0.38\% &
  \multicolumn{1}{r|}{0.7076} &
  -0.13\% &
  \multicolumn{1}{r|}{0.6005} &
  0.07\% &
  \multicolumn{1}{r|}{\cellcolor[RGB]{242, 206, 239}0.6612} &
  \cellcolor[RGB]{242, 206, 239}-1.83\% &
  \multicolumn{1}{r|}{\cellcolor[RGB]{242, 206, 239}0.7028} &
  \cellcolor[RGB]{242, 206, 239}-0.90\% &
  \multicolumn{1}{r|}{0.6515} &
  -1.20\% \\ \hline
\end{tabular}%
} \\

    \\
    (b) Recall \\
    \resizebox{\columnwidth}{!}{%
\begin{tabular}{l|l|c|rr|rr|rr|rr|rr|rr}
\hline
Data &
  \begin{tabular}[c]{@{}l@{}}Anonymization \\ Model\end{tabular} &
  \multicolumn{1}{l|}{$k$} &
  \multicolumn{2}{c|}{DT} &
  \multicolumn{2}{c|}{LR} &
  \multicolumn{2}{c|}{NB} &
  \multicolumn{2}{c|}{NN} &
  \multicolumn{2}{c|}{RF} &
  \multicolumn{2}{c}{SVM} \\ \hline
OR data &
   &
  \multicolumn{1}{l|}{} &
  \multicolumn{1}{r|}{0.5748} &
   &
  \multicolumn{1}{r|}{0.6197} &
   &
  \multicolumn{1}{r|}{0.4370} &
   &
  \multicolumn{1}{r|}{0.6160} &
   &
  \multicolumn{1}{r|}{0.6021} &
   &
  \multicolumn{1}{r|}{0.6097} &
   \\ \hline
 &
  $k$-anonymity &
   &
  \multicolumn{1}{r|}{0.5746} &
  -0.03\% &
  \multicolumn{1}{r|}{0.6160} &
  -0.60\% &
  \multicolumn{1}{r|}{0.4370} &
  0.00\% &
  \multicolumn{1}{r|}{0.6221} &
  0.99\% &
  \multicolumn{1}{r|}{0.6018} &
  -0.05\% &
  \multicolumn{1}{r|}{\cellcolor[RGB]{242, 206, 239}0.5205} &
  \cellcolor[RGB]{242, 206, 239}-14.63\% \\ \cline{2-2} \cline{4-15} 
 &
  Zheng et al &
   &
  \multicolumn{1}{r|}{0.5792} &
  0.77\% &
  \multicolumn{1}{r|}{0.6214} &
  0.27\% &
  \multicolumn{1}{r|}{0.4419} &
  1.12\% &
  \multicolumn{1}{r|}{0.6153} &
  -0.11\% &
  \multicolumn{1}{r|}{0.6006} &
  -0.25\% &
  \multicolumn{1}{r|}{0.6267} &
  2.79\% \\ \cline{2-2} \cline{4-15} 
 &
  MO-OBAM &
  \multirow{-3}{*}{5} &
  \multicolumn{1}{r|}{0.5760} &
  0.21\% &
  \multicolumn{1}{r|}{0.6158} &
  -0.63\% &
  \multicolumn{1}{r|}{0.4405} &
  0.80\% &
  \multicolumn{1}{r|}{0.6170} &
  0.16\% &
  \multicolumn{1}{r|}{0.6051} &
  0.50\% &
  \multicolumn{1}{r|}{0.5544} &
  -9.07\% \\ \cline{2-15} 
 &
  $k$-anonymity &
   &
  \multicolumn{1}{r|}{0.5759} &
  0.19\% &
  \multicolumn{1}{r|}{0.6180} &
  -0.27\% &
  \multicolumn{1}{r|}{0.4375} &
  0.11\% &
  \multicolumn{1}{r|}{0.6101} &
  -0.96\% &
  \multicolumn{1}{r|}{0.6035} &
  0.23\% &
  \multicolumn{1}{r|}{0.5525} &
  -9.38\% \\ \cline{2-2} \cline{4-15} 
 &
  Zheng et al &
   &
  \multicolumn{1}{r|}{0.5739} &
  -0.16\% &
  \multicolumn{1}{r|}{0.6206} &
  0.15\% &
  \multicolumn{1}{r|}{\cellcolor[RGB]{242, 206, 239}0.4442} &
  \cellcolor[RGB]{242, 206, 239}1.65\% &
  \multicolumn{1}{r|}{0.6059} &
  -1.64\% &
  \multicolumn{1}{r|}{0.5988} &
  -0.55\% &
  \multicolumn{1}{r|}{0.5752} &
  -5.66\% \\ \cline{2-2} \cline{4-15} 
 &
  MO-OBAM &
  \multirow{-3}{*}{10} &
  \multicolumn{1}{r|}{0.5706} &
  -0.73\% &
  \multicolumn{1}{r|}{0.6169} &
  -0.45\% &
  \multicolumn{1}{r|}{0.4424} &
  1.24\% &
  \multicolumn{1}{r|}{0.6239} &
  1.28\% &
  \multicolumn{1}{r|}{\cellcolor[RGB]{242, 206, 239}0.5974} &
  \cellcolor[RGB]{242, 206, 239}-0.78\% &
  \multicolumn{1}{r|}{\cellcolor[RGB]{242, 206, 239}0.5420} &
  \cellcolor[RGB]{242, 206, 239}-11.10\% \\ \cline{2-15} 
 &
  $k$-anonymity &
   &
  \multicolumn{1}{r|}{0.5727} &
  -0.37\% &
  \multicolumn{1}{r|}{0.6158} &
  -0.63\% &
  \multicolumn{1}{r|}{0.4454} &
  1.92\% &
  \multicolumn{1}{r|}{0.6141} &
  -0.31\% &
  \multicolumn{1}{r|}{0.6017} &
  -0.07\% &
  \multicolumn{1}{r|}{0.6194} &
  1.59\% \\ \cline{2-2} \cline{4-15} 
 &
  Zheng et al &
   &
  \multicolumn{1}{r|}{0.5775} &
  0.47\% &
  \multicolumn{1}{r|}{0.6210} &
  0.21\% &
  \multicolumn{1}{r|}{\cellcolor[RGB]{242, 206, 239}0.4486} &
  \cellcolor[RGB]{242, 206, 239}2.65\% &
  \multicolumn{1}{r|}{0.6225} &
  1.06\% &
  \multicolumn{1}{r|}{0.6054} &
  0.55\% &
  \multicolumn{1}{r|}{0.6266} &
  2.77\% \\ \cline{2-2} \cline{4-15} 
 &
  MO-OBAM &
  \multirow{-3}{*}{15} &
  \multicolumn{1}{r|}{0.5743} &
  -0.09\% &
  \multicolumn{1}{r|}{0.6194} &
  -0.05\% &
  \multicolumn{1}{r|}{0.4428} &
  1.33\% &
  \multicolumn{1}{r|}{0.6214} &
  0.88\% &
  \multicolumn{1}{r|}{0.6016} &
  -0.08\% &
  \multicolumn{1}{r|}{0.5777} &
  -5.25\% \\ \cline{2-15} 
 &
  $k$-anonymity &
   &
  \multicolumn{1}{r|}{0.5762} &
  0.24\% &
  \multicolumn{1}{r|}{0.6177} &
  -0.32\% &
  \multicolumn{1}{r|}{0.4435} &
  1.49\% &
  \multicolumn{1}{r|}{0.6136} &
  -0.39\% &
  \multicolumn{1}{r|}{0.6060} &
  0.65\% &
  \multicolumn{1}{r|}{0.6084} &
  -0.21\% \\ \cline{2-2} \cline{4-15} 
 &
  Zheng et al &
   &
  \multicolumn{1}{r|}{0.5731} &
  -0.30\% &
  \multicolumn{1}{r|}{0.6187} &
  -0.16\% &
  \multicolumn{1}{r|}{\cellcolor[RGB]{242, 206, 239}0.4455} &
  \cellcolor[RGB]{242, 206, 239}1.95\% &
  \multicolumn{1}{r|}{0.6232} &
  1.17\% &
  \multicolumn{1}{r|}{0.5979} &
  -0.70\% &
  \multicolumn{1}{r|}{\cellcolor[RGB]{242, 206, 239}0.5255} &
  \cellcolor[RGB]{242, 206, 239}-13.81\% \\ \cline{2-2} \cline{4-15} 
\multirow{-12}{*}{FA data} &
  MO-OBAM &
  \multirow{-3}{*}{20} &
  \multicolumn{1}{r|}{0.5700} &
  -0.84\% &
  \multicolumn{1}{r|}{0.6174} &
  -0.37\% &
  \multicolumn{1}{r|}{0.4374} &
  0.09\% &
  \multicolumn{1}{r|}{0.6181} &
  0.34\% &
  \multicolumn{1}{r|}{0.5995} &
  -0.43\% &
  \multicolumn{1}{r|}{0.6188} &
  1.49\% \\ \hline
\end{tabular}%
}
    \end{tabular}
\end{table}


\begin{table}[h!]
    \centering
    \caption{OR vs OR-NV-only}
    \label{tab:OR vs OR-NV-only in app}
    \begin{tabular}{c}
        (a) Precision \\
        \resizebox{\columnwidth}{!}{
    \begin{tabular}{cc}
    (1) Linkage attack with the risk level of $\tau=0.05$ & (2) Linkage attack with the risk level of $\tau=0.075$\\
\begin{tabular}{l|r|r|r}
\hline
Model &
  \multicolumn{1}{l|}{\begin{tabular}[c]{@{}l@{}}Average \\ (OR)\end{tabular}} &
  \multicolumn{1}{l|}{\begin{tabular}[c]{@{}l@{}}Average \\ (OR-NV-only)\end{tabular}} &
  \multicolumn{1}{l}{\begin{tabular}[c]{@{}l@{}}Percentage \\ Change\end{tabular}} \\ \hline
DT  & \cellcolor[RGB]{242, 206, 239}0.5793 & \cellcolor[RGB]{242, 206, 239}0.4208 & \cellcolor[RGB]{242, 206, 239}-27.36\% \\ \hline
LR  & \cellcolor[RGB]{242, 206, 239}0.7085 & \cellcolor[RGB]{242, 206, 239}0.5761 & \cellcolor[RGB]{242, 206, 239}-18.69\% \\ \hline
NB  & \cellcolor[RGB]{242, 206, 239}0.6001 & \cellcolor[RGB]{242, 206, 239}0.3459 & \cellcolor[RGB]{242, 206, 239}-42.36\% \\ \hline
NN  & \cellcolor[RGB]{242, 206, 239}0.6735 & \cellcolor[RGB]{242, 206, 239}0.5713 & \cellcolor[RGB]{242, 206, 239}-15.17\% \\ \hline
RF  & \cellcolor[RGB]{242, 206, 239}0.7092 & \cellcolor[RGB]{242, 206, 239}0.6228 & \cellcolor[RGB]{242, 206, 239}-12.18\% \\ \hline
SVM & \cellcolor[RGB]{242, 206, 239}0.6594 & \cellcolor[RGB]{242, 206, 239}0.3262 & \cellcolor[RGB]{242, 206, 239}-50.53\% \\ \hline
\end{tabular}
 &
\begin{tabular}{l|r|r|r}
\hline
Model &
  \multicolumn{1}{l|}{\begin{tabular}[c]{@{}l@{}}Average \\ (OR)\end{tabular}} &
  \multicolumn{1}{l|}{\begin{tabular}[c]{@{}l@{}}Average \\ (OR-NV-only)\end{tabular}} &
  \multicolumn{1}{l}{\begin{tabular}[c]{@{}l@{}}Percentage \\ Change\end{tabular}} \\ \hline
DT  & \cellcolor[RGB]{242, 206, 239}0.5793 & \cellcolor[RGB]{242, 206, 239}0.4590 & \cellcolor[RGB]{242, 206, 239}-20.77\% \\ \hline
LR  & \cellcolor[RGB]{242, 206, 239}0.7085 & \cellcolor[RGB]{242, 206, 239}0.6052 & \cellcolor[RGB]{242, 206, 239}-14.58\% \\ \hline
NB  & \cellcolor[RGB]{242, 206, 239}0.6001 & \cellcolor[RGB]{242, 206, 239}0.3093 & \cellcolor[RGB]{242, 206, 239}-48.46\% \\ \hline
NN  & \cellcolor[RGB]{242, 206, 239}0.6735 & \cellcolor[RGB]{242, 206, 239}0.6043 & \cellcolor[RGB]{242, 206, 239}-10.27\% \\ \hline
RF  & \cellcolor[RGB]{242, 206, 239}0.7092 & \cellcolor[RGB]{242, 206, 239}0.6418 & \cellcolor[RGB]{242, 206, 239}-9.50\%  \\ \hline
SVM & \cellcolor[RGB]{242, 206, 239}0.6594 & \cellcolor[RGB]{242, 206, 239}0.5140 & \cellcolor[RGB]{242, 206, 239}-22.05\% \\ \hline
\end{tabular}%
\\
    \\
    (3) Linkage attack with the risk level of $\tau=0.1$ & (4) Homogeneity attack \\
    \begin{tabular}{l|r|r|r}
\hline
Model &
  \multicolumn{1}{l|}{\begin{tabular}[c]{@{}l@{}}Average \\ (OR)\end{tabular}} &
  \multicolumn{1}{l|}{\begin{tabular}[c]{@{}l@{}}Average \\ (OR-NV-only)\end{tabular}} &
  \multicolumn{1}{l}{\begin{tabular}[c]{@{}l@{}}Percentage \\ Change\end{tabular}} \\ \hline
DT  & \cellcolor[RGB]{242, 206, 239}0.5793 & \cellcolor[RGB]{242, 206, 239}0.4666 & \cellcolor[RGB]{242, 206, 239}-19.45\% \\ \hline
LR  & \cellcolor[RGB]{242, 206, 239}0.7085 & \cellcolor[RGB]{242, 206, 239}0.6321 & \cellcolor[RGB]{242, 206, 239}-10.78\% \\ \hline
NB  & \cellcolor[RGB]{242, 206, 239}0.6001 & \cellcolor[RGB]{242, 206, 239}0.3735 & \cellcolor[RGB]{242, 206, 239}-37.76\% \\ \hline
NN  & \cellcolor[RGB]{242, 206, 239}0.6735 & \cellcolor[RGB]{242, 206, 239}0.6041 & \cellcolor[RGB]{242, 206, 239}-10.30\% \\ \hline
RF  & \cellcolor[RGB]{242, 206, 239}0.7092 & \cellcolor[RGB]{242, 206, 239}0.6414 & \cellcolor[RGB]{242, 206, 239}-9.56\%  \\ \hline
SVM & \cellcolor[RGB]{242, 206, 239}0.6594 & \cellcolor[RGB]{242, 206, 239}0.5488 & \cellcolor[RGB]{242, 206, 239}-16.77\% \\ \hline
\end{tabular} &
    \begin{tabular}{l|r|r|r}
\hline
Model &
  \multicolumn{1}{l|}{\begin{tabular}[c]{@{}l@{}}Average\\ (OR)\end{tabular}} &
  \multicolumn{1}{l|}{\begin{tabular}[c]{@{}l@{}}Average \\ (OR-NV-only)\end{tabular}} &
  \multicolumn{1}{l}{\begin{tabular}[c]{@{}l@{}}Percentage \\ Change\end{tabular}} \\ \hline
DT  & \cellcolor[RGB]{242, 206, 239}0.5793 & \cellcolor[RGB]{242, 206, 239}0.5636 & \cellcolor[RGB]{242, 206, 239}-2.71\% \\ \hline
LR  & \cellcolor[RGB]{242, 206, 239}0.7085 & \cellcolor[RGB]{242, 206, 239}0.6941 & \cellcolor[RGB]{242, 206, 239}-2.03\% \\ \hline
NB  & \cellcolor[RGB]{242, 206, 239}0.6001 & \cellcolor[RGB]{242, 206, 239}0.5635 & \cellcolor[RGB]{242, 206, 239}-6.10\% \\ \hline
NN  & 0.6735                         & 0.6766                         & 0.46\%                          \\ \hline
RF  & \cellcolor[RGB]{242, 206, 239}0.7092 & \cellcolor[RGB]{242, 206, 239}0.6920 & \cellcolor[RGB]{242, 206, 239}-2.43\% \\ \hline
SVM & 0.6594                         & 0.6598                         & 0.06\%                          \\ \hline
\end{tabular} 
    \end{tabular}
    } 
\end{tabular}
\end{table}
\begin{table}[h!]
    \centering
    \label{tab:my_label}
    \begin{tabular}{c}
        (b) Recall \\
        \resizebox{\columnwidth}{!}{
    \begin{tabular}{cc}
    (1) Linkage attack with the risk level of $\tau=0.05$ & (2) Linkage attack with the risk level of $\tau=0.075$\\
    \begin{tabular}{l|r|r|r}
\hline
Model &
  \multicolumn{1}{l|}{\begin{tabular}[c]{@{}l@{}}Average\\ (OR)\end{tabular}} &
  \multicolumn{1}{l|}{\begin{tabular}[c]{@{}l@{}}Average \\ (OR-NV-only)\end{tabular}} &
  \multicolumn{1}{l}{\begin{tabular}[c]{@{}l@{}}Percentage \\ Change\end{tabular}} \\ \hline
DT  & \cellcolor[RGB]{242, 206, 239}0.5748 & \cellcolor[RGB]{242, 206, 239}0.4214 & \cellcolor[RGB]{242, 206, 239}-26.69\% \\ \hline
LR  & \cellcolor[RGB]{242, 206, 239}0.6197 & \cellcolor[RGB]{242, 206, 239}0.3624 & \cellcolor[RGB]{242, 206, 239}-41.52\% \\ \hline
NB  & \cellcolor[RGB]{242, 206, 239}0.4370 & \cellcolor[RGB]{242, 206, 239}0.7608 & \cellcolor[RGB]{242, 206, 239}74.10\%  \\ \hline
NN  & \cellcolor[RGB]{242, 206, 239}0.6160 & \cellcolor[RGB]{242, 206, 239}0.4004 & \cellcolor[RGB]{242, 206, 239}-35.00\% \\ \hline
RF  & \cellcolor[RGB]{242, 206, 239}0.6021 & \cellcolor[RGB]{242, 206, 239}0.2446 & \cellcolor[RGB]{242, 206, 239}-59.38\% \\ \hline
SVM & \cellcolor[RGB]{242, 206, 239}0.6097 & \cellcolor[RGB]{242, 206, 239}0.3420 & \cellcolor[RGB]{242, 206, 239}-43.91\% \\ \hline
\end{tabular} &
    \begin{tabular}{l|r|r|r}
\hline
Model &
  \multicolumn{1}{l|}{\begin{tabular}[c]{@{}l@{}}Average\\ (OR)\end{tabular}} &
  \multicolumn{1}{l|}{\begin{tabular}[c]{@{}l@{}}Average \\ (OR-NV-only)\end{tabular}} &
  \multicolumn{1}{l}{\begin{tabular}[c]{@{}l@{}}Percentage \\ Change\end{tabular}} \\ \hline
DT  & \cellcolor[RGB]{242, 206, 239}0.5748 & \cellcolor[RGB]{242, 206, 239}0.4586 & \cellcolor[RGB]{242, 206, 239}-20.22\% \\ \hline
LR  & \cellcolor[RGB]{242, 206, 239}0.6197 & \cellcolor[RGB]{242, 206, 239}0.3893 & \cellcolor[RGB]{242, 206, 239}-37.18\% \\ \hline
NB  & \cellcolor[RGB]{242, 206, 239}0.4370 & \cellcolor[RGB]{242, 206, 239}0.6099 & \cellcolor[RGB]{242, 206, 239}39.57\%  \\ \hline
NN  & \cellcolor[RGB]{242, 206, 239}0.6160 & \cellcolor[RGB]{242, 206, 239}0.4154 & \cellcolor[RGB]{242, 206, 239}-32.56\% \\ \hline
RF  & \cellcolor[RGB]{242, 206, 239}0.6021 & \cellcolor[RGB]{242, 206, 239}0.2721 & \cellcolor[RGB]{242, 206, 239}-54.81\% \\ \hline
SVM & \cellcolor[RGB]{242, 206, 239}0.6097 & \cellcolor[RGB]{242, 206, 239}0.4021 & \cellcolor[RGB]{242, 206, 239}-34.05\% \\ \hline
\end{tabular}\\
    \\
    (3) Linkage attack with the risk level of $\tau=0.1$ & (4) Homogeneity attack \\
     \begin{tabular}{l|r|r|r}
\hline
Model &
  \multicolumn{1}{l|}{\begin{tabular}[c]{@{}l@{}}Average\\ (OR)\end{tabular}} &
  \multicolumn{1}{l|}{\begin{tabular}[c]{@{}l@{}}Average \\ (OR-NV-only)\end{tabular}} &
  \multicolumn{1}{l}{\begin{tabular}[c]{@{}l@{}}Percentage \\ Change\end{tabular}} \\ \hline
DT  & \cellcolor[RGB]{242, 206, 239}0.5748 & \cellcolor[RGB]{242, 206, 239}0.4606 & \cellcolor[RGB]{242, 206, 239}-19.87\% \\ \hline
LR  & \cellcolor[RGB]{242, 206, 239}0.6197 & \cellcolor[RGB]{242, 206, 239}0.4100 & \cellcolor[RGB]{242, 206, 239}-33.84\% \\ \hline
NB  & \cellcolor[RGB]{242, 206, 239}0.4370 & \cellcolor[RGB]{242, 206, 239}0.6481 & \cellcolor[RGB]{242, 206, 239}48.31\%  \\ \hline
NN  & \cellcolor[RGB]{242, 206, 239}0.6160 & \cellcolor[RGB]{242, 206, 239}0.4549 & \cellcolor[RGB]{242, 206, 239}-26.15\% \\ \hline
RF  & \cellcolor[RGB]{242, 206, 239}0.6021 & \cellcolor[RGB]{242, 206, 239}0.3211 & \cellcolor[RGB]{242, 206, 239}-46.67\% \\ \hline
SVM & \cellcolor[RGB]{242, 206, 239}0.6097 & \cellcolor[RGB]{242, 206, 239}0.4508 & \cellcolor[RGB]{242, 206, 239}-26.06\% \\ \hline
\end{tabular} &
     \begin{tabular}{l|r|r|r}
\hline
Model &
  \multicolumn{1}{l|}{\begin{tabular}[c]{@{}l@{}}Average\\ (OR)\end{tabular}} &
  \multicolumn{1}{l|}{\begin{tabular}[c]{@{}l@{}}Average \\ (OR-NV-only)\end{tabular}} &
  \multicolumn{1}{l}{\begin{tabular}[c]{@{}l@{}}Percentage \\ Change\end{tabular}} \\ \hline
DT  & \cellcolor[RGB]{242, 206, 239}0.5748 & \cellcolor[RGB]{242, 206, 239}0.5636 & \cellcolor[RGB]{242, 206, 239}-1.95\% \\ \hline
LR  & \cellcolor[RGB]{242, 206, 239}0.6197 & \cellcolor[RGB]{242, 206, 239}0.5744 & \cellcolor[RGB]{242, 206, 239}-7.31\% \\ \hline
NB  & \cellcolor[RGB]{242, 206, 239}0.4370 & \cellcolor[RGB]{242, 206, 239}0.4519 & \cellcolor[RGB]{242, 206, 239}3.41\%  \\ \hline
NN  & 0.6160                         & 0.5951                         & -3.39\%                         \\ \hline
RF  & \cellcolor[RGB]{242, 206, 239}0.6021 & \cellcolor[RGB]{242, 206, 239}0.5574 & \cellcolor[RGB]{242, 206, 239}-7.42\% \\ \hline
SVM & 0.6097                         & 0.5451                         & -10.60\%                        \\ \hline
\end{tabular}
    \end{tabular}
    }
    \end{tabular}
\end{table}

\begin{table}[h!]
    \centering
    \caption{OR-NV-only vs FA-NV-only (Precision)}
    \label{tab:OR-NV-only vs FA-NV-only (pre)}
    \begin{tabular}{c}
    (a) Linkage attack with the risk level of $\tau = 0.05$ \\
    \resizebox{\columnwidth}{!}{%
\begin{tabular}{l|l|c|rr|rr|rr|rr|rr|rr}
\hline
Data &
  \begin{tabular}[c]{@{}l@{}}Anonymization \\ Model\end{tabular} &
  \multicolumn{1}{l|}{$k$} &
  \multicolumn{2}{c|}{DT} &
  \multicolumn{2}{c|}{LR} &
  \multicolumn{2}{c|}{NB} &
  \multicolumn{2}{c|}{NN} &
  \multicolumn{2}{c|}{RF} &
  \multicolumn{2}{c}{SVM} \\ \hline
OR-NV-only data &
   &
  \multicolumn{1}{l|}{} &
  \multicolumn{1}{r|}{0.4208} &
   &
  \multicolumn{1}{r|}{0.5761} &
   &
  \multicolumn{1}{r|}{0.3459} &
   &
  \multicolumn{1}{r|}{0.5713} &
   &
  \multicolumn{1}{r|}{0.6228} &
   &
  \multicolumn{1}{r|}{0.3262} &
   \\ \hline
 &
  $k$-anonymity &
   &
  \multicolumn{1}{r|}{\cellcolor[RGB]{242, 206, 239}0.4838} &
  \cellcolor[RGB]{242, 206, 239}14.97\% &
  \multicolumn{1}{r|}{\cellcolor[RGB]{242, 206, 239}0.6567} &
  \cellcolor[RGB]{242, 206, 239}13.99\% &
  \multicolumn{1}{r|}{\cellcolor[RGB]{242, 206, 239}0.4349} &
  \cellcolor[RGB]{242, 206, 239}25.73\% &
  \multicolumn{1}{r|}{\cellcolor[RGB]{242, 206, 239}0.6031} &
  \cellcolor[RGB]{242, 206, 239}5.57\% &
  \multicolumn{1}{r|}{0.6446} &
  3.50\% &
  \multicolumn{1}{r|}{\cellcolor[RGB]{242, 206, 239}0.6081} &
  \cellcolor[RGB]{242, 206, 239}86.42\% \\ \cline{2-2} \cline{4-15} 
 &
  Zheng et al &
   &
  \multicolumn{1}{r|}{\cellcolor[RGB]{242, 206, 239}0.5237} &
  \cellcolor[RGB]{242, 206, 239}24.45\% &
  \multicolumn{1}{r|}{\cellcolor[RGB]{242, 206, 239}0.6703} &
  \cellcolor[RGB]{242, 206, 239}16.35\% &
  \multicolumn{1}{r|}{\cellcolor[RGB]{242, 206, 239}0.5029} &
  \cellcolor[RGB]{242, 206, 239}45.39\% &
  \multicolumn{1}{r|}{\cellcolor[RGB]{242, 206, 239}0.6306} &
  \cellcolor[RGB]{242, 206, 239}10.38\% &
  \multicolumn{1}{r|}{\cellcolor[RGB]{242, 206, 239}0.6754} &
  \cellcolor[RGB]{242, 206, 239}8.45\% &
  \multicolumn{1}{r|}{\cellcolor[RGB]{242, 206, 239}0.6550} &
  \cellcolor[RGB]{242, 206, 239}100.80\% \\ \cline{2-2} \cline{4-15} 
 &
  MO-OBAM &
  \multirow{-3}{*}{5} &
  \multicolumn{1}{r|}{\cellcolor[RGB]{242, 206, 239}0.5501} &
  \cellcolor[RGB]{242, 206, 239}30.73\% &
  \multicolumn{1}{r|}{\cellcolor[RGB]{242, 206, 239}0.6854} &
  \cellcolor[RGB]{242, 206, 239}18.97\% &
  \multicolumn{1}{r|}{\cellcolor[RGB]{242, 206, 239}0.5491} &
  \cellcolor[RGB]{242, 206, 239}58.75\% &
  \multicolumn{1}{r|}{\cellcolor[RGB]{242, 206, 239}0.6358} &
  \cellcolor[RGB]{242, 206, 239}11.29\% &
  \multicolumn{1}{r|}{\cellcolor[RGB]{242, 206, 239}0.6868} &
  \cellcolor[RGB]{242, 206, 239}10.28\% &
  \multicolumn{1}{r|}{\cellcolor[RGB]{242, 206, 239}0.6623} &
  \cellcolor[RGB]{242, 206, 239}103.03\% \\ \cline{2-15} 
 &
  $k$-anonymity &
   &
  \multicolumn{1}{r|}{\cellcolor[RGB]{242, 206, 239}0.5053} &
  \cellcolor[RGB]{242, 206, 239}20.08\% &
  \multicolumn{1}{r|}{\cellcolor[RGB]{242, 206, 239}0.6679} &
  \cellcolor[RGB]{242, 206, 239}15.93\% &
  \multicolumn{1}{r|}{\cellcolor[RGB]{242, 206, 239}0.4629} &
  \cellcolor[RGB]{242, 206, 239}33.82\% &
  \multicolumn{1}{r|}{\cellcolor[RGB]{242, 206, 239}0.6251} &
  \cellcolor[RGB]{242, 206, 239}9.42\% &
  \multicolumn{1}{r|}{\cellcolor[RGB]{242, 206, 239}0.6557} &
  \cellcolor[RGB]{242, 206, 239}5.28\% &
  \multicolumn{1}{r|}{\cellcolor[RGB]{242, 206, 239}0.6491} &
  \cellcolor[RGB]{242, 206, 239}98.99\% \\ \cline{2-2} \cline{4-15} 
 &
  Zheng et al &
   &
  \multicolumn{1}{r|}{\cellcolor[RGB]{242, 206, 239}0.5229} &
  \cellcolor[RGB]{242, 206, 239}24.26\% &
  \multicolumn{1}{r|}{\cellcolor[RGB]{242, 206, 239}0.6780} &
  \cellcolor[RGB]{242, 206, 239}17.69\% &
  \multicolumn{1}{r|}{\cellcolor[RGB]{242, 206, 239}0.5158} &
  \cellcolor[RGB]{242, 206, 239}49.12\% &
  \multicolumn{1}{r|}{\cellcolor[RGB]{242, 206, 239}0.6355} &
  \cellcolor[RGB]{242, 206, 239}11.24\% &
  \multicolumn{1}{r|}{\cellcolor[RGB]{242, 206, 239}0.6752} &
  \cellcolor[RGB]{242, 206, 239}8.41\% &
  \multicolumn{1}{r|}{\cellcolor[RGB]{242, 206, 239}0.6797} &
  \cellcolor[RGB]{242, 206, 239}108.37\% \\ \cline{2-2} \cline{4-15} 
 &
  MO-OBAM &
  \multirow{-3}{*}{10} &
  \multicolumn{1}{r|}{\cellcolor[RGB]{242, 206, 239}0.5684} &
  \cellcolor[RGB]{242, 206, 239}35.08\% &
  \multicolumn{1}{r|}{\cellcolor[RGB]{242, 206, 239}0.7017} &
  \cellcolor[RGB]{242, 206, 239}21.80\% &
  \multicolumn{1}{r|}{\cellcolor[RGB]{242, 206, 239}0.5831} &
  \cellcolor[RGB]{242, 206, 239}68.57\% &
  \multicolumn{1}{r|}{\cellcolor[RGB]{242, 206, 239}0.6568} &
  \cellcolor[RGB]{242, 206, 239}14.97\% &
  \multicolumn{1}{r|}{\cellcolor[RGB]{242, 206, 239}0.6975} &
  \cellcolor[RGB]{242, 206, 239}11.99\% &
  \multicolumn{1}{r|}{\cellcolor[RGB]{242, 206, 239}0.6925} &
  \cellcolor[RGB]{242, 206, 239}112.29\% \\ \cline{2-15} 
 &
  $k$-anonymity &
   &
  \multicolumn{1}{r|}{\cellcolor[RGB]{242, 206, 239}0.4861} &
  \cellcolor[RGB]{242, 206, 239}15.52\% &
  \multicolumn{1}{r|}{\cellcolor[RGB]{242, 206, 239}0.6605} &
  \cellcolor[RGB]{242, 206, 239}14.65\% &
  \multicolumn{1}{r|}{\cellcolor[RGB]{242, 206, 239}0.4203} &
  \cellcolor[RGB]{242, 206, 239}21.51\% &
  \multicolumn{1}{r|}{\cellcolor[RGB]{242, 206, 239}0.6384} &
  \cellcolor[RGB]{242, 206, 239}11.75\% &
  \multicolumn{1}{r|}{0.6443} &
  3.45\% &
  \multicolumn{1}{r|}{\cellcolor[RGB]{242, 206, 239}0.6466} &
  \cellcolor[RGB]{242, 206, 239}98.22\% \\ \cline{2-2} \cline{4-15} 
 &
  Zheng et al &
   &
  \multicolumn{1}{r|}{\cellcolor[RGB]{242, 206, 239}0.5233} &
  \cellcolor[RGB]{242, 206, 239}24.36\% &
  \multicolumn{1}{r|}{\cellcolor[RGB]{242, 206, 239}0.6779} &
  \cellcolor[RGB]{242, 206, 239}17.67\% &
  \multicolumn{1}{r|}{\cellcolor[RGB]{242, 206, 239}0.5092} &
  \cellcolor[RGB]{242, 206, 239}47.21\% &
  \multicolumn{1}{r|}{\cellcolor[RGB]{242, 206, 239}0.6325} &
  \cellcolor[RGB]{242, 206, 239}10.71\% &
  \multicolumn{1}{r|}{\cellcolor[RGB]{242, 206, 239}0.6793} &
  \cellcolor[RGB]{242, 206, 239}9.07\% &
  \multicolumn{1}{r|}{\cellcolor[RGB]{242, 206, 239}0.6480} &
  \cellcolor[RGB]{242, 206, 239}98.65\% \\ \cline{2-2} \cline{4-15} 
 &
  MO-OBAM &
  \multirow{-3}{*}{15} &
  \multicolumn{1}{r|}{\cellcolor[RGB]{242, 206, 239}0.5735} &
  \cellcolor[RGB]{242, 206, 239}36.29\% &
  \multicolumn{1}{r|}{\cellcolor[RGB]{242, 206, 239}0.7042} &
  \cellcolor[RGB]{242, 206, 239}22.24\% &
  \multicolumn{1}{r|}{\cellcolor[RGB]{242, 206, 239}0.6001} &
  \cellcolor[RGB]{242, 206, 239}73.49\% &
  \multicolumn{1}{r|}{\cellcolor[RGB]{242, 206, 239}0.6606} &
  \cellcolor[RGB]{242, 206, 239}15.63\% &
  \multicolumn{1}{r|}{\cellcolor[RGB]{242, 206, 239}0.7007} &
  \cellcolor[RGB]{242, 206, 239}12.51\% &
  \multicolumn{1}{r|}{\cellcolor[RGB]{242, 206, 239}0.6723} &
  \cellcolor[RGB]{242, 206, 239}106.10\% \\ \cline{2-15} 
 &
  $k$-anonymity &
   &
  \multicolumn{1}{r|}{\cellcolor[RGB]{242, 206, 239}0.5786} &
  \cellcolor[RGB]{242, 206, 239}37.50\% &
  \multicolumn{1}{r|}{\cellcolor[RGB]{242, 206, 239}0.7102} &
  \cellcolor[RGB]{242, 206, 239}23.28\% &
  \multicolumn{1}{r|}{\cellcolor[RGB]{242, 206, 239}0.6004} &
  \cellcolor[RGB]{242, 206, 239}73.58\% &
  \multicolumn{1}{r|}{\cellcolor[RGB]{242, 206, 239}0.6639} &
  \cellcolor[RGB]{242, 206, 239}16.21\% &
  \multicolumn{1}{r|}{\cellcolor[RGB]{242, 206, 239}0.7039} &
  \cellcolor[RGB]{242, 206, 239}13.02\% &
  \multicolumn{1}{r|}{\cellcolor[RGB]{242, 206, 239}0.6883} &
  \cellcolor[RGB]{242, 206, 239}111.01\% \\ \cline{2-2} \cline{4-15} 
 &
  Zheng et al &
   &
  \multicolumn{1}{r|}{\cellcolor[RGB]{242, 206, 239}0.5819} &
  \cellcolor[RGB]{242, 206, 239}38.28\% &
  \multicolumn{1}{r|}{\cellcolor[RGB]{242, 206, 239}0.7071} &
  \cellcolor[RGB]{242, 206, 239}22.74\% &
  \multicolumn{1}{r|}{\cellcolor[RGB]{242, 206, 239}0.6042} &
  \cellcolor[RGB]{242, 206, 239}74.67\% &
  \multicolumn{1}{r|}{\cellcolor[RGB]{242, 206, 239}0.6699} &
  \cellcolor[RGB]{242, 206, 239}17.26\% &
  \multicolumn{1}{r|}{\cellcolor[RGB]{242, 206, 239}0.7027} &
  \cellcolor[RGB]{242, 206, 239}12.83\% &
  \multicolumn{1}{r|}{\cellcolor[RGB]{242, 206, 239}0.6967} &
  \cellcolor[RGB]{242, 206, 239}113.58\% \\ \cline{2-2} \cline{4-15} 
\multirow{-12}{*}{FA-NV-only data} &
  MO-OBAM &
  \multirow{-3}{*}{20} &
  \multicolumn{1}{r|}{\cellcolor[RGB]{242, 206, 239}0.5765} &
  \cellcolor[RGB]{242, 206, 239}37.00\% &
  \multicolumn{1}{r|}{\cellcolor[RGB]{242, 206, 239}0.7055} &
  \cellcolor[RGB]{242, 206, 239}22.46\% &
  \multicolumn{1}{r|}{\cellcolor[RGB]{242, 206, 239}0.6006} &
  \cellcolor[RGB]{242, 206, 239}73.63\% &
  \multicolumn{1}{r|}{\cellcolor[RGB]{242, 206, 239}0.6678} &
  \cellcolor[RGB]{242, 206, 239}16.89\% &
  \multicolumn{1}{r|}{\cellcolor[RGB]{242, 206, 239}0.6995} &
  \cellcolor[RGB]{242, 206, 239}12.32\% &
  \multicolumn{1}{r|}{\cellcolor[RGB]{242, 206, 239}0.7082} &
  \cellcolor[RGB]{242, 206, 239}117.11\% \\ \hline
\end{tabular}%
} 
    \end{tabular}
\end{table}
\begin{table}[h!]
    \centering
    \begin{tabular}{c}
    (b) Linkage attack with the risk level of $\tau = 0.075$ \\
    \resizebox{\columnwidth}{!}{%
\begin{tabular}{l|l|c|rr|rr|rr|rr|rr|rr}
\hline
Data &
  \begin{tabular}[c]{@{}l@{}}Anonymization \\ Model\end{tabular} &
  \multicolumn{1}{l|}{$k$} &
  \multicolumn{2}{c|}{DT} &
  \multicolumn{2}{c|}{LR} &
  \multicolumn{2}{c|}{NB} &
  \multicolumn{2}{c|}{NN} &
  \multicolumn{2}{c|}{RF} &
  \multicolumn{2}{c}{SVM} \\ \hline
OR-NV-only data &
   &
  \multicolumn{1}{l|}{} &
  \multicolumn{1}{r|}{0.4590} &
   &
  \multicolumn{1}{r|}{0.6052} &
   &
  \multicolumn{1}{r|}{0.3093} &
   &
  \multicolumn{1}{r|}{0.6043} &
   &
  \multicolumn{1}{r|}{0.6418} &
   &
  \multicolumn{1}{r|}{0.5140} &
   \\ \hline
 &
  $k$-anonymity &
   &
  \multicolumn{1}{r|}{\cellcolor[RGB]{242, 206, 239}0.4800} &
  \cellcolor[RGB]{242, 206, 239}4.58\% &
  \multicolumn{1}{r|}{\cellcolor[RGB]{242, 206, 239}0.6580} &
  \cellcolor[RGB]{242, 206, 239}8.72\% &
  \multicolumn{1}{r|}{\cellcolor[RGB]{242, 206, 239}0.4511} &
  \cellcolor[RGB]{242, 206, 239}45.85\% &
  \multicolumn{1}{r|}{0.6076} &
  0.55\% &
  \multicolumn{1}{r|}{0.6263} &
  -2.42\% &
  \multicolumn{1}{r|}{\cellcolor[RGB]{242, 206, 239}0.6170} &
  \cellcolor[RGB]{242, 206, 239}20.04\% \\ \cline{2-2} \cline{4-15} 
 &
  Zheng et al &
   &
  \multicolumn{1}{r|}{\cellcolor[RGB]{242, 206, 239}0.5358} &
  \cellcolor[RGB]{242, 206, 239}16.73\% &
  \multicolumn{1}{r|}{\cellcolor[RGB]{242, 206, 239}0.6786} &
  \cellcolor[RGB]{242, 206, 239}12.13\% &
  \multicolumn{1}{r|}{\cellcolor[RGB]{242, 206, 239}0.5354} &
  \cellcolor[RGB]{242, 206, 239}73.10\% &
  \multicolumn{1}{r|}{\cellcolor[RGB]{242, 206, 239}0.6301} &
  \cellcolor[RGB]{242, 206, 239}4.27\% &
  \multicolumn{1}{r|}{\cellcolor[RGB]{242, 206, 239}0.6835} &
  \cellcolor[RGB]{242, 206, 239}6.50\% &
  \multicolumn{1}{r|}{\cellcolor[RGB]{242, 206, 239}0.6830} &
  \cellcolor[RGB]{242, 206, 239}32.88\% \\ \cline{2-2} \cline{4-15} 
 &
  MO-OBAM &
  \multirow{-3}{*}{5} &
  \multicolumn{1}{r|}{\cellcolor[RGB]{242, 206, 239}0.5621} &
  \cellcolor[RGB]{242, 206, 239}22.46\% &
  \multicolumn{1}{r|}{\cellcolor[RGB]{242, 206, 239}0.6967} &
  \cellcolor[RGB]{242, 206, 239}15.12\% &
  \multicolumn{1}{r|}{\cellcolor[RGB]{242, 206, 239}0.5668} &
  \cellcolor[RGB]{242, 206, 239}83.25\% &
  \multicolumn{1}{r|}{\cellcolor[RGB]{242, 206, 239}0.6552} &
  \cellcolor[RGB]{242, 206, 239}8.42\% &
  \multicolumn{1}{r|}{\cellcolor[RGB]{242, 206, 239}0.6889} &
  \cellcolor[RGB]{242, 206, 239}7.34\% &
  \multicolumn{1}{r|}{\cellcolor[RGB]{242, 206, 239}0.6632} &
  \cellcolor[RGB]{242, 206, 239}29.03\% \\ \cline{2-15} 
 &
  $k$-anonymity &
   &
  \multicolumn{1}{r|}{\cellcolor[RGB]{242, 206, 239}0.5044} &
  \cellcolor[RGB]{242, 206, 239}9.89\% &
  \multicolumn{1}{r|}{\cellcolor[RGB]{242, 206, 239}0.6720} &
  \cellcolor[RGB]{242, 206, 239}11.04\% &
  \multicolumn{1}{r|}{\cellcolor[RGB]{242, 206, 239}0.4643} &
  \cellcolor[RGB]{242, 206, 239}50.11\% &
  \multicolumn{1}{r|}{0.6245} &
  3.34\% &
  \multicolumn{1}{r|}{0.6558} &
  2.18\% &
  \multicolumn{1}{r|}{\cellcolor[RGB]{242, 206, 239}0.6123} &
  \cellcolor[RGB]{242, 206, 239}19.12\% \\ \cline{2-2} \cline{4-15} 
 &
  Zheng et al &
   &
  \multicolumn{1}{r|}{\cellcolor[RGB]{242, 206, 239}0.5372} &
  \cellcolor[RGB]{242, 206, 239}17.04\% &
  \multicolumn{1}{r|}{\cellcolor[RGB]{242, 206, 239}0.6818} &
  \cellcolor[RGB]{242, 206, 239}12.66\% &
  \multicolumn{1}{r|}{\cellcolor[RGB]{242, 206, 239}0.5344} &
  \cellcolor[RGB]{242, 206, 239}72.78\% &
  \multicolumn{1}{r|}{\cellcolor[RGB]{242, 206, 239}0.6359} &
  \cellcolor[RGB]{242, 206, 239}5.23\% &
  \multicolumn{1}{r|}{\cellcolor[RGB]{242, 206, 239}0.6817} &
  \cellcolor[RGB]{242, 206, 239}6.22\% &
  \multicolumn{1}{r|}{\cellcolor[RGB]{242, 206, 239}0.6651} &
  \cellcolor[RGB]{242, 206, 239}29.40\% \\ \cline{2-2} \cline{4-15} 
 &
  MO-OBAM &
  \multirow{-3}{*}{10} &
  \multicolumn{1}{r|}{\cellcolor[RGB]{242, 206, 239}0.5700} &
  \cellcolor[RGB]{242, 206, 239}24.18\% &
  \multicolumn{1}{r|}{\cellcolor[RGB]{242, 206, 239}0.7017} &
  \cellcolor[RGB]{242, 206, 239}15.95\% &
  \multicolumn{1}{r|}{\cellcolor[RGB]{242, 206, 239}0.5949} &
  \cellcolor[RGB]{242, 206, 239}92.34\% &
  \multicolumn{1}{r|}{\cellcolor[RGB]{242, 206, 239}0.6655} &
  \cellcolor[RGB]{242, 206, 239}10.13\% &
  \multicolumn{1}{r|}{\cellcolor[RGB]{242, 206, 239}0.6990} &
  \cellcolor[RGB]{242, 206, 239}8.91\% &
  \multicolumn{1}{r|}{\cellcolor[RGB]{242, 206, 239}0.6734} &
  \cellcolor[RGB]{242, 206, 239}31.01\% \\ \cline{2-15} 
 &
  $k$-anonymity &
   &
  \multicolumn{1}{r|}{\cellcolor[RGB]{242, 206, 239}0.5751} &
  \cellcolor[RGB]{242, 206, 239}25.29\% &
  \multicolumn{1}{r|}{\cellcolor[RGB]{242, 206, 239}0.7044} &
  \cellcolor[RGB]{242, 206, 239}16.39\% &
  \multicolumn{1}{r|}{\cellcolor[RGB]{242, 206, 239}0.6008} &
  \cellcolor[RGB]{242, 206, 239}94.25\% &
  \multicolumn{1}{r|}{\cellcolor[RGB]{242, 206, 239}0.6675} &
  \cellcolor[RGB]{242, 206, 239}10.46\% &
  \multicolumn{1}{r|}{\cellcolor[RGB]{242, 206, 239}0.7057} &
  \cellcolor[RGB]{242, 206, 239}9.96\% &
  \multicolumn{1}{r|}{\cellcolor[RGB]{242, 206, 239}0.6565} &
  \cellcolor[RGB]{242, 206, 239}27.72\% \\ \cline{2-2} \cline{4-15} 
 &
  Zheng et al &
   &
  \multicolumn{1}{r|}{\cellcolor[RGB]{242, 206, 239}0.5837} &
  \cellcolor[RGB]{242, 206, 239}27.17\% &
  \multicolumn{1}{r|}{\cellcolor[RGB]{242, 206, 239}0.7051} &
  \cellcolor[RGB]{242, 206, 239}16.51\% &
  \multicolumn{1}{r|}{\cellcolor[RGB]{242, 206, 239}0.5986} &
  \cellcolor[RGB]{242, 206, 239}93.53\% &
  \multicolumn{1}{r|}{\cellcolor[RGB]{242, 206, 239}0.6687} &
  \cellcolor[RGB]{242, 206, 239}10.66\% &
  \multicolumn{1}{r|}{\cellcolor[RGB]{242, 206, 239}0.7011} &
  \cellcolor[RGB]{242, 206, 239}9.24\% &
  \multicolumn{1}{r|}{\cellcolor[RGB]{242, 206, 239}0.6755} &
  \cellcolor[RGB]{242, 206, 239}31.42\% \\ \cline{2-2} \cline{4-15} 
 &
  MO-OBAM &
  \multirow{-3}{*}{15} &
  \multicolumn{1}{r|}{\cellcolor[RGB]{242, 206, 239}0.5761} &
  \cellcolor[RGB]{242, 206, 239}25.51\% &
  \multicolumn{1}{r|}{\cellcolor[RGB]{242, 206, 239}0.7068} &
  \cellcolor[RGB]{242, 206, 239}16.79\% &
  \multicolumn{1}{r|}{\cellcolor[RGB]{242, 206, 239}0.5973} &
  \cellcolor[RGB]{242, 206, 239}93.11\% &
  \multicolumn{1}{r|}{\cellcolor[RGB]{242, 206, 239}0.6629} &
  \cellcolor[RGB]{242, 206, 239}9.70\% &
  \multicolumn{1}{r|}{\cellcolor[RGB]{242, 206, 239}0.7085} &
  \cellcolor[RGB]{242, 206, 239}10.39\% &
  \multicolumn{1}{r|}{\cellcolor[RGB]{242, 206, 239}0.6926} &
  \cellcolor[RGB]{242, 206, 239}34.75\% \\ \cline{2-15} 
 &
  $k$-anonymity &
   &
  \multicolumn{1}{r|}{\cellcolor[RGB]{242, 206, 239}0.5804} &
  \cellcolor[RGB]{242, 206, 239}26.45\% &
  \multicolumn{1}{r|}{\cellcolor[RGB]{242, 206, 239}0.7058} &
  \cellcolor[RGB]{242, 206, 239}16.62\% &
  \multicolumn{1}{r|}{\cellcolor[RGB]{242, 206, 239}0.6030} &
  \cellcolor[RGB]{242, 206, 239}94.96\% &
  \multicolumn{1}{r|}{\cellcolor[RGB]{242, 206, 239}0.6623} &
  \cellcolor[RGB]{242, 206, 239}9.60\% &
  \multicolumn{1}{r|}{\cellcolor[RGB]{242, 206, 239}0.7079} &
  \cellcolor[RGB]{242, 206, 239}10.30\% &
  \multicolumn{1}{r|}{\cellcolor[RGB]{242, 206, 239}0.7009} &
  \cellcolor[RGB]{242, 206, 239}36.36\% \\ \cline{2-2} \cline{4-15} 
 &
  Zheng et al &
   &
  \multicolumn{1}{r|}{\cellcolor[RGB]{242, 206, 239}0.5810} &
  \cellcolor[RGB]{242, 206, 239}26.58\% &
  \multicolumn{1}{r|}{\cellcolor[RGB]{242, 206, 239}0.7067} &
  \cellcolor[RGB]{242, 206, 239}16.77\% &
  \multicolumn{1}{r|}{\cellcolor[RGB]{242, 206, 239}0.6013} &
  \cellcolor[RGB]{242, 206, 239}94.41\% &
  \multicolumn{1}{r|}{\cellcolor[RGB]{242, 206, 239}0.6668} &
  \cellcolor[RGB]{242, 206, 239}10.34\% &
  \multicolumn{1}{r|}{\cellcolor[RGB]{242, 206, 239}0.7004} &
  \cellcolor[RGB]{242, 206, 239}9.13\% &
  \multicolumn{1}{r|}{\cellcolor[RGB]{242, 206, 239}0.6680} &
  \cellcolor[RGB]{242, 206, 239}29.96\% \\ \cline{2-2} \cline{4-15} 
\multirow{-12}{*}{FA-NV-only data} &
  MO-OBAM &
  \multirow{-3}{*}{20} &
  \multicolumn{1}{r|}{\cellcolor[RGB]{242, 206, 239}0.5761} &
  \cellcolor[RGB]{242, 206, 239}25.51\% &
  \multicolumn{1}{r|}{\cellcolor[RGB]{242, 206, 239}0.7068} &
  \cellcolor[RGB]{242, 206, 239}16.79\% &
  \multicolumn{1}{r|}{\cellcolor[RGB]{242, 206, 239}0.6020} &
  \cellcolor[RGB]{242, 206, 239}94.63\% &
  \multicolumn{1}{r|}{\cellcolor[RGB]{242, 206, 239}0.6640} &
  \cellcolor[RGB]{242, 206, 239}9.88\% &
  \multicolumn{1}{r|}{\cellcolor[RGB]{242, 206, 239}0.6992} &
  \cellcolor[RGB]{242, 206, 239}8.94\% &
  \multicolumn{1}{r|}{\cellcolor[RGB]{242, 206, 239}0.6629} &
  \cellcolor[RGB]{242, 206, 239}28.97\% \\ \hline
\end{tabular}%
} 
    \end{tabular}
\end{table}
\begin{table}[h!]
    \centering
    \begin{tabular}{c}
    (c) Linkage attack with the risk level of $\tau = 0.1$ \\
    \resizebox{\columnwidth}{!}{%
\begin{tabular}{l|l|c|rr|rr|rr|rr|rr|rr}
\hline
Data &
  \begin{tabular}[c]{@{}l@{}}Anonymization \\ Model\end{tabular} &
  \multicolumn{1}{l|}{$k$} &
  \multicolumn{2}{c|}{DT} &
  \multicolumn{2}{c|}{LR} &
  \multicolumn{2}{c|}{NB} &
  \multicolumn{2}{c|}{NN} &
  \multicolumn{2}{c|}{RF} &
  \multicolumn{2}{c}{SVM} \\ \hline
OR-NV-only data &
   &
  \multicolumn{1}{l|}{} &
  \multicolumn{1}{r|}{0.4666} &
   &
  \multicolumn{1}{r|}{0.6321} &
   &
  \multicolumn{1}{r|}{0.3735} &
   &
  \multicolumn{1}{r|}{0.6041} &
   &
  \multicolumn{1}{r|}{0.6414} &
   &
  \multicolumn{1}{r|}{0.5488} &
   \\ \hline
 &
  $k$-anonymity &
   &
  \multicolumn{1}{r|}{\cellcolor[RGB]{242, 206, 239}0.4844} &
  \cellcolor[RGB]{242, 206, 239}3.81\% &
  \multicolumn{1}{r|}{\cellcolor[RGB]{242, 206, 239}0.6522} &
  \cellcolor[RGB]{242, 206, 239}3.18\% &
  \multicolumn{1}{r|}{\cellcolor[RGB]{242, 206, 239}0.4466} &
  \cellcolor[RGB]{242, 206, 239}19.57\% &
  \multicolumn{1}{r|}{\cellcolor[RGB]{242, 206, 239}0.6248} &
  \cellcolor[RGB]{242, 206, 239}3.43\% &
  \multicolumn{1}{r|}{0.6416} &
  0.03\% &
  \multicolumn{1}{r|}{0.6070} &
  10.60\% \\ \cline{2-2} \cline{4-15} 
 &
  Zheng et al &
   &
  \multicolumn{1}{r|}{\cellcolor[RGB]{242, 206, 239}0.5535} &
  \cellcolor[RGB]{242, 206, 239}18.62\% &
  \multicolumn{1}{r|}{\cellcolor[RGB]{242, 206, 239}0.6877} &
  \cellcolor[RGB]{242, 206, 239}8.80\% &
  \multicolumn{1}{r|}{\cellcolor[RGB]{242, 206, 239}0.5588} &
  \cellcolor[RGB]{242, 206, 239}49.61\% &
  \multicolumn{1}{r|}{\cellcolor[RGB]{242, 206, 239}0.6404} &
  \cellcolor[RGB]{242, 206, 239}6.01\% &
  \multicolumn{1}{r|}{\cellcolor[RGB]{242, 206, 239}0.6827} &
  \cellcolor[RGB]{242, 206, 239}6.44\% &
  \multicolumn{1}{r|}{\cellcolor[RGB]{242, 206, 239}0.6732} &
  \cellcolor[RGB]{242, 206, 239}22.67\% \\ \cline{2-2} \cline{4-15} 
 &
  MO-OBAM &
  \multirow{-3}{*}{5} &
  \multicolumn{1}{r|}{\cellcolor[RGB]{242, 206, 239}0.5763} &
  \cellcolor[RGB]{242, 206, 239}23.51\% &
  \multicolumn{1}{r|}{\cellcolor[RGB]{242, 206, 239}0.7045} &
  \cellcolor[RGB]{242, 206, 239}11.45\% &
  \multicolumn{1}{r|}{\cellcolor[RGB]{242, 206, 239}0.5924} &
  \cellcolor[RGB]{242, 206, 239}58.61\% &
  \multicolumn{1}{r|}{\cellcolor[RGB]{242, 206, 239}0.6656} &
  \cellcolor[RGB]{242, 206, 239}10.18\% &
  \multicolumn{1}{r|}{\cellcolor[RGB]{242, 206, 239}0.7027} &
  \cellcolor[RGB]{242, 206, 239}9.56\% &
  \multicolumn{1}{r|}{\cellcolor[RGB]{242, 206, 239}0.6882} &
  \cellcolor[RGB]{242, 206, 239}25.40\% \\ \cline{2-15} 
 &
  $k$-anonymity &
   &
  \multicolumn{1}{r|}{\cellcolor[RGB]{242, 206, 239}0.5809} &
  \cellcolor[RGB]{242, 206, 239}24.50\% &
  \multicolumn{1}{r|}{\cellcolor[RGB]{242, 206, 239}0.7063} &
  \cellcolor[RGB]{242, 206, 239}11.74\% &
  \multicolumn{1}{r|}{\cellcolor[RGB]{242, 206, 239}0.6017} &
  \cellcolor[RGB]{242, 206, 239}61.10\% &
  \multicolumn{1}{r|}{\cellcolor[RGB]{242, 206, 239}0.6697} &
  \cellcolor[RGB]{242, 206, 239}10.86\% &
  \multicolumn{1}{r|}{\cellcolor[RGB]{242, 206, 239}0.7093} &
  \cellcolor[RGB]{242, 206, 239}10.59\% &
  \multicolumn{1}{r|}{\cellcolor[RGB]{242, 206, 239}0.6593} &
  \cellcolor[RGB]{242, 206, 239}20.13\% \\ \cline{2-2} \cline{4-15} 
 &
  Zheng et al &
   &
  \multicolumn{1}{r|}{\cellcolor[RGB]{242, 206, 239}0.5791} &
  \cellcolor[RGB]{242, 206, 239}24.11\% &
  \multicolumn{1}{r|}{\cellcolor[RGB]{242, 206, 239}0.7046} &
  \cellcolor[RGB]{242, 206, 239}11.47\% &
  \multicolumn{1}{r|}{\cellcolor[RGB]{242, 206, 239}0.6053} &
  \cellcolor[RGB]{242, 206, 239}62.06\% &
  \multicolumn{1}{r|}{\cellcolor[RGB]{242, 206, 239}0.6668} &
  \cellcolor[RGB]{242, 206, 239}10.38\% &
  \multicolumn{1}{r|}{\cellcolor[RGB]{242, 206, 239}0.7050} &
  \cellcolor[RGB]{242, 206, 239}9.92\% &
  \multicolumn{1}{r|}{\cellcolor[RGB]{242, 206, 239}0.7073} &
  \cellcolor[RGB]{242, 206, 239}28.88\% \\ \cline{2-2} \cline{4-15} 
 &
  MO-OBAM &
  \multirow{-3}{*}{10} &
  \multicolumn{1}{r|}{\cellcolor[RGB]{242, 206, 239}0.5775} &
  \cellcolor[RGB]{242, 206, 239}23.77\% &
  \multicolumn{1}{r|}{\cellcolor[RGB]{242, 206, 239}0.7062} &
  \cellcolor[RGB]{242, 206, 239}11.72\% &
  \multicolumn{1}{r|}{\cellcolor[RGB]{242, 206, 239}0.6026} &
  \cellcolor[RGB]{242, 206, 239}61.34\% &
  \multicolumn{1}{r|}{\cellcolor[RGB]{242, 206, 239}0.6682} &
  \cellcolor[RGB]{242, 206, 239}10.61\% &
  \multicolumn{1}{r|}{\cellcolor[RGB]{242, 206, 239}0.7005} &
  \cellcolor[RGB]{242, 206, 239}9.21\% &
  \multicolumn{1}{r|}{\cellcolor[RGB]{242, 206, 239}0.6839} &
  \cellcolor[RGB]{242, 206, 239}24.62\% \\ \cline{2-15} 
 &
  $k$-anonymity &
   &
  \multicolumn{1}{r|}{\cellcolor[RGB]{242, 206, 239}0.5744} &
  \cellcolor[RGB]{242, 206, 239}23.10\% &
  \multicolumn{1}{r|}{\cellcolor[RGB]{242, 206, 239}0.7055} &
  \cellcolor[RGB]{242, 206, 239}11.61\% &
  \multicolumn{1}{r|}{\cellcolor[RGB]{242, 206, 239}0.6004} &
  \cellcolor[RGB]{242, 206, 239}60.75\% &
  \multicolumn{1}{r|}{\cellcolor[RGB]{242, 206, 239}0.6760} &
  \cellcolor[RGB]{242, 206, 239}11.90\% &
  \multicolumn{1}{r|}{\cellcolor[RGB]{242, 206, 239}0.7041} &
  \cellcolor[RGB]{242, 206, 239}9.78\% &
  \multicolumn{1}{r|}{\cellcolor[RGB]{242, 206, 239}0.6966} &
  \cellcolor[RGB]{242, 206, 239}26.93\% \\ \cline{2-2} \cline{4-15} 
 &
  Zheng et al &
   &
  \multicolumn{1}{r|}{\cellcolor[RGB]{242, 206, 239}0.5797} &
  \cellcolor[RGB]{242, 206, 239}24.24\% &
  \multicolumn{1}{r|}{\cellcolor[RGB]{242, 206, 239}0.7064} &
  \cellcolor[RGB]{242, 206, 239}11.75\% &
  \multicolumn{1}{r|}{\cellcolor[RGB]{242, 206, 239}0.5986} &
  \cellcolor[RGB]{242, 206, 239}60.27\% &
  \multicolumn{1}{r|}{\cellcolor[RGB]{242, 206, 239}0.6655} &
  \cellcolor[RGB]{242, 206, 239}10.16\% &
  \multicolumn{1}{r|}{\cellcolor[RGB]{242, 206, 239}0.7039} &
  \cellcolor[RGB]{242, 206, 239}9.74\% &
  \multicolumn{1}{r|}{\cellcolor[RGB]{242, 206, 239}0.6739} &
  \cellcolor[RGB]{242, 206, 239}22.80\% \\ \cline{2-2} \cline{4-15} 
 &
  MO-OBAM &
  \multirow{-3}{*}{15} &
  \multicolumn{1}{r|}{\cellcolor[RGB]{242, 206, 239}0.5785} &
  \cellcolor[RGB]{242, 206, 239}23.98\% &
  \multicolumn{1}{r|}{\cellcolor[RGB]{242, 206, 239}0.7015} &
  \cellcolor[RGB]{242, 206, 239}10.98\% &
  \multicolumn{1}{r|}{\cellcolor[RGB]{242, 206, 239}0.5995} &
  \cellcolor[RGB]{242, 206, 239}60.51\% &
  \multicolumn{1}{r|}{\cellcolor[RGB]{242, 206, 239}0.6728} &
  \cellcolor[RGB]{242, 206, 239}11.37\% &
  \multicolumn{1}{r|}{\cellcolor[RGB]{242, 206, 239}0.7076} &
  \cellcolor[RGB]{242, 206, 239}10.32\% &
  \multicolumn{1}{r|}{\cellcolor[RGB]{242, 206, 239}0.6815} &
  \cellcolor[RGB]{242, 206, 239}24.18\% \\ \cline{2-15} 
 &
  $k$-anonymity &
   &
  \multicolumn{1}{r|}{\cellcolor[RGB]{242, 206, 239}0.5787} &
  \cellcolor[RGB]{242, 206, 239}24.02\% &
  \multicolumn{1}{r|}{\cellcolor[RGB]{242, 206, 239}0.7064} &
  \cellcolor[RGB]{242, 206, 239}11.75\% &
  \multicolumn{1}{r|}{\cellcolor[RGB]{242, 206, 239}0.6021} &
  \cellcolor[RGB]{242, 206, 239}61.20\% &
  \multicolumn{1}{r|}{\cellcolor[RGB]{242, 206, 239}0.6729} &
  \cellcolor[RGB]{242, 206, 239}11.39\% &
  \multicolumn{1}{r|}{\cellcolor[RGB]{242, 206, 239}0.7055} &
  \cellcolor[RGB]{242, 206, 239}9.99\% &
  \multicolumn{1}{r|}{\cellcolor[RGB]{242, 206, 239}0.6798} &
  \cellcolor[RGB]{242, 206, 239}23.87\% \\ \cline{2-2} \cline{4-15} 
 &
  Zheng et al &
   &
  \multicolumn{1}{r|}{\cellcolor[RGB]{242, 206, 239}0.5841} &
  \cellcolor[RGB]{242, 206, 239}25.18\% &
  \multicolumn{1}{r|}{\cellcolor[RGB]{242, 206, 239}0.7052} &
  \cellcolor[RGB]{242, 206, 239}11.56\% &
  \multicolumn{1}{r|}{\cellcolor[RGB]{242, 206, 239}0.5992} &
  \cellcolor[RGB]{242, 206, 239}60.43\% &
  \multicolumn{1}{r|}{\cellcolor[RGB]{242, 206, 239}0.6656} &
  \cellcolor[RGB]{242, 206, 239}10.18\% &
  \multicolumn{1}{r|}{\cellcolor[RGB]{242, 206, 239}0.7015} &
  \cellcolor[RGB]{242, 206, 239}9.37\% &
  \multicolumn{1}{r|}{\cellcolor[RGB]{242, 206, 239}0.7118} &
  \cellcolor[RGB]{242, 206, 239}29.70\% \\ \cline{2-2} \cline{4-15} 
\multirow{-12}{*}{FA-NV-only data} &
  MO-OBAM &
  \multirow{-3}{*}{20} &
  \multicolumn{1}{r|}{\cellcolor[RGB]{242, 206, 239}0.5780} &
  \cellcolor[RGB]{242, 206, 239}23.87\% &
  \multicolumn{1}{r|}{\cellcolor[RGB]{242, 206, 239}0.7050} &
  \cellcolor[RGB]{242, 206, 239}11.53\% &
  \multicolumn{1}{r|}{\cellcolor[RGB]{242, 206, 239}0.6000} &
  \cellcolor[RGB]{242, 206, 239}60.64\% &
  \multicolumn{1}{r|}{\cellcolor[RGB]{242, 206, 239}0.6662} &
  \cellcolor[RGB]{242, 206, 239}10.28\% &
  \multicolumn{1}{r|}{\cellcolor[RGB]{242, 206, 239}0.7008} &
  \cellcolor[RGB]{242, 206, 239}9.26\% &
  \multicolumn{1}{r|}{\cellcolor[RGB]{242, 206, 239}0.6899} &
  \cellcolor[RGB]{242, 206, 239}25.71\% \\ \hline
\end{tabular}%
} \\ 
    \end{tabular}
\end{table}
\begin{table}[h!]
    \centering
    \begin{tabular}{c}
    (d) Homogeneity attack \\
    \resizebox{\columnwidth}{!}{%
\begin{tabular}{l|l|c|rr|rr|rr|rr|rr|rr}
\hline
Data &
  \begin{tabular}[c]{@{}l@{}}Anonymization \\ Model\end{tabular} &
  \multicolumn{1}{l|}{$k$} &
  \multicolumn{2}{c|}{DT} &
  \multicolumn{2}{c|}{LR} &
  \multicolumn{2}{c|}{NB} &
  \multicolumn{2}{c|}{NN} &
  \multicolumn{2}{c|}{RF} &
  \multicolumn{2}{c}{SVM} \\ \hline
OR-NV-only data &
   &
  \multicolumn{1}{l|}{} &
  \multicolumn{1}{r|}{0.5636} &
   &
  \multicolumn{1}{r|}{0.6941} &
   &
  \multicolumn{1}{r|}{0.5635} &
   &
  \multicolumn{1}{r|}{0.6766} &
   &
  \multicolumn{1}{r|}{0.6920} &
   &
  \multicolumn{1}{r|}{0.6598} &
   \\ \hline
 &
  $k$-anonymity &
   &
  \multicolumn{1}{r|}{\cellcolor[RGB]{242, 206, 239}0.5739} &
  \cellcolor[RGB]{242, 206, 239}1.83\% &
  \multicolumn{1}{r|}{\cellcolor[RGB]{242, 206, 239}0.7049} &
  \cellcolor[RGB]{242, 206, 239}1.56\% &
  \multicolumn{1}{r|}{\cellcolor[RGB]{242, 206, 239}0.5924} &
  \cellcolor[RGB]{242, 206, 239}5.13\% &
  \multicolumn{1}{r|}{0.6674} &
  -1.36\% &
  \multicolumn{1}{r|}{\cellcolor[RGB]{242, 206, 239}0.7027} &
  \cellcolor[RGB]{242, 206, 239}1.55\% &
  \multicolumn{1}{r|}{0.6871} &
  4.14\% \\ \cline{2-2} \cline{4-15} 
 &
  Zheng et al &
   &
  \multicolumn{1}{r|}{\cellcolor[RGB]{242, 206, 239}0.5794} &
  \cellcolor[RGB]{242, 206, 239}2.80\% &
  \multicolumn{1}{r|}{\cellcolor[RGB]{242, 206, 239}0.7102} &
  \cellcolor[RGB]{242, 206, 239}2.32\% &
  \multicolumn{1}{r|}{\cellcolor[RGB]{242, 206, 239}0.6035} &
  \cellcolor[RGB]{242, 206, 239}7.10\% &
  \multicolumn{1}{r|}{0.6750} &
  -0.24\% &
  \multicolumn{1}{r|}{\cellcolor[RGB]{242, 206, 239}0.7038} &
  \cellcolor[RGB]{242, 206, 239}1.71\% &
  \multicolumn{1}{r|}{0.6788} &
  2.88\% \\ \cline{2-2} \cline{4-15} 
 &
  MO-OBAM &
  \multirow{-3}{*}{5} &
  \multicolumn{1}{r|}{\cellcolor[RGB]{242, 206, 239}0.5831} &
  \cellcolor[RGB]{242, 206, 239}3.46\% &
  \multicolumn{1}{r|}{\cellcolor[RGB]{242, 206, 239}0.7068} &
  \cellcolor[RGB]{242, 206, 239}1.83\% &
  \multicolumn{1}{r|}{\cellcolor[RGB]{242, 206, 239}0.5992} &
  \cellcolor[RGB]{242, 206, 239}6.34\% &
  \multicolumn{1}{r|}{0.6727} &
  -0.58\% &
  \multicolumn{1}{r|}{\cellcolor[RGB]{242, 206, 239}0.7060} &
  \cellcolor[RGB]{242, 206, 239}2.02\% &
  \multicolumn{1}{r|}{\cellcolor[RGB]{242, 206, 239}0.7104} &
  \cellcolor[RGB]{242, 206, 239}7.67\% \\ \cline{2-15} 
 &
  $k$-anonymity &
   &
  \multicolumn{1}{r|}{\cellcolor[RGB]{242, 206, 239}0.5851} &
  \cellcolor[RGB]{242, 206, 239}3.81\% &
  \multicolumn{1}{r|}{\cellcolor[RGB]{242, 206, 239}0.7073} &
  \cellcolor[RGB]{242, 206, 239}1.90\% &
  \multicolumn{1}{r|}{\cellcolor[RGB]{242, 206, 239}0.6038} &
  \cellcolor[RGB]{242, 206, 239}7.15\% &
  \multicolumn{1}{r|}{0.6676} &
  -1.33\% &
  \multicolumn{1}{r|}{\cellcolor[RGB]{242, 206, 239}0.7054} &
  \cellcolor[RGB]{242, 206, 239}1.94\% &
  \multicolumn{1}{r|}{0.6962} &
  5.52\% \\ \cline{2-2} \cline{4-15} 
 &
  Zheng et al &
   &
  \multicolumn{1}{r|}{\cellcolor[RGB]{242, 206, 239}0.5806} &
  \cellcolor[RGB]{242, 206, 239}3.02\% &
  \multicolumn{1}{r|}{\cellcolor[RGB]{242, 206, 239}0.7045} &
  \cellcolor[RGB]{242, 206, 239}1.50\% &
  \multicolumn{1}{r|}{\cellcolor[RGB]{242, 206, 239}0.6030} &
  \cellcolor[RGB]{242, 206, 239}7.01\% &
  \multicolumn{1}{r|}{\cellcolor[RGB]{242, 206, 239}0.6673} &
  \cellcolor[RGB]{242, 206, 239}-1.37\% &
  \multicolumn{1}{r|}{\cellcolor[RGB]{242, 206, 239}0.7055} &
  \cellcolor[RGB]{242, 206, 239}1.95\% &
  \multicolumn{1}{r|}{0.6881} &
  4.29\% \\ \cline{2-2} \cline{4-15} 
 &
  MO-OBAM &
  \multirow{-3}{*}{10} &
  \multicolumn{1}{r|}{\cellcolor[RGB]{242, 206, 239}0.5768} &
  \cellcolor[RGB]{242, 206, 239}2.34\% &
  \multicolumn{1}{r|}{\cellcolor[RGB]{242, 206, 239}0.7070} &
  \cellcolor[RGB]{242, 206, 239}1.86\% &
  \multicolumn{1}{r|}{\cellcolor[RGB]{242, 206, 239}0.6002} &
  \cellcolor[RGB]{242, 206, 239}6.51\% &
  \multicolumn{1}{r|}{0.6733} &
  -0.49\% &
  \multicolumn{1}{r|}{\cellcolor[RGB]{242, 206, 239}0.7037} &
  \cellcolor[RGB]{242, 206, 239}1.69\% &
  \multicolumn{1}{r|}{\cellcolor[RGB]{242, 206, 239}0.7336} &
  \cellcolor[RGB]{242, 206, 239}11.19\% \\ \cline{2-15} 
 &
  $k$-anonymity &
   &
  \multicolumn{1}{r|}{\cellcolor[RGB]{242, 206, 239}0.5752} &
  \cellcolor[RGB]{242, 206, 239}2.06\% &
  \multicolumn{1}{r|}{\cellcolor[RGB]{242, 206, 239}0.7030} &
  \cellcolor[RGB]{242, 206, 239}1.28\% &
  \multicolumn{1}{r|}{\cellcolor[RGB]{242, 206, 239}0.5998} &
  \cellcolor[RGB]{242, 206, 239}6.44\% &
  \multicolumn{1}{r|}{\cellcolor[RGB]{242, 206, 239}0.6654} &
  \cellcolor[RGB]{242, 206, 239}-1.66\% &
  \multicolumn{1}{r|}{\cellcolor[RGB]{242, 206, 239}0.7043} &
  \cellcolor[RGB]{242, 206, 239}1.78\% &
  \multicolumn{1}{r|}{0.6724} &
  1.91\% \\ \cline{2-2} \cline{4-15} 
 &
  Zheng et al &
   &
  \multicolumn{1}{r|}{\cellcolor[RGB]{242, 206, 239}0.5795} &
  \cellcolor[RGB]{242, 206, 239}2.82\% &
  \multicolumn{1}{r|}{\cellcolor[RGB]{242, 206, 239}0.7103} &
  \cellcolor[RGB]{242, 206, 239}2.33\% &
  \multicolumn{1}{r|}{\cellcolor[RGB]{242, 206, 239}0.6010} &
  \cellcolor[RGB]{242, 206, 239}6.65\% &
  \multicolumn{1}{r|}{\cellcolor[RGB]{242, 206, 239}0.6621} &
  \cellcolor[RGB]{242, 206, 239}-2.14\% &
  \multicolumn{1}{r|}{\cellcolor[RGB]{242, 206, 239}0.7063} &
  \cellcolor[RGB]{242, 206, 239}2.07\% &
  \multicolumn{1}{r|}{0.6881} &
  4.29\% \\ \cline{2-2} \cline{4-15} 
 &
  MO-OBAM &
  \multirow{-3}{*}{15} &
  \multicolumn{1}{r|}{\cellcolor[RGB]{242, 206, 239}0.5790} &
  \cellcolor[RGB]{242, 206, 239}2.73\% &
  \multicolumn{1}{r|}{\cellcolor[RGB]{242, 206, 239}0.7082} &
  \cellcolor[RGB]{242, 206, 239}2.03\% &
  \multicolumn{1}{r|}{\cellcolor[RGB]{242, 206, 239}0.6038} &
  \cellcolor[RGB]{242, 206, 239}7.15\% &
  \multicolumn{1}{r|}{0.6695} &
  -1.05\% &
  \multicolumn{1}{r|}{\cellcolor[RGB]{242, 206, 239}0.7052} &
  \cellcolor[RGB]{242, 206, 239}1.91\% &
  \multicolumn{1}{r|}{0.6642} &
  0.67\% \\ \cline{2-15} 
 &
  $k$-anonymity &
   &
  \multicolumn{1}{r|}{\cellcolor[RGB]{242, 206, 239}0.5757} &
  \cellcolor[RGB]{242, 206, 239}2.15\% &
  \multicolumn{1}{r|}{\cellcolor[RGB]{242, 206, 239}0.7055} &
  \cellcolor[RGB]{242, 206, 239}1.64\% &
  \multicolumn{1}{r|}{\cellcolor[RGB]{242, 206, 239}0.6004} &
  \cellcolor[RGB]{242, 206, 239}6.55\% &
  \multicolumn{1}{r|}{\cellcolor[RGB]{242, 206, 239}0.6619} &
  \cellcolor[RGB]{242, 206, 239}-2.17\% &
  \multicolumn{1}{r|}{\cellcolor[RGB]{242, 206, 239}0.7033} &
  \cellcolor[RGB]{242, 206, 239}1.63\% &
  \multicolumn{1}{r|}{0.6926} &
  4.97\% \\ \cline{2-2} \cline{4-15} 
 &
  Zheng et al &
   &
  \multicolumn{1}{r|}{\cellcolor[RGB]{242, 206, 239}0.5831} &
  \cellcolor[RGB]{242, 206, 239}3.46\% &
  \multicolumn{1}{r|}{\cellcolor[RGB]{242, 206, 239}0.7065} &
  \cellcolor[RGB]{242, 206, 239}1.79\% &
  \multicolumn{1}{r|}{\cellcolor[RGB]{242, 206, 239}0.6017} &
  \cellcolor[RGB]{242, 206, 239}6.78\% &
  \multicolumn{1}{r|}{0.6752} &
  -0.21\% &
  \multicolumn{1}{r|}{\cellcolor[RGB]{242, 206, 239}0.7018} &
  \cellcolor[RGB]{242, 206, 239}1.42\% &
  \multicolumn{1}{r|}{0.6901} &
  4.59\% \\ \cline{2-2} \cline{4-15} 
\multirow{-12}{*}{FA-NV-only data} &
  MO-OBAM &
  \multirow{-3}{*}{20} &
  \multicolumn{1}{r|}{\cellcolor[RGB]{242, 206, 239}0.5779} &
  \cellcolor[RGB]{242, 206, 239}2.54\% &
  \multicolumn{1}{r|}{\cellcolor[RGB]{242, 206, 239}0.7060} &
  \cellcolor[RGB]{242, 206, 239}1.71\% &
  \multicolumn{1}{r|}{\cellcolor[RGB]{242, 206, 239}0.6041} &
  \cellcolor[RGB]{242, 206, 239}7.20\% &
  \multicolumn{1}{r|}{0.6697} &
  -1.02\% &
  \multicolumn{1}{r|}{\cellcolor[RGB]{242, 206, 239}0.7037} &
  \cellcolor[RGB]{242, 206, 239}1.69\% &
  \multicolumn{1}{r|}{0.6507} &
  -1.38\% \\ \hline
\end{tabular}%
}
    \end{tabular}
\end{table}
\begin{table}[h!]
    \centering
    \caption{OR-NV-only vs FA-NV-only (Recall)}
    \label{tab:OR-NV-only vs FA-NV-only (recall)}
    \begin{tabular}{c}
    (a) Linkage attack with the risk level of $\tau = 0.05$ \\
    \resizebox{\columnwidth}{!}{%
\begin{tabular}{l|l|c|ll|ll|ll|ll|ll|ll}
\hline
Data &
  \begin{tabular}[c]{@{}l@{}}Anonymization \\ Model\end{tabular} &
  \multicolumn{1}{l|}{$k$} &
  \multicolumn{2}{c|}{DT} &
  \multicolumn{2}{c|}{LR} &
  \multicolumn{2}{c|}{NB} &
  \multicolumn{2}{c|}{NN} &
  \multicolumn{2}{c|}{RF} &
  \multicolumn{2}{c}{SVM} \\ \hline
OR-NV-only data &
   &
  \multicolumn{1}{l|}{} &
  \multicolumn{1}{l|}{0.4214} &
   &
  \multicolumn{1}{l|}{0.3624} &
   &
  \multicolumn{1}{l|}{0.7608} &
   &
  \multicolumn{1}{l|}{0.4004} &
   &
  \multicolumn{1}{l|}{0.2446} &
   &
  \multicolumn{1}{l|}{0.3420} &
   \\ \hline
 &
  $k$-anonymity &
   &
  \multicolumn{1}{l|}{\cellcolor[RGB]{242, 206, 239}0.4750} &
  \cellcolor[RGB]{242, 206, 239}12.72\% &
  \multicolumn{1}{l|}{\cellcolor[RGB]{242, 206, 239}0.4585} &
  \cellcolor[RGB]{242, 206, 239}26.52\% &
  \multicolumn{1}{l|}{\cellcolor[RGB]{242, 206, 239}0.5037} &
  \cellcolor[RGB]{242, 206, 239}-33.79\% &
  \multicolumn{1}{l|}{\cellcolor[RGB]{242, 206, 239}0.4991} &
  \cellcolor[RGB]{242, 206, 239}24.65\% &
  \multicolumn{1}{l|}{\cellcolor[RGB]{242, 206, 239}0.3772} &
  \cellcolor[RGB]{242, 206, 239}54.21\% &
  \multicolumn{1}{l|}{\cellcolor[RGB]{242, 206, 239}0.4621} &
  \cellcolor[RGB]{242, 206, 239}35.12\% \\ \cline{2-2} \cline{4-15} 
 &
  Zheng et al &
   &
  \multicolumn{1}{l|}{\cellcolor[RGB]{242, 206, 239}0.5236} &
  \cellcolor[RGB]{242, 206, 239}24.25\% &
  \multicolumn{1}{l|}{\cellcolor[RGB]{242, 206, 239}0.5443} &
  \cellcolor[RGB]{242, 206, 239}50.19\% &
  \multicolumn{1}{l|}{\cellcolor[RGB]{242, 206, 239}0.4345} &
  \cellcolor[RGB]{242, 206, 239}-42.89\% &
  \multicolumn{1}{l|}{\cellcolor[RGB]{242, 206, 239}0.5467} &
  \cellcolor[RGB]{242, 206, 239}36.54\% &
  \multicolumn{1}{l|}{\cellcolor[RGB]{242, 206, 239}0.4982} &
  \cellcolor[RGB]{242, 206, 239}103.68\% &
  \multicolumn{1}{l|}{\cellcolor[RGB]{242, 206, 239}0.5137} &
  \cellcolor[RGB]{242, 206, 239}50.20\% \\ \cline{2-2} \cline{4-15} 
 &
  MO-OBAM &
  \multirow{-3}{*}{5} &
  \multicolumn{1}{l|}{\cellcolor[RGB]{242, 206, 239}0.5413} &
  \cellcolor[RGB]{242, 206, 239}28.45\% &
  \multicolumn{1}{l|}{\cellcolor[RGB]{242, 206, 239}0.5599} &
  \cellcolor[RGB]{242, 206, 239}54.50\% &
  \multicolumn{1}{l|}{\cellcolor[RGB]{242, 206, 239}0.4428} &
  \cellcolor[RGB]{242, 206, 239}-41.80\% &
  \multicolumn{1}{l|}{\cellcolor[RGB]{242, 206, 239}0.5731} &
  \cellcolor[RGB]{242, 206, 239}43.13\% &
  \multicolumn{1}{l|}{\cellcolor[RGB]{242, 206, 239}0.5325} &
  \cellcolor[RGB]{242, 206, 239}117.70\% &
  \multicolumn{1}{l|}{\cellcolor[RGB]{242, 206, 239}0.5323} &
  \cellcolor[RGB]{242, 206, 239}55.64\% \\ \cline{2-15} 
 &
  $k$-anonymity &
   &
  \multicolumn{1}{l|}{\cellcolor[RGB]{242, 206, 239}0.4971} &
  \cellcolor[RGB]{242, 206, 239}17.96\% &
  \multicolumn{1}{l|}{\cellcolor[RGB]{242, 206, 239}0.4771} &
  \cellcolor[RGB]{242, 206, 239}31.65\% &
  \multicolumn{1}{l|}{\cellcolor[RGB]{242, 206, 239}0.4425} &
  \cellcolor[RGB]{242, 206, 239}-41.84\% &
  \multicolumn{1}{l|}{\cellcolor[RGB]{242, 206, 239}0.5006} &
  \cellcolor[RGB]{242, 206, 239}25.02\% &
  \multicolumn{1}{l|}{\cellcolor[RGB]{242, 206, 239}0.4039} &
  \cellcolor[RGB]{242, 206, 239}65.13\% &
  \multicolumn{1}{l|}{0.4355} &
  27.34\% \\ \cline{2-2} \cline{4-15} 
 &
  Zheng et al &
   &
  \multicolumn{1}{l|}{\cellcolor[RGB]{242, 206, 239}0.5216} &
  \cellcolor[RGB]{242, 206, 239}23.78\% &
  \multicolumn{1}{l|}{\cellcolor[RGB]{242, 206, 239}0.5449} &
  \cellcolor[RGB]{242, 206, 239}50.36\% &
  \multicolumn{1}{l|}{\cellcolor[RGB]{242, 206, 239}0.4337} &
  \cellcolor[RGB]{242, 206, 239}-42.99\% &
  \multicolumn{1}{l|}{\cellcolor[RGB]{242, 206, 239}0.5643} &
  \cellcolor[RGB]{242, 206, 239}40.93\% &
  \multicolumn{1}{l|}{\cellcolor[RGB]{242, 206, 239}0.4991} &
  \cellcolor[RGB]{242, 206, 239}104.05\% &
  \multicolumn{1}{l|}{\cellcolor[RGB]{242, 206, 239}0.4752} &
  \cellcolor[RGB]{242, 206, 239}38.95\% \\ \cline{2-2} \cline{4-15} 
 &
  MO-OBAM &
  \multirow{-3}{*}{10} &
  \multicolumn{1}{l|}{\cellcolor[RGB]{242, 206, 239}0.5642} &
  \cellcolor[RGB]{242, 206, 239}33.89\% &
  \multicolumn{1}{l|}{\cellcolor[RGB]{242, 206, 239}0.6023} &
  \cellcolor[RGB]{242, 206, 239}66.20\% &
  \multicolumn{1}{l|}{\cellcolor[RGB]{242, 206, 239}0.4253} &
  \cellcolor[RGB]{242, 206, 239}-44.10\% &
  \multicolumn{1}{l|}{\cellcolor[RGB]{242, 206, 239}0.6028} &
  \cellcolor[RGB]{242, 206, 239}50.55\% &
  \multicolumn{1}{l|}{\cellcolor[RGB]{242, 206, 239}0.5785} &
  \cellcolor[RGB]{242, 206, 239}136.51\% &
  \multicolumn{1}{l|}{\cellcolor[RGB]{242, 206, 239}0.5484} &
  \cellcolor[RGB]{242, 206, 239}60.35\% \\ \cline{2-15} 
 &
  $k$-anonymity &
   &
  \multicolumn{1}{l|}{\cellcolor[RGB]{242, 206, 239}0.4769} &
  \cellcolor[RGB]{242, 206, 239}13.17\% &
  \multicolumn{1}{l|}{\cellcolor[RGB]{242, 206, 239}0.4651} &
  \cellcolor[RGB]{242, 206, 239}28.34\% &
  \multicolumn{1}{l|}{\cellcolor[RGB]{242, 206, 239}0.6782} &
  \cellcolor[RGB]{242, 206, 239}-10.86\% &
  \multicolumn{1}{l|}{\cellcolor[RGB]{242, 206, 239}0.4708} &
  \cellcolor[RGB]{242, 206, 239}17.58\% &
  \multicolumn{1}{l|}{\cellcolor[RGB]{242, 206, 239}0.3789} &
  \cellcolor[RGB]{242, 206, 239}54.91\% &
  \multicolumn{1}{l|}{\cellcolor[RGB]{242, 206, 239}0.4355} &
  \cellcolor[RGB]{242, 206, 239}27.34\% \\ \cline{2-2} \cline{4-15} 
 &
  Zheng et al &
   &
  \multicolumn{1}{l|}{\cellcolor[RGB]{242, 206, 239}0.5189} &
  \cellcolor[RGB]{242, 206, 239}23.14\% &
  \multicolumn{1}{l|}{\cellcolor[RGB]{242, 206, 239}0.5400} &
  \cellcolor[RGB]{242, 206, 239}49.01\% &
  \multicolumn{1}{l|}{\cellcolor[RGB]{242, 206, 239}0.4130} &
  \cellcolor[RGB]{242, 206, 239}-45.72\% &
  \multicolumn{1}{l|}{\cellcolor[RGB]{242, 206, 239}0.5567} &
  \cellcolor[RGB]{242, 206, 239}39.04\% &
  \multicolumn{1}{l|}{\cellcolor[RGB]{242, 206, 239}0.4974} &
  \cellcolor[RGB]{242, 206, 239}103.35\% &
  \multicolumn{1}{l|}{\cellcolor[RGB]{242, 206, 239}0.5318} &
  \cellcolor[RGB]{242, 206, 239}55.50\% \\ \cline{2-2} \cline{4-15} 
 &
  MO-OBAM &
  \multirow{-3}{*}{15} &
  \multicolumn{1}{l|}{\cellcolor[RGB]{242, 206, 239}0.5670} &
  \cellcolor[RGB]{242, 206, 239}34.55\% &
  \multicolumn{1}{l|}{\cellcolor[RGB]{242, 206, 239}0.6158} &
  \cellcolor[RGB]{242, 206, 239}69.92\% &
  \multicolumn{1}{l|}{\cellcolor[RGB]{242, 206, 239}0.4409} &
  \cellcolor[RGB]{242, 206, 239}-42.05\% &
  \multicolumn{1}{l|}{\cellcolor[RGB]{242, 206, 239}0.6169} &
  \cellcolor[RGB]{242, 206, 239}54.07\% &
  \multicolumn{1}{l|}{\cellcolor[RGB]{242, 206, 239}0.5930} &
  \cellcolor[RGB]{242, 206, 239}142.44\% &
  \multicolumn{1}{l|}{\cellcolor[RGB]{242, 206, 239}0.5888} &
  \cellcolor[RGB]{242, 206, 239}72.16\% \\ \cline{2-15} 
 &
  $k$-anonymity &
   &
  \multicolumn{1}{l|}{\cellcolor[RGB]{242, 206, 239}0.5749} &
  \cellcolor[RGB]{242, 206, 239}36.43\% &
  \multicolumn{1}{l|}{\cellcolor[RGB]{242, 206, 239}0.6173} &
  \cellcolor[RGB]{242, 206, 239}70.34\% &
  \multicolumn{1}{l|}{\cellcolor[RGB]{242, 206, 239}0.4389} &
  \cellcolor[RGB]{242, 206, 239}-42.31\% &
  \multicolumn{1}{l|}{\cellcolor[RGB]{242, 206, 239}0.6195} &
  \cellcolor[RGB]{242, 206, 239}54.72\% &
  \multicolumn{1}{l|}{\cellcolor[RGB]{242, 206, 239}0.5999} &
  \cellcolor[RGB]{242, 206, 239}145.26\% &
  \multicolumn{1}{l|}{\cellcolor[RGB]{242, 206, 239}0.5443} &
  \cellcolor[RGB]{242, 206, 239}59.15\% \\ \cline{2-2} \cline{4-15} 
 &
  Zheng et al &
   &
  \multicolumn{1}{l|}{\cellcolor[RGB]{242, 206, 239}0.5687} &
  \cellcolor[RGB]{242, 206, 239}34.95\% &
  \multicolumn{1}{l|}{\cellcolor[RGB]{242, 206, 239}0.6210} &
  \cellcolor[RGB]{242, 206, 239}71.36\% &
  \multicolumn{1}{l|}{\cellcolor[RGB]{242, 206, 239}0.4369} &
  \cellcolor[RGB]{242, 206, 239}-42.57\% &
  \multicolumn{1}{l|}{\cellcolor[RGB]{242, 206, 239}0.6091} &
  \cellcolor[RGB]{242, 206, 239}52.12\% &
  \multicolumn{1}{l|}{\cellcolor[RGB]{242, 206, 239}0.5976} &
  \cellcolor[RGB]{242, 206, 239}144.32\% &
  \multicolumn{1}{l|}{\cellcolor[RGB]{242, 206, 239}0.5560} &
  \cellcolor[RGB]{242, 206, 239}62.57\% \\ \cline{2-2} \cline{4-15} 
\multirow{-12}{*}{FA-NV-only data} &
  MO-OBAM &
  \multirow{-3}{*}{20} &
  \multicolumn{1}{l|}{\cellcolor[RGB]{242, 206, 239}0.5704} &
  \cellcolor[RGB]{242, 206, 239}35.36\% &
  \multicolumn{1}{l|}{\cellcolor[RGB]{242, 206, 239}0.6172} &
  \cellcolor[RGB]{242, 206, 239}70.31\% &
  \multicolumn{1}{l|}{\cellcolor[RGB]{242, 206, 239}0.4475} &
  \cellcolor[RGB]{242, 206, 239}-41.18\% &
  \multicolumn{1}{l|}{\cellcolor[RGB]{242, 206, 239}0.6085} &
  \cellcolor[RGB]{242, 206, 239}51.97\% &
  \multicolumn{1}{l|}{\cellcolor[RGB]{242, 206, 239}0.5972} &
  \cellcolor[RGB]{242, 206, 239}144.15\% &
  \multicolumn{1}{l|}{\cellcolor[RGB]{242, 206, 239}0.5206} &
  \cellcolor[RGB]{242, 206, 239}52.22\% \\ \hline
\end{tabular}%
} \\
    \end{tabular}
\end{table}
\begin{table}[h!]
    \centering
    \begin{tabular}{c}
    (b) Linkage attack with the risk level of $\tau = 0.075$ \\
    \resizebox{\columnwidth}{!}{%
\begin{tabular}{l|l|c|ll|ll|ll|ll|ll|ll}
\hline
Data &
  \begin{tabular}[c]{@{}l@{}}Anonymization \\ Model\end{tabular} &
  \multicolumn{1}{l|}{$k$} &
  \multicolumn{2}{c|}{DT} &
  \multicolumn{2}{c|}{LR} &
  \multicolumn{2}{c|}{NB} &
  \multicolumn{2}{c|}{NN} &
  \multicolumn{2}{c|}{RF} &
  \multicolumn{2}{c}{SVM} \\ \hline
OR-NV-only data &
   &
  \multicolumn{1}{l|}{} &
  \multicolumn{1}{l|}{0.4586} &
   &
  \multicolumn{1}{l|}{0.3893} &
   &
  \multicolumn{1}{l|}{0.6099} &
   &
  \multicolumn{1}{l|}{0.4154} &
   &
  \multicolumn{1}{l|}{0.2721} &
   &
  \multicolumn{1}{l|}{0.4021} &
   \\ \hline
 &
  $k$-anonymity &
   &
  \multicolumn{1}{l|}{\cellcolor[RGB]{242, 206, 239}0.4790} &
  \cellcolor[RGB]{242, 206, 239}4.45\% &
  \multicolumn{1}{l|}{\cellcolor[RGB]{242, 206, 239}0.4568} &
  \cellcolor[RGB]{242, 206, 239}17.34\% &
  \multicolumn{1}{l|}{\cellcolor[RGB]{242, 206, 239}0.4332} &
  \cellcolor[RGB]{242, 206, 239}-28.97\% &
  \multicolumn{1}{l|}{\cellcolor[RGB]{242, 206, 239}0.4961} &
  \cellcolor[RGB]{242, 206, 239}19.43\% &
  \multicolumn{1}{l|}{\cellcolor[RGB]{242, 206, 239}0.3729} &
  \cellcolor[RGB]{242, 206, 239}37.05\% &
  \multicolumn{1}{l|}{\cellcolor[RGB]{242, 206, 239}0.5145} &
  \cellcolor[RGB]{242, 206, 239}27.95\% \\ \cline{2-2} \cline{4-15} 
 &
  Zheng et al &
   &
  \multicolumn{1}{l|}{\cellcolor[RGB]{242, 206, 239}0.5362} &
  \cellcolor[RGB]{242, 206, 239}16.92\% &
  \multicolumn{1}{l|}{\cellcolor[RGB]{242, 206, 239}0.5632} &
  \cellcolor[RGB]{242, 206, 239}44.67\% &
  \multicolumn{1}{l|}{\cellcolor[RGB]{242, 206, 239}0.4173} &
  \cellcolor[RGB]{242, 206, 239}-31.58\% &
  \multicolumn{1}{l|}{\cellcolor[RGB]{242, 206, 239}0.5672} &
  \cellcolor[RGB]{242, 206, 239}36.54\% &
  \multicolumn{1}{l|}{\cellcolor[RGB]{242, 206, 239}0.5258} &
  \cellcolor[RGB]{242, 206, 239}93.24\% &
  \multicolumn{1}{l|}{0.4868} &
  21.06\% \\ \cline{2-2} \cline{4-15} 
 &
  MO-OBAM &
  \multirow{-3}{*}{5} &
  \multicolumn{1}{l|}{\cellcolor[RGB]{242, 206, 239}0.5554} &
  \cellcolor[RGB]{242, 206, 239}21.11\% &
  \multicolumn{1}{l|}{\cellcolor[RGB]{242, 206, 239}0.5922} &
  \cellcolor[RGB]{242, 206, 239}52.12\% &
  \multicolumn{1}{l|}{\cellcolor[RGB]{242, 206, 239}0.4679} &
  \cellcolor[RGB]{242, 206, 239}-23.28\% &
  \multicolumn{1}{l|}{\cellcolor[RGB]{242, 206, 239}0.5806} &
  \cellcolor[RGB]{242, 206, 239}39.77\% &
  \multicolumn{1}{l|}{\cellcolor[RGB]{242, 206, 239}0.5695} &
  \cellcolor[RGB]{242, 206, 239}109.30\% &
  \multicolumn{1}{l|}{\cellcolor[RGB]{242, 206, 239}0.5772} &
  \cellcolor[RGB]{242, 206, 239}43.55\% \\ \cline{2-15} 
 &
  $k$-anonymity &
   &
  \multicolumn{1}{l|}{\cellcolor[RGB]{242, 206, 239}0.5026} &
  \cellcolor[RGB]{242, 206, 239}9.59\% &
  \multicolumn{1}{l|}{\cellcolor[RGB]{242, 206, 239}0.4804} &
  \cellcolor[RGB]{242, 206, 239}23.40\% &
  \multicolumn{1}{l|}{\cellcolor[RGB]{242, 206, 239}0.4376} &
  \cellcolor[RGB]{242, 206, 239}-28.25\% &
  \multicolumn{1}{l|}{\cellcolor[RGB]{242, 206, 239}0.5089} &
  \cellcolor[RGB]{242, 206, 239}22.51\% &
  \multicolumn{1}{l|}{\cellcolor[RGB]{242, 206, 239}0.4065} &
  \cellcolor[RGB]{242, 206, 239}49.39\% &
  \multicolumn{1}{l|}{\cellcolor[RGB]{242, 206, 239}0.5109} &
  \cellcolor[RGB]{242, 206, 239}27.06\% \\ \cline{2-2} \cline{4-15} 
 &
  Zheng et al &
   &
  \multicolumn{1}{l|}{\cellcolor[RGB]{242, 206, 239}0.5344} &
  \cellcolor[RGB]{242, 206, 239}16.53\% &
  \multicolumn{1}{l|}{\cellcolor[RGB]{242, 206, 239}0.5585} &
  \cellcolor[RGB]{242, 206, 239}43.46\% &
  \multicolumn{1}{l|}{\cellcolor[RGB]{242, 206, 239}0.4341} &
  \cellcolor[RGB]{242, 206, 239}-28.82\% &
  \multicolumn{1}{l|}{\cellcolor[RGB]{242, 206, 239}0.5694} &
  \cellcolor[RGB]{242, 206, 239}37.07\% &
  \multicolumn{1}{l|}{\cellcolor[RGB]{242, 206, 239}0.5143} &
  \cellcolor[RGB]{242, 206, 239}89.01\% &
  \multicolumn{1}{l|}{\cellcolor[RGB]{242, 206, 239}0.5145} &
  \cellcolor[RGB]{242, 206, 239}27.95\% \\ \cline{2-2} \cline{4-15} 
 &
  MO-OBAM &
  \multirow{-3}{*}{10} &
  \multicolumn{1}{l|}{\cellcolor[RGB]{242, 206, 239}0.5672} &
  \cellcolor[RGB]{242, 206, 239}23.68\% &
  \multicolumn{1}{l|}{\cellcolor[RGB]{242, 206, 239}0.6113} &
  \cellcolor[RGB]{242, 206, 239}57.03\% &
  \multicolumn{1}{l|}{\cellcolor[RGB]{242, 206, 239}0.4341} &
  \cellcolor[RGB]{242, 206, 239}-28.82\% &
  \multicolumn{1}{l|}{\cellcolor[RGB]{242, 206, 239}0.6050} &
  \cellcolor[RGB]{242, 206, 239}45.64\% &
  \multicolumn{1}{l|}{\cellcolor[RGB]{242, 206, 239}0.5927} &
  \cellcolor[RGB]{242, 206, 239}117.82\% &
  \multicolumn{1}{l|}{\cellcolor[RGB]{242, 206, 239}0.5998} &
  \cellcolor[RGB]{242, 206, 239}49.17\% \\ \cline{2-15} 
 &
  $k$-anonymity &
   &
  \multicolumn{1}{l|}{\cellcolor[RGB]{242, 206, 239}0.5699} &
  \cellcolor[RGB]{242, 206, 239}24.27\% &
  \multicolumn{1}{l|}{\cellcolor[RGB]{242, 206, 239}0.6175} &
  \cellcolor[RGB]{242, 206, 239}58.62\% &
  \multicolumn{1}{l|}{\cellcolor[RGB]{242, 206, 239}0.4357} &
  \cellcolor[RGB]{242, 206, 239}-28.56\% &
  \multicolumn{1}{l|}{\cellcolor[RGB]{242, 206, 239}0.6140} &
  \cellcolor[RGB]{242, 206, 239}47.81\% &
  \multicolumn{1}{l|}{\cellcolor[RGB]{242, 206, 239}0.6037} &
  \cellcolor[RGB]{242, 206, 239}121.87\% &
  \multicolumn{1}{l|}{\cellcolor[RGB]{242, 206, 239}0.6219} &
  \cellcolor[RGB]{242, 206, 239}54.66\% \\ \cline{2-2} \cline{4-15} 
 &
  Zheng et al &
   &
  \multicolumn{1}{l|}{\cellcolor[RGB]{242, 206, 239}0.5784} &
  \cellcolor[RGB]{242, 206, 239}26.12\% &
  \multicolumn{1}{l|}{\cellcolor[RGB]{242, 206, 239}0.6190} &
  \cellcolor[RGB]{242, 206, 239}59.00\% &
  \multicolumn{1}{l|}{\cellcolor[RGB]{242, 206, 239}0.4417} &
  \cellcolor[RGB]{242, 206, 239}-27.58\% &
  \multicolumn{1}{l|}{\cellcolor[RGB]{242, 206, 239}0.6055} &
  \cellcolor[RGB]{242, 206, 239}45.76\% &
  \multicolumn{1}{l|}{\cellcolor[RGB]{242, 206, 239}0.6021} &
  \cellcolor[RGB]{242, 206, 239}121.28\% &
  \multicolumn{1}{l|}{\cellcolor[RGB]{242, 206, 239}0.5982} &
  \cellcolor[RGB]{242, 206, 239}48.77\% \\ \cline{2-2} \cline{4-15} 
 &
  MO-OBAM &
  \multirow{-3}{*}{15} &
  \multicolumn{1}{l|}{\cellcolor[RGB]{242, 206, 239}0.5758} &
  \cellcolor[RGB]{242, 206, 239}25.56\% &
  \multicolumn{1}{l|}{\cellcolor[RGB]{242, 206, 239}0.6209} &
  \cellcolor[RGB]{242, 206, 239}59.49\% &
  \multicolumn{1}{l|}{\cellcolor[RGB]{242, 206, 239}0.4452} &
  \cellcolor[RGB]{242, 206, 239}-27.00\% &
  \multicolumn{1}{l|}{\cellcolor[RGB]{242, 206, 239}0.6307} &
  \cellcolor[RGB]{242, 206, 239}51.83\% &
  \multicolumn{1}{l|}{\cellcolor[RGB]{242, 206, 239}0.5992} &
  \cellcolor[RGB]{242, 206, 239}120.21\% &
  \multicolumn{1}{l|}{\cellcolor[RGB]{242, 206, 239}0.5610} &
  \cellcolor[RGB]{242, 206, 239}39.52\% \\ \cline{2-15} 
 &
  $k$-anonymity &
   &
  \multicolumn{1}{l|}{\cellcolor[RGB]{242, 206, 239}0.5747} &
  \cellcolor[RGB]{242, 206, 239}25.32\% &
  \multicolumn{1}{l|}{\cellcolor[RGB]{242, 206, 239}0.6111} &
  \cellcolor[RGB]{242, 206, 239}56.97\% &
  \multicolumn{1}{l|}{\cellcolor[RGB]{242, 206, 239}0.4398} &
  \cellcolor[RGB]{242, 206, 239}-27.89\% &
  \multicolumn{1}{l|}{\cellcolor[RGB]{242, 206, 239}0.6237} &
  \cellcolor[RGB]{242, 206, 239}50.14\% &
  \multicolumn{1}{l|}{\cellcolor[RGB]{242, 206, 239}0.6047} &
  \cellcolor[RGB]{242, 206, 239}122.23\% &
  \multicolumn{1}{l|}{\cellcolor[RGB]{242, 206, 239}0.5181} &
  \cellcolor[RGB]{242, 206, 239}28.85\% \\ \cline{2-2} \cline{4-15} 
 &
  Zheng et al &
   &
  \multicolumn{1}{l|}{\cellcolor[RGB]{242, 206, 239}0.5716} &
  \cellcolor[RGB]{242, 206, 239}24.64\% &
  \multicolumn{1}{l|}{\cellcolor[RGB]{242, 206, 239}0.6203} &
  \cellcolor[RGB]{242, 206, 239}59.34\% &
  \multicolumn{1}{l|}{\cellcolor[RGB]{242, 206, 239}0.4426} &
  \cellcolor[RGB]{242, 206, 239}-27.43\% &
  \multicolumn{1}{l|}{\cellcolor[RGB]{242, 206, 239}0.6070} &
  \cellcolor[RGB]{242, 206, 239}46.12\% &
  \multicolumn{1}{l|}{\cellcolor[RGB]{242, 206, 239}0.5977} &
  \cellcolor[RGB]{242, 206, 239}119.66\% &
  \multicolumn{1}{l|}{\cellcolor[RGB]{242, 206, 239}0.6043} &
  \cellcolor[RGB]{242, 206, 239}50.29\% \\ \cline{2-2} \cline{4-15} 
\multirow{-12}{*}{FA-NV-only data} &
  MO-OBAM &
  \multirow{-3}{*}{20} &
  \multicolumn{1}{l|}{\cellcolor[RGB]{242, 206, 239}0.5721} &
  \cellcolor[RGB]{242, 206, 239}24.75\% &
  \multicolumn{1}{l|}{\cellcolor[RGB]{242, 206, 239}0.6185} &
  \cellcolor[RGB]{242, 206, 239}58.87\% &
  \multicolumn{1}{l|}{\cellcolor[RGB]{242, 206, 239}0.4374} &
  \cellcolor[RGB]{242, 206, 239}-28.28\% &
  \multicolumn{1}{l|}{\cellcolor[RGB]{242, 206, 239}0.6133} &
  \cellcolor[RGB]{242, 206, 239}47.64\% &
  \multicolumn{1}{l|}{\cellcolor[RGB]{242, 206, 239}0.5992} &
  \cellcolor[RGB]{242, 206, 239}120.21\% &
  \multicolumn{1}{l|}{\cellcolor[RGB]{242, 206, 239}0.6167} &
  \cellcolor[RGB]{242, 206, 239}53.37\% \\ \hline
\end{tabular}%
} \\
    \end{tabular}
\end{table}
\begin{table}[h!]
    \centering
    \begin{tabular}{c}
    (c) Linkage attack with the risk level of $\tau = 0.1$ \\
    \resizebox{\columnwidth}{!}{%
\begin{tabular}{l|l|c|ll|ll|ll|ll|ll|ll}
\hline
Data &
  \begin{tabular}[c]{@{}l@{}}Anonymization \\ Model\end{tabular} &
  \multicolumn{1}{l|}{$k$} &
  \multicolumn{2}{c|}{DT} &
  \multicolumn{2}{c|}{LR} &
  \multicolumn{2}{c|}{NB} &
  \multicolumn{2}{c|}{NN} &
  \multicolumn{2}{c|}{RF} &
  \multicolumn{2}{c}{SVM} \\ \hline
OR-NV-only data &
   &
  \multicolumn{1}{l|}{} &
  \multicolumn{1}{l|}{0.4606} &
   &
  \multicolumn{1}{l|}{0.4100} &
   &
  \multicolumn{1}{l|}{0.6481} &
   &
  \multicolumn{1}{l|}{0.4549} &
   &
  \multicolumn{1}{l|}{0.3211} &
   &
  \multicolumn{1}{l|}{0.4508} &
   \\ \hline
 &
  $k$-anonymity &
   &
  \multicolumn{1}{l|}{\cellcolor[RGB]{242, 206, 239}0.4829} &
  \cellcolor[RGB]{242, 206, 239}4.84\% &
  \multicolumn{1}{l|}{\cellcolor[RGB]{242, 206, 239}0.4597} &
  \cellcolor[RGB]{242, 206, 239}12.12\% &
  \multicolumn{1}{l|}{\cellcolor[RGB]{242, 206, 239}0.3950} &
  \cellcolor[RGB]{242, 206, 239}-39.05\% &
  \multicolumn{1}{l|}{\cellcolor[RGB]{242, 206, 239}0.4967} &
  \cellcolor[RGB]{242, 206, 239}9.19\% &
  \multicolumn{1}{l|}{\cellcolor[RGB]{242, 206, 239}0.4056} &
  \cellcolor[RGB]{242, 206, 239}26.32\% &
  \multicolumn{1}{l|}{0.5135} &
  13.91\% \\ \cline{2-2} \cline{4-15} 
 &
  Zheng et al &
   &
  \multicolumn{1}{l|}{\cellcolor[RGB]{242, 206, 239}0.5503} &
  \cellcolor[RGB]{242, 206, 239}19.47\% &
  \multicolumn{1}{l|}{\cellcolor[RGB]{242, 206, 239}0.5760} &
  \cellcolor[RGB]{242, 206, 239}40.49\% &
  \multicolumn{1}{l|}{\cellcolor[RGB]{242, 206, 239}0.4235} &
  \cellcolor[RGB]{242, 206, 239}-34.66\% &
  \multicolumn{1}{l|}{\cellcolor[RGB]{242, 206, 239}0.5738} &
  \cellcolor[RGB]{242, 206, 239}26.14\% &
  \multicolumn{1}{l|}{\cellcolor[RGB]{242, 206, 239}0.5518} &
  \cellcolor[RGB]{242, 206, 239}71.85\% &
  \multicolumn{1}{l|}{0.5280} &
  17.13\% \\ \cline{2-2} \cline{4-15} 
 &
  MO-OBAM &
  \multirow{-3}{*}{5} &
  \multicolumn{1}{l|}{\cellcolor[RGB]{242, 206, 239}0.5690} &
  \cellcolor[RGB]{242, 206, 239}23.53\% &
  \multicolumn{1}{l|}{\cellcolor[RGB]{242, 206, 239}0.6121} &
  \cellcolor[RGB]{242, 206, 239}49.29\% &
  \multicolumn{1}{l|}{\cellcolor[RGB]{242, 206, 239}0.4527} &
  \cellcolor[RGB]{242, 206, 239}-30.15\% &
  \multicolumn{1}{l|}{\cellcolor[RGB]{242, 206, 239}0.6080} &
  \cellcolor[RGB]{242, 206, 239}33.66\% &
  \multicolumn{1}{l|}{\cellcolor[RGB]{242, 206, 239}0.5949} &
  \cellcolor[RGB]{242, 206, 239}85.27\% &
  \multicolumn{1}{l|}{0.5494} &
  21.87\% \\ \cline{2-15} 
 &
  $k$-anonymity &
   &
  \multicolumn{1}{l|}{\cellcolor[RGB]{242, 206, 239}0.5773} &
  \cellcolor[RGB]{242, 206, 239}25.34\% &
  \multicolumn{1}{l|}{\cellcolor[RGB]{242, 206, 239}0.6168} &
  \cellcolor[RGB]{242, 206, 239}50.44\% &
  \multicolumn{1}{l|}{\cellcolor[RGB]{242, 206, 239}0.4401} &
  \cellcolor[RGB]{242, 206, 239}-32.09\% &
  \multicolumn{1}{l|}{\cellcolor[RGB]{242, 206, 239}0.6113} &
  \cellcolor[RGB]{242, 206, 239}34.38\% &
  \multicolumn{1}{l|}{\cellcolor[RGB]{242, 206, 239}0.6025} &
  \cellcolor[RGB]{242, 206, 239}87.64\% &
  \multicolumn{1}{l|}{\cellcolor[RGB]{242, 206, 239}0.6185} &
  \cellcolor[RGB]{242, 206, 239}37.20\% \\ \cline{2-2} \cline{4-15} 
 &
  Zheng et al &
   &
  \multicolumn{1}{l|}{\cellcolor[RGB]{242, 206, 239}0.5755} &
  \cellcolor[RGB]{242, 206, 239}24.95\% &
  \multicolumn{1}{l|}{\cellcolor[RGB]{242, 206, 239}0.6176} &
  \cellcolor[RGB]{242, 206, 239}50.63\% &
  \multicolumn{1}{l|}{\cellcolor[RGB]{242, 206, 239}0.4393} &
  \cellcolor[RGB]{242, 206, 239}-32.22\% &
  \multicolumn{1}{l|}{\cellcolor[RGB]{242, 206, 239}0.6131} &
  \cellcolor[RGB]{242, 206, 239}34.78\% &
  \multicolumn{1}{l|}{\cellcolor[RGB]{242, 206, 239}0.6018} &
  \cellcolor[RGB]{242, 206, 239}87.42\% &
  \multicolumn{1}{l|}{0.5218} &
  15.75\% \\ \cline{2-2} \cline{4-15} 
 &
  MO-OBAM &
  \multirow{-3}{*}{10} &
  \multicolumn{1}{l|}{\cellcolor[RGB]{242, 206, 239}0.5727} &
  \cellcolor[RGB]{242, 206, 239}24.34\% &
  \multicolumn{1}{l|}{\cellcolor[RGB]{242, 206, 239}0.6175} &
  \cellcolor[RGB]{242, 206, 239}50.61\% &
  \multicolumn{1}{l|}{\cellcolor[RGB]{242, 206, 239}0.4440} &
  \cellcolor[RGB]{242, 206, 239}-31.49\% &
  \multicolumn{1}{l|}{\cellcolor[RGB]{242, 206, 239}0.6172} &
  \cellcolor[RGB]{242, 206, 239}35.68\% &
  \multicolumn{1}{l|}{\cellcolor[RGB]{242, 206, 239}0.5990} &
  \cellcolor[RGB]{242, 206, 239}86.55\% &
  \multicolumn{1}{l|}{\cellcolor[RGB]{242, 206, 239}0.5772} &
  \cellcolor[RGB]{242, 206, 239}28.04\% \\ \cline{2-15} 
 &
  $k$-anonymity &
   &
  \multicolumn{1}{l|}{\cellcolor[RGB]{242, 206, 239}0.5689} &
  \cellcolor[RGB]{242, 206, 239}23.51\% &
  \multicolumn{1}{l|}{\cellcolor[RGB]{242, 206, 239}0.6152} &
  \cellcolor[RGB]{242, 206, 239}50.05\% &
  \multicolumn{1}{l|}{\cellcolor[RGB]{242, 206, 239}0.4447} &
  \cellcolor[RGB]{242, 206, 239}-31.38\% &
  \multicolumn{1}{l|}{\cellcolor[RGB]{242, 206, 239}0.6025} &
  \cellcolor[RGB]{242, 206, 239}32.45\% &
  \multicolumn{1}{l|}{\cellcolor[RGB]{242, 206, 239}0.5999} &
  \cellcolor[RGB]{242, 206, 239}86.83\% &
  \multicolumn{1}{l|}{0.5419} &
  20.21\% \\ \cline{2-2} \cline{4-15} 
 &
  Zheng et al &
   &
  \multicolumn{1}{l|}{\cellcolor[RGB]{242, 206, 239}0.5738} &
  \cellcolor[RGB]{242, 206, 239}24.58\% &
  \multicolumn{1}{l|}{\cellcolor[RGB]{242, 206, 239}0.6169} &
  \cellcolor[RGB]{242, 206, 239}50.46\% &
  \multicolumn{1}{l|}{\cellcolor[RGB]{242, 206, 239}0.4382} &
  \cellcolor[RGB]{242, 206, 239}-32.39\% &
  \multicolumn{1}{l|}{\cellcolor[RGB]{242, 206, 239}0.6175} &
  \cellcolor[RGB]{242, 206, 239}35.74\% &
  \multicolumn{1}{l|}{\cellcolor[RGB]{242, 206, 239}0.6026} &
  \cellcolor[RGB]{242, 206, 239}87.67\% &
  \multicolumn{1}{l|}{\cellcolor[RGB]{242, 206, 239}0.5942} &
  \cellcolor[RGB]{242, 206, 239}31.81\% \\ \cline{2-2} \cline{4-15} 
 &
  MO-OBAM &
  \multirow{-3}{*}{15} &
  \multicolumn{1}{l|}{\cellcolor[RGB]{242, 206, 239}0.5750} &
  \cellcolor[RGB]{242, 206, 239}24.84\% &
  \multicolumn{1}{l|}{\cellcolor[RGB]{242, 206, 239}0.6192} &
  \cellcolor[RGB]{242, 206, 239}51.02\% &
  \multicolumn{1}{l|}{\cellcolor[RGB]{242, 206, 239}0.4416} &
  \cellcolor[RGB]{242, 206, 239}-31.86\% &
  \multicolumn{1}{l|}{\cellcolor[RGB]{242, 206, 239}0.6087} &
  \cellcolor[RGB]{242, 206, 239}33.81\% &
  \multicolumn{1}{l|}{\cellcolor[RGB]{242, 206, 239}0.6001} &
  \cellcolor[RGB]{242, 206, 239}86.89\% &
  \multicolumn{1}{l|}{\cellcolor[RGB]{242, 206, 239}0.5741} &
  \cellcolor[RGB]{242, 206, 239}27.35\% \\ \cline{2-15} 
 &
  $k$-anonymity &
   &
  \multicolumn{1}{l|}{\cellcolor[RGB]{242, 206, 239}0.5759} &
  \cellcolor[RGB]{242, 206, 239}25.03\% &
  \multicolumn{1}{l|}{\cellcolor[RGB]{242, 206, 239}0.6181} &
  \cellcolor[RGB]{242, 206, 239}50.76\% &
  \multicolumn{1}{l|}{\cellcolor[RGB]{242, 206, 239}0.4420} &
  \cellcolor[RGB]{242, 206, 239}-31.80\% &
  \multicolumn{1}{l|}{\cellcolor[RGB]{242, 206, 239}0.6040} &
  \cellcolor[RGB]{242, 206, 239}32.78\% &
  \multicolumn{1}{l|}{\cellcolor[RGB]{242, 206, 239}0.6022} &
  \cellcolor[RGB]{242, 206, 239}87.54\% &
  \multicolumn{1}{l|}{\cellcolor[RGB]{242, 206, 239}0.5735} &
  \cellcolor[RGB]{242, 206, 239}27.22\% \\ \cline{2-2} \cline{4-15} 
 &
  Zheng et al &
   &
  \multicolumn{1}{l|}{\cellcolor[RGB]{242, 206, 239}0.5723} &
  \cellcolor[RGB]{242, 206, 239}24.25\% &
  \multicolumn{1}{l|}{\cellcolor[RGB]{242, 206, 239}0.6207} &
  \cellcolor[RGB]{242, 206, 239}51.39\% &
  \multicolumn{1}{l|}{\cellcolor[RGB]{242, 206, 239}0.4415} &
  \cellcolor[RGB]{242, 206, 239}-31.88\% &
  \multicolumn{1}{l|}{\cellcolor[RGB]{242, 206, 239}0.6190} &
  \cellcolor[RGB]{242, 206, 239}36.07\% &
  \multicolumn{1}{l|}{\cellcolor[RGB]{242, 206, 239}0.5985} &
  \cellcolor[RGB]{242, 206, 239}86.39\% &
  \multicolumn{1}{l|}{0.5279} &
  17.10\% \\ \cline{2-2} \cline{4-15} 
\multirow{-12}{*}{FA-NV-only data} &
  MO-OBAM &
  \multirow{-3}{*}{20} &
  \multicolumn{1}{l|}{\cellcolor[RGB]{242, 206, 239}0.5688} &
  \cellcolor[RGB]{242, 206, 239}23.49\% &
  \multicolumn{1}{l|}{\cellcolor[RGB]{242, 206, 239}0.6203} &
  \cellcolor[RGB]{242, 206, 239}51.29\% &
  \multicolumn{1}{l|}{\cellcolor[RGB]{242, 206, 239}0.4398} &
  \cellcolor[RGB]{242, 206, 239}-32.14\% &
  \multicolumn{1}{l|}{\cellcolor[RGB]{242, 206, 239}0.6165} &
  \cellcolor[RGB]{242, 206, 239}35.52\% &
  \multicolumn{1}{l|}{\cellcolor[RGB]{242, 206, 239}0.5956} &
  \cellcolor[RGB]{242, 206, 239}85.49\% &
  \multicolumn{1}{l|}{0.5568} &
  23.51\% \\ \hline
\end{tabular}%
} 
    \end{tabular}
\end{table}
\begin{table}[h!]
    \centering
    \begin{tabular}{c}
    (d) Homogeneity attack \\
    \resizebox{\columnwidth}{!}{%
\begin{tabular}{l|l|c|ll|ll|ll|ll|ll|ll}
\hline
Data &
  \begin{tabular}[c]{@{}l@{}}Anonymization \\ Model\end{tabular} &
  \multicolumn{1}{l|}{$k$} &
  \multicolumn{2}{c|}{DT} &
  \multicolumn{2}{c|}{LR} &
  \multicolumn{2}{c|}{NB} &
  \multicolumn{2}{c|}{NN} &
  \multicolumn{2}{c|}{RF} &
  \multicolumn{2}{c}{SVM} \\ \hline
OR-NV-only data &
   &
  \multicolumn{1}{l|}{} &
  \multicolumn{1}{l|}{0.5636} &
   &
  \multicolumn{1}{l|}{0.5744} &
   &
  \multicolumn{1}{l|}{0.4519} &
   &
  \multicolumn{1}{l|}{0.5951} &
   &
  \multicolumn{1}{l|}{0.5574} &
   &
  \multicolumn{1}{l|}{0.5451} &
   \\ \hline
 &
  $k$-anonymity &
   &
  \multicolumn{1}{l|}{\cellcolor[RGB]{242, 206, 239}0.5728} &
  \cellcolor[RGB]{242, 206, 239}1.63\% &
  \multicolumn{1}{l|}{\cellcolor[RGB]{242, 206, 239}0.6044} &
  \cellcolor[RGB]{242, 206, 239}5.22\% &
  \multicolumn{1}{l|}{0.4476} &
  -0.95\% &
  \multicolumn{1}{l|}{0.6060} &
  1.83\% &
  \multicolumn{1}{l|}{\cellcolor[RGB]{242, 206, 239}0.5893} &
  \cellcolor[RGB]{242, 206, 239}5.72\% &
  \multicolumn{1}{l|}{0.5349} &
  -1.87\% \\ \cline{2-2} \cline{4-15} 
 &
  Zheng et al &
   &
  \multicolumn{1}{l|}{\cellcolor[RGB]{242, 206, 239}0.5751} &
  \cellcolor[RGB]{242, 206, 239}2.04\% &
  \multicolumn{1}{l|}{\cellcolor[RGB]{242, 206, 239}0.6205} &
  \cellcolor[RGB]{242, 206, 239}8.03\% &
  \multicolumn{1}{l|}{\cellcolor[RGB]{242, 206, 239}0.4439} &
  \cellcolor[RGB]{242, 206, 239}-1.77\% &
  \multicolumn{1}{l|}{0.6031} &
  1.34\% &
  \multicolumn{1}{l|}{\cellcolor[RGB]{242, 206, 239}0.6001} &
  \cellcolor[RGB]{242, 206, 239}7.66\% &
  \multicolumn{1}{l|}{0.5896} &
  8.16\% \\ \cline{2-2} \cline{4-15} 
 &
  MO-OBAM &
  \multirow{-3}{*}{5} &
  \multicolumn{1}{l|}{\cellcolor[RGB]{242, 206, 239}0.5787} &
  \cellcolor[RGB]{242, 206, 239}2.68\% &
  \multicolumn{1}{l|}{\cellcolor[RGB]{242, 206, 239}0.6193} &
  \cellcolor[RGB]{242, 206, 239}7.82\% &
  \multicolumn{1}{l|}{\cellcolor[RGB]{242, 206, 239}0.4417} &
  \cellcolor[RGB]{242, 206, 239}-2.26\% &
  \multicolumn{1}{l|}{0.6088} &
  2.30\% &
  \multicolumn{1}{l|}{\cellcolor[RGB]{242, 206, 239}0.6009} &
  \cellcolor[RGB]{242, 206, 239}7.80\% &
  \multicolumn{1}{l|}{0.5272} &
  -3.28\% \\ \cline{2-15} 
 &
  $k$-anonymity &
   &
  \multicolumn{1}{l|}{\cellcolor[RGB]{242, 206, 239}0.5771} &
  \cellcolor[RGB]{242, 206, 239}2.40\% &
  \multicolumn{1}{l|}{\cellcolor[RGB]{242, 206, 239}0.6187} &
  \cellcolor[RGB]{242, 206, 239}7.71\% &
  \multicolumn{1}{l|}{\cellcolor[RGB]{242, 206, 239}0.4384} &
  \cellcolor[RGB]{242, 206, 239}-2.99\% &
  \multicolumn{1}{l|}{0.6150} &
  3.34\% &
  \multicolumn{1}{l|}{\cellcolor[RGB]{242, 206, 239}0.6080} &
  \cellcolor[RGB]{242, 206, 239}9.08\% &
  \multicolumn{1}{l|}{0.5439} &
  -0.22\% \\ \cline{2-2} \cline{4-15} 
 &
  Zheng et al &
   &
  \multicolumn{1}{l|}{\cellcolor[RGB]{242, 206, 239}0.5705} &
  \cellcolor[RGB]{242, 206, 239}1.22\% &
  \multicolumn{1}{l|}{\cellcolor[RGB]{242, 206, 239}0.6151} &
  \cellcolor[RGB]{242, 206, 239}7.09\% &
  \multicolumn{1}{l|}{0.4482} &
  -0.82\% &
  \multicolumn{1}{l|}{0.6115} &
  2.76\% &
  \multicolumn{1}{l|}{\cellcolor[RGB]{242, 206, 239}0.6023} &
  \cellcolor[RGB]{242, 206, 239}8.06\% &
  \multicolumn{1}{l|}{0.5740} &
  5.30\% \\ \cline{2-2} \cline{4-15} 
 &
  MO-OBAM &
  \multirow{-3}{*}{10} &
  \multicolumn{1}{l|}{\cellcolor[RGB]{242, 206, 239}0.5707} &
  \cellcolor[RGB]{242, 206, 239}1.26\% &
  \multicolumn{1}{l|}{\cellcolor[RGB]{242, 206, 239}0.6195} &
  \cellcolor[RGB]{242, 206, 239}7.85\% &
  \multicolumn{1}{l|}{0.4443} &
  -1.68\% &
  \multicolumn{1}{l|}{0.6122} &
  2.87\% &
  \multicolumn{1}{l|}{\cellcolor[RGB]{242, 206, 239}0.5978} &
  \cellcolor[RGB]{242, 206, 239}7.25\% &
  \multicolumn{1}{l|}{0.4787} &
  -12.18\% \\ \cline{2-15} 
 &
  $k$-anonymity &
   &
  \multicolumn{1}{l|}{\cellcolor[RGB]{242, 206, 239}0.5710} &
  \cellcolor[RGB]{242, 206, 239}1.31\% &
  \multicolumn{1}{l|}{\cellcolor[RGB]{242, 206, 239}0.6145} &
  \cellcolor[RGB]{242, 206, 239}6.98\% &
  \multicolumn{1}{l|}{\cellcolor[RGB]{242, 206, 239}0.4401} &
  \cellcolor[RGB]{242, 206, 239}-2.61\% &
  \multicolumn{1}{l|}{\cellcolor[RGB]{242, 206, 239}0.6216} &
  \cellcolor[RGB]{242, 206, 239}4.45\% &
  \multicolumn{1}{l|}{\cellcolor[RGB]{242, 206, 239}0.5976} &
  \cellcolor[RGB]{242, 206, 239}7.21\% &
  \multicolumn{1}{l|}{0.5722} &
  4.97\% \\ \cline{2-2} \cline{4-15} 
 &
  Zheng et al &
   &
  \multicolumn{1}{l|}{\cellcolor[RGB]{242, 206, 239}0.5765} &
  \cellcolor[RGB]{242, 206, 239}2.29\% &
  \multicolumn{1}{l|}{\cellcolor[RGB]{242, 206, 239}0.6191} &
  \cellcolor[RGB]{242, 206, 239}7.78\% &
  \multicolumn{1}{l|}{0.4440} &
  -1.75\% &
  \multicolumn{1}{l|}{0.6128} &
  2.97\% &
  \multicolumn{1}{l|}{\cellcolor[RGB]{242, 206, 239}0.6021} &
  \cellcolor[RGB]{242, 206, 239}8.02\% &
  \multicolumn{1}{l|}{0.5723} &
  4.99\% \\ \cline{2-2} \cline{4-15} 
 &
  MO-OBAM &
  \multirow{-3}{*}{15} &
  \multicolumn{1}{l|}{\cellcolor[RGB]{242, 206, 239}0.5711} &
  \cellcolor[RGB]{242, 206, 239}1.33\% &
  \multicolumn{1}{l|}{\cellcolor[RGB]{242, 206, 239}0.6190} &
  \cellcolor[RGB]{242, 206, 239}7.76\% &
  \multicolumn{1}{l|}{\cellcolor[RGB]{242, 206, 239}0.4386} &
  \cellcolor[RGB]{242, 206, 239}-2.94\% &
  \multicolumn{1}{l|}{0.6139} &
  3.16\% &
  \multicolumn{1}{l|}{\cellcolor[RGB]{242, 206, 239}0.6009} &
  \cellcolor[RGB]{242, 206, 239}7.80\% &
  \multicolumn{1}{l|}{0.6119} &
  12.25\% \\ \cline{2-15} 
 &
  $k$-anonymity &
   &
  \multicolumn{1}{l|}{\cellcolor[RGB]{242, 206, 239}0.5733} &
  \cellcolor[RGB]{242, 206, 239}1.72\% &
  \multicolumn{1}{l|}{\cellcolor[RGB]{242, 206, 239}0.6163} &
  \cellcolor[RGB]{242, 206, 239}7.29\% &
  \multicolumn{1}{l|}{\cellcolor[RGB]{242, 206, 239}0.4405} &
  \cellcolor[RGB]{242, 206, 239}-2.52\% &
  \multicolumn{1}{l|}{\cellcolor[RGB]{242, 206, 239}0.6200} &
  \cellcolor[RGB]{242, 206, 239}4.18\% &
  \multicolumn{1}{l|}{\cellcolor[RGB]{242, 206, 239}0.5952} &
  \cellcolor[RGB]{242, 206, 239}6.78\% &
  \multicolumn{1}{l|}{0.5444} &
  -0.13\% \\ \cline{2-2} \cline{4-15} 
 &
  Zheng et al &
   &
  \multicolumn{1}{l|}{\cellcolor[RGB]{242, 206, 239}0.5710} &
  \cellcolor[RGB]{242, 206, 239}1.31\% &
  \multicolumn{1}{l|}{\cellcolor[RGB]{242, 206, 239}0.6206} &
  \cellcolor[RGB]{242, 206, 239}8.04\% &
  \multicolumn{1}{l|}{\cellcolor[RGB]{242, 206, 239}0.4382} &
  \cellcolor[RGB]{242, 206, 239}-3.03\% &
  \multicolumn{1}{l|}{0.6083} &
  2.22\% &
  \multicolumn{1}{l|}{\cellcolor[RGB]{242, 206, 239}0.5949} &
  \cellcolor[RGB]{242, 206, 239}6.73\% &
  \multicolumn{1}{l|}{0.5587} &
  2.49\% \\ \cline{2-2} \cline{4-15} 
\multirow{-12}{*}{FA-NV-only data} &
  MO-OBAM &
  \multirow{-3}{*}{20} &
  \multicolumn{1}{l|}{0.5698} &
  1.10\% &
  \multicolumn{1}{l|}{\cellcolor[RGB]{242, 206, 239}0.6180} &
  \cellcolor[RGB]{242, 206, 239}7.59\% &
  \multicolumn{1}{l|}{\cellcolor[RGB]{242, 206, 239}0.4403} &
  \cellcolor[RGB]{242, 206, 239}-2.57\% &
  \multicolumn{1}{l|}{0.6096} &
  2.44\% &
  \multicolumn{1}{l|}{\cellcolor[RGB]{242, 206, 239}0.5970} &
  \cellcolor[RGB]{242, 206, 239}7.10\% &
  \multicolumn{1}{l|}{\cellcolor[RGB]{242, 206, 239}0.6420} &
  \cellcolor[RGB]{242, 206, 239}17.78\% \\ \hline
\end{tabular}%
}
    \end{tabular}
\end{table}

\begin{table}[h!]
    \centering
    \caption{FA vs FA-NV-only (Precision)}
    \label{tab:FA vs FA-NV-only-precision}
    \begin{tabular}{c}
         (a) Linkage attack with the risk level of $\tau=0.05$ \\
         \resizebox{\columnwidth}{!}{%
\begin{tabular}{l|c|rrr|rrr|rrr|rrr|rrr|rrr}
\hline
 &
  \multicolumn{1}{l|}{} &
  \multicolumn{3}{c|}{DT} &
  \multicolumn{3}{c|}{LR} &
  \multicolumn{3}{c|}{NB} &
  \multicolumn{3}{c|}{NN} &
  \multicolumn{3}{c|}{RF} &
  \multicolumn{3}{c}{SVM} \\ \hline
\begin{tabular}[c]{@{}l@{}}Anonymization \\ Model\end{tabular} &
  $k$ &
  \multicolumn{1}{l|}{\begin{tabular}[c]{@{}l@{}}Average \\ (FA)\end{tabular}} &
  \multicolumn{1}{l|}{\begin{tabular}[c]{@{}l@{}}Average \\ (FA-NV-only)\end{tabular}} &
  \multicolumn{1}{l|}{\begin{tabular}[c]{@{}l@{}}Percentage \\ Change\end{tabular}} &
  \multicolumn{1}{l|}{\begin{tabular}[c]{@{}l@{}}Average \\ (FA)\end{tabular}} &
  \multicolumn{1}{l|}{\begin{tabular}[c]{@{}l@{}}Average \\ (FA-NV-only)\end{tabular}} &
  \multicolumn{1}{l|}{\begin{tabular}[c]{@{}l@{}}Percentage \\ Change\end{tabular}} &
  \multicolumn{1}{l|}{\begin{tabular}[c]{@{}l@{}}Average \\ (FA)\end{tabular}} &
  \multicolumn{1}{l|}{\begin{tabular}[c]{@{}l@{}}Average \\ (FA-NV-only)\end{tabular}} &
  \multicolumn{1}{l|}{\begin{tabular}[c]{@{}l@{}}Percentage \\ Change\end{tabular}} &
  \multicolumn{1}{l|}{\begin{tabular}[c]{@{}l@{}}Average \\ (FA)\end{tabular}} &
  \multicolumn{1}{l|}{\begin{tabular}[c]{@{}l@{}}Average \\ (FA-NV-only)\end{tabular}} &
  \multicolumn{1}{l|}{\begin{tabular}[c]{@{}l@{}}Percentage \\ Change\end{tabular}} &
  \multicolumn{1}{l|}{\begin{tabular}[c]{@{}l@{}}Average \\ (FA)\end{tabular}} &
  \multicolumn{1}{l|}{\begin{tabular}[c]{@{}l@{}}Average \\ (FA-NV-only)\end{tabular}} &
  \multicolumn{1}{l|}{\begin{tabular}[c]{@{}l@{}}Percentage \\ Change\end{tabular}} &
  \multicolumn{1}{l|}{\begin{tabular}[c]{@{}l@{}}Average \\ (FA)\end{tabular}} &
  \multicolumn{1}{l|}{\begin{tabular}[c]{@{}l@{}}Average \\ (FA-NV-only)\end{tabular}} &
  \multicolumn{1}{l}{\begin{tabular}[c]{@{}l@{}}Percentage \\ Change\end{tabular}} \\ \hline
 &
  5 &
  \multicolumn{1}{r|}{\cellcolor[RGB]{242, 206, 239}0.5793} &
  \multicolumn{1}{r|}{\cellcolor[RGB]{242, 206, 239}0.4838} &
  \cellcolor[RGB]{242, 206, 239}-16.49\% &
  \multicolumn{1}{r|}{\cellcolor[RGB]{242, 206, 239}0.7027} &
  \multicolumn{1}{r|}{\cellcolor[RGB]{242, 206, 239}0.6567} &
  \cellcolor[RGB]{242, 206, 239}-6.55\% &
  \multicolumn{1}{r|}{\cellcolor[RGB]{242, 206, 239}0.6016} &
  \multicolumn{1}{r|}{\cellcolor[RGB]{242, 206, 239}0.4349} &
  \cellcolor[RGB]{242, 206, 239}-27.71\% &
  \multicolumn{1}{r|}{\cellcolor[RGB]{242, 206, 239}0.668} &
  \multicolumn{1}{r|}{\cellcolor[RGB]{242, 206, 239}0.6031} &
  \cellcolor[RGB]{242, 206, 239}-9.72\% &
  \multicolumn{1}{r|}{\cellcolor[RGB]{242, 206, 239}0.7045} &
  \multicolumn{1}{r|}{\cellcolor[RGB]{242, 206, 239}0.6446} &
  \cellcolor[RGB]{242, 206, 239}-8.50\% &
  \multicolumn{1}{r|}{\cellcolor[RGB]{242, 206, 239}0.7042} &
  \multicolumn{1}{r|}{\cellcolor[RGB]{242, 206, 239}0.6081} &
  \cellcolor[RGB]{242, 206, 239}-13.65\% \\ \cline{2-20} 
 &
  10 &
  \multicolumn{1}{r|}{\cellcolor[RGB]{242, 206, 239}0.5833} &
  \multicolumn{1}{r|}{\cellcolor[RGB]{242, 206, 239}0.5053} &
  \cellcolor[RGB]{242, 206, 239}-13.37\% &
  \multicolumn{1}{r|}{\cellcolor[RGB]{242, 206, 239}0.705} &
  \multicolumn{1}{r|}{\cellcolor[RGB]{242, 206, 239}0.6679} &
  \cellcolor[RGB]{242, 206, 239}-5.26\% &
  \multicolumn{1}{r|}{\cellcolor[RGB]{242, 206, 239}0.5995} &
  \multicolumn{1}{r|}{\cellcolor[RGB]{242, 206, 239}0.4629} &
  \cellcolor[RGB]{242, 206, 239}-22.79\% &
  \multicolumn{1}{r|}{\cellcolor[RGB]{242, 206, 239}0.6682} &
  \multicolumn{1}{r|}{\cellcolor[RGB]{242, 206, 239}0.6251} &
  \cellcolor[RGB]{242, 206, 239}-6.45\% &
  \multicolumn{1}{r|}{\cellcolor[RGB]{242, 206, 239}0.7053} &
  \multicolumn{1}{r|}{\cellcolor[RGB]{242, 206, 239}0.6557} &
  \cellcolor[RGB]{242, 206, 239}-7.03\% &
  \multicolumn{1}{r|}{0.6905} &
  \multicolumn{1}{r|}{0.6491} &
  -6.00\% \\ \cline{2-20} 
 &
  15 &
  \multicolumn{1}{r|}{\cellcolor[RGB]{242, 206, 239}0.5771} &
  \multicolumn{1}{r|}{\cellcolor[RGB]{242, 206, 239}0.4861} &
  \cellcolor[RGB]{242, 206, 239}-15.77\% &
  \multicolumn{1}{r|}{\cellcolor[RGB]{242, 206, 239}0.706} &
  \multicolumn{1}{r|}{\cellcolor[RGB]{242, 206, 239}0.6605} &
  \cellcolor[RGB]{242, 206, 239}-6.44\% &
  \multicolumn{1}{r|}{\cellcolor[RGB]{242, 206, 239}0.6037} &
  \multicolumn{1}{r|}{\cellcolor[RGB]{242, 206, 239}0.4203} &
  \cellcolor[RGB]{242, 206, 239}-30.38\% &
  \multicolumn{1}{r|}{\cellcolor[RGB]{242, 206, 239}0.6663} &
  \multicolumn{1}{r|}{\cellcolor[RGB]{242, 206, 239}0.6384} &
  \cellcolor[RGB]{242, 206, 239}-4.19\% &
  \multicolumn{1}{r|}{\cellcolor[RGB]{242, 206, 239}0.7024} &
  \multicolumn{1}{r|}{\cellcolor[RGB]{242, 206, 239}0.6443} &
  \cellcolor[RGB]{242, 206, 239}-8.27\% &
  \multicolumn{1}{r|}{0.661} &
  \multicolumn{1}{r|}{0.6466} &
  -2.18\% \\ \cline{2-20} 
\multirow{-4}{*}{$k$-anonymity} &
  20 &
  \multicolumn{1}{r|}{0.5793} &
  \multicolumn{1}{r|}{0.5786} &
  -0.12\% &
  \multicolumn{1}{r|}{\cellcolor[RGB]{242, 206, 239}0.7037} &
  \multicolumn{1}{r|}{\cellcolor[RGB]{242, 206, 239}0.7102} &
  \cellcolor[RGB]{242, 206, 239}0.92\% &
  \multicolumn{1}{r|}{0.6037} &
  \multicolumn{1}{r|}{0.6004} &
  -0.55\% &
  \multicolumn{1}{r|}{0.6673} &
  \multicolumn{1}{r|}{0.6639} &
  -0.51\% &
  \multicolumn{1}{r|}{0.704} &
  \multicolumn{1}{r|}{0.7039} &
  -0.01\% &
  \multicolumn{1}{r|}{0.6661} &
  \multicolumn{1}{r|}{0.6883} &
  3.33\% \\ \hline
 &
  5 &
  \multicolumn{1}{r|}{\cellcolor[RGB]{242, 206, 239}0.5822} &
  \multicolumn{1}{r|}{\cellcolor[RGB]{242, 206, 239}0.5237} &
  \cellcolor[RGB]{242, 206, 239}-10.05\% &
  \multicolumn{1}{r|}{\cellcolor[RGB]{242, 206, 239}0.709} &
  \multicolumn{1}{r|}{\cellcolor[RGB]{242, 206, 239}0.6703} &
  \cellcolor[RGB]{242, 206, 239}-5.46\% &
  \multicolumn{1}{r|}{\cellcolor[RGB]{242, 206, 239}0.5978} &
  \multicolumn{1}{r|}{\cellcolor[RGB]{242, 206, 239}0.5029} &
  \cellcolor[RGB]{242, 206, 239}-15.87\% &
  \multicolumn{1}{r|}{\cellcolor[RGB]{242, 206, 239}0.6736} &
  \multicolumn{1}{r|}{\cellcolor[RGB]{242, 206, 239}0.6306} &
  \cellcolor[RGB]{242, 206, 239}-6.38\% &
  \multicolumn{1}{r|}{\cellcolor[RGB]{242, 206, 239}0.7025} &
  \multicolumn{1}{r|}{\cellcolor[RGB]{242, 206, 239}0.6754} &
  \cellcolor[RGB]{242, 206, 239}-3.86\% &
  \multicolumn{1}{r|}{0.6576} &
  \multicolumn{1}{r|}{0.655} &
  -0.40\% \\ \cline{2-20} 
 &
  10 &
  \multicolumn{1}{r|}{\cellcolor[RGB]{242, 206, 239}0.5777} &
  \multicolumn{1}{r|}{\cellcolor[RGB]{242, 206, 239}0.5229} &
  \cellcolor[RGB]{242, 206, 239}-9.49\% &
  \multicolumn{1}{r|}{\cellcolor[RGB]{242, 206, 239}0.7045} &
  \multicolumn{1}{r|}{\cellcolor[RGB]{242, 206, 239}0.678} &
  \cellcolor[RGB]{242, 206, 239}-3.76\% &
  \multicolumn{1}{r|}{\cellcolor[RGB]{242, 206, 239}0.6005} &
  \multicolumn{1}{r|}{\cellcolor[RGB]{242, 206, 239}0.5158} &
  \cellcolor[RGB]{242, 206, 239}-14.10\% &
  \multicolumn{1}{r|}{\cellcolor[RGB]{242, 206, 239}0.6703} &
  \multicolumn{1}{r|}{\cellcolor[RGB]{242, 206, 239}0.6355} &
  \cellcolor[RGB]{242, 206, 239}-5.19\% &
  \multicolumn{1}{r|}{\cellcolor[RGB]{242, 206, 239}0.7049} &
  \multicolumn{1}{r|}{\cellcolor[RGB]{242, 206, 239}0.6752} &
  \cellcolor[RGB]{242, 206, 239}-4.21\% &
  \multicolumn{1}{r|}{0.6823} &
  \multicolumn{1}{r|}{0.6797} &
  -0.38\% \\ \cline{2-20} 
 &
  15 &
  \multicolumn{1}{r|}{\cellcolor[RGB]{242, 206, 239}0.5801} &
  \multicolumn{1}{r|}{\cellcolor[RGB]{242, 206, 239}0.5233} &
  \cellcolor[RGB]{242, 206, 239}-9.79\% &
  \multicolumn{1}{r|}{\cellcolor[RGB]{242, 206, 239}0.7102} &
  \multicolumn{1}{r|}{\cellcolor[RGB]{242, 206, 239}0.6779} &
  \cellcolor[RGB]{242, 206, 239}-4.55\% &
  \multicolumn{1}{r|}{\cellcolor[RGB]{242, 206, 239}0.6024} &
  \multicolumn{1}{r|}{\cellcolor[RGB]{242, 206, 239}0.5092} &
  \cellcolor[RGB]{242, 206, 239}-15.47\% &
  \multicolumn{1}{r|}{\cellcolor[RGB]{242, 206, 239}0.6648} &
  \multicolumn{1}{r|}{\cellcolor[RGB]{242, 206, 239}0.6325} &
  \cellcolor[RGB]{242, 206, 239}-4.86\% &
  \multicolumn{1}{r|}{\cellcolor[RGB]{242, 206, 239}0.7069} &
  \multicolumn{1}{r|}{\cellcolor[RGB]{242, 206, 239}0.6793} &
  \cellcolor[RGB]{242, 206, 239}-3.90\% &
  \multicolumn{1}{r|}{0.6589} &
  \multicolumn{1}{r|}{0.648} &
  -1.65\% \\ \cline{2-20} 
\multirow{-4}{*}{Zheng et al} &
  20 &
  \multicolumn{1}{r|}{0.5808} &
  \multicolumn{1}{r|}{0.5819} &
  0.19\% &
  \multicolumn{1}{r|}{0.7071} &
  \multicolumn{1}{r|}{0.7071} &
  0.00\% &
  \multicolumn{1}{r|}{0.6017} &
  \multicolumn{1}{r|}{0.6042} &
  0.42\% &
  \multicolumn{1}{r|}{0.6631} &
  \multicolumn{1}{r|}{0.6699} &
  1.03\% &
  \multicolumn{1}{r|}{0.7037} &
  \multicolumn{1}{r|}{0.7027} &
  -0.14\% &
  \multicolumn{1}{r|}{0.7063} &
  \multicolumn{1}{r|}{0.6967} &
  -1.36\% \\ \hline
 &
  5 &
  \multicolumn{1}{r|}{\cellcolor[RGB]{242, 206, 239}0.5809} &
  \multicolumn{1}{r|}{\cellcolor[RGB]{242, 206, 239}0.5501} &
  \cellcolor[RGB]{242, 206, 239}-5.30\% &
  \multicolumn{1}{r|}{\cellcolor[RGB]{242, 206, 239}0.711} &
  \multicolumn{1}{r|}{\cellcolor[RGB]{242, 206, 239}0.6854} &
  \cellcolor[RGB]{242, 206, 239}-3.60\% &
  \multicolumn{1}{r|}{\cellcolor[RGB]{242, 206, 239}0.6029} &
  \multicolumn{1}{r|}{\cellcolor[RGB]{242, 206, 239}0.5491} &
  \cellcolor[RGB]{242, 206, 239}-8.92\% &
  \multicolumn{1}{r|}{\cellcolor[RGB]{242, 206, 239}0.6671} &
  \multicolumn{1}{r|}{\cellcolor[RGB]{242, 206, 239}0.6358} &
  \cellcolor[RGB]{242, 206, 239}-4.69\% &
  \multicolumn{1}{r|}{\cellcolor[RGB]{242, 206, 239}0.7087} &
  \multicolumn{1}{r|}{\cellcolor[RGB]{242, 206, 239}0.6868} &
  \cellcolor[RGB]{242, 206, 239}-3.09\% &
  \multicolumn{1}{r|}{0.6999} &
  \multicolumn{1}{r|}{0.6623} &
  -5.37\% \\ \cline{2-20} 
 &
  10 &
  \multicolumn{1}{r|}{\cellcolor[RGB]{242, 206, 239}0.5777} &
  \multicolumn{1}{r|}{\cellcolor[RGB]{242, 206, 239}0.5684} &
  \cellcolor[RGB]{242, 206, 239}-1.61\% &
  \multicolumn{1}{r|}{\cellcolor[RGB]{242, 206, 239}0.7082} &
  \multicolumn{1}{r|}{\cellcolor[RGB]{242, 206, 239}0.7017} &
  \cellcolor[RGB]{242, 206, 239}-0.92\% &
  \multicolumn{1}{r|}{\cellcolor[RGB]{242, 206, 239}0.6006} &
  \multicolumn{1}{r|}{\cellcolor[RGB]{242, 206, 239}0.5831} &
  \cellcolor[RGB]{242, 206, 239}-2.91\% &
  \multicolumn{1}{r|}{\cellcolor[RGB]{242, 206, 239}0.669} &
  \multicolumn{1}{r|}{\cellcolor[RGB]{242, 206, 239}0.6568} &
  \cellcolor[RGB]{242, 206, 239}-1.82\% &
  \multicolumn{1}{r|}{\cellcolor[RGB]{242, 206, 239}0.7031} &
  \multicolumn{1}{r|}{\cellcolor[RGB]{242, 206, 239}0.6975} &
  \cellcolor[RGB]{242, 206, 239}-0.80\% &
  \multicolumn{1}{r|}{0.7087} &
  \multicolumn{1}{r|}{0.6925} &
  -2.29\% \\ \cline{2-20} 
 &
  15 &
  \multicolumn{1}{r|}{\cellcolor[RGB]{242, 206, 239}0.5785} &
  \multicolumn{1}{r|}{\cellcolor[RGB]{242, 206, 239}0.5735} &
  \cellcolor[RGB]{242, 206, 239}-0.86\% &
  \multicolumn{1}{r|}{0.7061} &
  \multicolumn{1}{r|}{0.7042} &
  -0.27\% &
  \multicolumn{1}{r|}{0.6037} &
  \multicolumn{1}{r|}{0.6001} &
  -0.60\% &
  \multicolumn{1}{r|}{0.6676} &
  \multicolumn{1}{r|}{0.6606} &
  -1.05\% &
  \multicolumn{1}{r|}{\cellcolor[RGB]{242, 206, 239}0.7061} &
  \multicolumn{1}{r|}{\cellcolor[RGB]{242, 206, 239}0.7007} &
  \cellcolor[RGB]{242, 206, 239}-0.76\% &
  \multicolumn{1}{r|}{0.6798} &
  \multicolumn{1}{r|}{0.6723} &
  -1.10\% \\ \cline{2-20} 
\multirow{-4}{*}{MO-OBAM} &
  20 &
  \multicolumn{1}{r|}{0.5771} &
  \multicolumn{1}{r|}{0.5765} &
  -0.10\% &
  \multicolumn{1}{r|}{0.7076} &
  \multicolumn{1}{r|}{0.7055} &
  -0.30\% &
  \multicolumn{1}{r|}{0.6005} &
  \multicolumn{1}{r|}{0.6006} &
  0.02\% &
  \multicolumn{1}{r|}{0.6612} &
  \multicolumn{1}{r|}{0.6678} &
  1.00\% &
  \multicolumn{1}{r|}{0.7028} &
  \multicolumn{1}{r|}{0.6995} &
  -0.47\% &
  \multicolumn{1}{r|}{\cellcolor[RGB]{242, 206, 239}0.6515} &
  \multicolumn{1}{r|}{\cellcolor[RGB]{242, 206, 239}0.7082} &
  \cellcolor[RGB]{242, 206, 239}8.70\% \\ \hline
\end{tabular}%
} \\
         \\
         (b) Linkage attack with the risk level of $\tau=0.075$ \\
         \resizebox{\columnwidth}{!}{%
\begin{tabular}{l|c|rrr|rrr|rrr|rrr|rrr|rrr}
\hline
 &
  \multicolumn{1}{l|}{} &
  \multicolumn{3}{c|}{DT} &
  \multicolumn{3}{c|}{LR} &
  \multicolumn{3}{c|}{NB} &
  \multicolumn{3}{c|}{NN} &
  \multicolumn{3}{c|}{RF} &
  \multicolumn{3}{c}{SVM} \\ \hline
\begin{tabular}[c]{@{}l@{}}Anonymization \\ Model\end{tabular} &
  $k$ &
  \multicolumn{1}{l|}{\begin{tabular}[c]{@{}l@{}}Average \\ (FA)\end{tabular}} &
  \multicolumn{1}{l|}{\begin{tabular}[c]{@{}l@{}}Average \\ (FA-NV-only)\end{tabular}} &
  \multicolumn{1}{l|}{\begin{tabular}[c]{@{}l@{}}Percentage \\ Change\end{tabular}} &
  \multicolumn{1}{l|}{\begin{tabular}[c]{@{}l@{}}Average \\ (FA)\end{tabular}} &
  \multicolumn{1}{l|}{\begin{tabular}[c]{@{}l@{}}Average \\ (FA-NV-only)\end{tabular}} &
  \multicolumn{1}{l|}{\begin{tabular}[c]{@{}l@{}}Percentage \\ Change\end{tabular}} &
  \multicolumn{1}{l|}{\begin{tabular}[c]{@{}l@{}}Average \\ (FA)\end{tabular}} &
  \multicolumn{1}{l|}{\begin{tabular}[c]{@{}l@{}}Average \\ (FA-NV-only)\end{tabular}} &
  \multicolumn{1}{l|}{\begin{tabular}[c]{@{}l@{}}Percentage \\ Change\end{tabular}} &
  \multicolumn{1}{l|}{\begin{tabular}[c]{@{}l@{}}Average \\ (FA)\end{tabular}} &
  \multicolumn{1}{l|}{\begin{tabular}[c]{@{}l@{}}Average \\ (FA-NV-only)\end{tabular}} &
  \multicolumn{1}{l|}{\begin{tabular}[c]{@{}l@{}}Percentage \\ Change\end{tabular}} &
  \multicolumn{1}{l|}{\begin{tabular}[c]{@{}l@{}}Average \\ (FA)\end{tabular}} &
  \multicolumn{1}{l|}{\begin{tabular}[c]{@{}l@{}}Average \\ (FA-NV-only)\end{tabular}} &
  \multicolumn{1}{l|}{\begin{tabular}[c]{@{}l@{}}Percentage \\ Change\end{tabular}} &
  \multicolumn{1}{l|}{\begin{tabular}[c]{@{}l@{}}Average \\ (FA)\end{tabular}} &
  \multicolumn{1}{l|}{\begin{tabular}[c]{@{}l@{}}Average \\ (FA-NV-only)\end{tabular}} &
  \multicolumn{1}{l}{\begin{tabular}[c]{@{}l@{}}Percentage \\ Change\end{tabular}} \\ \hline
 &
  5 &
  \multicolumn{1}{r|}{\cellcolor[RGB]{242, 206, 239}0.5793} &
  \multicolumn{1}{r|}{\cellcolor[RGB]{242, 206, 239}0.48} &
  \cellcolor[RGB]{242, 206, 239}-17.14\% &
  \multicolumn{1}{r|}{\cellcolor[RGB]{242, 206, 239}0.7027} &
  \multicolumn{1}{r|}{\cellcolor[RGB]{242, 206, 239}0.658} &
  \cellcolor[RGB]{242, 206, 239}-6.36\% &
  \multicolumn{1}{r|}{\cellcolor[RGB]{242, 206, 239}0.6016} &
  \multicolumn{1}{r|}{\cellcolor[RGB]{242, 206, 239}0.4511} &
  \cellcolor[RGB]{242, 206, 239}-25.02\% &
  \multicolumn{1}{r|}{\cellcolor[RGB]{242, 206, 239}0.668} &
  \multicolumn{1}{r|}{\cellcolor[RGB]{242, 206, 239}0.6076} &
  \cellcolor[RGB]{242, 206, 239}-9.04\% &
  \multicolumn{1}{r|}{\cellcolor[RGB]{242, 206, 239}0.7045} &
  \multicolumn{1}{r|}{\cellcolor[RGB]{242, 206, 239}0.6263} &
  \cellcolor[RGB]{242, 206, 239}-11.10\% &
  \multicolumn{1}{r|}{\cellcolor[RGB]{242, 206, 239}0.7042} &
  \multicolumn{1}{r|}{\cellcolor[RGB]{242, 206, 239}0.617} &
  \cellcolor[RGB]{242, 206, 239}-12.38\% \\ \cline{2-20} 
 &
  10 &
  \multicolumn{1}{r|}{\cellcolor[RGB]{242, 206, 239}0.5833} &
  \multicolumn{1}{r|}{\cellcolor[RGB]{242, 206, 239}0.5044} &
  \cellcolor[RGB]{242, 206, 239}-13.53\% &
  \multicolumn{1}{r|}{\cellcolor[RGB]{242, 206, 239}0.705} &
  \multicolumn{1}{r|}{\cellcolor[RGB]{242, 206, 239}0.672} &
  \cellcolor[RGB]{242, 206, 239}-4.68\% &
  \multicolumn{1}{r|}{\cellcolor[RGB]{242, 206, 239}0.5995} &
  \multicolumn{1}{r|}{\cellcolor[RGB]{242, 206, 239}0.4643} &
  \cellcolor[RGB]{242, 206, 239}-22.55\% &
  \multicolumn{1}{r|}{\cellcolor[RGB]{242, 206, 239}0.6682} &
  \multicolumn{1}{r|}{\cellcolor[RGB]{242, 206, 239}0.6245} &
  \cellcolor[RGB]{242, 206, 239}-6.54\% &
  \multicolumn{1}{r|}{\cellcolor[RGB]{242, 206, 239}0.7053} &
  \multicolumn{1}{r|}{\cellcolor[RGB]{242, 206, 239}0.6558} &
  \cellcolor[RGB]{242, 206, 239}-7.02\% &
  \multicolumn{1}{r|}{\cellcolor[RGB]{242, 206, 239}0.6905} &
  \multicolumn{1}{r|}{\cellcolor[RGB]{242, 206, 239}0.6123} &
  \cellcolor[RGB]{242, 206, 239}-11.33\% \\ \cline{2-20} 
 &
  15 &
  \multicolumn{1}{r|}{0.5771} &
  \multicolumn{1}{r|}{0.5751} &
  -0.35\% &
  \multicolumn{1}{r|}{0.706} &
  \multicolumn{1}{r|}{0.7044} &
  -0.23\% &
  \multicolumn{1}{r|}{0.6037} &
  \multicolumn{1}{r|}{0.6008} &
  -0.48\% &
  \multicolumn{1}{r|}{0.6663} &
  \multicolumn{1}{r|}{0.6675} &
  0.18\% &
  \multicolumn{1}{r|}{0.7024} &
  \multicolumn{1}{r|}{0.7057} &
  0.47\% &
  \multicolumn{1}{r|}{0.661} &
  \multicolumn{1}{r|}{0.6565} &
  -0.68\% \\ \cline{2-20} 
\multirow{-4}{*}{$k$-anonymity} &
  20 &
  \multicolumn{1}{r|}{0.5793} &
  \multicolumn{1}{r|}{0.5804} &
  0.19\% &
  \multicolumn{1}{r|}{0.7037} &
  \multicolumn{1}{r|}{0.7058} &
  0.30\% &
  \multicolumn{1}{r|}{0.6037} &
  \multicolumn{1}{r|}{0.603} &
  -0.12\% &
  \multicolumn{1}{r|}{0.6673} &
  \multicolumn{1}{r|}{0.6623} &
  -0.75\% &
  \multicolumn{1}{r|}{0.704} &
  \multicolumn{1}{r|}{0.7079} &
  0.55\% &
  \multicolumn{1}{r|}{0.6661} &
  \multicolumn{1}{r|}{0.7009} &
  5.22\% \\ \hline
 &
  5 &
  \multicolumn{1}{r|}{\cellcolor[RGB]{242, 206, 239}0.5822} &
  \multicolumn{1}{r|}{\cellcolor[RGB]{242, 206, 239}0.5358} &
  \cellcolor[RGB]{242, 206, 239}-7.97\% &
  \multicolumn{1}{r|}{\cellcolor[RGB]{242, 206, 239}0.709} &
  \multicolumn{1}{r|}{\cellcolor[RGB]{242, 206, 239}0.6786} &
  \cellcolor[RGB]{242, 206, 239}-4.29\% &
  \multicolumn{1}{r|}{\cellcolor[RGB]{242, 206, 239}0.5978} &
  \multicolumn{1}{r|}{\cellcolor[RGB]{242, 206, 239}0.5354} &
  \cellcolor[RGB]{242, 206, 239}-10.44\% &
  \multicolumn{1}{r|}{\cellcolor[RGB]{242, 206, 239}0.6736} &
  \multicolumn{1}{r|}{\cellcolor[RGB]{242, 206, 239}0.6301} &
  \cellcolor[RGB]{242, 206, 239}-6.46\% &
  \multicolumn{1}{r|}{\cellcolor[RGB]{242, 206, 239}0.7025} &
  \multicolumn{1}{r|}{\cellcolor[RGB]{242, 206, 239}0.6835} &
  \cellcolor[RGB]{242, 206, 239}-2.70\% &
  \multicolumn{1}{r|}{0.6576} &
  \multicolumn{1}{r|}{0.683} &
  3.86\% \\ \cline{2-20} 
 &
  10 &
  \multicolumn{1}{r|}{\cellcolor[RGB]{242, 206, 239}0.5777} &
  \multicolumn{1}{r|}{\cellcolor[RGB]{242, 206, 239}0.5372} &
  \cellcolor[RGB]{242, 206, 239}-7.01\% &
  \multicolumn{1}{r|}{\cellcolor[RGB]{242, 206, 239}0.7045} &
  \multicolumn{1}{r|}{\cellcolor[RGB]{242, 206, 239}0.6818} &
  \cellcolor[RGB]{242, 206, 239}-3.22\% &
  \multicolumn{1}{r|}{\cellcolor[RGB]{242, 206, 239}0.6005} &
  \multicolumn{1}{r|}{\cellcolor[RGB]{242, 206, 239}0.5344} &
  \cellcolor[RGB]{242, 206, 239}-11.01\% &
  \multicolumn{1}{r|}{\cellcolor[RGB]{242, 206, 239}0.6703} &
  \multicolumn{1}{r|}{\cellcolor[RGB]{242, 206, 239}0.6359} &
  \cellcolor[RGB]{242, 206, 239}-5.13\% &
  \multicolumn{1}{r|}{\cellcolor[RGB]{242, 206, 239}0.7049} &
  \multicolumn{1}{r|}{\cellcolor[RGB]{242, 206, 239}0.6817} &
  \cellcolor[RGB]{242, 206, 239}-3.29\% &
  \multicolumn{1}{r|}{0.6823} &
  \multicolumn{1}{r|}{0.6651} &
  -2.52\% \\ \cline{2-20} 
 &
  15 &
  \multicolumn{1}{r|}{0.5801} &
  \multicolumn{1}{r|}{0.5837} &
  0.62\% &
  \multicolumn{1}{r|}{\cellcolor[RGB]{242, 206, 239}0.7102} &
  \multicolumn{1}{r|}{\cellcolor[RGB]{242, 206, 239}0.7051} &
  \cellcolor[RGB]{242, 206, 239}-0.72\% &
  \multicolumn{1}{r|}{0.6024} &
  \multicolumn{1}{r|}{0.5986} &
  -0.63\% &
  \multicolumn{1}{r|}{0.6648} &
  \multicolumn{1}{r|}{0.6687} &
  0.59\% &
  \multicolumn{1}{r|}{\cellcolor[RGB]{242, 206, 239}0.7069} &
  \multicolumn{1}{r|}{\cellcolor[RGB]{242, 206, 239}0.7011} &
  \cellcolor[RGB]{242, 206, 239}-0.82\% &
  \multicolumn{1}{r|}{0.6589} &
  \multicolumn{1}{r|}{0.6755} &
  2.52\% \\ \cline{2-20} 
\multirow{-4}{*}{Zheng et al} &
  20 &
  \multicolumn{1}{r|}{0.5808} &
  \multicolumn{1}{r|}{0.581} &
  0.03\% &
  \multicolumn{1}{r|}{0.7071} &
  \multicolumn{1}{r|}{0.7067} &
  -0.06\% &
  \multicolumn{1}{r|}{0.6017} &
  \multicolumn{1}{r|}{0.6013} &
  -0.07\% &
  \multicolumn{1}{r|}{0.6631} &
  \multicolumn{1}{r|}{0.6668} &
  0.56\% &
  \multicolumn{1}{r|}{0.7037} &
  \multicolumn{1}{r|}{0.7004} &
  -0.47\% &
  \multicolumn{1}{r|}{\cellcolor[RGB]{242, 206, 239}0.7063} &
  \multicolumn{1}{r|}{\cellcolor[RGB]{242, 206, 239}0.668} &
  \cellcolor[RGB]{242, 206, 239}-5.42\% \\ \hline
 &
  5 &
  \multicolumn{1}{r|}{\cellcolor[RGB]{242, 206, 239}0.5809} &
  \multicolumn{1}{r|}{\cellcolor[RGB]{242, 206, 239}0.5621} &
  \cellcolor[RGB]{242, 206, 239}-3.24\% &
  \multicolumn{1}{r|}{\cellcolor[RGB]{242, 206, 239}0.711} &
  \multicolumn{1}{r|}{\cellcolor[RGB]{242, 206, 239}0.6967} &
  \cellcolor[RGB]{242, 206, 239}-2.01\% &
  \multicolumn{1}{r|}{\cellcolor[RGB]{242, 206, 239}0.6029} &
  \multicolumn{1}{r|}{\cellcolor[RGB]{242, 206, 239}0.5668} &
  \cellcolor[RGB]{242, 206, 239}-5.99\% &
  \multicolumn{1}{r|}{\cellcolor[RGB]{242, 206, 239}0.6671} &
  \multicolumn{1}{r|}{\cellcolor[RGB]{242, 206, 239}0.6552} &
  \cellcolor[RGB]{242, 206, 239}-1.78\% &
  \multicolumn{1}{r|}{\cellcolor[RGB]{242, 206, 239}0.7087} &
  \multicolumn{1}{r|}{\cellcolor[RGB]{242, 206, 239}0.6889} &
  \cellcolor[RGB]{242, 206, 239}-2.79\% &
  \multicolumn{1}{r|}{0.6999} &
  \multicolumn{1}{r|}{0.6632} &
  -5.24\% \\ \cline{2-20} 
 &
  10 &
  \multicolumn{1}{r|}{\cellcolor[RGB]{242, 206, 239}0.5777} &
  \multicolumn{1}{r|}{\cellcolor[RGB]{242, 206, 239}0.57} &
  \cellcolor[RGB]{242, 206, 239}-1.33\% &
  \multicolumn{1}{r|}{\cellcolor[RGB]{242, 206, 239}0.7082} &
  \multicolumn{1}{r|}{\cellcolor[RGB]{242, 206, 239}0.7017} &
  \cellcolor[RGB]{242, 206, 239}-0.92\% &
  \multicolumn{1}{r|}{\cellcolor[RGB]{242, 206, 239}0.6006} &
  \multicolumn{1}{r|}{\cellcolor[RGB]{242, 206, 239}0.5949} &
  \cellcolor[RGB]{242, 206, 239}-0.95\% &
  \multicolumn{1}{r|}{0.669} &
  \multicolumn{1}{r|}{0.6655} &
  -0.52\% &
  \multicolumn{1}{r|}{\cellcolor[RGB]{242, 206, 239}0.7031} &
  \multicolumn{1}{r|}{\cellcolor[RGB]{242, 206, 239}0.699} &
  \cellcolor[RGB]{242, 206, 239}-0.58\% &
  \multicolumn{1}{r|}{0.7087} &
  \multicolumn{1}{r|}{0.6734} &
  -4.98\% \\ \cline{2-20} 
 &
  15 &
  \multicolumn{1}{r|}{0.5785} &
  \multicolumn{1}{r|}{0.5761} &
  -0.41\% &
  \multicolumn{1}{r|}{0.7061} &
  \multicolumn{1}{r|}{0.7068} &
  0.10\% &
  \multicolumn{1}{r|}{\cellcolor[RGB]{242, 206, 239}0.6037} &
  \multicolumn{1}{r|}{\cellcolor[RGB]{242, 206, 239}0.5973} &
  \cellcolor[RGB]{242, 206, 239}-1.06\% &
  \multicolumn{1}{r|}{0.6676} &
  \multicolumn{1}{r|}{0.6629} &
  -0.70\% &
  \multicolumn{1}{r|}{0.7061} &
  \multicolumn{1}{r|}{0.7085} &
  0.34\% &
  \multicolumn{1}{r|}{0.6798} &
  \multicolumn{1}{r|}{0.6926} &
  1.88\% \\ \cline{2-20} 
\multirow{-4}{*}{MO-OBAM} &
  20 &
  \multicolumn{1}{r|}{0.5771} &
  \multicolumn{1}{r|}{0.5761} &
  -0.17\% &
  \multicolumn{1}{r|}{0.7076} &
  \multicolumn{1}{r|}{0.7068} &
  -0.11\% &
  \multicolumn{1}{r|}{0.6005} &
  \multicolumn{1}{r|}{0.602} &
  0.25\% &
  \multicolumn{1}{r|}{0.6612} &
  \multicolumn{1}{r|}{0.664} &
  0.42\% &
  \multicolumn{1}{r|}{0.7028} &
  \multicolumn{1}{r|}{0.6992} &
  -0.51\% &
  \multicolumn{1}{r|}{0.6515} &
  \multicolumn{1}{r|}{0.6629} &
  1.75\% \\ \hline
\end{tabular}%
} \\
         \\
         (c) Linkage attack with the risk level of $\tau=0.1$ \\
         \resizebox{\columnwidth}{!}{%
\begin{tabular}{l|c|rrr|rrr|rrr|rrr|rrr|rrr}
\hline
 &
  \multicolumn{1}{l|}{} &
  \multicolumn{3}{c|}{DT} &
  \multicolumn{3}{c|}{LR} &
  \multicolumn{3}{c|}{NB} &
  \multicolumn{3}{c|}{NN} &
  \multicolumn{3}{c|}{RF} &
  \multicolumn{3}{c}{SVM} \\ \hline
\begin{tabular}[c]{@{}l@{}}Anonymization \\ Model\end{tabular} &
  $k$ &
  \multicolumn{1}{l|}{\begin{tabular}[c]{@{}l@{}}Average \\ (FA)\end{tabular}} &
  \multicolumn{1}{l|}{\begin{tabular}[c]{@{}l@{}}Average \\ (FA-NV-only)\end{tabular}} &
  \multicolumn{1}{l|}{\begin{tabular}[c]{@{}l@{}}Percentage \\ Change\end{tabular}} &
  \multicolumn{1}{l|}{\begin{tabular}[c]{@{}l@{}}Average \\ (FA)\end{tabular}} &
  \multicolumn{1}{l|}{\begin{tabular}[c]{@{}l@{}}Average \\ (FA-NV-only)\end{tabular}} &
  \multicolumn{1}{l|}{\begin{tabular}[c]{@{}l@{}}Percentage \\ Change\end{tabular}} &
  \multicolumn{1}{l|}{\begin{tabular}[c]{@{}l@{}}Average \\ (FA)\end{tabular}} &
  \multicolumn{1}{l|}{\begin{tabular}[c]{@{}l@{}}Average \\ (FA-NV-only)\end{tabular}} &
  \multicolumn{1}{l|}{\begin{tabular}[c]{@{}l@{}}Percentage \\ Change\end{tabular}} &
  \multicolumn{1}{l|}{\begin{tabular}[c]{@{}l@{}}Average \\ (FA)\end{tabular}} &
  \multicolumn{1}{l|}{\begin{tabular}[c]{@{}l@{}}Average \\ (FA-NV-only)\end{tabular}} &
  \multicolumn{1}{l|}{\begin{tabular}[c]{@{}l@{}}Percentage \\ Change\end{tabular}} &
  \multicolumn{1}{l|}{\begin{tabular}[c]{@{}l@{}}Average \\ (FA)\end{tabular}} &
  \multicolumn{1}{l|}{\begin{tabular}[c]{@{}l@{}}Average \\ (FA-NV-only)\end{tabular}} &
  \multicolumn{1}{l|}{\begin{tabular}[c]{@{}l@{}}Percentage \\ Change\end{tabular}} &
  \multicolumn{1}{l|}{\begin{tabular}[c]{@{}l@{}}Average \\ (FA)\end{tabular}} &
  \multicolumn{1}{l|}{\begin{tabular}[c]{@{}l@{}}Average \\ (FA-NV-only)\end{tabular}} &
  \multicolumn{1}{l}{\begin{tabular}[c]{@{}l@{}}Percentage \\ Change\end{tabular}} \\ \hline
 &
  5 &
  \multicolumn{1}{r|}{\cellcolor[RGB]{242, 206, 239}0.5793} &
  \multicolumn{1}{r|}{\cellcolor[RGB]{242, 206, 239}0.4844} &
  \cellcolor[RGB]{242, 206, 239}-16.38\% &
  \multicolumn{1}{r|}{\cellcolor[RGB]{242, 206, 239}0.7027} &
  \multicolumn{1}{r|}{\cellcolor[RGB]{242, 206, 239}0.6522} &
  \cellcolor[RGB]{242, 206, 239}-7.19\% &
  \multicolumn{1}{r|}{\cellcolor[RGB]{242, 206, 239}0.6016} &
  \multicolumn{1}{r|}{\cellcolor[RGB]{242, 206, 239}0.4466} &
  \cellcolor[RGB]{242, 206, 239}-25.76\% &
  \multicolumn{1}{r|}{\cellcolor[RGB]{242, 206, 239}0.668} &
  \multicolumn{1}{r|}{\cellcolor[RGB]{242, 206, 239}0.6248} &
  \cellcolor[RGB]{242, 206, 239}-6.47\% &
  \multicolumn{1}{r|}{\cellcolor[RGB]{242, 206, 239}0.7045} &
  \multicolumn{1}{r|}{\cellcolor[RGB]{242, 206, 239}0.6416} &
  \cellcolor[RGB]{242, 206, 239}-8.93\% &
  \multicolumn{1}{r|}{\cellcolor[RGB]{242, 206, 239}0.7042} &
  \multicolumn{1}{r|}{\cellcolor[RGB]{242, 206, 239}0.607} &
  \cellcolor[RGB]{242, 206, 239}-13.80\% \\ \cline{2-20} 
 &
  10 &
  \multicolumn{1}{r|}{0.5833} &
  \multicolumn{1}{r|}{0.5809} &
  -0.41\% &
  \multicolumn{1}{r|}{0.705} &
  \multicolumn{1}{r|}{0.7063} &
  0.18\% &
  \multicolumn{1}{r|}{0.5995} &
  \multicolumn{1}{r|}{0.6017} &
  0.37\% &
  \multicolumn{1}{r|}{0.6682} &
  \multicolumn{1}{r|}{0.6697} &
  0.22\% &
  \multicolumn{1}{r|}{0.7053} &
  \multicolumn{1}{r|}{0.7093} &
  0.57\% &
  \multicolumn{1}{r|}{0.6905} &
  \multicolumn{1}{r|}{0.6593} &
  -4.52\% \\ \cline{2-20} 
 &
  15 &
  \multicolumn{1}{r|}{0.5771} &
  \multicolumn{1}{r|}{0.5744} &
  -0.47\% &
  \multicolumn{1}{r|}{0.706} &
  \multicolumn{1}{r|}{0.7055} &
  -0.07\% &
  \multicolumn{1}{r|}{0.6037} &
  \multicolumn{1}{r|}{0.6004} &
  -0.55\% &
  \multicolumn{1}{r|}{\cellcolor[RGB]{242, 206, 239}0.6663} &
  \multicolumn{1}{r|}{\cellcolor[RGB]{242, 206, 239}0.676} &
  \cellcolor[RGB]{242, 206, 239}1.46\% &
  \multicolumn{1}{r|}{0.7024} &
  \multicolumn{1}{r|}{0.7041} &
  0.24\% &
  \multicolumn{1}{r|}{0.661} &
  \multicolumn{1}{r|}{0.6966} &
  5.39\% \\ \cline{2-20} 
\multirow{-4}{*}{$k$-anonymity} &
  20 &
  \multicolumn{1}{r|}{0.5793} &
  \multicolumn{1}{r|}{0.5787} &
  -0.10\% &
  \multicolumn{1}{r|}{0.7037} &
  \multicolumn{1}{r|}{0.7064} &
  0.38\% &
  \multicolumn{1}{r|}{0.6037} &
  \multicolumn{1}{r|}{0.6021} &
  -0.27\% &
  \multicolumn{1}{r|}{0.6673} &
  \multicolumn{1}{r|}{0.6729} &
  0.84\% &
  \multicolumn{1}{r|}{0.704} &
  \multicolumn{1}{r|}{0.7055} &
  0.21\% &
  \multicolumn{1}{r|}{0.6661} &
  \multicolumn{1}{r|}{0.6798} &
  2.06\% \\ \hline
 &
  5 &
  \multicolumn{1}{r|}{\cellcolor[RGB]{242, 206, 239}0.5822} &
  \multicolumn{1}{r|}{\cellcolor[RGB]{242, 206, 239}0.5535} &
  \cellcolor[RGB]{242, 206, 239}-4.93\% &
  \multicolumn{1}{r|}{\cellcolor[RGB]{242, 206, 239}0.709} &
  \multicolumn{1}{r|}{\cellcolor[RGB]{242, 206, 239}0.6877} &
  \cellcolor[RGB]{242, 206, 239}-3.00\% &
  \multicolumn{1}{r|}{\cellcolor[RGB]{242, 206, 239}0.5978} &
  \multicolumn{1}{r|}{\cellcolor[RGB]{242, 206, 239}0.5588} &
  \cellcolor[RGB]{242, 206, 239}-6.52\% &
  \multicolumn{1}{r|}{\cellcolor[RGB]{242, 206, 239}0.6736} &
  \multicolumn{1}{r|}{\cellcolor[RGB]{242, 206, 239}0.6404} &
  \cellcolor[RGB]{242, 206, 239}-4.93\% &
  \multicolumn{1}{r|}{\cellcolor[RGB]{242, 206, 239}0.7025} &
  \multicolumn{1}{r|}{\cellcolor[RGB]{242, 206, 239}0.6827} &
  \cellcolor[RGB]{242, 206, 239}-2.82\% &
  \multicolumn{1}{r|}{0.6576} &
  \multicolumn{1}{r|}{0.6732} &
  2.37\% \\ \cline{2-20} 
 &
  10 &
  \multicolumn{1}{r|}{0.5777} &
  \multicolumn{1}{r|}{0.5791} &
  0.24\% &
  \multicolumn{1}{r|}{0.7045} &
  \multicolumn{1}{r|}{0.7046} &
  0.01\% &
  \multicolumn{1}{r|}{\cellcolor[RGB]{242, 206, 239}0.6005} &
  \multicolumn{1}{r|}{\cellcolor[RGB]{242, 206, 239}0.6053} &
  \cellcolor[RGB]{242, 206, 239}0.80\% &
  \multicolumn{1}{r|}{0.6703} &
  \multicolumn{1}{r|}{0.6668} &
  -0.52\% &
  \multicolumn{1}{r|}{0.7049} &
  \multicolumn{1}{r|}{0.705} &
  0.01\% &
  \multicolumn{1}{r|}{0.6823} &
  \multicolumn{1}{r|}{0.7073} &
  3.66\% \\ \cline{2-20} 
 &
  15 &
  \multicolumn{1}{r|}{0.5801} &
  \multicolumn{1}{r|}{0.5797} &
  -0.07\% &
  \multicolumn{1}{r|}{0.7102} &
  \multicolumn{1}{r|}{0.7064} &
  -0.54\% &
  \multicolumn{1}{r|}{0.6024} &
  \multicolumn{1}{r|}{0.5986} &
  -0.63\% &
  \multicolumn{1}{r|}{0.6648} &
  \multicolumn{1}{r|}{0.6655} &
  0.11\% &
  \multicolumn{1}{r|}{0.7069} &
  \multicolumn{1}{r|}{0.7039} &
  -0.42\% &
  \multicolumn{1}{r|}{0.6589} &
  \multicolumn{1}{r|}{0.6739} &
  2.28\% \\ \cline{2-20} 
\multirow{-4}{*}{Zheng et al} &
  20 &
  \multicolumn{1}{r|}{0.5808} &
  \multicolumn{1}{r|}{0.5841} &
  0.57\% &
  \multicolumn{1}{r|}{0.7071} &
  \multicolumn{1}{r|}{0.7052} &
  -0.27\% &
  \multicolumn{1}{r|}{0.6017} &
  \multicolumn{1}{r|}{0.5992} &
  -0.42\% &
  \multicolumn{1}{r|}{0.6631} &
  \multicolumn{1}{r|}{0.6656} &
  0.38\% &
  \multicolumn{1}{r|}{0.7037} &
  \multicolumn{1}{r|}{0.7015} &
  -0.31\% &
  \multicolumn{1}{r|}{0.7063} &
  \multicolumn{1}{r|}{0.7118} &
  0.78\% \\ \hline
 &
  5 &
  \multicolumn{1}{r|}{0.5809} &
  \multicolumn{1}{r|}{0.5763} &
  -0.79\% &
  \multicolumn{1}{r|}{\cellcolor[RGB]{242, 206, 239}0.711} &
  \multicolumn{1}{r|}{\cellcolor[RGB]{242, 206, 239}0.7045} &
  \cellcolor[RGB]{242, 206, 239}-0.91\% &
  \multicolumn{1}{r|}{\cellcolor[RGB]{242, 206, 239}0.6029} &
  \multicolumn{1}{r|}{\cellcolor[RGB]{242, 206, 239}0.5924} &
  \cellcolor[RGB]{242, 206, 239}-1.74\% &
  \multicolumn{1}{r|}{0.6671} &
  \multicolumn{1}{r|}{0.6656} &
  -0.22\% &
  \multicolumn{1}{r|}{\cellcolor[RGB]{242, 206, 239}0.7087} &
  \multicolumn{1}{r|}{\cellcolor[RGB]{242, 206, 239}0.7027} &
  \cellcolor[RGB]{242, 206, 239}-0.85\% &
  \multicolumn{1}{r|}{0.6999} &
  \multicolumn{1}{r|}{0.6882} &
  -1.67\% \\ \cline{2-20} 
 &
  10 &
  \multicolumn{1}{r|}{0.5777} &
  \multicolumn{1}{r|}{0.5775} &
  -0.03\% &
  \multicolumn{1}{r|}{0.7082} &
  \multicolumn{1}{r|}{0.7062} &
  -0.28\% &
  \multicolumn{1}{r|}{0.6006} &
  \multicolumn{1}{r|}{0.6026} &
  0.33\% &
  \multicolumn{1}{r|}{0.669} &
  \multicolumn{1}{r|}{0.6682} &
  -0.12\% &
  \multicolumn{1}{r|}{0.7031} &
  \multicolumn{1}{r|}{0.7005} &
  -0.37\% &
  \multicolumn{1}{r|}{0.7087} &
  \multicolumn{1}{r|}{0.6839} &
  -3.50\% \\ \cline{2-20} 
 &
  15 &
  \multicolumn{1}{r|}{0.5785} &
  \multicolumn{1}{r|}{0.5785} &
  0.00\% &
  \multicolumn{1}{r|}{0.7061} &
  \multicolumn{1}{r|}{0.7015} &
  -0.65\% &
  \multicolumn{1}{r|}{0.6037} &
  \multicolumn{1}{r|}{0.5995} &
  -0.70\% &
  \multicolumn{1}{r|}{0.6676} &
  \multicolumn{1}{r|}{0.6728} &
  0.78\% &
  \multicolumn{1}{r|}{0.7061} &
  \multicolumn{1}{r|}{0.7076} &
  0.21\% &
  \multicolumn{1}{r|}{0.6798} &
  \multicolumn{1}{r|}{0.6815} &
  0.25\% \\ \cline{2-20} 
\multirow{-4}{*}{MO-OBAM} &
  20 &
  \multicolumn{1}{r|}{0.5771} &
  \multicolumn{1}{r|}{0.578} &
  0.16\% &
  \multicolumn{1}{r|}{0.7076} &
  \multicolumn{1}{r|}{0.705} &
  -0.37\% &
  \multicolumn{1}{r|}{0.6005} &
  \multicolumn{1}{r|}{0.6} &
  -0.08\% &
  \multicolumn{1}{r|}{0.6612} &
  \multicolumn{1}{r|}{0.6662} &
  0.76\% &
  \multicolumn{1}{r|}{0.7028} &
  \multicolumn{1}{r|}{0.7008} &
  -0.28\% &
  \multicolumn{1}{r|}{0.6515} &
  \multicolumn{1}{r|}{0.6899} &
  5.89\% \\ \hline
\end{tabular}%
} \\
         \\
         (d) Homogeneity attack \\
         \resizebox{\columnwidth}{!}{%
\begin{tabular}{l|c|rrr|rrr|rrr|rrr|rrr|rrr}
\hline
 &
  \multicolumn{1}{l|}{} &
  \multicolumn{3}{c|}{DT} &
  \multicolumn{3}{c|}{LR} &
  \multicolumn{3}{c|}{NB} &
  \multicolumn{3}{c|}{NN} &
  \multicolumn{3}{c|}{RF} &
  \multicolumn{3}{c}{SVM} \\ \hline
\begin{tabular}[c]{@{}l@{}}Anonymization \\ Model\end{tabular} &
  $k$ &
  \multicolumn{1}{l|}{\begin{tabular}[c]{@{}l@{}}Average \\ (FA)\end{tabular}} &
  \multicolumn{1}{l|}{\begin{tabular}[c]{@{}l@{}}Average \\ (FA-NV-only)\end{tabular}} &
  \multicolumn{1}{l|}{\begin{tabular}[c]{@{}l@{}}Percentage \\ Change\end{tabular}} &
  \multicolumn{1}{l|}{\begin{tabular}[c]{@{}l@{}}Average \\ (FA)\end{tabular}} &
  \multicolumn{1}{l|}{\begin{tabular}[c]{@{}l@{}}Average \\ (FA-NV-only)\end{tabular}} &
  \multicolumn{1}{l|}{\begin{tabular}[c]{@{}l@{}}Percentage \\ Change\end{tabular}} &
  \multicolumn{1}{l|}{\begin{tabular}[c]{@{}l@{}}Average \\ (FA)\end{tabular}} &
  \multicolumn{1}{l|}{\begin{tabular}[c]{@{}l@{}}Average \\ (FA-NV-only)\end{tabular}} &
  \multicolumn{1}{l|}{\begin{tabular}[c]{@{}l@{}}Percentage \\ Change\end{tabular}} &
  \multicolumn{1}{l|}{\begin{tabular}[c]{@{}l@{}}Average \\ (FA)\end{tabular}} &
  \multicolumn{1}{l|}{\begin{tabular}[c]{@{}l@{}}Average \\ (FA-NV-only)\end{tabular}} &
  \multicolumn{1}{l|}{\begin{tabular}[c]{@{}l@{}}Percentage \\ Change\end{tabular}} &
  \multicolumn{1}{l|}{\begin{tabular}[c]{@{}l@{}}Average \\ (FA)\end{tabular}} &
  \multicolumn{1}{l|}{\begin{tabular}[c]{@{}l@{}}Average \\ (FA-NV-only)\end{tabular}} &
  \multicolumn{1}{l|}{\begin{tabular}[c]{@{}l@{}}Percentage \\ Change\end{tabular}} &
  \multicolumn{1}{l|}{\begin{tabular}[c]{@{}l@{}}Average \\ (FA)\end{tabular}} &
  \multicolumn{1}{l|}{\begin{tabular}[c]{@{}l@{}}Average \\ (FA-NV-only)\end{tabular}} &
  \multicolumn{1}{l}{\begin{tabular}[c]{@{}l@{}}Percentage \\ Change\end{tabular}} \\ \hline
 &
  5 &
  \multicolumn{1}{r|}{\cellcolor[RGB]{242, 206, 239}0.5793} &
  \multicolumn{1}{r|}{\cellcolor[RGB]{242, 206, 239}0.5739} &
  \cellcolor[RGB]{242, 206, 239}-0.93\% &
  \multicolumn{1}{r|}{0.7027} &
  \multicolumn{1}{r|}{0.7049} &
  0.31\% &
  \multicolumn{1}{r|}{\cellcolor[RGB]{242, 206, 239}0.6016} &
  \multicolumn{1}{r|}{\cellcolor[RGB]{242, 206, 239}0.5924} &
  \cellcolor[RGB]{242, 206, 239}-1.53\% &
  \multicolumn{1}{r|}{0.668} &
  \multicolumn{1}{r|}{0.6674} &
  -0.09\% &
  \multicolumn{1}{r|}{0.7045} &
  \multicolumn{1}{r|}{0.7027} &
  -0.26\% &
  \multicolumn{1}{r|}{0.7042} &
  \multicolumn{1}{r|}{0.6871} &
  -2.43\% \\ \cline{2-20} 
 &
  10 &
  \multicolumn{1}{r|}{0.5833} &
  \multicolumn{1}{r|}{0.5851} &
  0.31\% &
  \multicolumn{1}{r|}{0.705} &
  \multicolumn{1}{r|}{0.7073} &
  0.33\% &
  \multicolumn{1}{r|}{0.5995} &
  \multicolumn{1}{r|}{0.6038} &
  0.72\% &
  \multicolumn{1}{r|}{0.6682} &
  \multicolumn{1}{r|}{0.6676} &
  -0.09\% &
  \multicolumn{1}{r|}{0.7053} &
  \multicolumn{1}{r|}{0.7054} &
  0.01\% &
  \multicolumn{1}{r|}{0.6905} &
  \multicolumn{1}{r|}{0.6962} &
  0.83\% \\ \cline{2-20} 
 &
  15 &
  \multicolumn{1}{r|}{0.5771} &
  \multicolumn{1}{r|}{0.5752} &
  -0.33\% &
  \multicolumn{1}{r|}{0.706} &
  \multicolumn{1}{r|}{0.703} &
  -0.42\% &
  \multicolumn{1}{r|}{0.6037} &
  \multicolumn{1}{r|}{0.5998} &
  -0.65\% &
  \multicolumn{1}{r|}{0.6663} &
  \multicolumn{1}{r|}{0.6654} &
  -0.14\% &
  \multicolumn{1}{r|}{0.7024} &
  \multicolumn{1}{r|}{0.7043} &
  0.27\% &
  \multicolumn{1}{r|}{0.661} &
  \multicolumn{1}{r|}{0.6724} &
  1.72\% \\ \cline{2-20} 
\multirow{-4}{*}{$k$-anonymity} &
  20 &
  \multicolumn{1}{r|}{0.5793} &
  \multicolumn{1}{r|}{0.5757} &
  -0.62\% &
  \multicolumn{1}{r|}{0.7037} &
  \multicolumn{1}{r|}{0.7055} &
  0.26\% &
  \multicolumn{1}{r|}{0.6037} &
  \multicolumn{1}{r|}{0.6004} &
  -0.55\% &
  \multicolumn{1}{r|}{0.6673} &
  \multicolumn{1}{r|}{0.6619} &
  -0.81\% &
  \multicolumn{1}{r|}{0.704} &
  \multicolumn{1}{r|}{0.7033} &
  -0.10\% &
  \multicolumn{1}{r|}{0.6661} &
  \multicolumn{1}{r|}{0.6926} &
  3.98\% \\ \hline
 &
  5 &
  \multicolumn{1}{r|}{0.5822} &
  \multicolumn{1}{r|}{0.5794} &
  -0.48\% &
  \multicolumn{1}{r|}{0.709} &
  \multicolumn{1}{r|}{0.7102} &
  0.17\% &
  \multicolumn{1}{r|}{\cellcolor[RGB]{242, 206, 239}0.5978} &
  \multicolumn{1}{r|}{\cellcolor[RGB]{242, 206, 239}0.6035} &
  \cellcolor[RGB]{242, 206, 239}0.95\% &
  \multicolumn{1}{r|}{0.6736} &
  \multicolumn{1}{r|}{0.675} &
  0.21\% &
  \multicolumn{1}{r|}{0.7025} &
  \multicolumn{1}{r|}{0.7038} &
  0.19\% &
  \multicolumn{1}{r|}{0.6576} &
  \multicolumn{1}{r|}{0.6788} &
  3.22\% \\ \cline{2-20} 
 &
  10 &
  \multicolumn{1}{r|}{0.5777} &
  \multicolumn{1}{r|}{0.5806} &
  0.50\% &
  \multicolumn{1}{r|}{0.7045} &
  \multicolumn{1}{r|}{0.7045} &
  0.00\% &
  \multicolumn{1}{r|}{0.6005} &
  \multicolumn{1}{r|}{0.603} &
  0.42\% &
  \multicolumn{1}{r|}{0.6703} &
  \multicolumn{1}{r|}{0.6673} &
  -0.45\% &
  \multicolumn{1}{r|}{0.7049} &
  \multicolumn{1}{r|}{0.7055} &
  0.09\% &
  \multicolumn{1}{r|}{0.6823} &
  \multicolumn{1}{r|}{0.6881} &
  0.85\% \\ \cline{2-20} 
 &
  15 &
  \multicolumn{1}{r|}{0.5801} &
  \multicolumn{1}{r|}{0.5795} &
  -0.10\% &
  \multicolumn{1}{r|}{0.7102} &
  \multicolumn{1}{r|}{0.7103} &
  0.01\% &
  \multicolumn{1}{r|}{0.6024} &
  \multicolumn{1}{r|}{0.601} &
  -0.23\% &
  \multicolumn{1}{r|}{0.6648} &
  \multicolumn{1}{r|}{0.6621} &
  -0.41\% &
  \multicolumn{1}{r|}{0.7069} &
  \multicolumn{1}{r|}{0.7063} &
  -0.08\% &
  \multicolumn{1}{r|}{0.6589} &
  \multicolumn{1}{r|}{0.6881} &
  4.43\% \\ \cline{2-20} 
\multirow{-4}{*}{Zheng et al} &
  20 &
  \multicolumn{1}{r|}{0.5808} &
  \multicolumn{1}{r|}{0.5831} &
  0.40\% &
  \multicolumn{1}{r|}{0.7071} &
  \multicolumn{1}{r|}{0.7065} &
  -0.08\% &
  \multicolumn{1}{r|}{0.6017} &
  \multicolumn{1}{r|}{0.6017} &
  0.00\% &
  \multicolumn{1}{r|}{\cellcolor[RGB]{242, 206, 239}0.6631} &
  \multicolumn{1}{r|}{\cellcolor[RGB]{242, 206, 239}0.6752} &
  \cellcolor[RGB]{242, 206, 239}1.82\% &
  \multicolumn{1}{r|}{0.7037} &
  \multicolumn{1}{r|}{0.7018} &
  -0.27\% &
  \multicolumn{1}{r|}{0.7063} &
  \multicolumn{1}{r|}{0.6901} &
  -2.29\% \\ \hline
 &
  5 &
  \multicolumn{1}{r|}{0.5809} &
  \multicolumn{1}{r|}{0.5831} &
  0.38\% &
  \multicolumn{1}{r|}{0.711} &
  \multicolumn{1}{r|}{0.7068} &
  -0.59\% &
  \multicolumn{1}{r|}{0.6029} &
  \multicolumn{1}{r|}{0.5992} &
  -0.61\% &
  \multicolumn{1}{r|}{0.6671} &
  \multicolumn{1}{r|}{0.6727} &
  0.84\% &
  \multicolumn{1}{r|}{0.7087} &
  \multicolumn{1}{r|}{0.706} &
  -0.38\% &
  \multicolumn{1}{r|}{0.6999} &
  \multicolumn{1}{r|}{0.7104} &
  1.50\% \\ \cline{2-20} 
 &
  10 &
  \multicolumn{1}{r|}{0.5777} &
  \multicolumn{1}{r|}{0.5768} &
  -0.16\% &
  \multicolumn{1}{r|}{0.7082} &
  \multicolumn{1}{r|}{0.707} &
  -0.17\% &
  \multicolumn{1}{r|}{0.6006} &
  \multicolumn{1}{r|}{0.6002} &
  -0.07\% &
  \multicolumn{1}{r|}{0.669} &
  \multicolumn{1}{r|}{0.6733} &
  0.64\% &
  \multicolumn{1}{r|}{0.7031} &
  \multicolumn{1}{r|}{0.7037} &
  0.09\% &
  \multicolumn{1}{r|}{0.7087} &
  \multicolumn{1}{r|}{0.7336} &
  3.51\% \\ \cline{2-20} 
 &
  15 &
  \multicolumn{1}{r|}{0.5785} &
  \multicolumn{1}{r|}{0.579} &
  0.09\% &
  \multicolumn{1}{r|}{0.7061} &
  \multicolumn{1}{r|}{0.7082} &
  0.30\% &
  \multicolumn{1}{r|}{0.6037} &
  \multicolumn{1}{r|}{0.6038} &
  0.02\% &
  \multicolumn{1}{r|}{0.6676} &
  \multicolumn{1}{r|}{0.6695} &
  0.28\% &
  \multicolumn{1}{r|}{0.7061} &
  \multicolumn{1}{r|}{0.7052} &
  -0.13\% &
  \multicolumn{1}{r|}{0.6798} &
  \multicolumn{1}{r|}{0.6642} &
  -2.29\% \\ \cline{2-20} 
\multirow{-4}{*}{MO-OBAM} &
  20 &
  \multicolumn{1}{r|}{0.5771} &
  \multicolumn{1}{r|}{0.5779} &
  0.14\% &
  \multicolumn{1}{r|}{0.7076} &
  \multicolumn{1}{r|}{0.706} &
  -0.23\% &
  \multicolumn{1}{r|}{0.6005} &
  \multicolumn{1}{r|}{0.6041} &
  0.60\% &
  \multicolumn{1}{r|}{0.6612} &
  \multicolumn{1}{r|}{0.6697} &
  1.29\% &
  \multicolumn{1}{r|}{0.7028} &
  \multicolumn{1}{r|}{0.7037} &
  0.13\% &
  \multicolumn{1}{r|}{0.6515} &
  \multicolumn{1}{r|}{0.6507} &
  -0.12\% \\ \hline
\end{tabular}%
} \\
         \\
    \end{tabular}
\end{table}
\begin{table}[h!]
    \centering
    \caption{FA vs FA-NV-only (Recall)}
    \label{tab:FA vs FA-NV-only-recall in app}
    \begin{tabular}{c}
         (a) Linkage attack with the risk level of $\tau=0.05$ \\
          \resizebox{\columnwidth}{!}{%
\begin{tabular}{l|c|rrr|rrr|rrr|rrr|rrr|rrr}
\hline
 &
  \multicolumn{1}{l|}{} &
  \multicolumn{3}{c|}{DT} &
  \multicolumn{3}{c|}{LR} &
  \multicolumn{3}{c|}{NB} &
  \multicolumn{3}{c|}{NN} &
  \multicolumn{3}{c|}{RF} &
  \multicolumn{3}{c}{SVM} \\ \hline
\begin{tabular}[c]{@{}l@{}}Anonymization \\ Model\end{tabular} &
  $k$ &
  \multicolumn{1}{l|}{\begin{tabular}[c]{@{}l@{}}Average \\ (FA)\end{tabular}} &
  \multicolumn{1}{l|}{\begin{tabular}[c]{@{}l@{}}Average \\ (FA-NV-only)\end{tabular}} &
  \multicolumn{1}{l|}{\begin{tabular}[c]{@{}l@{}}Percentage \\ Change\end{tabular}} &
  \multicolumn{1}{l|}{\begin{tabular}[c]{@{}l@{}}Average \\ (FA)\end{tabular}} &
  \multicolumn{1}{l|}{\begin{tabular}[c]{@{}l@{}}Average \\ (FA-NV-only)\end{tabular}} &
  \multicolumn{1}{l|}{\begin{tabular}[c]{@{}l@{}}Percentage \\ Change\end{tabular}} &
  \multicolumn{1}{l|}{\begin{tabular}[c]{@{}l@{}}Average \\ (FA)\end{tabular}} &
  \multicolumn{1}{l|}{\begin{tabular}[c]{@{}l@{}}Average \\ (FA-NV-only)\end{tabular}} &
  \multicolumn{1}{l|}{\begin{tabular}[c]{@{}l@{}}Percentage \\ Change\end{tabular}} &
  \multicolumn{1}{l|}{\begin{tabular}[c]{@{}l@{}}Average \\ (FA)\end{tabular}} &
  \multicolumn{1}{l|}{\begin{tabular}[c]{@{}l@{}}Average \\ (FA-NV-only)\end{tabular}} &
  \multicolumn{1}{l|}{\begin{tabular}[c]{@{}l@{}}Percentage \\ Change\end{tabular}} &
  \multicolumn{1}{l|}{\begin{tabular}[c]{@{}l@{}}Average \\ (FA)\end{tabular}} &
  \multicolumn{1}{l|}{\begin{tabular}[c]{@{}l@{}}Average \\ (FA-NV-only)\end{tabular}} &
  \multicolumn{1}{l|}{\begin{tabular}[c]{@{}l@{}}Percentage \\ Change\end{tabular}} &
  \multicolumn{1}{l|}{\begin{tabular}[c]{@{}l@{}}Average \\ (FA)\end{tabular}} &
  \multicolumn{1}{l|}{\begin{tabular}[c]{@{}l@{}}Average \\ (FA-NV-only)\end{tabular}} &
  \multicolumn{1}{l}{\begin{tabular}[c]{@{}l@{}}Percentage \\ Change\end{tabular}} \\ \hline
 &
  5 &
  \multicolumn{1}{r|}{\cellcolor[RGB]{242, 206, 239}0.5746} &
  \multicolumn{1}{r|}{\cellcolor[RGB]{242, 206, 239}0.475} &
  \cellcolor[RGB]{242, 206, 239}-17.33\% &
  \multicolumn{1}{r|}{\cellcolor[RGB]{242, 206, 239}0.616} &
  \multicolumn{1}{r|}{\cellcolor[RGB]{242, 206, 239}0.4585} &
  \cellcolor[RGB]{242, 206, 239}-25.57\% &
  \multicolumn{1}{r|}{\cellcolor[RGB]{242, 206, 239}0.437} &
  \multicolumn{1}{r|}{\cellcolor[RGB]{242, 206, 239}0.5037} &
  \cellcolor[RGB]{242, 206, 239}15.26\% &
  \multicolumn{1}{r|}{\cellcolor[RGB]{242, 206, 239}0.6221} &
  \multicolumn{1}{r|}{\cellcolor[RGB]{242, 206, 239}0.4991} &
  \cellcolor[RGB]{242, 206, 239}-19.77\% &
  \multicolumn{1}{r|}{\cellcolor[RGB]{242, 206, 239}0.6018} &
  \multicolumn{1}{r|}{\cellcolor[RGB]{242, 206, 239}0.3772} &
  \cellcolor[RGB]{242, 206, 239}-37.32\% &
  \multicolumn{1}{r|}{0.5205} &
  \multicolumn{1}{r|}{0.4621} &
  -11.22\% \\ \cline{2-20} 
 &
  10 &
  \multicolumn{1}{r|}{\cellcolor[RGB]{242, 206, 239}0.5759} &
  \multicolumn{1}{r|}{\cellcolor[RGB]{242, 206, 239}0.4971} &
  \cellcolor[RGB]{242, 206, 239}-13.68\% &
  \multicolumn{1}{r|}{\cellcolor[RGB]{242, 206, 239}0.618} &
  \multicolumn{1}{r|}{\cellcolor[RGB]{242, 206, 239}0.4771} &
  \cellcolor[RGB]{242, 206, 239}-22.80\% &
  \multicolumn{1}{r|}{0.4375} &
  \multicolumn{1}{r|}{0.4425} &
  1.14\% &
  \multicolumn{1}{r|}{\cellcolor[RGB]{242, 206, 239}0.6101} &
  \multicolumn{1}{r|}{\cellcolor[RGB]{242, 206, 239}0.5006} &
  \cellcolor[RGB]{242, 206, 239}-17.95\% &
  \multicolumn{1}{r|}{\cellcolor[RGB]{242, 206, 239}0.6035} &
  \multicolumn{1}{r|}{\cellcolor[RGB]{242, 206, 239}0.4039} &
  \cellcolor[RGB]{242, 206, 239}-33.07\% &
  \multicolumn{1}{r|}{\cellcolor[RGB]{242, 206, 239}0.5525} &
  \multicolumn{1}{r|}{\cellcolor[RGB]{242, 206, 239}0.4355} &
  \cellcolor[RGB]{242, 206, 239}-21.18\% \\ \cline{2-20} 
 &
  15 &
  \multicolumn{1}{r|}{\cellcolor[RGB]{242, 206, 239}0.5727} &
  \multicolumn{1}{r|}{\cellcolor[RGB]{242, 206, 239}0.4769} &
  \cellcolor[RGB]{242, 206, 239}-16.73\% &
  \multicolumn{1}{r|}{\cellcolor[RGB]{242, 206, 239}0.6158} &
  \multicolumn{1}{r|}{\cellcolor[RGB]{242, 206, 239}0.4651} &
  \cellcolor[RGB]{242, 206, 239}-24.47\% &
  \multicolumn{1}{r|}{\cellcolor[RGB]{242, 206, 239}0.4454} &
  \multicolumn{1}{r|}{\cellcolor[RGB]{242, 206, 239}0.6782} &
  \cellcolor[RGB]{242, 206, 239}52.27\% &
  \multicolumn{1}{r|}{\cellcolor[RGB]{242, 206, 239}0.6141} &
  \multicolumn{1}{r|}{\cellcolor[RGB]{242, 206, 239}0.4708} &
  \cellcolor[RGB]{242, 206, 239}-23.33\% &
  \multicolumn{1}{r|}{\cellcolor[RGB]{242, 206, 239}0.6017} &
  \multicolumn{1}{r|}{\cellcolor[RGB]{242, 206, 239}0.3789} &
  \cellcolor[RGB]{242, 206, 239}-37.03\% &
  \multicolumn{1}{r|}{\cellcolor[RGB]{242, 206, 239}0.6194} &
  \multicolumn{1}{r|}{\cellcolor[RGB]{242, 206, 239}0.4355} &
  \cellcolor[RGB]{242, 206, 239}-29.69\% \\ \cline{2-20} 
\multirow{-4}{*}{$k$-anonymity} &
  20 &
  \multicolumn{1}{r|}{0.5762} &
  \multicolumn{1}{r|}{0.5749} &
  -0.23\% &
  \multicolumn{1}{r|}{0.6177} &
  \multicolumn{1}{r|}{0.6173} &
  -0.06\% &
  \multicolumn{1}{r|}{0.4435} &
  \multicolumn{1}{r|}{0.4389} &
  -1.04\% &
  \multicolumn{1}{r|}{0.6136} &
  \multicolumn{1}{r|}{0.6195} &
  0.96\% &
  \multicolumn{1}{r|}{\cellcolor[RGB]{242, 206, 239}0.606} &
  \multicolumn{1}{r|}{\cellcolor[RGB]{242, 206, 239}0.5999} &
  \cellcolor[RGB]{242, 206, 239}-1.01\% &
  \multicolumn{1}{r|}{0.6084} &
  \multicolumn{1}{r|}{0.5443} &
  -10.54\% \\ \hline
 &
  5 &
  \multicolumn{1}{r|}{\cellcolor[RGB]{242, 206, 239}0.5792} &
  \multicolumn{1}{r|}{\cellcolor[RGB]{242, 206, 239}0.5236} &
  \cellcolor[RGB]{242, 206, 239}-9.60\% &
  \multicolumn{1}{r|}{\cellcolor[RGB]{242, 206, 239}0.6214} &
  \multicolumn{1}{r|}{\cellcolor[RGB]{242, 206, 239}0.5443} &
  \cellcolor[RGB]{242, 206, 239}-12.41\% &
  \multicolumn{1}{r|}{0.4419} &
  \multicolumn{1}{r|}{0.4345} &
  -1.67\% &
  \multicolumn{1}{r|}{\cellcolor[RGB]{242, 206, 239}0.6153} &
  \multicolumn{1}{r|}{\cellcolor[RGB]{242, 206, 239}0.5467} &
  \cellcolor[RGB]{242, 206, 239}-11.15\% &
  \multicolumn{1}{r|}{\cellcolor[RGB]{242, 206, 239}0.6006} &
  \multicolumn{1}{r|}{\cellcolor[RGB]{242, 206, 239}0.4982} &
  \cellcolor[RGB]{242, 206, 239}-17.05\% &
  \multicolumn{1}{r|}{\cellcolor[RGB]{242, 206, 239}0.6267} &
  \multicolumn{1}{r|}{\cellcolor[RGB]{242, 206, 239}0.5137} &
  \cellcolor[RGB]{242, 206, 239}-18.03\% \\ \cline{2-20} 
 &
  10 &
  \multicolumn{1}{r|}{\cellcolor[RGB]{242, 206, 239}0.5739} &
  \multicolumn{1}{r|}{\cellcolor[RGB]{242, 206, 239}0.5216} &
  \cellcolor[RGB]{242, 206, 239}-9.11\% &
  \multicolumn{1}{r|}{\cellcolor[RGB]{242, 206, 239}0.6206} &
  \multicolumn{1}{r|}{\cellcolor[RGB]{242, 206, 239}0.5449} &
  \cellcolor[RGB]{242, 206, 239}-12.20\% &
  \multicolumn{1}{r|}{\cellcolor[RGB]{242, 206, 239}0.4442} &
  \multicolumn{1}{r|}{\cellcolor[RGB]{242, 206, 239}0.4337} &
  \cellcolor[RGB]{242, 206, 239}-2.36\% &
  \multicolumn{1}{r|}{\cellcolor[RGB]{242, 206, 239}0.6059} &
  \multicolumn{1}{r|}{\cellcolor[RGB]{242, 206, 239}0.5643} &
  \cellcolor[RGB]{242, 206, 239}-6.87\% &
  \multicolumn{1}{r|}{\cellcolor[RGB]{242, 206, 239}0.5988} &
  \multicolumn{1}{r|}{\cellcolor[RGB]{242, 206, 239}0.4991} &
  \cellcolor[RGB]{242, 206, 239}-16.65\% &
  \multicolumn{1}{r|}{\cellcolor[RGB]{242, 206, 239}0.5752} &
  \multicolumn{1}{r|}{\cellcolor[RGB]{242, 206, 239}0.4752} &
  \cellcolor[RGB]{242, 206, 239}-17.39\% \\ \cline{2-20} 
 &
  15 &
  \multicolumn{1}{r|}{\cellcolor[RGB]{242, 206, 239}0.5775} &
  \multicolumn{1}{r|}{\cellcolor[RGB]{242, 206, 239}0.5189} &
  \cellcolor[RGB]{242, 206, 239}-10.15\% &
  \multicolumn{1}{r|}{\cellcolor[RGB]{242, 206, 239}0.621} &
  \multicolumn{1}{r|}{\cellcolor[RGB]{242, 206, 239}0.54} &
  \cellcolor[RGB]{242, 206, 239}-13.04\% &
  \multicolumn{1}{r|}{\cellcolor[RGB]{242, 206, 239}0.4486} &
  \multicolumn{1}{r|}{\cellcolor[RGB]{242, 206, 239}0.413} &
  \cellcolor[RGB]{242, 206, 239}-7.94\% &
  \multicolumn{1}{r|}{\cellcolor[RGB]{242, 206, 239}0.6225} &
  \multicolumn{1}{r|}{\cellcolor[RGB]{242, 206, 239}0.5567} &
  \cellcolor[RGB]{242, 206, 239}-10.57\% &
  \multicolumn{1}{r|}{\cellcolor[RGB]{242, 206, 239}0.6054} &
  \multicolumn{1}{r|}{\cellcolor[RGB]{242, 206, 239}0.4974} &
  \cellcolor[RGB]{242, 206, 239}-17.84\% &
  \multicolumn{1}{r|}{\cellcolor[RGB]{242, 206, 239}0.6266} &
  \multicolumn{1}{r|}{\cellcolor[RGB]{242, 206, 239}0.5318} &
  \cellcolor[RGB]{242, 206, 239}-15.13\% \\ \cline{2-20} 
\multirow{-4}{*}{Zheng et al} &
  20 &
  \multicolumn{1}{r|}{0.5731} &
  \multicolumn{1}{r|}{0.5687} &
  -0.77\% &
  \multicolumn{1}{r|}{0.6187} &
  \multicolumn{1}{r|}{0.621} &
  0.37\% &
  \multicolumn{1}{r|}{\cellcolor[RGB]{242, 206, 239}0.4455} &
  \multicolumn{1}{r|}{\cellcolor[RGB]{242, 206, 239}0.4369} &
  \cellcolor[RGB]{242, 206, 239}-1.93\% &
  \multicolumn{1}{r|}{0.6232} &
  \multicolumn{1}{r|}{0.6091} &
  -2.26\% &
  \multicolumn{1}{r|}{0.5979} &
  \multicolumn{1}{r|}{0.5976} &
  -0.05\% &
  \multicolumn{1}{r|}{0.5255} &
  \multicolumn{1}{r|}{0.556} &
  5.80\% \\ \hline
 &
  5 &
  \multicolumn{1}{r|}{\cellcolor[RGB]{242, 206, 239}0.576} &
  \multicolumn{1}{r|}{\cellcolor[RGB]{242, 206, 239}0.5413} &
  \cellcolor[RGB]{242, 206, 239}-6.02\% &
  \multicolumn{1}{r|}{\cellcolor[RGB]{242, 206, 239}0.6158} &
  \multicolumn{1}{r|}{\cellcolor[RGB]{242, 206, 239}0.5599} &
  \cellcolor[RGB]{242, 206, 239}-9.08\% &
  \multicolumn{1}{r|}{0.4405} &
  \multicolumn{1}{r|}{0.4428} &
  0.52\% &
  \multicolumn{1}{r|}{\cellcolor[RGB]{242, 206, 239}0.617} &
  \multicolumn{1}{r|}{\cellcolor[RGB]{242, 206, 239}0.5731} &
  \cellcolor[RGB]{242, 206, 239}-7.12\% &
  \multicolumn{1}{r|}{\cellcolor[RGB]{242, 206, 239}0.6051} &
  \multicolumn{1}{r|}{\cellcolor[RGB]{242, 206, 239}0.5325} &
  \cellcolor[RGB]{242, 206, 239}-12.00\% &
  \multicolumn{1}{r|}{0.5544} &
  \multicolumn{1}{r|}{0.5323} &
  -3.99\% \\ \cline{2-20} 
 &
  10 &
  \multicolumn{1}{r|}{\cellcolor[RGB]{242, 206, 239}0.5706} &
  \multicolumn{1}{r|}{\cellcolor[RGB]{242, 206, 239}0.5642} &
  \cellcolor[RGB]{242, 206, 239}-1.12\% &
  \multicolumn{1}{r|}{\cellcolor[RGB]{242, 206, 239}0.6169} &
  \multicolumn{1}{r|}{\cellcolor[RGB]{242, 206, 239}0.6023} &
  \cellcolor[RGB]{242, 206, 239}-2.37\% &
  \multicolumn{1}{r|}{\cellcolor[RGB]{242, 206, 239}0.4424} &
  \multicolumn{1}{r|}{\cellcolor[RGB]{242, 206, 239}0.4253} &
  \cellcolor[RGB]{242, 206, 239}-3.87\% &
  \multicolumn{1}{r|}{\cellcolor[RGB]{242, 206, 239}0.6239} &
  \multicolumn{1}{r|}{\cellcolor[RGB]{242, 206, 239}0.6028} &
  \cellcolor[RGB]{242, 206, 239}-3.38\% &
  \multicolumn{1}{r|}{\cellcolor[RGB]{242, 206, 239}0.5974} &
  \multicolumn{1}{r|}{\cellcolor[RGB]{242, 206, 239}0.5785} &
  \cellcolor[RGB]{242, 206, 239}-3.16\% &
  \multicolumn{1}{r|}{0.542} &
  \multicolumn{1}{r|}{0.5484} &
  1.18\% \\ \cline{2-20} 
 &
  15 &
  \multicolumn{1}{r|}{\cellcolor[RGB]{242, 206, 239}0.5743} &
  \multicolumn{1}{r|}{\cellcolor[RGB]{242, 206, 239}0.567} &
  \cellcolor[RGB]{242, 206, 239}-1.27\% &
  \multicolumn{1}{r|}{0.6194} &
  \multicolumn{1}{r|}{0.6158} &
  -0.58\% &
  \multicolumn{1}{r|}{0.4428} &
  \multicolumn{1}{r|}{0.4409} &
  -0.43\% &
  \multicolumn{1}{r|}{0.6214} &
  \multicolumn{1}{r|}{0.6169} &
  -0.72\% &
  \multicolumn{1}{r|}{\cellcolor[RGB]{242, 206, 239}0.6016} &
  \multicolumn{1}{r|}{\cellcolor[RGB]{242, 206, 239}0.593} &
  \cellcolor[RGB]{242, 206, 239}-1.43\% &
  \multicolumn{1}{r|}{0.5777} &
  \multicolumn{1}{r|}{0.5888} &
  1.92\% \\ \cline{2-20} 
\multirow{-4}{*}{MO-OBAM} &
  20 &
  \multicolumn{1}{r|}{0.57} &
  \multicolumn{1}{r|}{0.5704} &
  0.07\% &
  \multicolumn{1}{r|}{0.6174} &
  \multicolumn{1}{r|}{0.6172} &
  -0.03\% &
  \multicolumn{1}{r|}{\cellcolor[RGB]{242, 206, 239}0.4374} &
  \multicolumn{1}{r|}{\cellcolor[RGB]{242, 206, 239}0.4475} &
  \cellcolor[RGB]{242, 206, 239}2.31\% &
  \multicolumn{1}{r|}{0.6181} &
  \multicolumn{1}{r|}{0.6085} &
  -1.55\% &
  \multicolumn{1}{r|}{0.5995} &
  \multicolumn{1}{r|}{0.5972} &
  -0.38\% &
  \multicolumn{1}{r|}{\cellcolor[RGB]{242, 206, 239}0.6188} &
  \multicolumn{1}{r|}{\cellcolor[RGB]{242, 206, 239}0.5206} &
  \cellcolor[RGB]{242, 206, 239}-15.87\% \\ \hline
\end{tabular}%
}\\
    \end{tabular}
\end{table}
\begin{table}[h!]
    \centering
    \begin{tabular}{c}
         \\
         (b) Linkage attack with the risk level of $\tau=0.075$ \\
         \resizebox{\columnwidth}{!}{%
\begin{tabular}{l|c|rrr|rrr|rrr|rrr|rrr|rrr}
\hline
 &
  \multicolumn{1}{l|}{} &
  \multicolumn{3}{c|}{DT} &
  \multicolumn{3}{c|}{LR} &
  \multicolumn{3}{c|}{NB} &
  \multicolumn{3}{c|}{NN} &
  \multicolumn{3}{c|}{RF} &
  \multicolumn{3}{c}{SVM} \\ \hline
\begin{tabular}[c]{@{}l@{}}Anonymization \\ Model\end{tabular} &
  $k$ &
  \multicolumn{1}{l|}{\begin{tabular}[c]{@{}l@{}}Average \\ (FA)\end{tabular}} &
  \multicolumn{1}{l|}{\begin{tabular}[c]{@{}l@{}}Average \\ (FA-NV-only)\end{tabular}} &
  \multicolumn{1}{l|}{\begin{tabular}[c]{@{}l@{}}Percentage \\ Change\end{tabular}} &
  \multicolumn{1}{l|}{\begin{tabular}[c]{@{}l@{}}Average \\ (FA)\end{tabular}} &
  \multicolumn{1}{l|}{\begin{tabular}[c]{@{}l@{}}Average \\ (FA-NV-only)\end{tabular}} &
  \multicolumn{1}{l|}{\begin{tabular}[c]{@{}l@{}}Percentage \\ Change\end{tabular}} &
  \multicolumn{1}{l|}{\begin{tabular}[c]{@{}l@{}}Average \\ (FA)\end{tabular}} &
  \multicolumn{1}{l|}{\begin{tabular}[c]{@{}l@{}}Average \\ (FA-NV-only)\end{tabular}} &
  \multicolumn{1}{l|}{\begin{tabular}[c]{@{}l@{}}Percentage \\ Change\end{tabular}} &
  \multicolumn{1}{l|}{\begin{tabular}[c]{@{}l@{}}Average \\ (FA)\end{tabular}} &
  \multicolumn{1}{l|}{\begin{tabular}[c]{@{}l@{}}Average \\ (FA-NV-only)\end{tabular}} &
  \multicolumn{1}{l|}{\begin{tabular}[c]{@{}l@{}}Percentage \\ Change\end{tabular}} &
  \multicolumn{1}{l|}{\begin{tabular}[c]{@{}l@{}}Average \\ (FA)\end{tabular}} &
  \multicolumn{1}{l|}{\begin{tabular}[c]{@{}l@{}}Average \\ (FA-NV-only)\end{tabular}} &
  \multicolumn{1}{l|}{\begin{tabular}[c]{@{}l@{}}Percentage \\ Change\end{tabular}} &
  \multicolumn{1}{l|}{\begin{tabular}[c]{@{}l@{}}Average \\ (FA)\end{tabular}} &
  \multicolumn{1}{l|}{\begin{tabular}[c]{@{}l@{}}Average \\ (FA-NV-only)\end{tabular}} &
  \multicolumn{1}{l}{\begin{tabular}[c]{@{}l@{}}Percentage \\ Change\end{tabular}} \\ \hline
 &
  5 &
  \multicolumn{1}{r|}{\cellcolor[RGB]{242, 206, 239}0.5746} &
  \multicolumn{1}{r|}{\cellcolor[RGB]{242, 206, 239}0.479} &
  \cellcolor[RGB]{242, 206, 239}-16.64\% &
  \multicolumn{1}{r|}{\cellcolor[RGB]{242, 206, 239}0.616} &
  \multicolumn{1}{r|}{\cellcolor[RGB]{242, 206, 239}0.4568} &
  \cellcolor[RGB]{242, 206, 239}-25.84\% &
  \multicolumn{1}{r|}{\cellcolor[RGB]{242, 206, 239}0.437} &
  \multicolumn{1}{r|}{\cellcolor[RGB]{242, 206, 239}0.4332} &
  \cellcolor[RGB]{242, 206, 239}-0.87\% &
  \multicolumn{1}{r|}{\cellcolor[RGB]{242, 206, 239}0.6221} &
  \multicolumn{1}{r|}{\cellcolor[RGB]{242, 206, 239}0.4961} &
  \cellcolor[RGB]{242, 206, 239}-20.25\% &
  \multicolumn{1}{r|}{\cellcolor[RGB]{242, 206, 239}0.6018} &
  \multicolumn{1}{r|}{\cellcolor[RGB]{242, 206, 239}0.3729} &
  \cellcolor[RGB]{242, 206, 239}-38.04\% &
  \multicolumn{1}{r|}{0.5205} &
  \multicolumn{1}{r|}{0.5145} &
  -1.15\% \\ \cline{2-20} 
 &
  10 &
  \multicolumn{1}{r|}{\cellcolor[RGB]{242, 206, 239}0.5759} &
  \multicolumn{1}{r|}{\cellcolor[RGB]{242, 206, 239}0.5026} &
  \cellcolor[RGB]{242, 206, 239}-12.73\% &
  \multicolumn{1}{r|}{\cellcolor[RGB]{242, 206, 239}0.618} &
  \multicolumn{1}{r|}{\cellcolor[RGB]{242, 206, 239}0.4804} &
  \cellcolor[RGB]{242, 206, 239}-22.27\% &
  \multicolumn{1}{r|}{0.4375} &
  \multicolumn{1}{r|}{0.4376} &
  0.02\% &
  \multicolumn{1}{r|}{\cellcolor[RGB]{242, 206, 239}0.6101} &
  \multicolumn{1}{r|}{\cellcolor[RGB]{242, 206, 239}0.5089} &
  \cellcolor[RGB]{242, 206, 239}-16.59\% &
  \multicolumn{1}{r|}{\cellcolor[RGB]{242, 206, 239}0.6035} &
  \multicolumn{1}{r|}{\cellcolor[RGB]{242, 206, 239}0.4065} &
  \cellcolor[RGB]{242, 206, 239}-32.64\% &
  \multicolumn{1}{r|}{0.5525} &
  \multicolumn{1}{r|}{0.5109} &
  -7.53\% \\ \cline{2-20} 
 &
  15 &
  \multicolumn{1}{r|}{0.5727} &
  \multicolumn{1}{r|}{0.5699} &
  -0.49\% &
  \multicolumn{1}{r|}{0.6158} &
  \multicolumn{1}{r|}{0.6175} &
  0.28\% &
  \multicolumn{1}{r|}{\cellcolor[RGB]{242, 206, 239}0.4454} &
  \multicolumn{1}{r|}{\cellcolor[RGB]{242, 206, 239}0.4357} &
  \cellcolor[RGB]{242, 206, 239}-2.18\% &
  \multicolumn{1}{r|}{0.6141} &
  \multicolumn{1}{r|}{0.614} &
  -0.02\% &
  \multicolumn{1}{r|}{0.6017} &
  \multicolumn{1}{r|}{0.6037} &
  0.33\% &
  \multicolumn{1}{r|}{0.6194} &
  \multicolumn{1}{r|}{0.6219} &
  0.40\% \\ \cline{2-20} 
\multirow{-4}{*}{$k$-anonymity} &
  20 &
  \multicolumn{1}{r|}{0.5762} &
  \multicolumn{1}{r|}{0.5747} &
  -0.26\% &
  \multicolumn{1}{r|}{\cellcolor[RGB]{242, 206, 239}0.6177} &
  \multicolumn{1}{r|}{\cellcolor[RGB]{242, 206, 239}0.6111} &
  \cellcolor[RGB]{242, 206, 239}-1.07\% &
  \multicolumn{1}{r|}{0.4435} &
  \multicolumn{1}{r|}{0.4398} &
  -0.83\% &
  \multicolumn{1}{r|}{0.6136} &
  \multicolumn{1}{r|}{0.6237} &
  1.65\% &
  \multicolumn{1}{r|}{0.606} &
  \multicolumn{1}{r|}{0.6047} &
  -0.21\% &
  \multicolumn{1}{r|}{\cellcolor[RGB]{242, 206, 239}0.6084} &
  \multicolumn{1}{r|}{\cellcolor[RGB]{242, 206, 239}0.5181} &
  \cellcolor[RGB]{242, 206, 239}-14.84\% \\ \hline
 &
  5 &
  \multicolumn{1}{r|}{\cellcolor[RGB]{242, 206, 239}0.5792} &
  \multicolumn{1}{r|}{\cellcolor[RGB]{242, 206, 239}0.5362} &
  \cellcolor[RGB]{242, 206, 239}-7.42\% &
  \multicolumn{1}{r|}{\cellcolor[RGB]{242, 206, 239}0.6214} &
  \multicolumn{1}{r|}{\cellcolor[RGB]{242, 206, 239}0.5632} &
  \cellcolor[RGB]{242, 206, 239}-9.37\% &
  \multicolumn{1}{r|}{\cellcolor[RGB]{242, 206, 239}0.4419} &
  \multicolumn{1}{r|}{\cellcolor[RGB]{242, 206, 239}0.4173} &
  \cellcolor[RGB]{242, 206, 239}-5.57\% &
  \multicolumn{1}{r|}{\cellcolor[RGB]{242, 206, 239}0.6153} &
  \multicolumn{1}{r|}{\cellcolor[RGB]{242, 206, 239}0.5672} &
  \cellcolor[RGB]{242, 206, 239}-7.82\% &
  \multicolumn{1}{r|}{\cellcolor[RGB]{242, 206, 239}0.6006} &
  \multicolumn{1}{r|}{\cellcolor[RGB]{242, 206, 239}0.5258} &
  \cellcolor[RGB]{242, 206, 239}-12.45\% &
  \multicolumn{1}{r|}{\cellcolor[RGB]{242, 206, 239}0.6267} &
  \multicolumn{1}{r|}{\cellcolor[RGB]{242, 206, 239}0.4868} &
  \cellcolor[RGB]{242, 206, 239}-22.32\% \\ \cline{2-20} 
 &
  10 &
  \multicolumn{1}{r|}{\cellcolor[RGB]{242, 206, 239}0.5739} &
  \multicolumn{1}{r|}{\cellcolor[RGB]{242, 206, 239}0.5344} &
  \cellcolor[RGB]{242, 206, 239}-6.88\% &
  \multicolumn{1}{r|}{\cellcolor[RGB]{242, 206, 239}0.6206} &
  \multicolumn{1}{r|}{\cellcolor[RGB]{242, 206, 239}0.5585} &
  \cellcolor[RGB]{242, 206, 239}-10.01\% &
  \multicolumn{1}{r|}{\cellcolor[RGB]{242, 206, 239}0.4442} &
  \multicolumn{1}{r|}{\cellcolor[RGB]{242, 206, 239}0.4341} &
  \cellcolor[RGB]{242, 206, 239}-2.27\% &
  \multicolumn{1}{r|}{\cellcolor[RGB]{242, 206, 239}0.6059} &
  \multicolumn{1}{r|}{\cellcolor[RGB]{242, 206, 239}0.5694} &
  \cellcolor[RGB]{242, 206, 239}-6.02\% &
  \multicolumn{1}{r|}{\cellcolor[RGB]{242, 206, 239}0.5988} &
  \multicolumn{1}{r|}{\cellcolor[RGB]{242, 206, 239}0.5143} &
  \cellcolor[RGB]{242, 206, 239}-14.11\% &
  \multicolumn{1}{r|}{0.5752} &
  \multicolumn{1}{r|}{0.5145} &
  -10.55\% \\ \cline{2-20} 
 &
  15 &
  \multicolumn{1}{r|}{0.5775} &
  \multicolumn{1}{r|}{0.5784} &
  0.16\% &
  \multicolumn{1}{r|}{0.621} &
  \multicolumn{1}{r|}{0.619} &
  -0.32\% &
  \multicolumn{1}{r|}{0.4486} &
  \multicolumn{1}{r|}{0.4417} &
  -1.54\% &
  \multicolumn{1}{r|}{0.6225} &
  \multicolumn{1}{r|}{0.6055} &
  -2.73\% &
  \multicolumn{1}{r|}{0.6054} &
  \multicolumn{1}{r|}{0.6021} &
  -0.55\% &
  \multicolumn{1}{r|}{0.6266} &
  \multicolumn{1}{r|}{0.5982} &
  -4.53\% \\ \cline{2-20} 
\multirow{-4}{*}{Zheng et al} &
  20 &
  \multicolumn{1}{r|}{0.5731} &
  \multicolumn{1}{r|}{0.5716} &
  -0.26\% &
  \multicolumn{1}{r|}{0.6187} &
  \multicolumn{1}{r|}{0.6203} &
  0.26\% &
  \multicolumn{1}{r|}{0.4455} &
  \multicolumn{1}{r|}{0.4426} &
  -0.65\% &
  \multicolumn{1}{r|}{0.6232} &
  \multicolumn{1}{r|}{0.607} &
  -2.60\% &
  \multicolumn{1}{r|}{0.5979} &
  \multicolumn{1}{r|}{0.5977} &
  -0.03\% &
  \multicolumn{1}{r|}{\cellcolor[RGB]{242, 206, 239}0.5255} &
  \multicolumn{1}{r|}{\cellcolor[RGB]{242, 206, 239}0.6043} &
  \cellcolor[RGB]{242, 206, 239}15.00\% \\ \hline
 &
  5 &
  \multicolumn{1}{r|}{\cellcolor[RGB]{242, 206, 239}0.576} &
  \multicolumn{1}{r|}{\cellcolor[RGB]{242, 206, 239}0.5554} &
  \cellcolor[RGB]{242, 206, 239}-3.58\% &
  \multicolumn{1}{r|}{\cellcolor[RGB]{242, 206, 239}0.6158} &
  \multicolumn{1}{r|}{\cellcolor[RGB]{242, 206, 239}0.5922} &
  \cellcolor[RGB]{242, 206, 239}-3.83\% &
  \multicolumn{1}{r|}{\cellcolor[RGB]{242, 206, 239}0.4405} &
  \multicolumn{1}{r|}{\cellcolor[RGB]{242, 206, 239}0.4679} &
  \cellcolor[RGB]{242, 206, 239}6.22\% &
  \multicolumn{1}{r|}{\cellcolor[RGB]{242, 206, 239}0.617} &
  \multicolumn{1}{r|}{\cellcolor[RGB]{242, 206, 239}0.5806} &
  \cellcolor[RGB]{242, 206, 239}-5.90\% &
  \multicolumn{1}{r|}{\cellcolor[RGB]{242, 206, 239}0.6051} &
  \multicolumn{1}{r|}{\cellcolor[RGB]{242, 206, 239}0.5695} &
  \cellcolor[RGB]{242, 206, 239}-5.88\% &
  \multicolumn{1}{r|}{0.5544} &
  \multicolumn{1}{r|}{0.5772} &
  4.11\% \\ \cline{2-20} 
 &
  10 &
  \multicolumn{1}{r|}{0.5706} &
  \multicolumn{1}{r|}{0.5672} &
  -0.60\% &
  \multicolumn{1}{r|}{\cellcolor[RGB]{242, 206, 239}0.6169} &
  \multicolumn{1}{r|}{\cellcolor[RGB]{242, 206, 239}0.6113} &
  \cellcolor[RGB]{242, 206, 239}-0.91\% &
  \multicolumn{1}{r|}{\cellcolor[RGB]{242, 206, 239}0.4424} &
  \multicolumn{1}{r|}{\cellcolor[RGB]{242, 206, 239}0.4341} &
  \cellcolor[RGB]{242, 206, 239}-1.88\% &
  \multicolumn{1}{r|}{\cellcolor[RGB]{242, 206, 239}0.6239} &
  \multicolumn{1}{r|}{\cellcolor[RGB]{242, 206, 239}0.605} &
  \cellcolor[RGB]{242, 206, 239}-3.03\% &
  \multicolumn{1}{r|}{\cellcolor[RGB]{242, 206, 239}0.5974} &
  \multicolumn{1}{r|}{\cellcolor[RGB]{242, 206, 239}0.5927} &
  \cellcolor[RGB]{242, 206, 239}-0.79\% &
  \multicolumn{1}{r|}{0.542} &
  \multicolumn{1}{r|}{0.5998} &
  10.66\% \\ \cline{2-20} 
 &
  15 &
  \multicolumn{1}{r|}{0.5743} &
  \multicolumn{1}{r|}{0.5758} &
  0.26\% &
  \multicolumn{1}{r|}{0.6194} &
  \multicolumn{1}{r|}{0.6209} &
  0.24\% &
  \multicolumn{1}{r|}{0.4428} &
  \multicolumn{1}{r|}{0.4452} &
  0.54\% &
  \multicolumn{1}{r|}{0.6214} &
  \multicolumn{1}{r|}{0.6307} &
  1.50\% &
  \multicolumn{1}{r|}{0.6016} &
  \multicolumn{1}{r|}{0.5992} &
  -0.40\% &
  \multicolumn{1}{r|}{0.5777} &
  \multicolumn{1}{r|}{0.561} &
  -2.89\% \\ \cline{2-20} 
\multirow{-4}{*}{MO-OBAM} &
  20 &
  \multicolumn{1}{r|}{0.57} &
  \multicolumn{1}{r|}{0.5721} &
  0.37\% &
  \multicolumn{1}{r|}{0.6174} &
  \multicolumn{1}{r|}{0.6185} &
  0.18\% &
  \multicolumn{1}{r|}{0.4374} &
  \multicolumn{1}{r|}{0.4374} &
  0.00\% &
  \multicolumn{1}{r|}{0.6181} &
  \multicolumn{1}{r|}{0.6133} &
  -0.78\% &
  \multicolumn{1}{r|}{0.5995} &
  \multicolumn{1}{r|}{0.5992} &
  -0.05\% &
  \multicolumn{1}{r|}{0.6188} &
  \multicolumn{1}{r|}{0.6167} &
  -0.34\% \\ \hline
\end{tabular}%
}\\
    \end{tabular}
\end{table}
\begin{table}[h!]
    \centering
    \begin{tabular}{c}
         \\
         (c) Linkage attack with the risk level of $\tau=0.1$ \\
          \resizebox{\columnwidth}{!}{%
\begin{tabular}{l|c|rrr|rrr|rrr|rrr|rrr|rrr}
\hline
 &
  \multicolumn{1}{l|}{} &
  \multicolumn{3}{c|}{DT} &
  \multicolumn{3}{c|}{LR} &
  \multicolumn{3}{c|}{NB} &
  \multicolumn{3}{c|}{NN} &
  \multicolumn{3}{c|}{RF} &
  \multicolumn{3}{c}{SVM} \\ \hline
\begin{tabular}[c]{@{}l@{}}Anonymization \\ Model\end{tabular} &
  $k$ &
  \multicolumn{1}{l|}{\begin{tabular}[c]{@{}l@{}}Average \\ (FA)\end{tabular}} &
  \multicolumn{1}{l|}{\begin{tabular}[c]{@{}l@{}}Average \\ (FA-NV-only)\end{tabular}} &
  \multicolumn{1}{l|}{\begin{tabular}[c]{@{}l@{}}Percentage \\ Change\end{tabular}} &
  \multicolumn{1}{l|}{\begin{tabular}[c]{@{}l@{}}Average \\ (FA)\end{tabular}} &
  \multicolumn{1}{l|}{\begin{tabular}[c]{@{}l@{}}Average \\ (FA-NV-only)\end{tabular}} &
  \multicolumn{1}{l|}{\begin{tabular}[c]{@{}l@{}}Percentage \\ Change\end{tabular}} &
  \multicolumn{1}{l|}{\begin{tabular}[c]{@{}l@{}}Average \\ (FA)\end{tabular}} &
  \multicolumn{1}{l|}{\begin{tabular}[c]{@{}l@{}}Average \\ (FA-NV-only)\end{tabular}} &
  \multicolumn{1}{l|}{\begin{tabular}[c]{@{}l@{}}Percentage \\ Change\end{tabular}} &
  \multicolumn{1}{l|}{\begin{tabular}[c]{@{}l@{}}Average \\ (FA)\end{tabular}} &
  \multicolumn{1}{l|}{\begin{tabular}[c]{@{}l@{}}Average \\ (FA-NV-only)\end{tabular}} &
  \multicolumn{1}{l|}{\begin{tabular}[c]{@{}l@{}}Percentage \\ Change\end{tabular}} &
  \multicolumn{1}{l|}{\begin{tabular}[c]{@{}l@{}}Average \\ (FA)\end{tabular}} &
  \multicolumn{1}{l|}{\begin{tabular}[c]{@{}l@{}}Average \\ (FA-NV-only)\end{tabular}} &
  \multicolumn{1}{l|}{\begin{tabular}[c]{@{}l@{}}Percentage \\ Change\end{tabular}} &
  \multicolumn{1}{l|}{\begin{tabular}[c]{@{}l@{}}Average \\ (FA)\end{tabular}} &
  \multicolumn{1}{l|}{\begin{tabular}[c]{@{}l@{}}Average \\ (FA-NV-only)\end{tabular}} &
  \multicolumn{1}{l}{\begin{tabular}[c]{@{}l@{}}Percentage \\ Change\end{tabular}} \\ \hline
 &
  5 &
  \multicolumn{1}{r|}{\cellcolor[RGB]{242, 206, 239}0.5746} &
  \multicolumn{1}{r|}{\cellcolor[RGB]{242, 206, 239}0.4829} &
  \cellcolor[RGB]{242, 206, 239}-15.96\% &
  \multicolumn{1}{r|}{\cellcolor[RGB]{242, 206, 239}0.616} &
  \multicolumn{1}{r|}{\cellcolor[RGB]{242, 206, 239}0.4597} &
  \cellcolor[RGB]{242, 206, 239}-25.37\% &
  \multicolumn{1}{r|}{\cellcolor[RGB]{242, 206, 239}0.437} &
  \multicolumn{1}{r|}{\cellcolor[RGB]{242, 206, 239}0.395} &
  \cellcolor[RGB]{242, 206, 239}-9.61\% &
  \multicolumn{1}{r|}{\cellcolor[RGB]{242, 206, 239}0.6221} &
  \multicolumn{1}{r|}{\cellcolor[RGB]{242, 206, 239}0.4967} &
  \cellcolor[RGB]{242, 206, 239}-20.16\% &
  \multicolumn{1}{r|}{\cellcolor[RGB]{242, 206, 239}0.6018} &
  \multicolumn{1}{r|}{\cellcolor[RGB]{242, 206, 239}0.4056} &
  \cellcolor[RGB]{242, 206, 239}-32.60\% &
  \multicolumn{1}{r|}{0.5205} &
  \multicolumn{1}{r|}{0.5135} &
  -1.34\% \\ \cline{2-20} 
 &
  10 &
  \multicolumn{1}{r|}{0.5759} &
  \multicolumn{1}{r|}{0.5773} &
  0.24\% &
  \multicolumn{1}{r|}{0.618} &
  \multicolumn{1}{r|}{0.6168} &
  -0.19\% &
  \multicolumn{1}{r|}{0.4375} &
  \multicolumn{1}{r|}{0.4401} &
  0.59\% &
  \multicolumn{1}{r|}{0.6101} &
  \multicolumn{1}{r|}{0.6113} &
  0.20\% &
  \multicolumn{1}{r|}{0.6035} &
  \multicolumn{1}{r|}{0.6025} &
  -0.17\% &
  \multicolumn{1}{r|}{0.5525} &
  \multicolumn{1}{r|}{0.6185} &
  11.95\% \\ \cline{2-20} 
 &
  15 &
  \multicolumn{1}{r|}{0.5727} &
  \multicolumn{1}{r|}{0.5689} &
  -0.66\% &
  \multicolumn{1}{r|}{0.6158} &
  \multicolumn{1}{r|}{0.6152} &
  -0.10\% &
  \multicolumn{1}{r|}{0.4454} &
  \multicolumn{1}{r|}{0.4447} &
  -0.16\% &
  \multicolumn{1}{r|}{0.6141} &
  \multicolumn{1}{r|}{0.6025} &
  -1.89\% &
  \multicolumn{1}{r|}{0.6017} &
  \multicolumn{1}{r|}{0.5999} &
  -0.30\% &
  \multicolumn{1}{r|}{0.6194} &
  \multicolumn{1}{r|}{0.5419} &
  -12.51\% \\ \cline{2-20} 
\multirow{-4}{*}{$k$-anonymity} &
  20 &
  \multicolumn{1}{r|}{0.5762} &
  \multicolumn{1}{r|}{0.5759} &
  -0.05\% &
  \multicolumn{1}{r|}{0.6177} &
  \multicolumn{1}{r|}{0.6181} &
  0.06\% &
  \multicolumn{1}{r|}{0.4435} &
  \multicolumn{1}{r|}{0.442} &
  -0.34\% &
  \multicolumn{1}{r|}{0.6136} &
  \multicolumn{1}{r|}{0.604} &
  -1.56\% &
  \multicolumn{1}{r|}{0.606} &
  \multicolumn{1}{r|}{0.6022} &
  -0.63\% &
  \multicolumn{1}{r|}{0.6084} &
  \multicolumn{1}{r|}{0.5735} &
  -5.74\% \\ \hline
 &
  5 &
  \multicolumn{1}{r|}{\cellcolor[RGB]{242, 206, 239}0.5792} &
  \multicolumn{1}{r|}{\cellcolor[RGB]{242, 206, 239}0.5503} &
  \cellcolor[RGB]{242, 206, 239}-4.99\% &
  \multicolumn{1}{r|}{\cellcolor[RGB]{242, 206, 239}0.6214} &
  \multicolumn{1}{r|}{\cellcolor[RGB]{242, 206, 239}0.576} &
  \cellcolor[RGB]{242, 206, 239}-7.31\% &
  \multicolumn{1}{r|}{\cellcolor[RGB]{242, 206, 239}0.4419} &
  \multicolumn{1}{r|}{\cellcolor[RGB]{242, 206, 239}0.4235} &
  \cellcolor[RGB]{242, 206, 239}-4.16\% &
  \multicolumn{1}{r|}{\cellcolor[RGB]{242, 206, 239}0.6153} &
  \multicolumn{1}{r|}{\cellcolor[RGB]{242, 206, 239}0.5738} &
  \cellcolor[RGB]{242, 206, 239}-6.74\% &
  \multicolumn{1}{r|}{\cellcolor[RGB]{242, 206, 239}0.6006} &
  \multicolumn{1}{r|}{\cellcolor[RGB]{242, 206, 239}0.5518} &
  \cellcolor[RGB]{242, 206, 239}-8.13\% &
  \multicolumn{1}{r|}{\cellcolor[RGB]{242, 206, 239}0.6267} &
  \multicolumn{1}{r|}{\cellcolor[RGB]{242, 206, 239}0.528} &
  \cellcolor[RGB]{242, 206, 239}-15.75\% \\ \cline{2-20} 
 &
  10 &
  \multicolumn{1}{r|}{0.5739} &
  \multicolumn{1}{r|}{0.5755} &
  0.28\% &
  \multicolumn{1}{r|}{0.6206} &
  \multicolumn{1}{r|}{0.6176} &
  -0.48\% &
  \multicolumn{1}{r|}{0.4442} &
  \multicolumn{1}{r|}{0.4393} &
  -1.10\% &
  \multicolumn{1}{r|}{0.6059} &
  \multicolumn{1}{r|}{0.6131} &
  1.19\% &
  \multicolumn{1}{r|}{0.5988} &
  \multicolumn{1}{r|}{0.6018} &
  0.50\% &
  \multicolumn{1}{r|}{0.5752} &
  \multicolumn{1}{r|}{0.5218} &
  -9.28\% \\ \cline{2-20} 
 &
  15 &
  \multicolumn{1}{r|}{0.5775} &
  \multicolumn{1}{r|}{0.5738} &
  -0.64\% &
  \multicolumn{1}{r|}{\cellcolor[RGB]{242, 206, 239}0.621} &
  \multicolumn{1}{r|}{\cellcolor[RGB]{242, 206, 239}0.6169} &
  \cellcolor[RGB]{242, 206, 239}-0.66\% &
  \multicolumn{1}{r|}{\cellcolor[RGB]{242, 206, 239}0.4486} &
  \multicolumn{1}{r|}{\cellcolor[RGB]{242, 206, 239}0.4382} &
  \cellcolor[RGB]{242, 206, 239}-2.32\% &
  \multicolumn{1}{r|}{0.6225} &
  \multicolumn{1}{r|}{0.6175} &
  -0.80\% &
  \multicolumn{1}{r|}{0.6054} &
  \multicolumn{1}{r|}{0.6026} &
  -0.46\% &
  \multicolumn{1}{r|}{0.6266} &
  \multicolumn{1}{r|}{0.5942} &
  -5.17\% \\ \cline{2-20} 
\multirow{-4}{*}{Zheng et al} &
  20 &
  \multicolumn{1}{r|}{0.5731} &
  \multicolumn{1}{r|}{0.5723} &
  -0.14\% &
  \multicolumn{1}{r|}{0.6187} &
  \multicolumn{1}{r|}{0.6207} &
  0.32\% &
  \multicolumn{1}{r|}{0.4455} &
  \multicolumn{1}{r|}{0.4415} &
  -0.90\% &
  \multicolumn{1}{r|}{0.6232} &
  \multicolumn{1}{r|}{0.619} &
  -0.67\% &
  \multicolumn{1}{r|}{0.5979} &
  \multicolumn{1}{r|}{0.5985} &
  0.10\% &
  \multicolumn{1}{r|}{0.5255} &
  \multicolumn{1}{r|}{0.5279} &
  0.46\% \\ \hline
 &
  5 &
  \multicolumn{1}{r|}{\cellcolor[RGB]{242, 206, 239}0.576} &
  \multicolumn{1}{r|}{\cellcolor[RGB]{242, 206, 239}0.569} &
  \cellcolor[RGB]{242, 206, 239}-1.22\% &
  \multicolumn{1}{r|}{0.6158} &
  \multicolumn{1}{r|}{0.6121} &
  -0.60\% &
  \multicolumn{1}{r|}{\cellcolor[RGB]{242, 206, 239}0.4405} &
  \multicolumn{1}{r|}{\cellcolor[RGB]{242, 206, 239}0.4527} &
  \cellcolor[RGB]{242, 206, 239}2.77\% &
  \multicolumn{1}{r|}{0.617} &
  \multicolumn{1}{r|}{0.608} &
  -1.46\% &
  \multicolumn{1}{r|}{\cellcolor[RGB]{242, 206, 239}0.6051} &
  \multicolumn{1}{r|}{\cellcolor[RGB]{242, 206, 239}0.5949} &
  \cellcolor[RGB]{242, 206, 239}-1.69\% &
  \multicolumn{1}{r|}{0.5544} &
  \multicolumn{1}{r|}{0.5494} &
  -0.90\% \\ \cline{2-20} 
 &
  10 &
  \multicolumn{1}{r|}{0.5706} &
  \multicolumn{1}{r|}{0.5727} &
  0.37\% &
  \multicolumn{1}{r|}{0.6169} &
  \multicolumn{1}{r|}{0.6175} &
  0.10\% &
  \multicolumn{1}{r|}{0.4424} &
  \multicolumn{1}{r|}{0.444} &
  0.36\% &
  \multicolumn{1}{r|}{0.6239} &
  \multicolumn{1}{r|}{0.6172} &
  -1.07\% &
  \multicolumn{1}{r|}{0.5974} &
  \multicolumn{1}{r|}{0.599} &
  0.27\% &
  \multicolumn{1}{r|}{0.542} &
  \multicolumn{1}{r|}{0.5772} &
  6.49\% \\ \cline{2-20} 
 &
  15 &
  \multicolumn{1}{r|}{0.5743} &
  \multicolumn{1}{r|}{0.575} &
  0.12\% &
  \multicolumn{1}{r|}{0.6194} &
  \multicolumn{1}{r|}{0.6192} &
  -0.03\% &
  \multicolumn{1}{r|}{0.4428} &
  \multicolumn{1}{r|}{0.4416} &
  -0.27\% &
  \multicolumn{1}{r|}{0.6214} &
  \multicolumn{1}{r|}{0.6087} &
  -2.04\% &
  \multicolumn{1}{r|}{0.6016} &
  \multicolumn{1}{r|}{0.6001} &
  -0.25\% &
  \multicolumn{1}{r|}{0.5777} &
  \multicolumn{1}{r|}{0.5741} &
  -0.62\% \\ \cline{2-20} 
\multirow{-4}{*}{MO-OBAM} &
  20 &
  \multicolumn{1}{r|}{0.57} &
  \multicolumn{1}{r|}{0.5688} &
  -0.21\% &
  \multicolumn{1}{r|}{0.6174} &
  \multicolumn{1}{r|}{0.6203} &
  0.47\% &
  \multicolumn{1}{r|}{0.4374} &
  \multicolumn{1}{r|}{0.4398} &
  0.55\% &
  \multicolumn{1}{r|}{0.6181} &
  \multicolumn{1}{r|}{0.6165} &
  -0.26\% &
  \multicolumn{1}{r|}{0.5995} &
  \multicolumn{1}{r|}{0.5956} &
  -0.65\% &
  \multicolumn{1}{r|}{0.6188} &
  \multicolumn{1}{r|}{0.5568} &
  -10.02\% \\ \hline
\end{tabular}%
}\\
    \end{tabular}
\end{table}
\begin{table}[h!]
    \centering
    \begin{tabular}{c}
         \\
         (d) Homogeneity attack \\
          \resizebox{\columnwidth}{!}{%
\begin{tabular}{l|c|rrr|rrr|rrr|rrr|rrr|rrr}
\hline
 &
  \multicolumn{1}{l|}{} &
  \multicolumn{3}{c|}{DT} &
  \multicolumn{3}{c|}{LR} &
  \multicolumn{3}{c|}{NB} &
  \multicolumn{3}{c|}{NN} &
  \multicolumn{3}{c|}{RF} &
  \multicolumn{3}{c}{SVM} \\ \hline
\begin{tabular}[c]{@{}l@{}}Anonymization \\ Model\end{tabular} &
  $k$ &
  \multicolumn{1}{l|}{\begin{tabular}[c]{@{}l@{}}Average \\ (FA)\end{tabular}} &
  \multicolumn{1}{l|}{\begin{tabular}[c]{@{}l@{}}Average \\ (FA-NV-only)\end{tabular}} &
  \multicolumn{1}{l|}{\begin{tabular}[c]{@{}l@{}}Percentage \\ Change\end{tabular}} &
  \multicolumn{1}{l|}{\begin{tabular}[c]{@{}l@{}}Average \\ (FA)\end{tabular}} &
  \multicolumn{1}{l|}{\begin{tabular}[c]{@{}l@{}}Average \\ (FA-NV-only)\end{tabular}} &
  \multicolumn{1}{l|}{\begin{tabular}[c]{@{}l@{}}Percentage \\ Change\end{tabular}} &
  \multicolumn{1}{l|}{\begin{tabular}[c]{@{}l@{}}Average \\ (FA)\end{tabular}} &
  \multicolumn{1}{l|}{\begin{tabular}[c]{@{}l@{}}Average \\ (FA-NV-only)\end{tabular}} &
  \multicolumn{1}{l|}{\begin{tabular}[c]{@{}l@{}}Percentage \\ Change\end{tabular}} &
  \multicolumn{1}{l|}{\begin{tabular}[c]{@{}l@{}}Average \\ (FA)\end{tabular}} &
  \multicolumn{1}{l|}{\begin{tabular}[c]{@{}l@{}}Average \\ (FA-NV-only)\end{tabular}} &
  \multicolumn{1}{l|}{\begin{tabular}[c]{@{}l@{}}Percentage \\ Change\end{tabular}} &
  \multicolumn{1}{l|}{\begin{tabular}[c]{@{}l@{}}Average \\ (FA)\end{tabular}} &
  \multicolumn{1}{l|}{\begin{tabular}[c]{@{}l@{}}Average \\ (FA-NV-only)\end{tabular}} &
  \multicolumn{1}{l|}{\begin{tabular}[c]{@{}l@{}}Percentage \\ Change\end{tabular}} &
  \multicolumn{1}{l|}{\begin{tabular}[c]{@{}l@{}}Average \\ (FA)\end{tabular}} &
  \multicolumn{1}{l|}{\begin{tabular}[c]{@{}l@{}}Average \\ (FA-NV-only)\end{tabular}} &
  \multicolumn{1}{l}{\begin{tabular}[c]{@{}l@{}}Percentage \\ Change\end{tabular}} \\ \hline
 &
  5 &
  \multicolumn{1}{r|}{0.5746} &
  \multicolumn{1}{r|}{0.5728} &
  -0.31\% &
  \multicolumn{1}{r|}{\cellcolor[RGB]{242, 206, 239}0.616} &
  \multicolumn{1}{r|}{\cellcolor[RGB]{242, 206, 239}0.6044} &
  \cellcolor[RGB]{242, 206, 239}-1.88\% &
  \multicolumn{1}{r|}{\cellcolor[RGB]{242, 206, 239}0.437} &
  \multicolumn{1}{r|}{\cellcolor[RGB]{242, 206, 239}0.4476} &
  \cellcolor[RGB]{242, 206, 239}2.43\% &
  \multicolumn{1}{r|}{0.6221} &
  \multicolumn{1}{r|}{0.606} &
  -2.59\% &
  \multicolumn{1}{r|}{\cellcolor[RGB]{242, 206, 239}0.6018} &
  \multicolumn{1}{r|}{\cellcolor[RGB]{242, 206, 239}0.5893} &
  \cellcolor[RGB]{242, 206, 239}-2.08\% &
  \multicolumn{1}{r|}{0.5205} &
  \multicolumn{1}{r|}{0.5349} &
  2.77\% \\ \cline{2-20} 
 &
  10 &
  \multicolumn{1}{r|}{0.5759} &
  \multicolumn{1}{r|}{0.5771} &
  0.21\% &
  \multicolumn{1}{r|}{0.618} &
  \multicolumn{1}{r|}{0.6187} &
  0.11\% &
  \multicolumn{1}{r|}{0.4375} &
  \multicolumn{1}{r|}{0.4384} &
  0.21\% &
  \multicolumn{1}{r|}{0.6101} &
  \multicolumn{1}{r|}{0.615} &
  0.80\% &
  \multicolumn{1}{r|}{0.6035} &
  \multicolumn{1}{r|}{0.608} &
  0.75\% &
  \multicolumn{1}{r|}{0.5525} &
  \multicolumn{1}{r|}{0.5439} &
  -1.56\% \\ \cline{2-20} 
 &
  15 &
  \multicolumn{1}{r|}{0.5727} &
  \multicolumn{1}{r|}{0.571} &
  -0.30\% &
  \multicolumn{1}{r|}{0.6158} &
  \multicolumn{1}{r|}{0.6145} &
  -0.21\% &
  \multicolumn{1}{r|}{0.4454} &
  \multicolumn{1}{r|}{0.4401} &
  -1.19\% &
  \multicolumn{1}{r|}{0.6141} &
  \multicolumn{1}{r|}{0.6216} &
  1.22\% &
  \multicolumn{1}{r|}{0.6017} &
  \multicolumn{1}{r|}{0.5976} &
  -0.68\% &
  \multicolumn{1}{r|}{0.6194} &
  \multicolumn{1}{r|}{0.5722} &
  -7.62\% \\ \cline{2-20} 
\multirow{-4}{*}{$k$-anonymity} &
  20 &
  \multicolumn{1}{r|}{0.5762} &
  \multicolumn{1}{r|}{0.5733} &
  -0.50\% &
  \multicolumn{1}{r|}{0.6177} &
  \multicolumn{1}{r|}{0.6163} &
  -0.23\% &
  \multicolumn{1}{r|}{0.4435} &
  \multicolumn{1}{r|}{0.4405} &
  -0.68\% &
  \multicolumn{1}{r|}{0.6136} &
  \multicolumn{1}{r|}{0.62} &
  1.04\% &
  \multicolumn{1}{r|}{\cellcolor[RGB]{242, 206, 239}0.606} &
  \multicolumn{1}{r|}{\cellcolor[RGB]{242, 206, 239}0.5952} &
  \cellcolor[RGB]{242, 206, 239}-1.78\% &
  \multicolumn{1}{r|}{0.6084} &
  \multicolumn{1}{r|}{0.5444} &
  -10.52\% \\ \hline
 &
  5 &
  \multicolumn{1}{r|}{0.5792} &
  \multicolumn{1}{r|}{0.5751} &
  -0.71\% &
  \multicolumn{1}{r|}{0.6214} &
  \multicolumn{1}{r|}{0.6205} &
  -0.14\% &
  \multicolumn{1}{r|}{0.4419} &
  \multicolumn{1}{r|}{0.4439} &
  0.45\% &
  \multicolumn{1}{r|}{0.6153} &
  \multicolumn{1}{r|}{0.6031} &
  -1.98\% &
  \multicolumn{1}{r|}{0.6006} &
  \multicolumn{1}{r|}{0.6001} &
  -0.08\% &
  \multicolumn{1}{r|}{0.6267} &
  \multicolumn{1}{r|}{0.5896} &
  -5.92\% \\ \cline{2-20} 
 &
  10 &
  \multicolumn{1}{r|}{0.5739} &
  \multicolumn{1}{r|}{0.5705} &
  -0.59\% &
  \multicolumn{1}{r|}{\cellcolor[RGB]{242, 206, 239}0.6206} &
  \multicolumn{1}{r|}{\cellcolor[RGB]{242, 206, 239}0.6151} &
  \cellcolor[RGB]{242, 206, 239}-0.89\% &
  \multicolumn{1}{r|}{0.4442} &
  \multicolumn{1}{r|}{0.4482} &
  0.90\% &
  \multicolumn{1}{r|}{0.6059} &
  \multicolumn{1}{r|}{0.6115} &
  0.92\% &
  \multicolumn{1}{r|}{0.5988} &
  \multicolumn{1}{r|}{0.6023} &
  0.58\% &
  \multicolumn{1}{r|}{0.5752} &
  \multicolumn{1}{r|}{0.574} &
  -0.21\% \\ \cline{2-20} 
 &
  15 &
  \multicolumn{1}{r|}{0.5775} &
  \multicolumn{1}{r|}{0.5765} &
  -0.17\% &
  \multicolumn{1}{r|}{0.621} &
  \multicolumn{1}{r|}{0.6191} &
  -0.31\% &
  \multicolumn{1}{r|}{0.4486} &
  \multicolumn{1}{r|}{0.444} &
  -1.03\% &
  \multicolumn{1}{r|}{0.6225} &
  \multicolumn{1}{r|}{0.6128} &
  -1.56\% &
  \multicolumn{1}{r|}{0.6054} &
  \multicolumn{1}{r|}{0.6021} &
  -0.55\% &
  \multicolumn{1}{r|}{0.6266} &
  \multicolumn{1}{r|}{0.5723} &
  -8.67\% \\ \cline{2-20} 
\multirow{-4}{*}{Zheng et al} &
  20 &
  \multicolumn{1}{r|}{0.5731} &
  \multicolumn{1}{r|}{0.571} &
  -0.37\% &
  \multicolumn{1}{r|}{0.6187} &
  \multicolumn{1}{r|}{0.6206} &
  0.31\% &
  \multicolumn{1}{r|}{0.4455} &
  \multicolumn{1}{r|}{0.4382} &
  -1.64\% &
  \multicolumn{1}{r|}{0.6232} &
  \multicolumn{1}{r|}{0.6083} &
  -2.39\% &
  \multicolumn{1}{r|}{0.5979} &
  \multicolumn{1}{r|}{0.5949} &
  -0.50\% &
  \multicolumn{1}{r|}{0.5255} &
  \multicolumn{1}{r|}{0.5587} &
  6.32\% \\ \hline
 &
  5 &
  \multicolumn{1}{r|}{0.576} &
  \multicolumn{1}{r|}{0.5787} &
  0.47\% &
  \multicolumn{1}{r|}{0.6158} &
  \multicolumn{1}{r|}{0.6193} &
  0.57\% &
  \multicolumn{1}{r|}{0.4405} &
  \multicolumn{1}{r|}{0.4417} &
  0.27\% &
  \multicolumn{1}{r|}{0.617} &
  \multicolumn{1}{r|}{0.6088} &
  -1.33\% &
  \multicolumn{1}{r|}{0.6051} &
  \multicolumn{1}{r|}{0.6009} &
  -0.69\% &
  \multicolumn{1}{r|}{0.5544} &
  \multicolumn{1}{r|}{0.5272} &
  -4.91\% \\ \cline{2-20} 
 &
  10 &
  \multicolumn{1}{r|}{0.5706} &
  \multicolumn{1}{r|}{0.5707} &
  0.02\% &
  \multicolumn{1}{r|}{0.6169} &
  \multicolumn{1}{r|}{0.6195} &
  0.42\% &
  \multicolumn{1}{r|}{0.4424} &
  \multicolumn{1}{r|}{0.4443} &
  0.43\% &
  \multicolumn{1}{r|}{0.6239} &
  \multicolumn{1}{r|}{0.6122} &
  -1.88\% &
  \multicolumn{1}{r|}{0.5974} &
  \multicolumn{1}{r|}{0.5978} &
  0.07\% &
  \multicolumn{1}{r|}{0.542} &
  \multicolumn{1}{r|}{0.4787} &
  -11.68\% \\ \cline{2-20} 
 &
  15 &
  \multicolumn{1}{r|}{0.5743} &
  \multicolumn{1}{r|}{0.5711} &
  -0.56\% &
  \multicolumn{1}{r|}{0.6194} &
  \multicolumn{1}{r|}{0.619} &
  -0.06\% &
  \multicolumn{1}{r|}{0.4428} &
  \multicolumn{1}{r|}{0.4386} &
  -0.95\% &
  \multicolumn{1}{r|}{0.6214} &
  \multicolumn{1}{r|}{0.6139} &
  -1.21\% &
  \multicolumn{1}{r|}{0.6016} &
  \multicolumn{1}{r|}{0.6009} &
  -0.12\% &
  \multicolumn{1}{r|}{0.5777} &
  \multicolumn{1}{r|}{0.6119} &
  5.92\% \\ \cline{2-20} 
\multirow{-4}{*}{MO-OBAM} &
  20 &
  \multicolumn{1}{r|}{0.57} &
  \multicolumn{1}{r|}{0.5698} &
  -0.04\% &
  \multicolumn{1}{r|}{0.6174} &
  \multicolumn{1}{r|}{0.618} &
  0.10\% &
  \multicolumn{1}{r|}{0.4374} &
  \multicolumn{1}{r|}{0.4403} &
  0.66\% &
  \multicolumn{1}{r|}{0.6181} &
  \multicolumn{1}{r|}{0.6096} &
  -1.38\% &
  \multicolumn{1}{r|}{0.5995} &
  \multicolumn{1}{r|}{0.597} &
  -0.42\% &
  \multicolumn{1}{r|}{0.6188} &
  \multicolumn{1}{r|}{0.642} &
  3.75\% \\ \hline
\end{tabular}%
}\\
         \\
    \end{tabular}
\end{table}

\end{document}